\pdfoutput=1
\documentclass[groupeaddress,aps,article,nofootinbib,10pt]{revtex4-1}
\usepackage{array}
\usepackage{titlesec}
\titleformat{\paragraph}
{\normalfont\normalsize\bfseries}{\theparagraph}{1em}{}
\titlespacing*{\paragraph}
{0pt}{3.25ex plus 1ex minus .2ex}{1.5ex plus .2ex}
\renewcommand*{\theparagraph}{\roman{paragraph})}

\usepackage{amssymb, amsmath,amsthm,bm}    
\usepackage{graphicx}   
\usepackage{verbatim}   
\usepackage{color}      
\usepackage[labelformat=simple]{subcaption}
\usepackage{mathrsfs} 
\usepackage[linktocpage = true]{hyperref}  
\usepackage{cancel}
\usepackage{graphicx}
\usepackage{caption} 
\usepackage{tikz}
\usepackage{tcolorbox}
\usepackage{pgflibraryarrows}
\usepackage{pgflibrarysnakes}
\usepackage{yfonts}
\usepackage{multirow}
\usetikzlibrary{decorations.markings}
\usetikzlibrary{decorations.footprints}
\definecolor{green}{RGB}{35,142,35}
\usetikzlibrary{decorations.pathmorphing}
\usepackage{float}

\usepackage{empheq} 
 
\usepackage{ulem}

\usepackage[framemethod=tikz]{mdframed}

\hypersetup{
    colorlinks   = true,
    linkcolor = blue,
    citecolor    = red
}  
\makeatletter
\def\p@subsection{}
\makeatother

\usepackage{cleveref}

\allowdisplaybreaks

\label{eq:partition_function_general2}

\newtcolorbox{mymathbox}[1][]{colback=white, #1}

\usepackage[framemethod=tikz]{mdframed}
\renewcommand{\FR}[2]{\frac{#1}{#2}}
\newcommand{\PFR}[2]{\left(\frac{#1}{#2}\right)}

\newcommand{\mc}{\mathcal}

\newcommand{\di}{\partial}

\newcommand{\ncmd}{\newcommand}
\ncmd{\lt}{\left}
\ncmd{\rt}{\right}
\newcommand{\eq}[1]{Eq. \eqref{#1}}
\newcommand{\fig}[1]{Fig. \ref{#1}}
\newcommand{\tr}[1]{\mbox{Tr}\lt[{#1}\rt]}

\ncmd{\kF}{$k_F$ }
\ncmd{\Lf}{$\Lambda_f$ }
\ncmd{\Lb}{$\Lambda_b$ }
\ncmd{\KF}{k_{\mathrm{F}}}

\newcommand{\murg}{\Lambda}

\definecolor{darkblue}{RGB}{10,10,150}
\DeclareMathOperator{\sgn}{sgn}

\newcommand{\ellthreekkp}{\ell^{(1L)}_{k',k}}

\newcommand{\ellthreezerokkp}{\ell^{(3,0)}_{k',k}}

\newcommand{\ellonek}{\ell^{(2L)}_{k}}

\newcommand{\ellonekp}{\ell^{(2L)}_{k'}}
\newcommand{\elltwok}{\ell^{(1L)}_{k}}
\newcommand{\elltwokp}{\ell^{(1L)}_{k'}}
\newcommand{\elltwokk}{\ell^{(1L)}_{k,k}}
\newcommand{\elltwokkp}{\ell^{(1L)}_{k,k'}}

\newcommand{\ellonekf}{{ L}^{(2L)}(k;\ell)}
\newcommand{\elltwokf}{{ L}^{(1L)}(k;\ell)}
\newcommand{\ellonekpf}{{ L}^{(2L)}(k';\ell)}
\newcommand{\elltwokpf}{{ L}^{(1L)}(k';\ell)}

\newcommand{\ellOneLoopX}{\elltwok}
\newcommand{\ellTwoLoopX}{\ellonek}
\newcommand{\ellOneLoopXX}{\elltwokk}

\newcommand{\lo}{\ell_\omega}
\newcommand{\lio}{\ell_{-i\omega}}

\newcommand{\ellthreekkpf}{{ L}^{(1L)}(k',k;\ell)}

\newcommand{\lk}{\ell_{k_0}}

\newcommand{\Eonex}{E^{(1L)}_k}
\newcommand{\Etwox}{E^{(2L)}_k}
\newcommand{\Eonexpx}{E^{(1L)}_{k',k}}

\newcommand{\eell}{e^{\ell}}
\newcommand{\emell}{e^{-\ell}}

\newcommand{\dd}{\mathrm{d}}

\newcommand{\vle}{v^{\textrm{l.e.}}}
\newcommand{\mfZ} {\mathcal{Z}}
\newcommand{\mfVF}{\mathcal{V}_{\mathrm{F}}}
\newcommand{\mfVFk}{\mathcal{V}_{\mathrm{F,k}}}

\newcommand{\whv}{\hat{v}}
\newcommand{\whV}{\hat{V}}
\newcommand{\whg}{\hat{g}}
\newcommand{\whl}{\hat{\lambda}}
\newcommand{\whkF}{\hat k_{\mathrm{F}}}
\newcommand{\whVF}{\hat V_{\mathrm{F}}}

\renewcommand{\textswab}{\mathcal}

\newcommand{\mfw}{\widetilde w}
\newcommand{\mfg}{\widetilde \gamma}

\renewcommand{\textswab}{\mathfrak}

\newcommand{\wh}{\hat}

\makeatletter
\newcommand*{\rom}[1]{\expandafter\@slowromancap\romannumeral #1@}
\makeatother

\newcommand{\td}{\tilde}

\makeatletter
\renewcommand\theequation{{\color{blue} \theequation@prefix \arabic{equation}}}
\makeatother

\makeatletter
\def\p@subsection{}
\makeatother
\newcommand{\s}[1]{\mathsf{#1}}

\newcommand{\bqa}{\begin{eqnarray}} 
\newcommand{\eqa}{\end{eqnarray}}
\newcommand{\nn}{\nonumber \\}
\newcommand{\ffc}{four-fermion couplings }

\newcommand{\pmqty}[1]{\begin{pmatrix}#1\end{pmatrix}}
\newcommand{\spmqty}[1]{\begin{psmallmatrix}#1\end{psmallmatrix}}

\newcommand{\sbmqty}[1]{\begin{bsmallmatrix}#1\end{bsmallmatrix}}

\newcommand{\abs}[1]{\left| #1 \right|}

\newcommand{\ImG}{\mathrm{Im}(G_{\mathrm{R}}(\omega,\vec{k})^{-1})}
\newcommand{\newelltwokN}       {\ell^{(1L)}_{k}}

\newcommand{\ellThree}{\ell^{(3)}_k}
\renewcommand{\lor}{\ell_{\omega}^{\mathrm{Re}}}

\begin{document}

\title{
Field-theoretic functional renormalization group formalism 
 for non-Fermi liquids and its application to 
the antiferromagnetic quantum critical metal in two dimensions
}

        \author{Francisco Borges$^{1,2}$ }
        \author{Anton Borissov$^{1,2}$ }
        \author{Ashutosh Singh$^{1,3}$\footnote{Current address:
        Department of Physics \& Astronomy, Texas A\&M University, College Station Texas 77843, United States}}
        \author{Andr\'es  Schlief$^{1,2,4}$\footnote{Current address: Deutsche Bank, Berlin, Germany }}
        \author{Sung-Sik Lee$^{1,2}$\footnote{slee@mcmaster.ca}}
        \affiliation{$^{1}$Department of Physics \& Astronomy, McMaster University, Hamilton ON L8S 4M1, Canada}
        \affiliation{$^{2}$Perimeter Institute for Theoretical Physics, Waterloo ON N2L 2Y5, Canada}
        \affiliation{$^{3}$
        Department of Physics,
        National Tsing Hua University,
        Hsinchu 300013, Taiwan 
        }
        \affiliation{$^{4}$Max-Planck-Institut f\"ur Physik Komplexer Systeme,
 01187,  Dresden, Germany}
    
        \date{\today}
        
\begin{abstract}

To capture the universal low-energy physics of metals within effective field theories, one has to generalize the usual notion of scale invariance and renormalizable field theory due to the presence of intrinsic scales (Fermi momenta). 
In this paper, we develop a field-theoretic functional renormalization group formalism for full low-energy effective field theories of non-Fermi liquids that include all gapless modes around the Fermi surface. 
%
%
The formalism is applied to the non-Fermi liquid that arises at the antiferromagnetic quantum critical point in two space dimensions. 
In the space of coupling functions, 
an interacting fixed point arises at a point with momentum-independent couplings and vanishing nesting angle.
In theories deformed with non-zero nesting angles,
coupling functions acquire 
universal momentum profiles 
controlled by the bare nesting angles at low energies before flowing to 
superconducting states
in the low-energy limit.
The superconducting instability is  unavoidable because
lukewarm electrons that
are coherent enough to be susceptible to pairing 
end up being subject to 
a renormalized attractive interaction 
with its minimum strength  
set by the nesting angle irrespective of the bare  four-fermion coupling.
%
%
%
Despite the inevitable superconducting instability, 
theories with small bare nesting angles and 
bare four-fermion couplings 
that are repulsive 
or weakly attractive 
must pass through the region with slow RG flow due to the proximity to the non-Fermi liquid fixed point.
The bottleneck region 
controls 
the scaling behaviours
of the normal state
and the quasi-universal pathway 
 from the non-Fermi liquid to superconductivity.
In the limit that 
the nesting angle is small,
the non-Fermi liquid scaling 
dictates the physics over 
a large window of energy scale 
above the superconducting transition temperature. 

\end{abstract}
\maketitle
\tableofcontents
\onecolumngrid

\newpage

\section*{Notation}

\begin{enumerate}

\item scales
\begin{itemize}
\item $\Lambda$ : UV energy cutoff
\item $k_F$ : size of Fermi surface
\item $\mu$ : floating energy scale at which renormalization conditions are imposed
\end{itemize}

\item momentum
\begin{itemize}
\item $\vec k = (k_x, k_y)$ : two-dimensional momentum measured with respect to the hot spot in each patch
\item ${\bf k} = ( k_0, \vec k)$ : 
three-momentum vector made of 
Matsubara frequency and two-dimensional momentum  
\item $\vec Q_{AF}$ : the antiferromagnetic ordering vector
\item $k_N$ : the component of $\vec k$ that is perpendicular to $\vec Q_{AF}$ 
\item $k$ : abbreviation for $k_N$ used when the associated hot spot index is obvious
\item $k_\perp$ : 
the component of $\vec k$ that is parallel to $\vec Q_{AF}$
\item $K = k e^{\ell}$ :  rescaled momentum associated with the logarithmic length scale $\ell$
\end{itemize}

\item coupling functions
\begin{itemize}

\item$V_{F,k}$ : the momentum dependent Fermi velocity parallel to the antiferromagnetic ordering vector
\item $v_k$ : the momentum dependent local nesting angle  
\item $c$ : the speed of the overdamped collective spin fluctuation
\item $w_k \equiv v_k/c$
($w \equiv v_0/c$)
\item $g_{k,p}$ : the momentum dependent fermion-boson coupling function
\item ${\lambda}^{\spmqty{N_1 & N_2 \\ N_4 & N_3};\spmqty{\sigma_1 & \sigma_2 \\ \sigma_4 & \sigma_3}}_{\spmqty{k_1 & k_2 \\ k_4 & k_3 }}$ : the momentum dependent four-fermion coupling function
\item $u_a$ ($a=1,2$) : quartic boson couplings
\end{itemize}



\end{enumerate}

\newpage

\section{Introduction}

One of the major goals in condensed matter physics is to understand
universal relations that low-energy observables satisfy through effective theories.
While the richness of condensed matter physics comes from  
the great variety of low-energy effective theories 
that can emerge\cite{XGW,
SUBIRBOOK,
Fradkin:2013sab,
tsvelik_2003,
coleman_2015,
burgess_2020},
the power of low-energy effective theories is derived from
their insensitivity to microscopic details\cite{KADANOFFRG,WILSONRG1,WILSONRG2}.
In no small part, the remarkable progress in modern condensed matter physics
is attributed to the balance between the diversity and the universality
that our nature granted.

Crudely, different types of low-energy effective theories
can be classified in terms of the number of low-energy modes.
In the trivial insulator, there is no low-energy degree of freedom,
and the theory is empty  in the infrared limit. 
In topological states, there can be low-energy modes
but their number scales sub-extensively with the volume of system due to the non-local nature of excitations.
In states with gapless local excitations,
the number of low-energy modes is extensive.  
Gapless states can be further divided into sub-classes 
in terms of the dimension of the manifold 
where gapless single particle excitations reside
in momentum space.
In systems with emergent Lorentz symmetry,
zero-energy modes reside at discrete points 
in momentum space.
In those theories,
the universal low-energy physics is encoded in 
a discrete set of operators 
classified in terms of their scaling dimensions and  charges\cite{DIFRANCESCO,1126-6708-2008-12-031}. 
In metals, on the other hand,
there are infinitely many gapless modes 
residing on extended Fermi surfaces,
and one needs function's worth  of data to fully characterize the universal low-energy physics.

In Fermi liquids\cite{LANDAU,SHANKAR,POLCHINSKI1}, 
the low-energy physics is specified by
the topology and geometric shape of the Fermi surface,
the Berry phase around the Fermi surface\cite{PhysRevLett.93.206602},
the angle-dependent effective mass 
and the forward scattering amplitude.
The simplicity of the low-energy effective theory of Fermi liquids
is due to the existence of
well-defined quasiparticles\footnote{
Well-defined quasiparticles imply a large emergent symmetry group\cite{2005cond.mat..5529H}, but the converse is not necessarily true.
It has been recently pointed out that a large emergent symmetry group  can be a general feature of compressible states beyond the quasiparticle paradigm\cite{PhysRevX.11.021005}. 
}.
In non-Fermi liquids, however,
the quasiparticle picture is not valid\cite{STEWART,SCHOFIELD},
and extracting dynamical information out of  effective field theories is significantly harder\cite{
	HOLSTEIN,
	HERTZ,
	PLEE1,
	REIZER,
	PLEE2,
	MILLIS,
	ALTSHULER,
	YBKIM,
	NAYAK,
	POLCHINSKI2,
	ABANOV1,
	ABANOV3,
	ABANOV2,
	LOH,
	SENTHIL,
	SSLEE,
	MROSS,
	MAX0,
	MAX2,
	HARTNOLL,
	ABRAHAMS,
	JIANG,
	FITZPATRICK,
	DENNIS,
	STRACK2,
	SHOUVIK2,
	PATEL2,
	SHOUVIK,
	RIDGWAY,
	HOLDER,
	PATEL,
	VARMA2,
	EBERLEIN,
	SCHATTNER,
	SHOUVIK3,
	LIU,
	XU,
	HONGLIU,
	CHOWDHURY,
	VARMA3,
	BERGLEDERER,
	YANGMENG}\footnote{
For recent progress in  non-Fermi liquid theories with random couplings, 
see Refs
\cite{
PhysRevLett.70.3339,
SYK_Kitaev,
PhysRevD.94.106002,
PhysRevB.98.165117,
PhysRevB.103.235129}.
}.
Among the additional data that are needed  
for non-Fermi liquids are
the singularity of the critical Fermi surface
characterized by anomalous dimensions of fermions, 
dynamical critical exponents,
and scaling properties of other critical modes
that are present in the system\cite{SENTHIL}.
Since the momentum along the Fermi surface
plays the role of a continuous flavour,
critical exponents 
(such as the scaling dimension of fermions) can 
depend on the momentum along the Fermi surface.

One approach to non-Fermi liquids
that has proven to be useful is the patch theory\footnote{
See \cite{SUNGSIKREVIEW} for a recent review.}.
The goal of a patch theory is to `divide and conquer' 
the full theory
by considering only a minimal set of patches 
of Fermi surface at a time.
The patch description is valid if
large-momentum scatterings that connect different parts of Fermi surface are negligible.
When the interactions mediated by critical collective modes
are sufficiently singular at low momenta,
large-angle scatterings
 are dynamically suppressed.
In two space dimensions,
two fermions on the Fermi surface can 
stay close to the Fermi surface while exchanging a boson with a small momentum
only when those fermions are in patches 
that are parallel or anti-parallel\cite{POLCHINSKI2}.
In this case, 
regions of Fermi surface with different orientations are not strongly coupled,
and the patch theory is a good description
at least within a range of energy scales in which inter-patch scatterings can be ignored.

However, the patch theory has its limitations.
In metals that support two-dimensional Fermi surface,
the patch description is not valid at any energy scale
as fermions 
in any two patches can be scattered along 
a common tangential direction
while staying close to the Fermi surface. 
Even for one-dimensional Fermi surfaces\footnote{
This includes metals in two space dimensions and semi-metals with line nodes in three dimensions.
},
the patch theory breaks down
if large-momentum scatterings are not suppressed.
This may happen in marginal Fermi liquids\cite{VARMALI}
 where the interaction mediated by a gapless collective mode is not singular enough to suppress 
 large-angle scatterings of electrons 
 along the Fermi surface\cite{PhysRevLett.128.106402}.
Inter-patch couplings become 
also important in the presence of strong superconducting fluctuations\cite{
PhysRevB.102.024524,
PhysRevB.102.024525,
PhysRevB.102.094516,
PhysRevB.103.024522,
PhysRevB.103.184508,
PhysRevB.104.144509,
PhysRevB.95.165137,
PhysRevB.92.205104,
PhysRevB.94.115138}
 or 
impurities\cite{2021arXiv211202562J} 
 that scatter electrons around the Fermi surface.
Furthermore, 
the patch theory is not ideal for
describing momentum-dependent scaling properties of fermions 
because it includes only a fraction of Fermi surface.
Although one can `patch' multiple patch theories to 
capture global aspects of Fermi surface\cite{PhysRevD.59.094019,PhysRevB.91.115111},
it is desirable to have one unified low-energy effective field theory
description that includes all gapless degrees of freedom on the equal footing.
It is also more convenient to include all low-energy modes within one theory to keep track of the anomaly\cite{PhysRevX.11.021005,2022arXiv220407585D}
because the anomaly is guaranteed to be preserved under the renormalization group flow only for the entire system\cite{Hooft1980}.

To capture the low-energy physics of the whole Fermi surface, 
the patch theory has to be extended to a theory that includes all gapless modes around the Fermi surface.
Such theories are characterized by couplings that are functions of momentum along the Fermi surface.
The universal low-energy data should be encoded in fixed points 
that arise in the space of coupling functions.
Ultimately, one would like to  identify the space of fixed points and extract the universal data associated with each fixed point.
The natural theoretical framework 
for this is the functional renormalization group method\cite{
POLCHINSKIFRG,
WETTERICH}
\cite{
MORRIS,
REUTER,
ROSA,
HOFLING,
HONERKAMP,
GIES,
GIES2,
BRAUN,
METZNER,
doi:10.1080/00018732.2013.862020,
PhysRevB.61.13609,
PhysRevLett.102.047005,
PhysRevB.61.7364,
SCHERER,
JANSSEN2,
MESTERHAZY,
PhysRevB.87.045104,
EBERLEINMETZNER,
PLATT,
WANGEBERLEIN,
JANSSEN,
MAIEREBERLEIN,
EBERLEIN2,
EBERLEIN3,
JAKUBCZYK,
MAIER,
TORRES}.
The functional renormalization group flow describes how the momentum dependent vertex function evolves as a function of an energy scale.
However, the exact renormalization group equation for the full vertex function is usually too difficult to solve for interacting theories.
As a result, some forms of truncation, which are often uncontrolled,
are employed to make the flow equation manageable.
Fortunately, one does not need to know 
the full momentum-dependent vertex function to characterize the universal low-energy physics.
Because gapless modes are residing on Fermi surface with a dimension lower than the space dimension,
one should be able to throw away 
a great deal of non-universal information associated with modes away from the Fermi surface.
For relativistic field theories, 
there exists a systematic way of achieving this : renormalizable field theory.
Born out of the locality principle 
and the gradient expansion,
a renormalizable field theory is the minimal theory that captures the low-energy physics shared by all theories within one universality class.
They are simple enough that one can   in principle study them with pen and paper,
yet powerful enough to produce among the most accurate predictions in the physical sciences\cite{PhysRevLett.97.030802}.
Then, it is natural to combine the functional renormalization group formalism with the notion of renormalizable field theory 
to capture the universal low-energy physics of metals.
The purpose of this work is to achieve this goal
\footnote{
For alternative approaches based on  bosonization,
see Refs.
\cite{PhysRevB.49.10877,doi:10.1080/000187300243363,2022arXiv220305004D}.
}.
In this paper, we develop a field-theoretic functional renormalization group formalism for 
general interacting theories of Fermi surface,
and apply it to the non-Fermi liquid that arises
at the antiferromagnetic quantum critical point
in two space dimensions.

The antiferromagnetic quantum critical metal is
potentially relevant to electron doped cuprates\cite{HELM}, 
iron pnictides\cite{HASHIMOTO}, 
and heavy fermion compounds\cite{PARK}. 
The theory has been intensively studied both in field theoretical and numerical approaches\cite{
ABANOV1,
ABANOV3,
ABANOV2,
HARTNOLL,
ABRAHAMS,
PhysRevB.87.045104,
VANUILDO,
DECARVALHO,
PATEL,
PATEL2,
VARMA2,
MAIER,
VARMA3,
MAX2,
SHOUVIK,
SHOUVIK3,
SUNGSIKREVIEW,
MAX1,
LIHAI2,
SCHATTNER2,
GERLACH,
LIHAI,
WANG2}. 
For the antiferromagnetic quantum critical metal in two dimensions,
the minimal patch description includes the critical spin fluctuations
and electrons residing near the hot spots.
Here the hot spots refer to points on the Fermi surface 
connected by the antiferromagnetic wave vector 
and remain strongly coupled with spin fluctuations at low energies. 
Although the theory in two dimensions is strongly coupled at low energies, 
a recent study has revealed 
a strongly coupled fixed point where 
exact critical exponents can be obtained\cite{SCHLIEF}.
The fixed point is characterized by 
the anomalous dimension of the spin fluctuations $1$,
the dynamical critical exponent $z=1$ 
and an emergent nesting of Fermi surface near the hot spots.
Due to a slow flow of the nesting angle under the RG flow, 
at finite length scales one expects to see scaling behaviours controlled by transient exponents that depend on the nesting angle\cite{LUNTS}.
A recent quantum Monte Carlo study\cite{2022arXiv220414241L}
that employs a sign problem-free lattice regularization\cite{MAX1,GERLACH,BERGLEDERER,SCHATTNER2}
shows scaling behaviours that are in qualitative agreement with the predictions.

However,
the hot-spot theory is not complete
because the antiferromagnetic quantum critical metal 
hosts both incoherent electrons near the hot spots 
and electrons with long lifetime away from the hot spots. 
The hot-spot theory is not capable of describing
the momentum dependent universal properties
of electrons on the Fermi surface
and potentially important interplay between hot and cold
electrons for superconductivity\cite{PhysRevB.95.174520}.
The patch theory does not include the four-fermion coupling either
because it is irrelevant in the scaling that leaves the patch theory invariant.
However, the four-fermion coupling should play an important role for superconductivity, 
which is another indication that the patch theory is incomplete.
If the four-fermion coupling
can give rise to infrared singularity,  
it is actually  {\it  relevant} for the low-energy physics. 
Superconducting fluctuations 
captured by the four-fermion coupling
can be intrinsic features of non-Fermi 
liquid states.
In this work, we include all gapless modes around the Fermi surface and the four-fermion coupling.
For the full low-energy effective field theory, we answer the following questions:

\begin{itemize}
\item 
{\bf Q}. How do quasiparticles in cold region gradually lose their coherence
as one approaches the hot spots  in the momentum space
\cite{ABANOV3,PhysRevB.87.045104}?

{\bf A}. 
Electrons at the hot spots 
are incoherent as they remain coupled with spin fluctuations at all energy scales.
On the other hand, 
electrons away from the hot spots decouple from spin fluctuations below a crossover energy scale that increases as one moves away from the hot spots.
This creates a critical fan for the electron spectral function in the momentum space (\fig{fig:dlogim}).
At a fixed momentum along the Fermi surface,
the single-electron decay rate exhibits 
a crossover from a non-Fermi liquid to the Fermi liquid behaviours 
as the momentum perpendicular to the Fermi surface (energy) is lowered.
The crossover, in turn, determines the momentum-dependent quasiparticle weight of the electrons on the Fermi surface 
and the shape of the deformed Fermi surface.
As one approaches the hot spots  on the Fermi surface,
the quasiparticle weight decays toward zero (\fig{figRENORMVF})
and the shape of the Fermi surface is deformed in such a way that the nesting is enhanced near the hot spots (\fig{figRENORMFERMISURFACE}).

\item
{\bf Q}. How do spin fluctuations make electrons incoherent
yet glue them into Cooper pairs?
Is the pairing glue strong enough to overcome the pair breaking effect of the gapless collective mode\cite{
PhysRevB.102.024524,
PhysRevB.102.024525,
PhysRevB.102.094516,
PhysRevB.103.024522,
PhysRevB.103.184508,
PhysRevB.104.144509,
PhysRevB.95.165137,
PhysRevB.92.205104,
PhysRevB.94.115138}?

{\bf A}. 
The gapless spin fluctuations provide the pairing glue, and at the same time they make electrons incoherent, which has the pair-breaking effect.
Although the pair-breaking effect is significant near the hot spots, pairing instability prevails at sufficiently low energies.
This is because 
(1) electrons away from the hot spots eventually become coherent at sufficiently low energies, 
(2) the high-energy spin fluctuations generate 
an attractive four-fermion interaction that coherent electrons become subject to at low energies,
and
(3) the minimum attractive four-fermion interaction is set by the bare nesting angle
formed by the patches of Fermi surface  connected by the antiferromagnetic ordering wave vector.
The largely coherent electrons  subject to the attractive interaction do not suffer from significant pair-breaking effect.

\item
{\bf Q}. 
If pairing instabilities are present,
do electrons near the hot spots that are subject to strong interactions
but small in number drive superconductivity, or
do abundant but cold electrons lead the instability?

{\bf A}. Electrons near the hot spots are too incoherent to drive superconducting instability by themselves despite the fact that they are subject to the strongest attractive interaction mediated by spin fluctuations.
Cold electrons that are too far away from the hot spots are not subject to strong pairing instability either due to weak attractive interactions.
Pairing instability is mainly driven by  `lukewarm' electrons that are incoherent at high energies but emerge as coherent excitations as they decouple from spin fluctuations at low energies.
The universal attractive interaction  generated for those lukewarm electrons
from high-energy spin fluctuations 
 is the main driving force of the superconducting instability.
The instability is strongest in the 
spin-singlet d-wave channel with
the pairing wavefunction extended far beyond the hot spot region 
defined at the scale of superconducting transition temperature.
Therefore, the hot spot theory 
can not capture the superconducting instability that involves large-angle scatterings.

\item
{\bf Q}. Is there a sizable energy window within which non-Fermi liquid scaling can be observed above the superconducting transition temperature
\cite{
CHUBUKOVSC,
PhysRevB.91.115111,
PhysRevB.92.205104}?

{\bf A}. 
For a smaller nesting angle, 
the screening of the interaction 
by particle-hole excitations becomes stronger.
Accordingly, the universal attractive interaction that drives the superconducting instability becomes weaker.
In the limit that the bare nesting angle is small, the superconducting transition temperature is suppressed, which gives rise to  a large window of energy for the non-Fermi liquid scaling as far as the bare four-fermion coupling is weaker than the interaction mediated by the spin fluctuations.
In this case, the superconducting transition temperature is  exponentially suppressed in a fractional power of the inverse of the bare nesting angle (\eq{eq:ellc}).

\item
{\bf Q}. What controls the universal behaviours of the normal state 
and the way superconductivity emerges at low energies\cite{ABANOV3}? 

{\bf A}. 
Universal scaling behaviours of the normal state and the pathway to the superconducting state is controlled by the nesting angle.
Although the nesting angle becomes smaller under the renormalization group flow, its flow becomes increasingly slower for smaller nesting angle.
In the limit that the bare nesting angle is small,
the flow of the nesting angle can be ignored 
above the superconducting transition temperature.
Consequently, the nesting angle acts as an exactly marginal parameter that controls the scaling behaviour of the normal state and the superconducting instability.

\end{itemize}
The above conclusions can be drawn in a controlled manner in the limit that the bare nesting angle is small. 
To answer these physics questions,
we develop a field-theoretic functional renormalization group formalism
and address the following theoretical issues 
that can be relevant to 
 general field theories with continuously many gapless modes\cite{
PhysRevB.66.054526,
PhysRevB.98.125105,
PhysRevB.100.024519,
PhysRevB.104.014517}. 
\begin{itemize}
\setcounter{enumi}{4}
\item
{\bf Q}. How do we construct renormalizable field theories 
that minimally capture universal low-energy physics
of systems with continuously many gapless modes?

{\bf A}. 
In the presence of Fermi surface, operators that are nominally irrelevant in terms of their scaling dimensions can give rise to infrared singularities.
This can happen because the Fermi momentum that determines the phase space of low-energy excitations enhances IR singularities for virtual processes that involve low-energy electrons across the full Fermi surface.
For this reason, the four-fermion operators that are `irrelevant' under the  patch scaling are actually marginally relevant in the pairing channel.
Renormalizable field theories must include all such operators.

\item
{\bf Q}. 
How is the notion of scale invariance defined 
for theories with intrinsic scales such as Fermi momentum
that do not decouple in the infrared (IR)?

{\bf A}. 
Fermi momentum in general does not decouple from low-energy observables.
Since it flows to infinity relative to the decreasing energy scale of the renormalization group flow,
theories with Fermi surface do not have the true sense of scale invariance.
Because the scaling equation for physical observables include the running Fermi momentum,
their scaling dimensions alone do not fix how they actually scale with probing energy in general.

\item 
{\bf Q}. 
In the presence of large momentum scales,
how does a low-energy  effective field theory protect  its predictability 
from high-energy physics that infiltrates into the low-energy realm?

{\bf A}. 
Because mixing among operators with large differences in momenta along the Fermi surface can be controlled by high-energy physics, 
some one-particle irreducible vertex functions can not be determined entirely in terms of other low-energy observables.
However, there still exist a set of low-energy observables that can be predicted within the low-energy effective field theory without resorting to `unknown' high-energy physics.

\item
{\bf Q}.
What are the strategies for understanding quantum field theories 
 that have continuously many gapless degrees of freedom?
What do RG flow and fixed points look like 
in the infinite dimensional space of coupling functions?

{\bf A}. 
In the presence of continuously many gapless modes in the momentum space,
the universal low-energy physics is captured by a set of coupling functions  in terms of which all low-energy observables of the theory can be expressed.
The RG flow of the coupling functions is captured by the beta functionals, 
which are functionals of coupling functions.
At fixed points, the coupling functions acquire universal momentum profiles.
While understanding the full RG flow requires a functional analysis, 
the flow projected onto a finite dimensional subspace of coupling functions provides an important insight into the full RG flow.
The projected RG flow can exhibit qualitatively different behaviours as the subspace is rotated within the infinite dimensional space of coupling functions.

\end{itemize}
Given that the paper is long,
we first provide a high-level summary of it
as a guidance for the more in-depth reading.

\section{Summary}
	\label{Sec:AFQCM}

	\begin{figure*}[htbp]
	\centering
	\begin{subfigure}[b]{0.49\linewidth}
		\centering
		\includegraphics[scale=0.8]{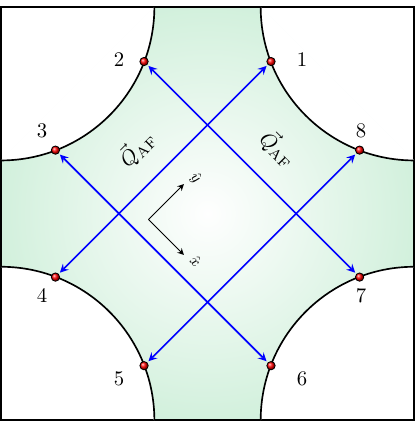}
		\caption{\label{fig:FermiSurface}}
	\end{subfigure}
	\begin{subfigure}[b]{0.49\linewidth}
		\centering
		\includegraphics[scale=0.8]{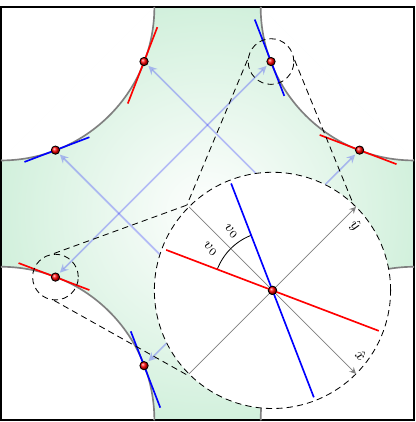}
		\caption{\label{fig:FermiSurfaceLinear}}
	\end{subfigure}
	\caption{({\color{blue}$a$}) 
	A $C_4$-symmetric Fermi surface in two dimensions. 
	The dots on the Fermi surface represent the hot spots connected by the antiferromagnetic  wave vector, 
	$\vec{Q}_{\mathrm{AF}}$.
	$\hat x$ ($\hat y$) is chosen to be 
	perpendicular (parallel) to $\vec Q_{AF}$ 
	that ends at hot spot $1$.
	({\color{blue}$b$}) 
	Near hot spot $1$, the Fermi surface is denoted as $k_y = - v_{k_x} k_x$,
	where $v_k$ is generally a function of a component of momentum along the Fermi surface.
	In the hot spot theory, $v_k$ is expanded around $k=0$ and only the leading order term is kept.
	In the small $v$ limit, the patches of Fermi surface connected by $\vec Q_{AF}$ become locally nested.
	}
\end{figure*}

Although the field theoretic functional renormalization group  formalism is general, we set it up through the concrete example.
Our starting point is the hot-spot theory
for the antiferromagnetic quantum critical point
in a two-dimensional metal with the $C_4$ symmetry 
(\eq{eq:Action1}).
We consider a simply connected Fermi surface (\fig{fig:FermiSurface})
that undergoes a collinear antiferromagnetic quantum phase transition
with a commensurate magnetic ordering vector $\vec Q_{AF}$, 
where $2 \vec Q_{AF}$ is equivalent to a reciprocal vector.
Generically, there are eight hot spots
on the Fermi surface that are connected by $\vec Q_{AF}$.
Only those electrons at the hot spots
remain strongly coupled with spin fluctuations
down to the zero-energy limit.
The hot-spot theory includes  
electrons near the eight hot spots
and the bosonic collective mode associated with critical spin fluctuations.
For generality, we consider fermions in the fundamental representation of
the $SU(N_f)$ flavour symmetry
and the $SU(N_c)$ spin rotational symmetry,
where the collective mode 
in the adjoint representation of $SU(N_c)$ 
is minimally coupled with electrons 
through a Yukawa coupling.
However, everything discussed in this paper remains valid for any $N_f \geq 1$ and $N_c \geq 2$.

In the patch theory,
parameters of the theory are expanded around the hot spots,
and only the leading-order constant terms are kept.
The Fermi surface is considered to be locally straight near the hot spots
whose orientation is parameterized by the nesting angle $v$.
If $v=0$, 
each pair of the hot spots 
connected by $\vec Q_{AF}$ are  anti-parallel		(\fig{fig:FermiSurfaceLinear}). 
However, the bare value of the nesting angle is nonzero without fine tuning.
The Fermi velocity along the direction of $\vec Q_{AF}$
can be set to be one through momentum scaling.
The Yukawa coupling constant
describes the interaction between electrons near the hot spots
and spin fluctuations.
The most prominent effect of the coupling is
to renormalize the dynamics of spin fluctuations 
through mixing with gapless particle-hole excitations.
The strong coupling gives rise to an $O(1)$ anomalous dimension of the boson,
and the entire bare kinetic term of the boson becomes irrelevant at low energies.
As a result, it is enough to consider the theory that includes 
only the kinetic term of the electrons and the Yukawa coupling.
The large anomalous dimension of the boson implies
that the scale associated with the Yukawa coupling is absorbed
into the boson field,
and the only dimensionless parameter of the hot-spot theory is $v$.
While the boson is strongly renormalized by particle-hole fluctuations,
its feedback to the fermionic sector is weak in the small $v$ limit.
For a small but non-zero $v$, 
quantum corrections 
make electrons incoherent near the hot spots
and generate the logarithmic RG flow of $v$ toward zero.

\begin{figure}[h]
	\centering
	\includegraphics[scale=0.8]{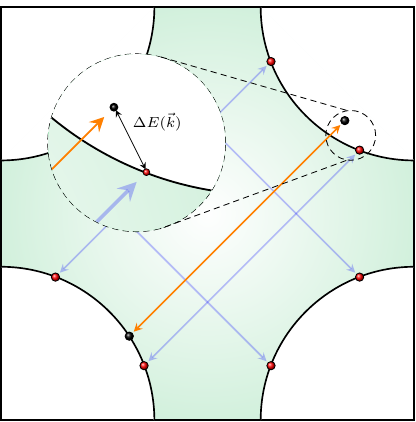}
	\caption{
By absorbing/emitting a boson with zero energy,
an electron on the Fermi surface (black dot) is scattered into a state away from the Fermi surface 
if the initial momentum is away from the hot spots.
Alternatively, the electron must absorb/emit a boson with non-zero energy to scatter onto the Fermi surface.
The minimum energy that virtual particles have to carry within a loop gives rise to a crossover energy scale below which electrons decouple from spin fluctuations at each point on the Fermi surface.
Electrons closer to the hot spots
remain coupled with spin fluctuations
down to lower energy scales,
which gives rise to a momentum dependent life time of electrons
that gradually vanishes as one approaches hot spots.
	\label{fig:Scatterings}
	}
\end{figure}

In the full theory that includes the whole Fermi surface (\eq{eq:LukewarmAction}),
the nesting angle, the Fermi velocity and the Yukawa coupling
are promoted to functions
of momenta along the Fermi surface.
The four-fermion coupling functions are also introduced as they give rise to IR singularities.
These couplings of the theory will be collectively referred to as {\it coupling functions} 
when there is no need to specify one. 
Even if one starts with momentum independent coupling functions,
they acquire non-trivial momentum dependences at low energies
through momentum dependent quantum corrections.
To understand how coupling functions acquire momentum dependence,
it is useful to introduce what is called {\it the space of IR singularity}.
It denotes a set of external momenta at which quantum corrections to 
each vertex function persist
down to the zero energy limit.
%
%
If external momenta are outside the space of IR singularity,
a non-zero energy scale that depends on external momenta cuts off IR divergences
(\fig{fig:space_of_IR_singularity}). 
For example, consider the fermion self-energy that renormalizes the kinetic term of electrons.
At the lowest order, 
an electron on the Fermi surface with momentum $k$ away from a hot spot 
can emit a virtual boson to be scattered to the other patch 
before it absorbs the boson to come back to the original state.
The quantum correction 
gives rise to an infrared divergence 
in the fermion self-energy 
and destroys the coherence of single-particle excitation 
at the hot spots.
However, the infrared singularity is cut off for electrons away from hot spots.
For $v \neq 0$, the patches connected by  the antiferromagnetic ordering vector, $\vec Q_{AF}$ is not perfectly nested,
and the electron away from the hot spots ($k \neq 0$) has to be scattered to an intermediate state 
away from the Fermi surface
unless it creates a boson with  a non-zero energy 
as is illustrated in \fig{fig:Scatterings}.
This necessitates creating virtual particles with non-zero energies 
whose minimum is proportional to $k$.
Due to the uncertainty principle, 
those virtual states can only last for a finite time period,
and the quantum fluctuation becomes unimportant
at time scales larger than a characteristic time scale given by the inverse of the infrared energy cutoff.
Since the quantum effects that renormalize the kinetic term of electron away from the hot spots turn off at sufficiently low energies,
the space of IR singularity for the electron self-energy is simply the hot spots.
Electrons at different locations on the Fermi surface
decouple from spin fluctuations at different energy scales,
and the renormalized coupling functions acquire
non-trivial momentum dependence.
Similarly, the momentum dependent renormalization
of the Yukawa coupling and the four-fermion coupling function
can be extracted from their spaces of IR singularity 
and the momentum dependent energy scales
that cut off IR divergences away from the spaces of IR singularity.
The Yukawa coupling function has the zero dimensional space of IR singularity.
However, the dimension of the space of IR singularity 
for the four-fermion coupling function is greater than zero.
The Yukawa coupling and the four-fermion coupling
represent the interaction energy that tend to localize particles in the real space,
and the nesting angle and the Fermi velocity represent
the strength of the kinetic term that tend to delocalize particles\cite{SHOUVIK}.
Consequently, these two groups of coupling functions acquire opposite momentum profiles :
as one approaches the hot spots,
the Yukawa coupling and the four-fermion coupling functions 
become stronger 
while the nesting angle and the Fermi velocity become smaller
(\fig{fig:vVFg}
and
\fig{fig:4f0_1111_plot}).

\begin{figure}[htbp]
	\centering
		\includegraphics[scale=0.65]{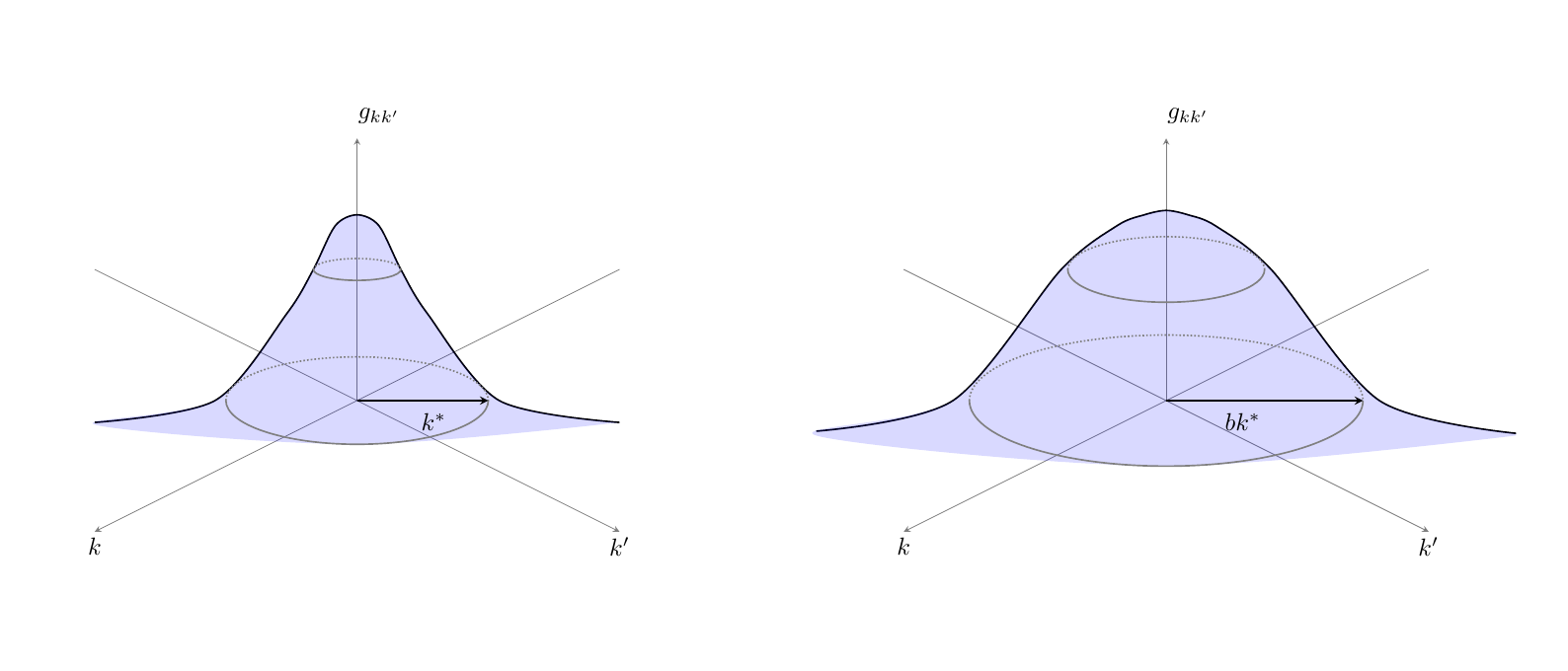}
	\caption{
Under the tree-level scaling, 
the momentum along the Fermi surface is rescaled, which causes
the momentum profiles of the coupling functions to be stretched out 
under the renormalization group flow.
}	
\label{fig:momentum_dilatation}
\end{figure}

The best way to understand the ultimate fate of a theory in the low-energy limit
is to examine its renormalization group (RG) flow\cite{WILSONRG1,WILSONRG2,WETTERICH,POLCHINSKIFRG,SHANKAR}.
Fixed points of the RG flow are of particular importance 
as they represent scale invariant states and 
control the low-energy physics 
that define universality classes.
Needless to say, the notion of scale invariance 
is defined with respect to a specific scale transformation.
In Fermi liquids, 
the fixed-point theory is invariant 
under the transformation that scales the frequency and the component of momentum perpendicular
to the Fermi surface\cite{POLCHINSKI1,SHANKAR}.
The component of momentum along the Fermi surface is treated as a dimensionless flavour 
that labels gapless modes on the Fermi surface.
This scaling leaves the forward scattering amplitudes invariant for which the momentum along the Fermi surface does not need to be compared with any scale. 
However, this scaling is not suitable for non-Fermi liquids
because the momentum along the Fermi surface plays an additional role:
it not only labels gapless modes along the Fermi surface but also carries a scale that needs to be measured against 
the momentum of the critical boson.
By emitting or absorbing a boson with a non-zero momentum,
electrons can change their  directions.
As a result, the momentum along the Fermi surface needs to be scaled
in the same way the boson momentum is scaled 
to keep the fermion-boson coupling invariant.
In the present theory, the $C_4$ symmetry forces momentum
to be scaled  isotropically.
Under this scale transformation, 
the region close to the Fermi surface is magnified,
which allows us to probe low-energy physics close to the Fermi surface.
At the same time, the momentum along the Fermi surface is stretched out.
This causes the size of Fermi surface 
measured in the unit of a running energy scale
to increase indefinitely as the low-energy limit is taken.
This is expected because gapless modes 
on the Fermi surface are not integrated out under a coarse graining.
The expanding Fermi surface reflects the fact that at lower energies 
two points on the Fermi surface become effectively farther apart
 compared to the momentum carried by critical boson.
Consequently, 
the couplings defined as functions of the momentum along the Fermi surface
are dilated under the scale transformation (\fig{fig:momentum_dilatation}).
With stretching,  bumps and wrinkles in the coupling functions tend to get flattened.
On the other hand, the coupling functions receive momentum 
dependent quantum corrections under the RG flow.
Combined, these two effects determine the universal profiles of the coupling functions 
 that emerge at a fixed point.

Promoting coupling constants to coupling functions amount
to incorporating infinitely many derivative terms when expanded around  a point on the Fermi surface (say a hot spot).
Higher derivative terms of the coupling functions have larger negative scaling dimensions,
and they are superficially irrelevant.
The irrelevancy indicated by the scaling dimension means that 
through small-angle scatterings 
those derivative terms do not affect the low-energy physics of electrons 
at a particular location on the Fermi surface 
around which the coupling functions are expanded.
However, we can not discard 
such higher derivative terms 
 all together even if they are irrelevant by power-counting.
This is because the derivative terms along the Fermi surface can describe the variation of the universal low-energy properties  of gapless electrons along the Fermi surface.
Such information is a part of the data that needs to be kept
within the low-energy effective field theory.
If there are large angle scatterings generated by superconducting fluctuations,
it is impossible to understand
even one part of Fermi surface 
without knowing what happens to the whole.
Therefore, we keep the full momentum dependence of the coupling functions
along the Fermi surface.  

Just as the relevancy of couplings can not be simply determined
from their scaling dimensions, the notion of renormalizable theory becomes subtle in field theories with continuously many gapless modes.
We define the renormalizable theory to be the minimal theory that can be used to
express all low-energy observables
in terms of those couplings included in the theory
within errors that vanish in powers of energy.
In relativistic field theories, renormalizable theories
only include couplings 
with non-negative scaling dimensions.
In the presence of Fermi surface, this simple rule does not apply.
For example, the four-fermion couplings have scaling dimension $-1$.
Nonetheless, they can give rise to IR singularities
that are responsible for superconducting instabilities.
Clearly, we can not capture the correct low-energy physics
within a power-law accuracy
if we drop the four-fermion coupling.
This discrepancy between 
the scaling dimension
and the actual relevancy of the coupling in the IR
arises because the scale associated with the size of Fermi surface
can change the degree of IR singularity 
of quantum corrections. 
Actual IR divergences depend not just on the dimension of
the coupling but also on the phase space available at low energies.
Through the four-fermion interaction in the pairing channel,
a pair of electrons with opposite momenta
can be scattered to anywhere on the Fermi surface.
Although the four-fermion coupling has a negative scaling dimension,
such processes can give rise to logarithmic singularities
because the four-fermion couplings 
combined with the size of Fermi surface which carries a positive scaling dimension
are effectively promoted to marginal couplings.
For this reason, the low-energy effective theory must include
the four-fermion coupling function 
even if it has the negative scaling dimension.
The general rule is that within a renormalizable theory
we have to include all couplings that can give rise to IR singularities 
irrespective of their scaling dimensions.
In the present theory, 
we don't need to include higher order couplings
beyond the quartic fermion couplings.
The sixth or higher order fermion couplings,
whose scaling dimensions are less than $-1$,
are too irrelevant to 
affect the low-energy physics
in a significant way
even with the help of the Fermi surface volume.

Physically, the coupling functions correspond to  
momentum dependent one-particle irreducible
vertex function of the theory measured at an energy scale.
The goal of the field theoretic functional RG is to extract the flow of 
the minimal set of momentum dependent coupling functions 
 needed to fully characterize the low-energy physics.
Ideally, one would want to use a scheme
in which the coupling functions are identical to
the physical vertex functions.
However, this `total subtraction scheme' is impractical 
because it requires computing
entire quantum corrections including finite parts.
To capture the universal low-energy physics in the minimal way,
we adopt a minimal subtraction scheme
in which we only require that the quantum effective action 
expressed in terms of the renormalized coupling functions
is regular in the zero energy limit.
In general, there are scheme dependent non-singular differences between
the actual quantum effective action and the coupling functions.
However, not knowing the precise relation between them is fine 
for the purpose of extracting scaling behaviours
as far as one is determined from the other 
through finite relations in the low-energy limit.
This minimal subtraction scheme can be readily implemented
for the marginal and relevant couplings 
in the standard way
in which one only adds counter terms to remove IR divergences.
However, there is a subtlety associated with the four-fermion couplings.
Having scaling dimension $-1$, quantum corrections to the four-fermion coupling exhibit $1/\mu$ singularity in energy $\mu$ 
when all external momenta are less than $\mu$ relative to hot spots.
If the IR divergence is cut off by external momenta larger than $\mu$,
the contribution of the power-law singularity to dimensionless physical observables remain finite
in the small $\mu$ limit 
due to the vanishingly small phase space of the external momenta. 
For this reason, we don't need to add counter terms to remove such IR divergences.
However, there can be 
IR singularities 
that are not cut off even when the external momenta
are far away from the hot spots.
In the channels whose spaces of IR singularity 
have dimensions greater than zero,
a dimensionless observable constructed
from a momentum integration of the vertex function
along the extended space
of IR singularity  can become singular 
in the low-energy limit.
An example of such physical observable is the energy of a Cooper pair determined from an integration of the  four-fermion coupling function 
over the momentum along the Fermi surface weighted with a pair wavefunction.
In order to express such physical observables 
in terms of the coupling functions 
through a non-singular relation,
one must introduce counter terms
to absorb those singularities 
into the coupling functions. 
Only then, the RG flow of the coupling functions
faithfully represent the behaviours of physical observables.
Therefore, one should impose renormalization conditions on the four-point vertex functions 
integrated over one-dimensional manifolds 
in the space of external momenta.
We call this {\it extended minimal subtraction scheme}.

As the next step toward computing the beta functional,
we need to compute the counter terms
as functionals of the coupling functions. 
In the present theory,
we use the nesting angle $v$
as a small parameter to organize the computation.
To the leading order in $v$,
one can identify an infinite set of diagrams 
that renormalize the boson propagator.
Those diagrams can be non-perturbatively summed over 
through the Schwinger-Dyson equation in \fig{fig:SDEq}.
It describes the processes in which
particle-hole fluctuations give 
the spin fluctuations
an over-damped dynamics with 
speed $c \sim \sqrt{v_0 \log 1/v_0}$,
where $v_0$ corresponds to the nesting angle at the hot spots.
The speed of the collective mode depends only on the nesting angle measured at the hot spots
because low-energy spin fluctuations are mainly renormalized
by electrons near the hot spots.
On the other hand, 
the feedback of the dressed boson to the fermion can be computed perturbatively
in powers of $v$ in the small $v$ limit.\footnote{
To be precise, the expansion is organized
in powers of $\sqrt{v}$ and $1/\log (1/v)$.}

The quantum effective action is a functional of the coupling  functions.
In general, quantum corrections are written as 
integrals over momentum along the Fermi surface.
Without knowing momentum profiles of the coupling functions,
we can not compute the loop integrations of quantum corrections in closed forms.
However, a simplification arises for quantum corrections 
that do not involve nested scatterings.
Here, the nested scatterings refer to the processes 
in which the phase space available for internal particles in the loops is not limited by the energy at which the scattering is probed.
Conversely, in non-nested processes, momenta of internal particles are limited by external energies. 
Thanks to the locality in the real space,
renormalized coupling functions at energy scale $\mu$
can not vary significantly over the momentum scale 
that is proportional to $\mu$.
In quantum corrections
that involve only non-nested scatterings,
loop momenta are dynamically cut off by a momentum that is proportional to $\mu$.
This allows us to treat the coupling functions within those  loops
as constants 
and replace them with the ones evaluated at the external momenta.
Renormalized coupling functions 
can vary significantly
across the entire Fermi surface. 
But, as far as they do not vary appreciably over the scale 
that is proportional to $\mu$, 
one can use the `instantaneous' values of the coupling functions
evaluated at the external momentum.
Using this {\it adiabaticity},
one can express the counter terms for non-nested quantum corrections in  terms of the couplings evaluated at the external momentum.

\begin{figure*}
	\centering
		\includegraphics[scale=0.8]{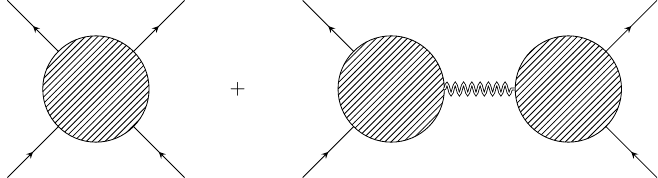}
	\caption{ 
The net two-body electron-electron interaction
composed of the one-particle irreducible (1PI) four-point function and two 1PI three-point functions connected with the dressed boson propagator.
}
	\label{fig:4pt3pts}
\end{figure*}

There are, however, nested quantum corrections 
that can not be computed by invoking adiabaticity.
For example, in the pairing channel, 
the four-fermion couplings with 
large differences in momentum 
along the Fermi surface can mix with each other
as Cooper pairs can be scattered  around the Fermi surface even at low energies.
For those nested scatterings, 
the momentum along the Fermi surface is not necessarily
bounded by the external energy,
and quantum corrections are generally expressed 
as integrals of coupling functions along the Fermi surface.
If the coupling functions do not decay fast enough at large momenta,
the contribution from large momenta along the Fermi surface
can be important.
In particular, the critical boson with large momenta
contributes to the mixing between Cooper pairs with
 different relative momenta even in the low-energy limit.
The fact that operators defined on different parts 
of the Fermi surface can mix with each other 
is not surprising in that they all describe gapless degrees of freedom.
What is peculiar though is the fact that 
the mixing between low-energy operators with large differences
in momentum along the Fermi surface can be influenced by
the high-energy physics of the critical boson.
Due to this ultra-violet/infrared (UV/IR) mixing, 
the four-point vertex function itself
can not be extracted from the low-energy effective field theory.
Although the predictability of the low-energy effective theory seems at risk,
there are still observables
protected from the UV/IR mixing.
The `protected' low-energy observable that describes the fermionic four-point function is the 
net two-body electron-electron interaction
given by the sum of the one-particle irreducible (1PI) four-point vertex function
and the tree graph formed by connecting 
two 1PI three-point function with the boson propagator
(see \fig{fig:4pt3pts}).
The net interaction determines physical
correlation functions at low energies,  
and it can be determined within the low-energy effective field theory
 without reference to high-energy physics. 
The RG flow of the net two-body interaction 
is insensitive to the ultra-violet (UV) physics.

	\begin{figure*}[htbp]
	\centering
	\begin{subfigure}[b]{0.49\linewidth}
		\centering
		\includegraphics[scale=1.0]{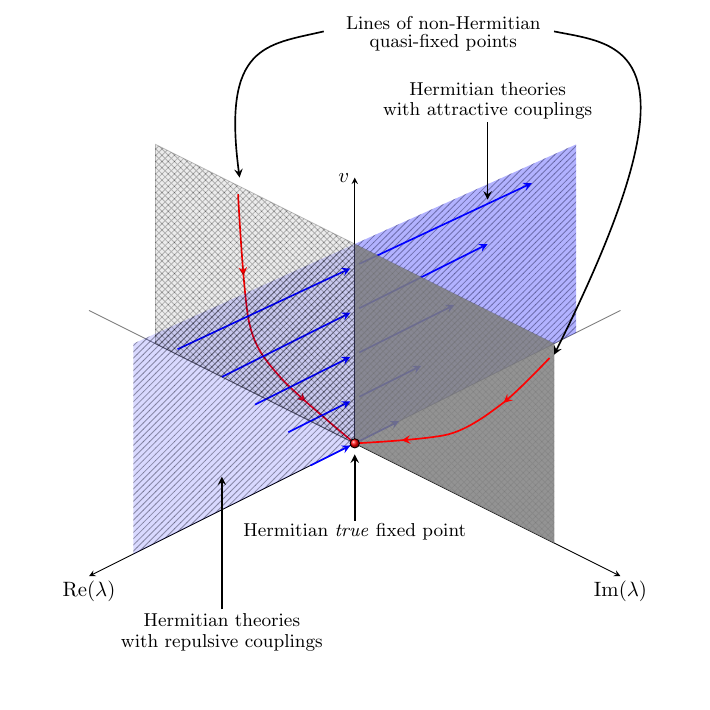}
		\caption{\label{fig:non_Hermitian_fp_a}}
	\end{subfigure}
	\begin{subfigure}[b]{0.49\linewidth}
		\centering
		\includegraphics[scale=0.9]{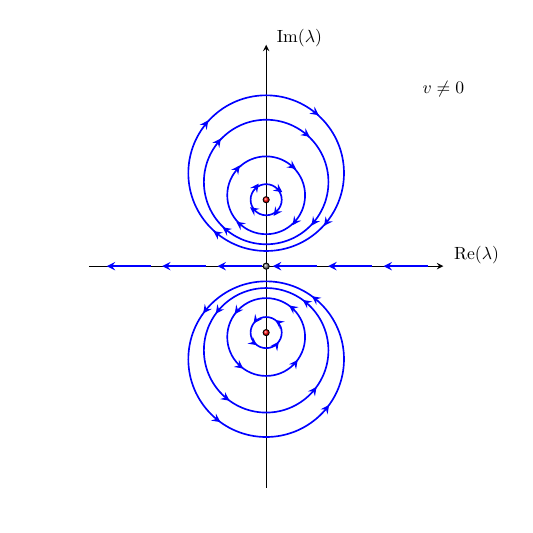}
		\caption{\label{fig:non_Hermitian_fp_b}}
	\end{subfigure}
	\caption{({\color{blue}$a$}) 
	The schematic functional renormalization (RG) group flow depicted in the space of the complexified four-fermion coupling function ($\lambda$) and the nesting angle ($v$).
For each nesting angle $v$, there exist a pair of non-Hermitian fixed points for the RG flow projected to the space of coupling functions with fixed $v$. 
These are called quasi-fixed points
and 
the pairs of non-Hermitian quasi-fixed points are related to each other through the Hermitian conjugation.
As $v$ decreases, the non-Hermitian quasi-fixed points merge to the true Hermitian fixed point located at $v=0$.
While $v$ flows towards zero under the full RG flow,
a Hermitian theory with a non-zero nesting angle 
necessarily flows to the superconducting state before the nesting angle changes significantly.
	({\color{blue}$b$}) 
A schematic functional RG flow projected to the space of complex four-fermion coupling function at a non-zero $v$.
The proximity of the non-Hermitian quasi-fixed points to the space of Hermitian theories creates a bottleneck region with constricted RG flow for physical theories with small nesting angles.	
	}
	\label{fig:non_Hermitian_fp}
\end{figure*}

Based on this theoretical formalism,
we extract the following physics 
for the two-dimensional
antiferromagnetic quantum critical metal.

\begin{enumerate}
\item 
In the space of coupling functions with small nesting angles,
there exists a unique interacting fixed point (\eq{eq:true_fp}). 
At the fixed point, the coupling functions
are momentum independent
with vanishing nesting angle 
and the dynamical critical exponent $1$.
The fixed point arises as the end point of 
one-parameter families of 
{\it quasi-fixed points} labeled by  momentum independent $v$.
The lines of quasi-fixed points,
which are non-Hermitian for $v>0$,
are the fixed points of the beta functionals 
projected to the space 
with a fixed nesting angle.
Although the nesting angle 
near the hot spots 
flows to zero logarithmically
under the full RG flow, 
the quasi-fixed points act as  approximate fixed points 
over a large window of distance scales for small $v$.
While those non-Hermitian quasi-fixed points are unphysical 
by themselves,
they come close to the space of
Hermitian theories in the small $v$ limit.
The proximity of the non-Hermitian quasi-fixed points
to the space of Hermitian theories
creates a bottleneck with slow RG flow speed for Hermitian theories.
This is illustrated in \fig{fig:non_Hermitian_fp}.

\item
The gapless spin fluctuations make electrons near the hot spots incoherent,
which has the pair-breaking effect.
However, the spin fluctuations provide an attractive glue that is strong enough to overcome the pair breaking effect and drive the system to superconducting states.
The following two groups of UV theories, which are divided by the bottleneck region,
exhibit superconductivity in qualitatively different ways.

\begin{enumerate}
    \item 
In UV theories in which the bare four-fermion coupling is attractive 
and stronger in magnitude 
than the interaction mediated by the spin fluctuations 
in one or more pairing channel
(called theories with attractive couplings),
superconductivity arises 
through the usual BCS pairing instability as in Fermi liquids.
The superconducting transition temperature 
and the pairing wavefunction 
are sensitive to the bare four-fermion coupling,
and the nearby non-Fermi liquid fixed point has
little to do with superconductivity.

\item
Theories whose
bare four-fermion interaction is repulsive in all pairing channels 
or attractive but weaker than the interaction mediated 
by the spin fluctuations
(called theories with repulsive UV couplings)
necessarily flow through the bottleneck region 
with constricted RG flow, 
exhibiting quasi-universal features \cite{Gorbenko:2018vt,PhysRevX.7.031051}.
In this case, 
normal state properties, 
the superconducting transition temperature
and the pairing wavefunction
are all controlled by 
the nesting angle
and
the scale 
at which the theory passes through 
the bottleneck region.
Superconducting instability occurs through two distinct stages.
In the first stage,
the gapless spin fluctuations generate attractive interactions  for  electrons 
in the vicinity of the hot spots
and make those very electrons incoherent.
The interaction generated by the spin fluctuations
is strongest in the d-wave spin singlet channel\cite{SCALAPINO}
with a universal momentum profile near the hot spots.
With increasing RG time (decreasing energy scale),
the region affected by spin fluctuations becomes increasingly localized near the hot spots.
At sufficiently low energies, 
most of the Fermi surface is left with
lukewarm electrons,
that is, electrons 
that were once strongly renormalized by spin fluctuations at high energies 
but are now largely decoupled from them.
The lukewarm electrons are coherent 
and no longer suffer from 
the strong pair breaking effect.
Nonetheless, they are still subject to an attractive four-fermion interaction that has been accumulated from the high energy scales.
The resulting attractive interaction is insensitive to the bare four-fermion coupling in the small $v$ limit
because the UV information 
is largely erased while the theory goes through the bottleneck region.
Below a crossover scale,
the accumulated four-fermion coupling for the lukewarm electrons become larger than the interaction mediated by the critical spin fluctuations.
This marks the beginning of the second stage in which 
the further growth of the  pairing interaction  is mainly driven by the  lukewarm electrons subject to the universal four-fermion coupling of the coherent electrons 
rather than the small number of hot electrons that are still  subject to strong renormalization from spin fluctuations.


\end{enumerate}

\item
In the normal state of the theories with repulsive couplings,
the shape of the renormalized Fermi surface,
the Fermi velocity,
the Yukawa coupling function
and the four-fermion coupling functions
all acquire universal momentum profiles near the hot spots (\fig{figRENORMFERMISURFACE}
and
\fig{figRENORMVF}).
At the hot spots, 
the electron spectral function is incoherent,
exhibiting a non-Fermi liquid behaviour at all available energy scales above the superconducting transition temperature.
Away from the hot spots, there exists a crossover 
from the high-energy non-Fermi liquid behaviour 
to the low-energy Fermi liquid behaviour.
The crossover energy scale increases with momentum away from the hot spots, 
creating a quantum critical fan with
the momentum along the Fermi surface playing the role of a tuning parameter for reaching `criticality'
within the critical theory (\fig{fig:dlogim}).
The size of the energy window for the non-Fermi liquid scaling above the superconducting transition temperature depends on the bare nesting angle.
For theories with repulsive couplings,
the window for the critical scaling becomes larger 
as the bare nesting angle becomes small.

\end{enumerate}

Here is the outline for the rest of the paper.
In Sec. \ref{Sec:Hot}, we review 
the hot-spot theory and 
the non-perturbative solution
obtained in the small $v$ limit.
In Sec. \ref{Sec:Lukewarm},
the hot-spot theory is generalized to 
the theory that includes the whole Fermi surface along with the four-fermion coupling.
Coupling constants of the hot spot theory are promoted to functions of momentum along the Fermi surface.
Sec. \ref{sec:FRG} lays out the theoretical foundation
for the field-theoretic functional renormalization group formalism.
The concept of 
renormalizable theory, 
minimal subtraction scheme
and scale invariance
are generalized for theories with continuously many gapless modes.
To facilitate the explicit calculations 
in the following sections,
we also introduce the ideas of
adiabaticity
and the space of IR singularity. 
In Sec. \ref{sec:beta_functionals},
we derive the beta functionals 
and identify the fixed point of the theory.
Sec. \ref{sec:fixedpoint}
discusses the fate of theories that are tuned 
away from the fixed point
with non-zero nesting angles.
Due to the slow flow of 
the nesting angle,
we consider 
the beta functionals projected
to the subspace of coupling functions 
with a fixed nesting angle
and examine its RG flow.
It is proven that the quasi-fixed points of the projected beta functional are necessarily non-Hermitian for a non-zero nesting angle.
We also discuss the channel dependent superconducting instability caused by the interplay between 
the pair forming effect 
of spin fluctuations
and the incoherence-induced
pair breaking effect. 
In Sec. \ref{sec:super},
we estimate the superconducting transition temperature 
by solving the beta functional  for the four-fermion coupling  in the pairing channel.
In Sec. \ref{sec:normal},
the normal state properties of the theory are discussed.
We  compute the electron spectral function
and extract momentum dependent single-particle properties of electron across the Fermi surface.
We end with concluding remarks in Sec. \ref{sec:conclusion}.
Each section begins with a brief summary of the main ideas put forward in that section.

\section{Review of the  hot spot theory}
\label{Sec:Hot}

	\fbox{\begin{minipage}{48em}
{\it
\begin{itemize}
\item
The hot spot theory describes gapless antiferromagnetic spin fluctuations
and electrons in the vicinity of the hot spots.
While the effective coupling, 
given by the ratio between the interaction strength and 
the smallest velocity in the theory, becomes $O(1)$ at low energies,
the theory is solvable 
thanks to the emergence of a hierarchy of velocities
in the limit that the nesting angle is small.  
\item
Non-perturbative quantum corrections 
that survive in the small nesting-angle limit
can be fully taken into account through 
the Schwinger-Dyson equation for the boson propagator, 
and the rest of the quantum corrections can be 
included perturbatively in powers of the nesting angle.

\item
The leading-order quantum corrections to fermions make the nesting angle flow toward zero at low energies,
and the small nesting-angle expansion becomes asymptotically exact in the low-energy limit.
However, the RG flow is expected to be eventually cut off by a superconducting instability driven by the four-fermion coupling, 
which is irrelevant by power-counting and not included in the hot spot theory.
\end{itemize}
}
\end{minipage}}
\vspace{0.5cm}

The hot spot theory that describes 
the gapless spin fluctuations
and the electrons near the hot spots
is written as\cite{SCHLIEF}
	\begin{align}
	\label{eq:Action1}
\begin{split}
S &=\sum^{8}_{N=1} \sum^{N_{c}}_{\sigma=1}\sum^{N_{f}}_{j=1}
\int \dd {\bf k}~ \psi^{\dagger}_{N,\sigma, j}({\bf k})\left[ i k_0 + e_{N}(\vec{k};v)\right] \psi_{N,\sigma, j}({\bf k})\\
& + \frac{1}{4} \int \dd {\bf q}~ \left( q_0^2 + c_0^{2} | \vec q |^2 \right) \tr{ \Phi({\bf q}) \Phi(-{\bf q}) } \\
&+\frac{g}{\sqrt{N_{f}}}\sum^{8}_{N=1}\sum^{N_{c}}_{\sigma\sigma'=1}\sum^{N_{f}}_{j=1}\int\dd {\bf k} ~\int \dd {\bf q} ~ \psi^\dagger_{N,\sigma' ,j}({\bf k}+{\bf q})\Phi_{\sigma'\sigma}({\bf q})\psi_{\bar{N},\sigma, j}({\bf k}).
\end{split}
\end{align}
\noindent 
Here ${\bf k}=(k_0,\vec{k})$ denotes the three momentum
that includes the  Matsubara frequency  $k_0$
and the two-dimensional momentum $\vec k$,
and
$\dd {\bf k} \equiv \frac{ \dd k_0 \dd k_x \dd k_y}{(2\pi)^3}$.
We consider a $C_4$-symmetric Fermi surface 
that supports eight hot spots labeled 
by $N=1,2,\dots,8$ 
as shown in \fig{fig:FermiSurface}. 
$\psi_{N,\sigma,j}({\bf k})$ represents the electron field 
near hot spot $N$ with spin $\sigma=1,2,\dots, N_c$
and flavour  $j=1,2,\dots,N_f$,
where $\vec{k}$ is measured relative to hot spot $N$. 
The electron is in the fundamental representation of 
spin $SU(N_c)$ and flavour $SU(N_f)$ groups.
The case that is most relevant to experiments is $N_c=2$ and $N_f=1$,
but we keep $N_c$ and $N_f$ general.
All results discussed in this paper
hold for any $N_c \geq 2$ and $N_f \geq 1$. 
The electron dispersion expanded to the linear order 
in momentum away from each hot spot is  
\begin{equation}
\begin{aligned}
e_{1}(\vec{k};v) =-e_5(\vec{k};v) =& vk_{x}+k_y, \\
e_2(\vec{k};v)=-e_{6}(\vec{k};v) =& -k_{x}-vk_y, \\
e_{3}(\vec{k};v) = -e_{7}(\vec{k};v) =& -k_{x}+vk_y, \\
e_4(\vec{k};v)=-e_8(\vec{k};v) =& vk_{x}-k_y.
\end{aligned}
\label{eq:Fermiondispersion}
\end{equation}
The coordinate is chosen so that 
$\vec{Q}_{\mathrm{AFM}}$ is parallel to 
$\hat y$ at hot spot $1$. 
The component of the Fermi velocity along $\vec{Q}_{\mathrm{AFM}}$ is set to be $1$ through a choice of momentum scale,
and $v$ denotes the  dimensionless ratio between the component of the Fermi velocity perpendicular to 
$\vec{Q}_{\mathrm{AFM}}$ 
and the component parallel to 
$\vec{Q}_{\mathrm{AFM}}$. 
The patches of Fermi surface 
connected by $\vec{Q}_{\mathrm{AFM}}$ 
have relative slope $2v$,
and $v$ is referred to as 
{\it nesting angle} (see \fig{fig:FermiSurfaceLinear}). 
The collective antiferromagnetic spin fluctuations are represented by a bosonic field in the adjoint representation of $SU(N_c)$, 
$\Phi({\bf q}) = \sum^{N^2_c-1}_{a=1}\phi^a({\bf q})\tau^a$, 
where $\tau^a$'s denote the $N_c \times N_c$ generators of SU($N_c$) 
with $\mathrm{Tr}[\tau^a\tau^b]=2\delta_{ab}$ 
and $\phi^a({\bf q})=\phi^a(-{\bf q})^{*}$. 
Momentum $\vec q$ of the boson is measured relative to $\vec{Q}_{\mathrm{AFM}}$. 
$c_0$ is the bare speed of the boson.
Finally, $g$ denotes the Yukawa coupling between the boson 
and electrons near the hot spots. 
The cubic vertex describes the processes 
where an electron near hot spot $\bar{N}$ 
is scattered to hot spot $N$ 
by absorbing or emitting a boson, 
where $\bar{1}=4,\bar{2}=7,\bar{3}=6$, $\bar{4}=1$,  
$\bar{5}=8$, $\bar{6}=3$, $\bar{7}=2$ and $\bar{8}=5$.

%

\begin{figure}[htbp]
	\centering
	\includegraphics[scale=1]{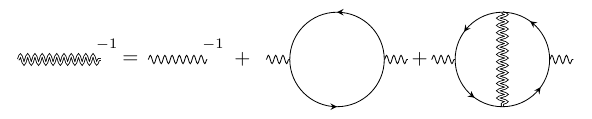}
	\caption{
The truncated Schwinger-Dyson equation that becomes exact in the small $v$ limit with $v\ll c(v)\ll 1$. 
The wiggly and solid lines denote the bare boson and fermion propagators, respectively.
The double wiggly line represents the dressed boson propagator.
\label{fig:SDEq}}
\end{figure}

Under the Gaussian scaling in which 
the kinetic terms are kept invariant,
the fields have dimension
$[\Psi]=-2$, 
$[\Phi] =-5/2$
and the Yukawa coupling has dimension $[g]=1/2$.
The four-fermion coupling, which is not included in the hot-spot theory,
has dimension $-1$.
The usual perturbative expansion 
in which physical observables are expressed
in powers of dimensionless coupling 
$g/E^{1/2}$ at energy scale $E$
is bound to fail at low energies.
Even if $g$ is small compared to the UV cutoff,
non-perturbative effects become important
at low energies.
Fortunately, the theory is solvable 
in the limit that the nesting angle $v$ is small.
If $v$ is non-zero but small at a UV scale,
it dynamically flows toward zero in the low energy limit,
and the solution obtained in the small $v$ limit becomes
 asymptotically exact 
in the low energy limit.
This makes it possible to extract the 
exact critical exponents 
at the infrared fixed point with vanishing $v$\cite{SCHLIEF,LUNTS,SCHLIEF2}\footnote{
It is in principle possible that there exist other fixed points 
with large nesting angle.
However, alternative perturbative analysis based on a dimensional regularization
which is under control for any value of nesting angle 
near three dimensions\cite{SHOUVIK,LUNTS}
indicates that there is no other fixed point besides the $z=1$ fixed point
that continuously evolves to the non-perturbative fixed point found 
in two dimensions\cite{SCHLIEF2}.}.
At the fixed-point,
both $g$ and $v$ vanish with $g^2/v \sim O(1)$,
where the anomalous dimension of the boson is controlled by $g^2/v$.
This interacting two-dimensional fixed point is distinct both from 
the Gaussian fixed point with $g^2/v=0$
and the one-dimensional  Fermi surface with the perfect nesting, $g^2/v=\infty$\cite{2018arXiv181101103C}.

To the leading order in $v$, 
the dynamics of the boson is 
generated by the infinite set of diagrams
included in \fig{fig:SDEq}.
At low energies, 
the solution to the self-consistent Schwinger-Dyson equation is given by\cite{SCHLIEF, SCHLIEF2}
\begin{align}
\label{eq:BosProp}
D({\bf q})^{-1} =
\frac{2g^2}{\pi v}
 \Bigl[
 |q_0| + c(v)(|q_{x}|+|q_y|) 
 \Bigr],
\end{align}
\noindent  where
\begin{align}\label{eq:CofV}
c(v) = \sqrt{\frac{v}{8N_c N_f}\log\left(\frac{1}{v}\right)}
\end{align}
\noindent 
is the speed of the over-damped collective mode\footnote{
When $q_0 \neq 0$,
$|q_{x}|+|q_y|$ in \eq{eq:BosProp} 
should be replaced with a function $f(q_0, \vec q)$ 
that approaches 
$f(q_0, \vec q) \approx |q_{x}|+|q_y|$
for $| \vec q | \gg |q_0|$.
For a small $c(v)$, 
$c(v)(|q_{x}|+|q_y|)$ is important only
for $| \vec q| \gg |q_0|$.
Therefore,
\eq{eq:BosProp} holds
for all $\vec q$ and $q_0$
to the leading order in $c(v)$
in the small $v$ limit.
}.
Interestingly, \eq{eq:CofV} is only a function of $v$ and 
independent of the bare speed of the boson ($c_0$). 
This is because the renormalization generated from gapless
particle-hole excitations is more singular than the local kinetic term 
at low momenta and energies.
The bare kinetic term, which is irrelevant, can be dropped from Eq. (\ref{eq:Action1})  at low energies.
Without the boson kinetic term,
one can rescale the boson field as
$\Phi \rightarrow \sqrt{\frac{\pi v}{2g^2}} \Phi$
so that the dressed boson propagator 
has the canonical normalization,
which gives rise to anomalous dimension $1$ for the boson.
After this rescaling, 
the Yukawa coupling becomes 
$\sqrt{\frac{\pi v}{2}}$.
Physically, this implies that the Yukawa coupling
and the nesting angle become dynamically 
related to each other at low energies.

\begin{figure}[t]
	\centering
	\begin{subfigure}[b]{0.49\linewidth}
		\centering
		\includegraphics[scale=1]{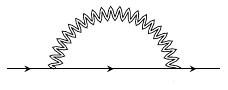}
		\caption{\label{fig:1LFSE}}
	\end{subfigure}
	\begin{subfigure}[b]{0.49\linewidth}
		\centering
		\includegraphics[scale=1]{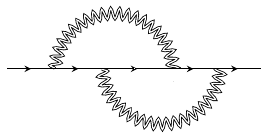}
		\caption{\label{fig:2LSE}}
	\end{subfigure}
		\begin{subfigure}[b]{0.49\linewidth}
		\centering
		\includegraphics[scale=1]{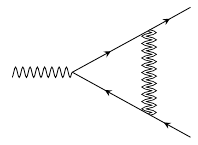}
		\caption{\label{fig:yukawa}}
	\end{subfigure}
	\caption{
The	 leading order 
fermion self-energy and
the vertex correction 
in the small $v$ limit.
\label{fig:SE}
}	
\end{figure}

While the boson is strongly dressed by particle-hole fluctuations,
its feedback to electrons is weak in the small $v$ limit.
The magnitude of a general $L$-loop quantum correction 
with $E$ external legs and $L_{f}$ fermion loops  
computed with the renormalized boson propagator is bounded by
\begin{align}\label{eq:UpperBound}
\mathcal{G}(L,L_{f},E) \leq  v^\frac{E-2}{2} w^{L-L_{f}}
\end{align}
up to logarithms of $v$,
where $w = v/c$\cite{SCHLIEF}.
%
%
%
%
%
According to Eq. (\ref{eq:UpperBound}), only \fig{fig:1LFSE} 
can potentially give the leading order contribution 
to the fermion self-energy that renormalizes $v$.
However, Eq. (\ref{eq:UpperBound}) is only an upper-bound,
and the actual correction to $v$ generated by \fig{fig:1LFSE} is further suppressed in $c$.
This is because \fig{fig:1LFSE} depends on external momentum $\vec k$ 
only through the combination, $c \vec k$ as
the external momentum can be directed to flow only through the boson propagator.
As a consequence, 
\fig{fig:1LFSE} 
becomes of the same order as \fig{fig:2LSE}
which saturates the inequality in Eq. (\ref{eq:UpperBound}),
and we have to include both \fig{fig:1LFSE}  and \fig{fig:2LSE} 
as the leading order correction to $v$.
The quantum corrections in \fig{fig:SE} renormalize the nesting angle at the hot spots,
and give rise to the beta function,
\bqa
\frac{dv}{d\ell}
=
- \frac{2(N^2_c-1)}{\pi^2 N_c N_f} v^2 \log \left( \frac{1}{v}\right),
\eqa
where $\ell$ is the logarithmic length scale.
The solution of the beta function is written as
\begin{align}\label{eq:Eqv}
\mathrm{Ei}\left[\log\left(\frac{1}{v(\ell)}\right)\right] = \mathrm{Ei}\left[\log\left(\frac{1}{v(0)}\right)\right]+\frac{2(N^2_c-1)}{\pi^2 N_c N_f}\ell,
\end{align}
\noindent where 
$\ell$ is the logarithmic length scale,
$v(0)$ is the value of $v$ measured at a UV scale set by $\ell=0$
and
$\mathrm{Ei}(x)$ is the exponential integral function which goes as 
$\mathrm{Ei}(x) = e^{x}(1/x+\mathcal{O}(1/x^2))$ for $x\gg 1$. 
For $v(0)\ll 1$, the solution becomes
\begin{align}\label{eq:ZeroMomentumslope}
v(\ell) = \frac{\pi^2N_cN_f}{2(N^2_c-1)}\frac{1}{(\ell+\ell_0)\log(\ell+\ell_0)},
\end{align}
where 
\begin{align}\label{eq:L0}
\ell_0 = \frac{\pi^2N_cN_f}{2(N^2_c-1)}\frac{1}{v(0)\log(1/v(0))}
\end{align}
is the crossover scale associated with the bare nesting angle, $v(0)$.
For $\ell \ll \ell_0$, the flow of $v$ can be ignored,
while $v$ flows to zero logarithmically for $\ell \gg \ell_0$.

For $\ell_0 \gg 1$, there is a large window of scales,
$1 < \ell \ll \ell_0$ in which the flow of $v$ can be ignored.
Within this window of length scales, 
physical observables obey approximate scaling relations
that are controlled by a set of {\it transient critical exponents},
\begin{align}
z &= 1+\frac{(N^2_c-1)}{2\pi N_c N_f}w,\label{eq:zhot}\\
[\Psi] &=-2 -\frac{(N^2_c-1)}{4\pi N_c N_f}w,\\
[\Phi] &=-2+\frac{1}{2\pi N_c N_f}w\log\left(\frac{1}{w}\right)\label{eq:Phi}.
\end{align}
Here $w= v(\ell)/c(\ell)$
(note that $c$ depends on $\ell$ through $v$).
$z$ is the dynamical critical exponent.
$[\Psi]$ and $[\Phi]$ denote the scaling dimension of the fermion and boson fields in the momentum space. 
If $v$ was independent of $\ell$,
these exponents would control the power-law scaling 
of correlation functions in the low energy limit,
and
the one-parameter family of theories labeled by $v$ 
would form a line of fixed points.
In reality, $v$ is not an exactly marginal parameter,
and it flows to zero logarithmically for $\ell \gg \ell_0$.
Still, these transient exponents control the scaling behaviors
over a finite window of length scales $1 < \ell \ll \ell_0$
in which the flow of $v$ can be ignored.
We call the one-parameter family of theories labeled by $v$
{\it quasi-fixed points } 
as they only act as approximate fixed points
in the intermediate energy scale.
If the RG flow was not cut off by an instability driven by the four-fermion coupling,
the theory would flow to the true fixed point with 
$v = 0$\cite{SCHLIEF,LUNTS, SCHLIEF2}
at length scales bigger than $\ell_0$.
At the true fixed point, 
the critical exponents become
$z=1$ and $[\Psi] = [\Phi] = -2$.\footnote{
It turns out that these exact critical exponents 
can be extracted from the {\it interaction-driven scaling} 
in which the Yukawa coupling and the fermion kinetic term are kept marginal\cite{SHOUVIK2,SCHLIEF}. }
The crossover created by the flow of $v$ manifests itself in physical observables.
For example, 
the spectral function 
of electrons at the hot spots 
($ \mathcal{A}(\omega)$)
and 
the dynamical spin susceptibility 
at the antiferromagnetic ordering vector 
($\chi^{''}(\omega)$)
take  different scaling forms at high and low energies as
\begin{equation}
\begin{aligned}
\mathcal{A}(\omega)
\sim & 
\left\{
\begin{array}{lll}
\omega^{- \left(  1 - \frac{(N^2_c-1)}{2\pi N_c N_f}w(0)    \right) } & 
 \mbox{for} & ~~ 
 \omega_0
 \ll \omega \ll \Lambda , \\
\left[
\omega  ~
e^{ 2 \sqrt{N_c^2-1} \frac{ \left( \log \frac{  1 }{\omega}  \right)^{1/2}}{ \log \log \frac{ 1}{\omega} }}
~\left( \log \frac{1}{\omega} \right)^{1/2} 
~\log \log \frac{1}{\omega} 
\right]^{-1}
  & 
 \mbox{for} &~~ \omega \ll 
 \omega_0
 \end{array}
\right., \\
\chi^{''}(\omega) \sim & 
\left\{ 
\begin{array}{lll}
\omega^{-\left( 1-\frac{1}{\pi N_c N_f}w(0) \log\frac{1}{w(0)} \right) }
 & 
 \mbox{for} & ~~
 \omega_0 \ll \omega \ll \Lambda , \\
\left[ \omega  ~
e^{
\frac{2}{\sqrt{N_c^2-1}}  
\sqrt{ \log \frac{1}{\omega} }
}
\right]^{-1}
& \mbox{for} & ~~ \omega \ll 
 \omega_0
\end{array}
\right..
\end{aligned}
\label{eq:SpectralHotspots}
\end{equation}
Here,
$\Lambda$ is the UV cutoff scale
and
$\omega_0 = \Lambda e^{-z(0) \ell_0}$ is the crossover energy scale,
where 
$z(0) = 1+\frac{(N^2_c-1)}{2\pi N_c N_f}w(0)$ is
the transient dynamical critical exponent 
defined at high energy.
At energies higher than $\omega_0$,
the flow of $v$ can be ignored,
and the spectral function decays in a power-law
controlled by the transient exponent that depends on $v(0)$.
At low energies, the spectral function is controlled by
the true fixed point with logarithmic corrections 
generated from the flow of $v$\cite{SCHLIEF,SCHLIEF2}.

Despite the success of the hot-spot theory
in explaining scaling properties of the critical spin fluctuations
and electrons at the hot spots, there are two important open issues.
First, the four-fermion coupling has not been included in the hot-spot theory.
While the four-fermion couplings  have scaling dimension $-1$ at the fixed point with $v=0$,
it can not be ignored 
if it gives rise to IR singularities,
which, for example, are responsible
for superconducting instabilities.
In priori,
both hot and cold electrons can play important roles 
in superconducting instabilities
because Cooper pairs from hot spots can be scattered to anywhere 
on the Fermi surface (and vice versa).
To capture such superconducting fluctuations,
it is crucial to include all gapless degrees of freedom on the equal footing.
If superconducting instabilities are indeed present,
as is seen ubiquitously in many quantum critical metals,
the flow of $v$ is cut off before the theory
flows to the true fixed point located at $v=0$.
Second, the hot spot theory does not capture the universal low-energy properties
that vary along the Fermi surface.
The antiferromagnetic quantum critical metal hosts both 
Fermi liquid away from the hot spots
and non-Fermi liquid at the hot spots
within one physical system.
Eq. (\ref{eq:SpectralHotspots}) describes the spectral properties of the electrons right at the hot spots.
The spectral function has no quasiparticle peak at the hot spots
because the gapless spin fluctuations remain coupled 
with electrons down to zero energy.
On the other hand, electrons away from the hot spots decouple from the low-energy spin fluctuations
at sufficiently low energies, 
and  they should be described by the Fermi liquid theory in the low-energy limit.
As the hot spots are approached,
the energy scale below which the electrons decouple from spin fluctuations
is lowered,
and 
the quasiparticle gradually loses coherence.
For the same reason, 
all other electronic properties such as 
the nesting angle,
Fermi velocities,
the quasiparticle weight and
Landau parameters
are expected to acquire singular momentum profiles near the hot spots.
In order to understand such momentum dependent critical properties of the system,
we have to go beyond the patch theory
and include all  
gapless modes 
within our effective field theory.

\section{The theory of the full Fermi surface}
\label{Sec:Lukewarm}

\fbox{\begin{minipage}{48em}
{\it
\begin{itemize}
\item
Full low-energy effective theories of metals should include all gapless modes around the Fermi surface, 
and they are characterized by couplings that are functions of momentum along the Fermi surface.
While the four-fermion coupling has scaling dimension $-1$, 
it should be included in the theory because it can give rise to IR singularities.

%

\item
Under the scale transformation that leaves the dynamics 
 of the critical collective mode invariant at low energies, 
momentum is rescaled in all directions.
Consequently, the beta functionals that vanish at fixed points
must include momentum dilatation that  
stretches out the Fermi surface under the RG flow.
\end{itemize}
}
\end{minipage}}
\vspace{0.5cm}

%
%

The full theory that includes all gapless modes
and the four-fermion coupling 
is written as
\begin{align}
\label{eq:LukewarmAction}
\begin{split}
&S
= \sum^{8}_{N=1}\sum^{N_c}_{\sigma =1}\sum^{N_f}_{j=1}\int\dd {\bf k}~\psi^{\dagger}_{N,\sigma,j}({\bf k})\left\{ik_0
+V^{(N)}_{F,k_N}  e_N[\vec k, v^{(N)}_{k_N}]
\right\}\psi_{N,\sigma,j}({\bf k})\\
&+\frac{1}{\sqrt{N}_f}\sum^{8}_{N=1}\sum^{N_c}_{\sigma\sigma'=1}\sum^{N_f}_{j=1}
\int \dd {\bf k}  \dd {\bf q} ~ 
g^{(N)}_{k_N+q_N, k_N}
\psi^\dagger_{N,\sigma',j}({\bf k}+{\bf q})
\Phi_{\sigma'\sigma}({\bf q})
\psi_{\bar{N},\sigma,j}({\bf k})\\
&+\frac{1}{4 \mu}\sum^{8}_{\{N_i=1\}}\sum^{N_c}_{\{\sigma_i=1\}}\sum^{N_f}_{\{j_i=1\}}
\int \prod^{4}_{i=1} 
\dd {\bf k}_i~~
\delta_{1+2,3+4}
\lambda^{\spmqty{N_1 & N_2 \\ N_4 & N_3};\spmqty{\sigma_1 & \sigma_2 \\ \sigma_4 & \sigma_3}}_{\spmqty{k_{1;N_1} & k_{2;N_2} \\ k_{4,N_4} & k_{3;N_3}}}
\psi^\dagger_{N_1,\sigma_1,j_1}({\bf k}_1)\psi^\dagger_{N_2,\sigma_2,j_2}({\bf k}_2)\psi_{N_3,\sigma_3,j_2}({\bf k}_3)\psi_{N_4,\sigma_4,j_1}({\bf k}_4) \\
& + \frac{1}{4 \mu} 
\int  \prod^{4}_{i=1} \dd {\bf k}_i ~~
\delta_{1+2+3+4,0}\Bigl\{
    u_1  \tr{ \Phi({\bf k}_1)  \Phi({\bf k}_2) } \tr{ \Phi({\bf k}_3)  \Phi({\bf k}_4) } 
 + u_2 \tr{ \Phi({\bf k}_1)  \Phi({\bf k}_2)  \Phi({\bf k}_3)  \Phi({\bf k}_4) }
\Bigr\}.
\end{split}
\end{align}
Here, the Fermi surface is still divided into eight disjoint patches each of which includes one hot spot as in the hot spot theory.
However,  unlike in the hot spot theory,
the union of those patches cover the entire Fermi surface
and the size of each patch is order of the Fermi momentum.
Therefore, the coupling  constants are promoted to general {\it coupling functions} that  
depend on momentum along the Fermi surface.
$\mu$ is the floating energy scale at which physical observables are related to the coupling functions.
Couplings that carry non-zero dimensions under the interaction driven scaling are expressed in the unit of $\mu$.
$\delta_{1+2,3+4} \equiv
(2\pi)^3\delta({\bf k}_1+{\bf k}_2-{\bf k}_3-{\bf k}_4)$
and
$\delta_{1+2+3+4,0} \equiv
(2\pi)^3\delta({\bf k}_1+{\bf k}_2+{\bf k}_3+{\bf k}_4)$.
$k_N$ denotes the component of momentum that labels the Fermi surface near hot spot $N$
(see Fig. \ref{fig:FermiSurface} for the choice of coordinate system),
\bqa
k_N &=& 
\left\{ 
\begin{array}{r}
   k_x ~~~~~ \mbox{ for $N=1,4,5,8$} \nn
   k_y ~~~~~~\mbox{for $N=2,3,6,7$} 
 \end{array}
\right..
\eqa
Although the $\hat x$ and $\hat y$ directions are not 
perfectly parallel to the Fermi surface in general,
there is one-to-one correspondence 
between $k_N$ and a point on the Fermi surface 
near hot spot $N$.
We call $k_N$ momentum `along' the Fermi surface near hot spot $N$.
$V^{(N)}_{F,k_N}$ is the momentum dependent Fermi velocity in the direction that is parallel to $\vec Q_{AF}$ near hot spot $N$.
$e_N[\vec k, v^{(N)}_{k_N}]$,
which determines the shape of the Fermi surface  
near each hot spot, 
is written as
\bqa
\begin{array}{llll}
 e_{1}[\vec{k};v^{(1)}_{k_x}] =  v^{(1)}_{k_x} k_x + k_y, &
 e_{2}[\vec{k};v^{(2)}_{k_y}] = - v^{(2)}_{k_y} k_y - k_x, &
 e_{3}[\vec{k};v^{(3)}_{k_y}] = v^{(3)}_{k_y} k_y - k_x, &
 e_{4}[\vec{k};v^{(4)}_{k_x}] =  v^{(4)}_{k_x} k_x - k_y,  \\
 e_{5}[\vec{k};v^{(5)}_{k_x}] =  -v^{(5)}_{k_x} k_x - k_y, &
 e_{6}[\vec{k};v^{(6)}_{k_y}] =  v^{(6)}_{k_y} k_y + k_x, &
 e_{7}[\vec{k};v^{(7)}_{k_y}] = - v^{(7)}_{k_y} k_y + k_x, &
 e_{8}[\vec{k};v^{(8)}_{k_x}] =  -v^{(8)}_{k_x} k_x + k_y,
 \end{array}
 \label{eq:e1vk2}
\eqa
where the nesting angle 
in \eq{eq:Fermiondispersion}
is promoted to functions.
The set of points that satisfy 
$e_N[\vec k, v^{(N)}_{k_N}]=0$
forms the Fermi surface
of a general shape.
%
$u_1$ and $u_2$ represent quartic couplings between the collective modes.
For $N_c=2$, the terms with $u_1$ and $u_2$ are not independent,
and one can set $u_2=0$ without loss of generality.
The momentum-dependent Yukawa coupling is denoted as $g^{(N)}_{k', k}$. 
Unlike $u_i$, which are coupling constants,
$g^{(N)}_{k', k}$ is a function that depends 
on  two momenta along the Fermi surface.
Similarly, $\lambda^{\spmqty{N_1 & N_2 \\ N_4 & N_3};\spmqty{\sigma_1 & \sigma_2 \\ \sigma_4 & \sigma_3}}_{
\spmqty{k_{1} & k_{2} \\ k_{4} & k_{3}}
}
$
($\lambda^{\{N_i\}; \{\sigma_i\}}_{\{k_{i}\}}$ in short)
denotes the short-range four-fermion interactions 
labeled by momenta of electrons on the Fermi surface.

Due to the $C_4$ symmetry, 
$v^{(N)}_{k}$, 
$V^{(N)}_{F,k}$ 
and 
$g^{(N)}_{k',k}$ 
can be represented 
in terms of just three coupling functions
$v_k$, $V_{F,k}$ and $g_{k',k}$ as
\bqa
\left( v^{(N)}_k, ~V^{(N)}_{F,k}, ~g^{(N)}_{k',k}  \right)
= \left\{ \begin{array}{ll}
\left( v_k, ~V_{F,k}, ~g_{k',k}  \right)
, & N=1,3,4,6 \\
\left( v_{-k}, ~V_{F,-k}, ~g_{-k',-k}  \right)
, & N=2,5,7,8 
\end{array}
\right..
\label{eq:vVgN_vVg}
\eqa
Similarly, four-fermion coupling functions 
that are mapped to each other under the $C_4$ symmetry are related.
We set the coefficient of the $ik_0$ term 
in the fermion kinetic term to $1$ 
by choosing the scaling of the fermion fields.
The relative scale between frequency and momentum is chosen to set 
the Fermi velocity along $\vec Q_{AFM}$ to be $1$ 
at the hot spots,
and
the normalization of the bosonic field 
is chosen so that the Yukawa coupling 
at the hot spots is tied to $v_0$,
\begin{align}\label{eq:ZeroMomParam}
V_{\mathrm{F},0} =1,\quad 
g_{0,0} = \sqrt{\frac{\pi v_0}{2}}.
\end{align}

\renewcommand{\arraystretch}{1.5}
\begin{table}
\begin{center}
\begin{minipage}{0.4\textwidth}
	\begin{tabular}{|c|c|c|c|c|}
		\hline
&   allowed channels  \\
		\hline
~Group 1 ~&  
$\sbmqty{1 & 1 \\ 1 & 1}_p$, 
 $\sbmqty{1 & 4 \\ 1 & 4}_p$,  
 $\sbmqty{4 & 4 \\ 1 & 1}$
 \\
 \hline		
~Group 2 ~ &  
 $\sbmqty{1 & 5 \\ 1 & 5}_p$,
  $\sbmqty{1 & 8 \\ 1 & 8}_p$,
  $\sbmqty{4 & 8 \\ 1 & 5}$
 $\sbmqty{1 & 8 \\ 4 & 5}$
 \\
 \hline		
~Group 3~  &  
~~~ $\sbmqty{1 & 2 \\ 1 & 2}_p$,  
 $\sbmqty{1 & 3 \\ 1 & 3}_p$, 
 $\sbmqty{1 & 6 \\ 1 & 6}_p$,
  $\sbmqty{1 & 2 \\ 4 & 7}$,  

  $\sbmqty{1 & 3 \\ 4 & 6}$,
  $\sbmqty{1 & 6 \\ 3 & 4}$,  
$\sbmqty{1 & 7 \\ 4 & 2}$ ~~~
 \\
 		\hline
	~Group 4 ~ &    
 $\sbmqty{1 & 5 \\ 2 & 6}$,  
 $\sbmqty{1 & 5 \\ 3 & 7}$,
 $\sbmqty{1 & 8 \\ 2 & 3}$ 
 \\
 		\hline	
	\end{tabular}
\end{minipage}
\end{center}
\caption{
The primary and secondary channels for the four-fermion coupling,
where the primary channels are the ones generated from the spin fluctuations at the leading order and the secondary channels are the ones generated from the primary channels through the linear mixing.
The ones with subscript $p$  denote primary channels.}
\label{table:PrimarySecondary}
\end{table}

The allowed four-fermion couplings are 
constrained by the crystal momentum conservation 
because the hot spots are located at different points in the momentum space.\footnote{
For example, the coupling function 
$\lambda^{\spmqty{1 & 5 \\ 4 & 8};\spmqty{\sigma_1 & \sigma_2 \\ \sigma_4 & \sigma_3}}_{
\spmqty{k_{1} & k_{2} \\ k_{4} & k_{3}}
}$
with $k_i \approx 0$ is allowed  
because a pair of electrons on hot spots $1$ and $5$ carry the same total momentum 
as the pair made of electrons on hot spots $4$ and $8$. 
On the other hand, the coupling function  
$\lambda^{\spmqty{1 & 1 \\ 2 & 5};\spmqty{\sigma_1 & \sigma_2 \\ \sigma_4 & \sigma_3}}_{
\spmqty{k_{1} & k_{2} \\ k_{4} & k_{3}}
}$
with $k_i \approx 0$
is not allowed because of momentum mismatch.}
Even if the four-fermion coupling is zero at a UV scale,
the Yukawa coupling generates 
four-fermion couplings.
To the leading order in the Yukawa coupling,
the diagrams in \fig{fig:4f0}
generate the four-fermion couplings
in channels $\spmqty{N & M \\ N & M}$ 
for $1 \leq N,M \leq 8$.
Due to the $C_4$ symmetry, we can focus on those channels with 
$N=1$ without loss of generality.
We call those couplings that are generated from the Yukawa coupling at the leading order {\it primary couplings}.
Once the primary couplings are generated,
secondary couplings are further generated through the linear mixing 
shown in \fig{fig:4f1}.
Because a set of coupling functions that forms a closed set of
beta functionals has common primary couplings,
it is convenient to group the four-fermion couplings 
according to their primary couplings.
%
Group 1 includes 
the primary couplings generated by the Yukawa coupling 
in  $(N,M)=(1,1), (1,4)$, 
and the secondary couplings that are further generated from mixing.
The couplings in group 2
represent 
the primary ones generated in channels $(N,M)=(1,5), (1,8)$ 
and the associated secondary couplings. 
Group 3
includes 
the primary couplings for  $(N,M)=(1,2), (1,3), (1,6)$, 
and their secondary couplings.
Those in group 4 
have no primary couplings.
The couplings in group 4 can be present 
only when there exists a bare short-range
four-fermion couplings at a UV scale.
These couplings are listed 
in Table \ref{table:PrimarySecondary}.
To avoid clutter in the table, 
we show only one channel among the ones
that are related to each other 
through the $C_4$ symmetry,
permutation of two incoming particles (or two outgoing particles),
and Hermitian conjugate.
Namely, each entry with the square brackets
in Table \ref{table:PrimarySecondary} represents a group of channels 
obtained by
the $C_4$ transformations,
the Hermitian conjugate and
the permutations between 
two incoming/outgoing particles,
\bqa 
 \sbmqty{N_1 & N_2 \\ N_4 & N_3}
=
\left\{
\left.
\begin{array}{cc}
{\bf R}_{N_1N^{'}_1}\cdots {\bf R}_{N_4N'_4} 
\spmqty{N_1' & N_2' \\ N_4' & N_3'}, &
{\bf R}_{N_1N^{'}_1}\cdots {\bf R}_{N_4N'_4} 
\spmqty{N_4' & N_3' \\ N_1' & N_2'}, \\
{\bf R}_{N_1N^{'}_1}\cdots {\bf R}_{N_4N'_4} 
\spmqty{N_2' & N_1' \\ N_4' & N_3'}, &
{\bf R}_{N_1N^{'}_1}\cdots {\bf R}_{N_4N'_4} 
\spmqty{N_3' & N_4' \\ N_1' & N_2'}
\end{array}
\right|
 {\bf R}_{N_i'N_i}
 \in C_4
\right\},
\eqa
where the repeated hot spot indices are summed over 
and ${\bf R}$ is the $8$-dimensional 
representation of the $C_4$ group that acts as permutations on hot spot indices.
For example, for
the $\pi/2$ rotation we have
${\bf R}^{\pi/2}_{NN'} =\delta_{N,[N'+2]_8}$, 
where $ [x]_{8}= x$ mod 8.
$\spmqty{1 & 7 \\ 1 & 7}$ is related to 
$\spmqty{1 & 3 \\ 1 & 3}$
through the $\pi/2$ rotation,
and it is not separately shown in group 3.
In the table, channels with subscript $p$  
denotes the ones that include primary couplings.

The theory has the  $U(1)$ charge, 
the  $SU(N_f)$ flavour
and
the $SU(N_c)$ spin rotational symmetry.
Due to the spin rotational symmetry, the four-fermion coupling function can be decomposed into two channels as
$
\lambda^{\spmqty{N_1 & N_2 \\ N_4 & N_3};\spmqty{\sigma_1 & \sigma_2 \\ \sigma_4 & \sigma_3}}_{
\spmqty{k_{1} & k_{2} \\ k_{4} & k_{3}}} 
=
\lambda^{\spmqty{N_1 & N_2 \\ N_4 & N_3}}_{D,\spmqty{k_{1} & k_{2} \\ k_{4} & k_{3}}} 
\delta_{\sigma_1 \sigma_4}\delta_{\sigma_2 \sigma_3} 
+
\lambda^{\spmqty{N_1 & N_2 \\ N_4 & N_3}}_{E,\spmqty{k_{1} & k_{2} \\ k_{4} & k_{3}}} 
\delta_{\sigma_1 \sigma_3}\delta_{\sigma_2 \sigma_4}$.
The action in Eq. (\ref{eq:LukewarmAction}) is also invariant under the  particle-hole (PH) transformation,
\begin{align}
\begin{split}\label{eq:PH}
\psi_{N,\sigma,j}(k)\longrightarrow \psi^{\dagger}_{N,\sigma,j}(-k), ~~~
\Phi(q)\longrightarrow - \Phi(q)^{\mathrm{T}} 
\end{split}
\end{align}
if the coupling functions satisfy  
\bqa
&& v_{-k} = v_k, ~~~
V_{\mathrm{F},-k} = V_{\mathrm{F},k},~~~
g_{-{k}',-k} = g_{k,k'}, ~~
\lambda^{\spmqty{N_1 & N_2 \\ N_4 & N_3};\spmqty{\sigma_1 & \sigma_2 \\ \sigma_4 & \sigma_3}}_{
\spmqty{k_{1} & k_{2} \\ k_{4} & k_{3}}} 
=
\lambda^{\spmqty{N_4 & N_3 \\ N_1 & N_2};\spmqty{\sigma_4 & \sigma_3 \\ \sigma_1 & \sigma_2}}_{
\spmqty{-k_{4} & -k_{3} \\ -k_{1} & -k_{2}}}.
\label{eq:PHcouplings}
\eqa
For Fermi surfaces with general shapes, 
the PH symmetry is absent.
In this paper, we are going to focus on the general cases
without the PH symmetry.

\begin{figure*}
	\centering
		\includegraphics[scale=0.75]{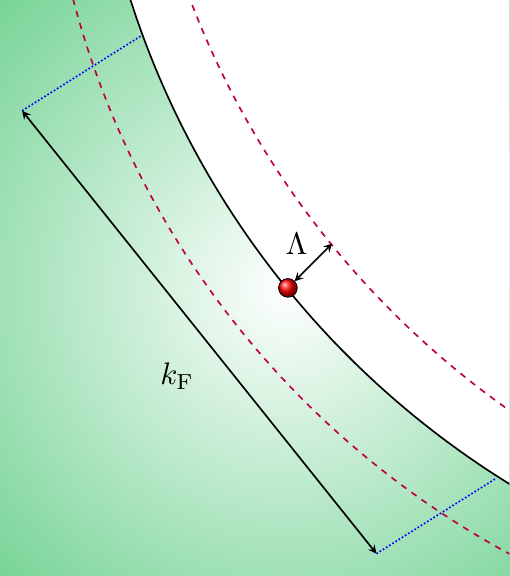}
	\caption{
Energy and momentum cutoffs.
$\Lambda$ is the energy cutoff,
and $k_F$ denotes the size of the patch 
near each hot spot.
The full Fermi surface consists of the union of the eight disjoint patches and the size of each patch is comparable to the Fermi momentum.
	}
\label{fig:cutoffs}
\end{figure*}

The theory has two cutoff scales.
The first is $k_F$ that represents the size of each patch.
The second is the energy cutoff $\Lambda$.
It sets the momentum cutoff of boson
and the momentum of electrons in the direction perpendicular 
to the Fermi surface.
Naturally, $k_F$ is the largest momentum scale.

Under the interaction-driven scaling in which frequency and momentum are 
rescaled by a factor $\s{b}>1$, 
the fields are transformed as
\bqa
\psi({\bf k}) =\s{b}^2 \psi'({\bf k}'), ~~
\Phi({\bf k}) =\s{b}^2 \Phi'({\bf k}')
\eqa
with ${\bf k} = \s{b}^{-1}{\bf k}'$.
Under this transformation, the coupling functions are transformed as
\bqa
v'_{k} = v_{ \s{b}^{-1} k}, ~~~~~
V'_{\mathrm{F}, {k}}=V_{\mathrm{F}, \s{b}^{-1} {k}}, ~~~~~
g'_{k+q, k}=g_{ \s{b}^{-1} (k+q), \s{b}^{-1} k},
~~~~~
\lambda^{'\{N_i\};\{\sigma_i\}}_{\{k_{i}\}} =\s{b}^{-1}\lambda^{\{N_i\}; \{\sigma_i\}}_{\{\s{b}^{-1}k_{i}\}}, ~~~~~
u_n' = \s{b}^{-1} u_n.
\label{eq:IDscaling}
\eqa
Here $k$, $q$ and $k_i$ represent the momentum along the Fermi surface. 
According to \eq{eq:IDscaling},
both fermionic and bosonic quartic interactions are irrelevant by power counting.
Unlike the pure $\phi^4$ theory in $2+1$ dimension, 
the bosonic quartic coupling is irrelevant 
because the boson acquires a large anomalous dimension 
due to the strong coupling 
with the fermions near the hot spots.
Indeed, loop corrections that involve $u_n$ are IR finite.
Therefore, we can drop the bosonic quartic coupling in the low-energy limit.

\begin{figure}
\centering
\includegraphics[scale=0.75]{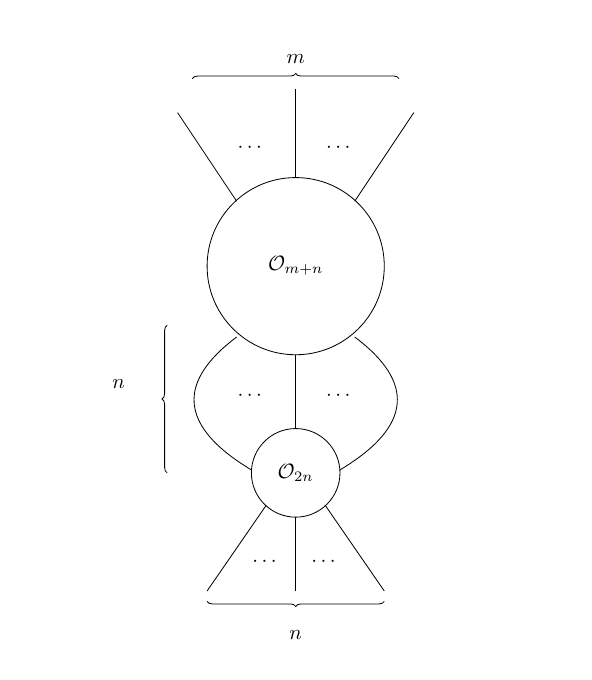}
\caption{
A loop correction in which an $m+n$-fermion operator is dressed with $2n$-fermion operator that results in an anomalous dimension of the $m+n$-fermion operator.
}
\label{fig:lambda_2n}
\end{figure}

On the contrary, one can not drop the four-fermion coupling 
because it can give rise to IR singularities through loop corrections.
The disagreement between what is expected from the power-counting
and the actual degree of IR divergence is caused by the scale
associated with the size of Fermi surface.
To see this,
let us consider a process in which a $2n$-fermion operator fuses with another operator, generating an anomalous dimension for the latter.
To the leading order in the perturbative expansion, this is represented by an $(n-1)$-loop process (see \fig{fig:lambda_2n}).
The scaling dimension 
of $\lambda_{2n}$ is $3-2n$,
and no IR divergence 
is expected for $n \geq 2$ 
from the power-counting.
However, the actual degree of IR divergence can be enhanced by the extended 
phase space for gapless fermions.
If fermions in the loop 
can stay on the Fermi surface within a manifold with dimension $\alpha_{2n}$ 
in the space of internal momenta,
the effective coupling that contributes to the quantum correction becomes 
$\lambda_{2n} k_F^{\alpha_{2n}}$
because the loop momenta 
within the manifold give 
the volume of the phase space, $k_F^{\alpha_{2n}}$.
For one-dimensional Fermi surfaces,
$\alpha_{2n} \leq (n-1)$. 
The upper bound is saturated if every momentum along the Fermi surface contributes a factor of $k_F$, 
which happens when the Fermi surface is straight.
For Fermi surfaces with generic shapes, $\alpha_{2n}$ becomes smaller than $(n-1)$ as the perfect nesting is destroyed by curvature of Fermi surface.
However, $n=2$ is special in that 
the upper bound is saturated 
in the pairing channel 
as far as the time-reversal symmetry is present.
Since a pair of fermions with zero center of mass momentum can be placed anywhere on the Fermi surface in the one-dimensional Fermi surface, $\alpha_4=1$.
This implies that the effective scaling dimension of the four-fermion coupling becomes zero, and the four-fermion coupling  should be included within the low-energy effective theory. 
Indeed, the BCS scattering processes that involve the four-fermion coupling give rise to logarithmic divergences in the low-energy limit.
On the other hand, $\lambda_{2n}$'s with $n >2$ are too irrelevant to create IR singularities even with the help of the enhancement from Fermi surface.
Therefore, we only keep the four-fermion coupling among the couplings that are irrelevant by power-counting.

Given that the action in Eq. (\ref{eq:LukewarmAction}) is local, all coupling functions can be expanded in the Taylor series of momentum along the Fermi surface as
\begin{align}
\label{eq:Parameters}
\begin{split}
v_{k}=\sum^{\infty}_{n=0} \frac{v^{[n]}}{n!}k^n,&~~~~~
V_{\mathrm{F}, k} =\sum^{\infty}_{n=0} \frac{V_F^{[n]}}{n!}k^n,\\
g_{k+q,k}=\sum^{\infty}_{m,n=0}\frac{g^{[m,n]}}{m! n!} k^{ m} q^n,&~~~~~
\lambda^{\{N_i\}; \{\sigma_i\}}_{ \{k_{i}\} } = \sum^{\infty}_{\{l_i\}=0}\frac{\lambda^{[l_1,..,l_4]; \{N_i\};\{\sigma_i\}}}{l_1!l_2!l_3!l_4!}k^{l_1}_{1}k^{l_2}_{2}k^{l_3}_{3}k^{l_4}_{4}.
\end{split}
\end{align}
The interaction driven scaling fixes the scaling dimensions of the coefficients in Eq. (\ref{eq:Parameters}) to
$[v^{[n]}]=[V_F^{[n]}]=-n$,
$[g^{[m, n]}]=-(m+n)$ and $\left[\lambda^{[l_1,..,l_4]; \{N_i\};\{\sigma_i\}}\right]=-(1+l_1+l_2+l_3+l_4)$. 
Formally, allowing the general momentum dependence in the coupling functions
amounts to introducing an infinite tower of coupling constants. 
Although the high-order coupling constants are highly `irrelevant' in terms of their scaling dimensions,
they are necessary to characterize the whole Fermi surface.
This rather unusual role of irrelevant couplings is due to the fact that 
the momentum along the Fermi surface 
not only acts as a scale 
but also as a label for the gapless electronic degrees of freedom. 
In particular, 
the momentum dependence in coupling functions is important
in understanding superconductivity that arises through
an interplay between hot and cold electrons.

In \eq{eq:Parameters}, 
$g_{k+q,k}$ denotes the strength
of the interaction 
in which an electron is scattered from momentum $k$ to $k+q$ 
near the Fermi surface by absorbing or emitting boson with momentum $q$ in magnitude. 
In relativistic quantum field theories,
scatterings that involve high-energy particles
are not important at low energies.
For this reason, 
one may just keep the leading order term in the expansion in $q$.
In the presence of Fermi surface, however,
the processes in which 
electrons are scattered 
by high-energy bosons 
within the Fermi surface
can give rise to
IR singularities.
The same mechanism is responsible for the logarithmic singularity associated with
the BCS instability
caused by short-range interactions
mediated by a massive boson
in Fermi liquids. 
For this reason, we include the Yukawa coupling 
with general $k$ and $q$ within the theory.
The fact that the coupling associated with
high-energy bosons should be included within the theory raises important questions on 
what constitute low-energy observables and what information the low-energy effective theory should include to be predictive.
Later, we will see that the predictions of the theory do not depend on UV physics 
if we choose the right observables.
For now, we proceed with the general Yukawa coupling function
as an intermediate step toward identifying
the universal observables 
that do not depend on UV physics.

\section{The field-theoretic functional renormalization group formalism}
\label{sec:FRG}

\fbox{\begin{minipage}{48em}
{\it
\begin{itemize}
\item
A renormalizable theory consists of the minimal set of coupling functions
in terms of which all low-energy observables 
can be expressed within errors
that vanish in powers of energy.
\item
A renormalizable theory 
must include all operators 
that give rise to IR divergences 
even if they are irrelevant in terms of their scaling dimensions.
In the presence of Fermi surfaces, 
 couplings with negative scaling dimensions can give rise to IR singularities 
 with the help of the  Fermi momentum.
The RG subtraction scheme
should be generalized so that those additional IR divergences 
are taken into account.
\item
Due to the Fermi momentum that runs under the RG flow,
general interacting theories of Fermi surfaces do not have the true sense of scale invariance even at fixed points.
Consequently,
the scaling relation alone does not fix the energy and momentum dependence of physical observables.
\item
The renormalization group flow of coupling functions 
are driven by IR singularities of quantum corrections
that depend on momentum along the Fermi surface.
In general, quantum corrections are functionals of coupling functions.
However, 
for quantum corrections
that only involve fermions in non-nested   parts of the Fermi surface,  
the locality allows one to extract singular parts
of quantum corrections 
in terms of coupling functions evaluated at external momenta.
\end{itemize}
}
\end{minipage}}
\vspace{0.5cm}

In this section we discuss the fundamentals
of the field theoretic functional RG scheme
that is used throughout the paper. 
For other functional RG approaches, 
see Refs. \cite{
	POLCHINSKIFRG,
	WETTERICH,
	MORRIS,
	REUTER,
	ROSA,
	HOFLING,
	HONERKAMP,
	GIES,
	GIES2,
	BRAUN,
	METZNER,
	SCHERER, 
	JANSSEN2,
	MESTERHAZY,
	PhysRevB.87.045104,
	EBERLEINMETZNER,
	PLATT,
	WANGEBERLEIN,
	JANSSEN,
	MAIEREBERLEIN,
	EBERLEIN2,
	EBERLEIN3,
	JAKUBCZYK,
	MAIER,
	TORRES2,
	TORRES}.

\subsection{Renormalizability}\label{sec:Renormalizability}

Explaining or predicting experiments 
is usually done in the following steps.
One first identifies relevant degrees of freedom and symmetry
to constructs a model that generally 
includes a set of free parameters.
After physical observables are computed from the model,
the parameters of the model are fixed from existing experimental data.
Once the parameters are fixed, one can make predictions 
for new observables.
Since there is freedom in choosing 
which observables are used to fix the parameters 
and which ones are used as predictions,
what a theory captures is the relation 
among physical observables.
A theory has stronger  predictive power 
if more observables 
 are fixed by fewer other observables.

In field theories, 
one aims to find relations among
low-energy observables 
measured at energy scales
much smaller than microscopic energy scales.
While individual low-energy observables  
can sensitively depend on microscopic details,
field theories can capture their relations  
that are independent of the microscopic details.
To achieve this goal, 
it is most convenient to use 
renormalizable field theories.
A renormalizable theory contains
a minimal set of couplings
in terms of which all other low-energy observables of the theory
can be expressed  
with errors that vanish in powers of $\mu/\Lambda$,
where $\mu$ is the energy scale at which observables
are probed and $\Lambda$ is a UV cutoff.
Although microscopic systems in general 
include more parameters,
one can use renormalizable theories
to extract universal relations 
among low-energy observables.
Two theories which differ 
by irrelevant couplings
give rise to the same relations among 
low-energy observables
within the power-law accuracy.

In the presence of  Fermi surface, a low-energy theory includes
momentum-dependent coupling functions.
Once expanded around a point (say a hot spot) on the Fermi surface, 
the momentum-dependent coupling functions 
in Eq. (\ref{eq:LukewarmAction}) can be viewed
as an infinite set of coupling constants.
Under a transformation that rescales the momentum relative to the hot spots,
the couplings associated with positive powers of momentum 
along the Fermi surface are formally irrelevant.
However, this does not imply that those higher order terms in momentum 
are unimportant for all low-energy observables.
Even if the higher order terms in \eq{eq:Parameters}
may not be needed for understanding the low-energy behaviours of electrons at the hot spots,
they are still important for electrons on the Fermi surface far away from the hot spots.
The higher order terms can be important 
even for electrons at the hot spots
if large momentum-scatterings are not suppressed at low energies.
Therefore, it is necessary to keep the full momentum-dependent 
coupling functions in order to characterize the low-energy physics
of the entire system.
Because the Fermi surface supports an infinite number of gapless modes,
the amount of universal low-energy data is in general infinite\footnote{
The low-energy data, while being infinite,
is still much smaller than the full information a microscopic theory can carry.
This is because the low-energy effective theory only keeps track of
the momentum dependence of the coupling functions along the Fermi surface.}.

The goal of low-energy effective theories for Fermi surface is to identify
the minimal set of functions' worth of low-energy data,
in terms of which all low-energy observables can be determined.
To achieve this, 
we use \eq{eq:LukewarmAction} 
to compute a set of physical observables 
as functionals of the coupling functions.
Local counter terms are added to the action 
so that those physical observables become 
what we set them to be 
as functions of momentum along the Fermi surface at an energy scale.
Once the bare theory that  includes the counter terms is fixed, 
it gives rise to the functional 
Callan-Symanzik equation 
that describes how the momentum dependent physical observables run 
as functions of energy.
From the flow equations,
we identify the set of low-energy observables
in terms of which
all other low-energy observables
can be expressed 
without resorting to unknown 
high-energy physics.
In particular, the RG flow of the coupling functions identified as low-energy observables should be captured solely in terms of those coupling functions  themselves.
Given that we don't know in priori what constitute universal low-energy observables, we first include all coupling functions
$\{
v_{k},
V_{\mathrm{F}, k},
g_{k+q,k},
\lambda^{\{N_i\}; \{\sigma_i\}}_{ \{k_{i}\} } \}$
that can be potentially needed in characterizing all low-energy observables.
From this, we isolate
the minimal subset whose RG flow can be extracted solely from those couplings in the minimal subset without resorting to any unknown UV physics.
The two-point function of fermion on the Fermi surface,
which is related to the momentum dependent nesting angle and Fermi velocity,
are low-energy observables.
While the forward cubic vertex function related to  $g_{k,k}$ is in the minimal set of low-energy observables,
$g_{k+q,k}$ with a non-zero $q$ is not  
because  the off-diagonal Yukawa coupling function 
with large momentum transfer
encodes the dynamics of 
the high-energy boson.
Remarkably, the 
one-particle irreducible  (1PI) 
four-fermion vertex function strictly 
defined on the Fermi surface 
does not belong to the minimal set of low-energy observables either.
This is because the flow of
$\lambda^{\{N_i\}; \{\sigma_i\}}_{ \{k_{i}\} } $
for general $k_i$'s
can not be determined 
within the low-energy effective field theory :
quartic fermion operators  defined on different points on the Fermi surface
can mix with each other at low energies  by exchanging high-energy bosons.
Nonetheless, the RG flow of the net two-body interaction that combines 
the 1PI four-fermion vertex function and 
the tree-diagram associated with two 1PI three-point vertex functions 
connected by the renormalized boson propagator 
can be understood within the low-energy effective theory (see \fig{fig:4pt3pts}).
This will be shown explicitly in  Sec. 
\ref{sec:four_fermion_fp2}.

In the present theory, there are two cutoff scales, $k_F$ and $\Lambda$.
Naively one might expect that the relations between 
observables at one scale, $\mu_1$,
and observables at another scale, $\mu_2$, 
should be independent of  
all of those short distance scales for $\mu_1, \mu_2 \ll \Lambda, k_F$.
This amounts to requiring that  divergences in any of those large momentum scales 
can be removed by adding local counter terms. 
However, it is in general impossible to remove \kF dependences in all low-energy observables\footnote{
For example, thermodynamic quantities such as the specific heat
are proportional to \kF in the low temperature limit.
\kF also determines the phase space of a pair of electrons with zero total momentum and energy,
and controls the mixing between quartic fermion operators in the pairing channel.}.
This is because  \kF
is a part of the low-energy data that
reflects the `number' of gapless modes in the system.
As a result, the beta functionals for the \ffc may explicitly depend on \kF,
and \kF needs to be included in characterizing  low-energy physics.
On the other hand,
$\Lambda$ represents the energy cutoffs,
and it can be removed 
from low-energy observables
by adding local counter terms.
In field theories of Fermi surface,
the validity of the low-energy effective field theory boils down to the question of 
whether one can remove $\Lambda$
but not necessarily \kF
in the relations among low-energy observables.

\subsection{Extended minimal subtraction scheme}
\label{sec:minimal}

To understand the low-energy physics of the theory, 
the quantum effective action is computed order by order in $v$
from the classical action in \eq{eq:LukewarmAction}.  
Since the Yukawa coupling is marginal under the interaction-driven scaling,
quantum corrections to the leading-order solution of the non-perturbative Schwinger-Dyson equation 
are logarithmically divergent in general.
While the four-fermion coupling is irrelevant under the interaction driven scaling,
it also gives rise to IR singularities
as will be shown later.
To capture how those singular corrections modify physical observables in the low-energy limit,
we express the vertex functions 
in terms of the coupling functions,
and keep track of the RG flow 
of the coupling functions
as the energy scale is lowered.
Since gapless electrons can be anywhere 
on the extended Fermi surface,
the low-energy vertex functions 
and the couplings are functions 
of momentum along the Fermi surface
and the energy scale.

The relation between the vertex functions and the coupling functions
is set by a set of renormalization conditions,
which is enforced by adding counter terms to  \eq{eq:LukewarmAction}.
The renormalization conditions are written as
\bqa
	Re \varGamma^{(2,0)}_{1}({\bf k})\bigg|_{{\bf k}=(\mu, k_x, -v_{k_x} k_x)} &=& 
0,
\label{eq:RG1}\\
 \frac{\partial }{\partial k_y} Re \varGamma^{(2,0)}_{1}({\bf k})\bigg|_{{\bf k}=(\mu, k_x, -v_{k_x} k_x)} &=& V_{\mathrm{F}, k_{x}}+\mathscr{F}_{1, k_{x}},\label{eq:RG2}\\
 -i \frac{\partial }{\partial k_0}Im \varGamma^{(2,0)}_{1}({\bf k})\bigg|_{{\bf k}=(\mu, k_x, -v_{k_x} k_x)} &=& 1+\mathscr{F}_{2, k_{x}},\label{eq:RG3}\\
 \varGamma^{(2,1)}_{1}({\bf k}',{\bf k})\bigg|_{
 \scriptsize
   \begin{array}{l}
  {\bf k}' =(2 \mu, k'_{x},-v_{k'_{x}} k'_{x}) \\
  {\bf k}=( \mu, k_{x},v_{k_{x}} k_{x})  
 \end{array}
 } 
 &=& \frac{g_{k'_{x},k_{x}}}{\sqrt{N_f}}+\mathscr{F}_{3,(k'_{x},k_{x})},\label{eq:RG4}\\
 \varGamma^{(4,0); \{N_i\};\{\sigma_i\}}
 (\{{\bf k}_i\})\bigg|_{{\bf k}_i={\bf k}^{*}_i} &=& 
 \frac{1}{4\mu} 
 \left[
 {\lambda}^{\spmqty{N_1 & N_2 \\ N_4 & N_3};\spmqty{\sigma_1 & \sigma_2 \\ \sigma_4 & \sigma_3}}_{\spmqty{k_1 & k_2 \\ k_4 & k_3 }}
+  {\mathscr F}^{\spmqty{N_1 & N_2 \\ N_4 & N_3};\spmqty{\sigma_1 & \sigma_2 \\ \sigma_4 & \sigma_3}}_{4,\spmqty{k_1 & k_2 \\ k_4 & k_3 }}
 \right].
 \label{eq:RG5}
\eqa
Here, $\varGamma^{(2,0)}_{N}({\bf k})$  is the two-point function of electrons near hot spot $N$.
$\varGamma^{(2,1)}_{N}({\bf k}',{\bf k})$ is the electron-boson vertex function 
that describes scattering of an electron from  three-momentum ${\bf k}$ near hot spot $\bar N$ 
to ${\bf k}'$ near hot spot $N$.
$\varGamma^{(4,0); \{N_i\};\{\sigma_i\}}(\{{\bf k}_i\})$ is the electron four-point function,
where the $i$-th external electron is near hot spot $N_i$, spin $\sigma_i$ and three-momentum ${\bf k}_i$.
%
%
Eq.
(\ref{eq:RG1}) is the defining equation for $v_k$
that specifies the renormalized Fermi surface.
Near hot spot $1$,
the renormalized Fermi surface at scale $\mu$
is given by the set of
$(k_x, - v_{k_x} k_x)$
at which the real part of the two-point function vanishes.
Eq. (\ref{eq:RG2}) defines the momentum dependent Fermi velocity, $V_{F,k}$.
$\mathscr{F}_{1,k_{x}}$ corresponds to a scheme dependent function that is regular in the small $\mu$ limit.
We choose the relative scale 
between frequency and spatial momentum 
to set
\bqa
V_{F,0}=1
\label{eq:VF0equalto1}
\eqa
at the hot spots.
To impose \eq{eq:VF0equalto1} 
at all  $\mu$,
the relative scale between frequency and momentum should be chosen in an energy dependent way, which gives rise to
a dynamical critical exponent different from $1$ in general.
The Fermi velocity away from the hot spots is in general different from that of the hot spots.
\eq{eq:RG3} determines the momentum dependent scaling of the fermion field :
it fixes the frequency dependent kinetic term 
of the fermion to be of the canonical form at all energy scales
up to a regular  correction, $\mathscr{F}_{2,k_{x}}$.
Finally,
 \eq{eq:RG4}
and
\eq{eq:RG5}
define the momentum dependent Yukawa coupling function
and the four-fermion coupling functions, respectively.
The renormalization condition 
for the four-point function is imposed at
\bqa
{\bf k}^{*}_{1}=(3\mu,\vec{k}^{*}_{1}), ~~
{\bf k}^{*}_2=(-\mu,\vec{k}^{*}_{2}), ~~
{\bf k}^{*}_3=(\mu,\vec{k}^{*}_{3}), ~~
{\bf k}^{*}_4=(\mu,\vec{k}^{*}_{4}),
\label{eq:4fmomenta}
\eqa
where $\vec{k}^{*}_{i}$'s are the spatial momenta 
that are near the Fermi surface\footnote{ To be precise, the energy associated with each momentum should be at most order of $\mu$. } 
and satisfy the momentum conservation.
For generic shapes of Fermi surface,  which is the main focus of our paper,  we can focus on the forward scattering and the pairing channels. 
Here, the external frequencies are chosen so that the energy that flows through the vertex function is $2 \mu$ in magnitude in all $s$, $t$, $u$ channels.

Ideally, one would want to choose the counter terms  
so that they completely cancel the quantum corrections.
In this total subtraction scheme, 
 $\mathscr{F}_i =0$,
and the coupling functions at scale $\mu$ coincides 
with the vertex functions measured at that energy.
However, the total subtraction scheme is rather impractical 
because it requires computing the full quantum corrections 
including finite parts.
In this paper, we use a minimal subtraction scheme,
where counter terms remove only divergent contributions to 
the quantum effective action
in the  small $\mu/\Lambda$ limit.
While the vertex functions do not exactly match the coupling functions
in the minimal subtraction scheme,
one can in principle infer one from the other as
they are related to each other through 
 relations that are regular 
in the  small $\mu$ limit.
For the purpose of extracting scaling behaviours, it suffices to know the existence of such finite functions but not their explicit forms.
As far as IR singularities in all physical observables are encoded in the coupling functions, any instability  of the system can be inferred from the RG flow of the coupling functions.

The minimal subtraction scheme is straightforward to implement
for dimensionless couplings.
Counter terms are added to remove singular corrections to 
the two and three-point functions as 
in Eqs. (\ref{eq:RG1})-(\ref{eq:RG4}).
In the minimal subtraction scheme,
$\mathscr{F}_{1,2,3}$ are generally non-zero,
but they stay finite in the small $\mu$ limit.
This guarantees that 
the dimensionless physical observables are
related to $v_{k}, V_{F,k}, g_{k',k}$
through non-singular relations. 
%
%

Implementing the minimal subtraction scheme for the four-fermion couplings is more subtle.
The four-fermion coupling is irrelevant, and 
$\varGamma^{(4,0)}$ has engineering scaling dimension $-1$.
A renormalization condition should be imposed on dimensionless quantities
constructed out of $\varGamma^{(4,0)}$ and a scale. 
$\mu \varGamma^{(4,0)}$ is one such dimensionless quantity.
Naively, one may only require that
$\mu \varGamma^{(4,0)}$ coincides with the 
dimensionless coupling function $\lambda$ 
up to any non-singular correction.
If this was the case,
${\mathscr F}_4$ in  \eq{eq:RG5} could be an arbitrary finite function of momenta.
However, this subtraction scheme is `too minimal' in that
$\lambda$ does not capture  
 all IR singularities of physical observables.
This is because
dimensionless observables
constructed out of integrations of
$\varGamma^{(4,0)}$ over momenta can exhibit singularities even if
$\mu \varGamma^{(4,0)}$ is finite in the small $\mu$ limit.
For example, the strength of the pairing interaction
at energy scale $\mu$  is measured by eigenvalue $E_\mu$ defined through
\bqa
\sum_{N_4} \sum_{\sigma_4, \sigma_3} \int dp~
\varGamma^{(4,0) \spmqty{N_1 & [N_1+4]_8 \\ N_4 & [N_4+4]_8 };\spmqty{\sigma_1 & \sigma_2 \\ \sigma_4 & \sigma_3}}_{\spmqty{{\bf q}+{\bf k} & -{\bf k} \\ {\bf p} & {\bf q}-{\bf p} }}
f^{N_4;(\sigma_4,\sigma_3)}_p
= 
E_\mu
f^{N_1;(\sigma_1,\sigma_2)}_k.
\label{eq:eigenGamma}
\eqa    
Here  ${\bf q}=(2 \mu,0,0)$ is the three momentum of a Cooper pair.
The frequencies of ${\bf k}$ and ${\bf p}$ are  set to be $\mu$.
$p$ and $k$ label the components of $\vec p$ and $\vec k$ 
along the Fermi surface.
The other components of the spatial momentum are chosen
so that $\vec p$ and $\vec k$ are on the Fermi surface.
$f^{N_1;(\sigma_1,\sigma_2)}_k$
is an eigen-wavefunction of the Cooper pair. 
Even if 
${\mathscr F}^{\spmqty{N_1 & N_2 \\ N_4 & N_3};\spmqty{\sigma_1 & \sigma_2 \\ \sigma_4 & \sigma_3}}_{4,\spmqty{k_1 & k_2 \\ k_4 & k_3 }}$ 
is finite at every $k_i$,
its contribution to the eigenvalue 
may diverge in the small $\mu$ limit
if it has an extended support in the momentum space.
For example, 
${\mathscr F}^{ \spmqty{N_1 & [N_1+4]_8 \\ N_4 & [N_4+4]_8 };\spmqty{\sigma_1 & \sigma_2 \\ \sigma_4 & \sigma_3}}_{4,\spmqty{k & -k \\ p & -p }} \sim \frac{\mu}{\sqrt{(k-p)^2 + \mu^2} }$ gives rise to a divergent correction to the eigenvalue $E_\mu$ although its element is finite in the small $\mu$ limit.
Without subtracting such divergent contribution,
the coupling function does not capture the IR singularity associated with the divergent pairing interaction.
To remove any singular discrepancy
between the eigenvalues 
of the vertex function 
and the coupling function,
we need to impose a more stringent condition on the finite part :
we require that not only
${\mathscr F}^{\spmqty{N_1 & N_2 \\ N_4 & N_3};\spmqty{\sigma_1 & \sigma_2 \\ \sigma_4 & \sigma_3}}_{4, \spmqty{k_1 & k_2 \\ k_4 & k_3 }}$ is finite at all momenta but also
\bqa
\lim_{\mu \rightarrow 0} 
\int_C 
 \frac{dk}{\mu}
~
\left| 
{\mathscr F}^{\spmqty{N_1 & N_2 \\ N_4 & N_3};\spmqty{\sigma_1 & \sigma_2 \\ \sigma_4 & \sigma_3}}_{4, \spmqty{k_1 & k_2 \\ k_4 & k_3 }}
\right| 
= \mbox{finite}
\label{eq:stringentF5}
\eqa
for any one-dimensional manifold $C$ in the space of $k_i$'s.
With this condition,  
the eigenvalue of 
${\mathscr F}^{\spmqty{N_1 & N_2 \\ N_4 & N_3};\spmqty{\sigma_1 & \sigma_2 \\ \sigma_4 & \sigma_3}}_{4, \spmqty{k_1 & k_2 \\ k_4 & k_3 }}
$ is non-divergent
in the small $\mu$ limit\footnote{
To see this, we consider matrix,
${\cal M}_{ij} = 
\frac{\Delta k}{\mu} 
{\mathscr F}^{ \spmqty{N_i & [N_i+4]_8 \\ N_j & [N_j+4]_8 };\spmqty{\sigma_i & \sigma'_i \\ \sigma_j & \sigma'_j}}_{4,\spmqty{k_i & -k_i \\ k_j & -k_j }}
$,
where the matrix indices $i,j$ label hot spot index, spin and discretized momentum of a Cooper pair, 
and $\Delta k$ denotes the mesh size of the discrete momentum.
\eq{eq:stringentF5} implies that sum of the absolute values of elements in any row of ${\cal M}$ is finite
in the small $\mu$ limit.
Gershgorin's circle theorem implies that all eigenvalues
are finite as well.}.
In this extended minimal subtraction scheme, 
IR singularities of the four-point function
are fully captured by $\lambda$.


%
%

\subsection{Scale invariance and the lack of thereof}

The local counter term action 
for \eq{eq:LukewarmAction} 
is written as
	\begin{align}\label{eq:CTAction}
	&\hspace{-1.5cm}S_{\mathrm{C.T}} =
	\sum^{8}_{N=1} \sum^{N_{c}}_{\sigma=1}\sum^{N_{f}}_{j=1}\int\dd {\bf k} ~
	\psi^{\dagger}_{N,\sigma, j}({\bf k})\left\{
i A_1^{(N)} (k_N) k_0 
+ A_3^{(N)} (k_N)V^{(N)}_{\mathrm{F},k_N}
e_{N}\left[\vec{k};\frac{A_2^{(N)} (k_N)}{A_3^{(N)} (k_N)}v^{(N)}_{k_N}\right]\right\}\psi_{N,\sigma, j}({\bf k})\notag\\
	&+\frac{1}{\sqrt{N_{f}}}\sum^{8}_{N=1}\sum^{N_{c}}_{\sigma\sigma'=1}\sum^{N_{f}}_{j=1}\int\dd {\bf k}'\int \dd {\bf k} ~ A^{(N)}_4 (k'_N,k_{\bar N}) g^{(N)}_{k'_N,k_{\bar N}}
	 \psi^\dagger_{N,\sigma' ,j}({\bf k}')
	 \Phi_{\sigma' \sigma}({\bf k}'-{\bf k})
	 \psi_{\bar{N},\sigma, j}({\bf k})\notag \\
		\begin{split}
	&+\frac{1}{4\mu}\sum^{8}_{\{N_i=1\}}\sum^{N_c}_{\{\sigma_i=1\}}\sum^{N_f}_{\{j_i=1\}}\left[\prod^{4}_{i=1}\int\dd {\bf k}_i\right]
	\left\{A^{\{N_i\};\{\sigma_i\}}(\{k_{i, N_i}\})
	\lambda^{\{N_i\}; \{\sigma_i\}}_{ \{k_{i,N_i}\} } 
	\delta_{1+2,3+4}
	\right.\\
	&\left. \times 
	\psi^\dagger_{N_1,\sigma_1,j_1}({\bf k}_1)
	\psi^\dagger_{N_2,\sigma_2,j_2}({\bf k}_2)
	\psi_{N_3,\sigma_3,j_2}({\bf k}_3)
	\psi_{N_4,\sigma_4,j_1}({\bf k}_4)
\right\} \\
&
	+ 
M_{CT}  
	\sum^{8}_{N=1} \sum^{N_{c}}_{\sigma=1}\sum^{N_{f}}_{j=1} ~
\int\dd {\bf k} ~ 
\psi^{\dagger}_{N,\sigma, j}({\bf k})
\psi_{N,\sigma, j}({\bf k})
	+ \frac{m_{CT}}{4} \int\dd {\bf k} ~ 
	\tr{
	\Phi({\bf k})
	\Phi(-{\bf k})
	}.
	\end{split}
	\end{align}
  Here, $A^{(N)}_i(k_N)$ with $i=1,2,3$, 
$A^{(N)}_4 (k',k)$ and 
$A^{\ \{N_i\}; \{\sigma_i\}}(\{k_{i}\})$
 are momentum-dependent local counter terms 
 which are functionals of the coupling functions.
They are determined from the quantum corrections 
so that the
renormalization conditions in 
Eqs.
\eqref{eq:RG1}
-
\eqref{eq:RG5}
are satisfied.
Due to the $C_4$ symmetry, 
$A^{(N)}_i(k)$ with $i=1,2,3$
and  $A^{(N)}_4(k',k)$ 
can be represented in terms of 
four counter term functions
$A_i(k)$ with $i=1,2,3$
and  $A_4(k',k)$ as
\bqa
\left( A_i^{(N)}(k), ~A_4^{(N)}(k',k)  \right)
= \left\{ \begin{array}{ll}
\left( A_i(k),  ~A_4(k',k)  \right)
, & N=1,3,4,6 \\
\left( A_i(-k),  ~A_4(-k',-k)  \right)
, & N=2,5,7,8 
\end{array}
\right.. 
\label{eq:ANi_symmetry}
\eqa
$M_{CT}$ is the counter term
that is needed to 
make sure that the hot spots 
are located at ${\bf k}=0$
in the fully renormalized Fermi surface.
$m_{CT}$ is the mass counter term for the boson,
which is needed to keep the system 
at the quantum critical point.

Adding Eqs. (\ref{eq:CTAction}) and (\ref{eq:LukewarmAction}) yields the renormalized action 
\begin{align}\label{eq:RenAction}
&	S_{\mathrm{Ren}}
=
\sum^{8}_{N=1} \sum^{N_{c}}_{\sigma=1}\sum^{N_{f}}_{j=1}\int\dd {\bf k}^{\mathsf{B}}~ 
\psi^{\mathsf{B} \dagger}_{N,\sigma, j}({\bf k}^{\mathsf{B}})
\left\{
i k^{\mathsf{B}}_0 + 
V^{\mathsf{B}(N)}_{\mathrm{F}, k^{\mathsf{B}}_{N}}
e_{N}\left[\vec{k}^{\mathsf{B}};v^{\mathsf{B}(N)}_{k^{\mathsf{B}}_{N}}\right]
\right\}
\psi^\mathsf{B}_{N,\sigma, j}({\bf k}^{\mathsf{B}})\notag\\
	&+\frac{1}{\sqrt{N_{f}}}\sum^{8}_{N=1}\sum^{N_{c}}_{\sigma\sigma'=1}\sum^{N_{f}}_{j=1}
	\int\dd {\bf k}^{'\mathsf{B}}
	\int \dd {\bf k}^{\mathsf{B}}~ 
	g^{\mathsf{B}(N)}_{ {k'}^{\mathsf{B}}_{N},k^{\mathsf{B}}_{\bar N}}
	\psi^{\mathsf{B} \dagger}_{N,\sigma' ,j}({\bf k}^{' \mathsf{B}} )
	\Phi^{\mathsf{B}}_{\sigma' \sigma}({\bf k}^{'\mathsf{B}}- {\bf k}^{\mathsf{B}})
	\psi^{\mathsf{B}}_{\bar{N},\sigma, j}({\bf k}^{\mathsf{B}})\notag\\
	& +
	\frac{1}{4}\sum^{8}_{\{N_i=1\}}\sum^{N_c}_{\{\sigma_i=1\}}\sum^{N_f}_{\{j_i=1\}}
	\left[\prod^{4}_{i=1}\int\dd {\bf k}^{\s{B}}_i\right]
	\left\{
	\lambda^{B \{ N_i \} ; \{ \sigma_i \} }_{ \{ k^\s{B}_{i;N_i} \} }
	\delta_{1^B+2^B,3^B+4^B}
	 \psi^{\s{B} \dagger}_{N_1,\sigma_1,j_1}({\bf k}^\s{B}_1)
	 \psi^{\s{B} \dagger}_{N_2,\sigma_2,j_2}({\bf k}^\s{B}_2)
	 \psi^{\s{B}}_{N_3,\sigma_3,j_2}({\bf k}^\s{B}_3)
	 \psi^{\s{B}}_{N_4,\sigma_4,j_1}({\bf k}^\s{B}_4)
\right\}  \notag \\
&+ 
M^{B}  
\sum^{8}_{N=1} \sum^{N_{c}}_{\sigma=1}\sum^{N_{f}}_{j=1} ~
\int\dd {\bf k}^{\s{B}} ~ 
\psi^{  \s{B} \dagger}_{N,\sigma, j}({\bf k}^B)
\psi^{\s{B}}_{N,\sigma, j}({\bf k}^{\s{B}})
	+ \frac{m^{\s{B}}}{4} \int\dd {\bf k}^{\s{B}} ~ 
	\tr{
	\Phi^{\s{\s{B}}}({\bf k}^{\s{B}})
	\Phi^{\s{\s{B}}}(-{\bf k}^{\s{B}})
	},
	\end{align}
where	
	$\delta_{1^B+2^B,3^B+4^B}
\equiv	 (2\pi)^3\delta({\bf k}^\s{B}_1+{\bf k}^\s{B}_2-{\bf k}^\s{B}_3-{\bf k}^\s{B}_4)$,
and
\begin{align}
\begin{split}\label{eq:RenormalizedQuantities}
&
k^{\mathsf{B}}_0 = Z_{\tau}k_0,\quad 
\vec{k}^{\mathsf{B}}= \vec{k},\quad  
k_F^B = \mu \td k_F, \quad
\Lambda^B = \mu \td  \Lambda, \quad
\\
& 
\psi^{\mathsf{B}}_{N,\sigma,j}({\bf k}^{\mathsf{B}}) =\sqrt{Z^{(\psi,N)}(k_N)} \psi_{N,\sigma,j}({\bf k}),\quad
\Phi^{\mathsf{B}}_{\sigma' \sigma}({\bf q}^{\mathsf{B}}) = \sqrt{Z^{(\Phi)}}\Phi_{\sigma' \sigma}({\bf q}), \\
&
v^{\mathsf{B}(N)}_{k^\mathsf{B}} = \frac{Z^{(N)}_{2}(k)}{Z^{(N)}_{3}(k)}v^{(N)}_{k},\quad 
V^{\mathsf{B}(N)}_{\mathrm{F}, k^{\mathsf{B}}} = Z_{\tau}\frac{Z^{(N)}_{3}(k)}{Z^{(N)}_{1}(k)}V^{(N)}_{\mathrm{F},k},\\
& 
g^{\mathsf{B}(N)}_{ {k'}^{\mathsf{B}},{k}^{\mathsf{B}} } =\frac{Z_1(0)}{Z_4(0,0)}\sqrt{\frac{Z_2(0)}{Z_3(0)}}\frac{Z^{(N)}_{4}(k',k)}{
\sqrt{
Z^{(N)}_{1}(k')
Z^{(\bar N)}_{1}(k)
}}
g^{(N)}_{k',k},\\ &
\lambda^{B \{N_i\};\{\sigma_i\}}_{\{k^{\s{B}}_{i}\}} = \mu^{-1}Z^{-3}_\tau
\left[\prod^{4}_{i=1}
Z^{(\psi,N_i)}(k_{i})
\right]^{-\frac{1}{2}}
Z^{\{N_i\};\{\sigma_i\}}(\{k_{i}\})
\lambda^{\{N_i\}; \{\sigma_i\}}_{ \{k_{i}\} }
	\end{split}
	\end{align}
with
\bqa
Z^{(\psi,N)}(k)  = \frac{Z_{1}^{(N)}(k)}{Z^2_\tau}, \quad
Z^{(\Phi)}  =   
\frac{Z^2_4(0,0)Z_3(0)}{Z^2_1(0)Z_2(0)}, \quad
Z_\tau = \frac{Z_1(0)}{Z_3(0)}.
\eqa 
$Z^{(N)}_{i}(k)\equiv 1+A^{(N)}_{i}(k)$ with $i=1,2,3$, 
$Z^{(N)}_{4}(k',k) \equiv 1+A^{(N)}_{4} (k',k)$ and 
$Z^{ \{N_i\};\{\sigma_i\}}(\{k_{i}\})\equiv 1+{A}^{\{N_i\};\{\sigma_i\}}(\{k_{i}\})$ 
are the momentum-dependent multiplicative renormalization factors.
$Z_i(0)=Z_i^{(N)}(0)$ for any $N$ due to the $C_4$ symmetry.
%
The field renormalization of the electron depends on momentum
because gapless electronic modes are labeled by the momentum along the Fermi surface. 
On the contrary, the bosonic field is rescaled in a momentum-independent way 
because the boson has zero energy only at one point in the momentum space.
The frequency is also rescaled with the momentum-independent scaling factor, $Z_\tau$.
If we keep only the momentum independent pieces in the Taylor series of the coupling functions,
these expressions reduce to those for the hot spot theory\cite{SCHLIEF}. 
$\td k_F$ and $\td \Lambda$ represent the dimensionless size of Fermi surface and the UV energy cutoff measured in the unit of $\mu$, respectively.

We denote the renormalized vertex function for 
$2m$ fermions at hot spots $\{ N_i\}$ and $n$ bosons as
\bqa
\varGamma^{(2m,n);\{N_j\} }
\left({\bf k}_i;\left[v,g,V_{\mathrm{F}},\lambda^{\{M_i\};\{\sigma_i\}}\right];\td k_F, \td \Lambda;\mu\right).
\label{eq:Gamma2mn}
\eqa
The vertex function depends on all external three-momenta, $\{ {\bf k}_i \}$.
It is also a functional of the coupling functions,
$v_{k},V_{\mathrm{F},k}, g_{k',k}, \lambda^{\{M_i\};\{\sigma_i\}}_{ \{k_i\}}$.
In general, the vertex function at a set of external momenta can depend on coupling functions at different momenta.
The vertex function can also depend on $\tilde \Lambda$ and $\tilde k_F$.
Although $\tilde \Lambda$ does not play any important role at low energies,
for now both $\td \Lambda$ and $\td k_F$  are kept in \eq{eq:Gamma2mn}
to contrast their different roles.
It also depends on scale $\mu$ at which 
the coupling functions are defined in terms of the vertex functions.
Using the facts that the bare vertex function is independent of $\mu$ and the vertex function has the scaling dimension $(3-2m-n)$ at the tree-level,
we obtain the RG equation, 
	\begin{equation}
	\begin{aligned}
	&	\left\{
	(2m + n-1)z - 2	+n\eta^{(\Phi)} +\sum^{2m}_{j=1}  
	\eta_{k_{N_{j}}}^{(\psi,N_j)} 
	+ \sum^{2m+n-1}_{j=1}\left[zk_{j;0}\frac{\partial}{\partial k_{j;0}}+\vec{k}_{j}\cdot\frac{\partial}{\partial\vec{k}_{j}}\right]
-\beta_{\td k_F}\frac{\partial}{\partial \td k_F}
-\beta_{\td \Lambda}\frac{\partial}{\partial \td \Lambda}
\right.\\
	&\left.  -\int \dd p 
	\left( 
	\left[p\frac{\partial v_{p}}{\partial p}
	+\beta^{(v)}_{p}\right]\frac{\delta}{\delta v_{p}}
	+\left[p\frac{\partial V_{\mathrm{F}, p}}{\partial p}
	+\beta^{(V_{\mathrm{F}})}_{p}\right]\frac{\delta}{\delta V_{\mathrm{F}, p}}\right.\right) 
	- \int \dd p_1 \dd p_2 \left( p_1\frac{\partial g_{p_1,p_2}}{\partial p_1}+p_2\frac{\partial g_{p_1,p_2}}{\partial p_2}+
	\beta^{(g)}_{p_1,p_2}\right)\frac{\delta}{\delta g_{p_1,p_2}} \\
	&\left.  
	-  
	\sum_{ \{M_i\} }
	\sum_{\{\sigma_i\}}
	\sum_{\{j_i\}}
	\int \dd p_1 \dd p_2 \dd p_3  
	\left(
	\sum_{\{p_i\}} p_i
	\frac{\partial \lambda^{\{M_i\};\{\sigma_i\}}_{\{p_i\}}}{\partial p_i} 
	+
	\beta^{(\lambda);\{M_i\};\{\sigma_i\}}_{\{p_i\}}
	\right)
	\frac{\delta}{\delta \lambda^{\{M_i\}; \{\sigma_i\}}_{\{p_i\}}}
	 \right\}  \times
	 \\
	&	\hspace{9cm}
	\varGamma^{(2m,n);  \{N_j\} }
	(\{{\bf k}_i\};[V_{\mathrm{F}},v,g,\lambda];\td k_F, \td \Lambda;\mu)=0.
	\end{aligned}
	\label{eq:RGEq2Point}
	\end{equation}
$\delta/\delta \mathsf{A}$ 
denotes the functional derivative with respect to $\mathsf{A}$ with $\mathsf{A}$ denoting the momentum-dependent coupling functions. 
The dynamical critical exponent, 
the anomalous scaling dimensions,
the beta functionals of the coupling functions, 
and the `beta functions' for $\td k_F$ and $\td \Lambda$  are defined by 
\bqa
\label{eq:Betas}
z = 1+ \frac{\dd \log Z_{\tau}}{\dd\log\mu}, 
~~~~~& \beta^{(\tilde{k}_F)} = \dfrac{\dd \td k_F}{\dd\log\mu},  &
~~~~~ \beta^{(\td \Lambda)} = \frac{\dd \td \Lambda}{\dd\log\mu},  
\nn
\eta_{k}^{(\psi,N)} = \frac{1}{2} \frac{\dd \log Z^{(\psi,N)}(k)}{\dd\log\mu}, 
 ~~~~~&
\eta^{(\Phi)} = \dfrac{1}{2} \dfrac{\dd \log Z^{(\Phi)}}{\dd\log\mu}, 
 &~~~~~
\beta^{(v)}_{k} =\frac{\dd v_{k}}{\dd\log\mu}, \nn
\beta^{(V_{\mathrm{F}})}_{k} =\frac{\dd V_{\mathrm{F},k}}{\dd\log\mu}, ~~~~~&
\beta^{(g)}_{k_1,k_2} = \dfrac{\dd g_{k_1,k_2}}{\dd\log\mu}, &~~~~~~
\beta^{(\lambda);\{N_i\};\{\sigma_i\}}_{\{k_i\}} = \dfrac{\dd \lambda^{\{N_i\};\{\sigma_i\}}_{\{k_i\}} }{\dd\log\mu}, 
\eqa
where the bare parameters are fixed in the derivatives.
It is noted that 
due to the $C_4$ symmetry,
we only need to keep track of 
one coupling function 
for each of the nesting angle,
Fermi velocity and the Yukawa coupling
as is shown in \eq{eq:vVgN_vVg}.
Furthermore, the fermion anomalous dimensions at different hot spots can be written in terms of one function as
\bqa
\eta_{k}^{(\psi,N)} 
= 
\left\{ \begin{array}{ll}
\eta_{k}^{(\psi)}, 
& N=1,3,4,6 \\
\eta_{-k}^{(\psi)}, 
& N=2,5,7,8 \end{array} \right..
\label{eq:etapsiNeta}
\eqa

From \eq{eq:RenormalizedQuantities},
one can express the beta functionals in terms of the counter terms as
\begin{align}
\beta^{(v)}_{k}  &=  v_k    
 \left( 
 \FR{\dd  \log  Z_{3}(k)}{\dd \log \mu} 
 - \FR{\dd  \log Z_{2}(k)}{\dd\log\mu} 
 \right),  
\label{eq:newZ1}\\
\beta^{(V_{\mathrm{F}})}_{k} &=
 V_{F,k}  \left( 
  \FR{\dd  \log Z_{1}(k)}{\dd \log \mu} 
 - \FR{\dd  \log Z_{3}(k)}{\dd\log\mu} 
 - \FR{\dd  \log Z_{1}(0)}{\dd\log\mu} 
 + \FR{\dd  \log Z_{3}(0)}{\dd\log\mu}
 \right),\label{eq:newZ2}\\
\beta^{(g)}_{k',k}  &=  g_{k',k}  \left( 
\left[
 \FR{\dd  \log Z_{4}(0,0)}{\dd\log\mu} 
 -\FR{\dd  \log Z_{1}(0)}{\dd \log \mu} 
 - \FR{1}{2}\FR{\dd  \log Z_{2}(0)}{\dd \log \mu} 
 + \FR{1}{2}\FR{\dd  \log Z_{3}(0)}{\dd\log\mu} 
 \right]
 \right.  \nn &\hspace{2cm} \left.
-\FR{\dd  \log Z_{4}(k',k)}{\dd \log \mu} 
+\FR{1}{2}\FR{\dd  \log Z_{1}(k')}{\dd \log \mu} 
+ \FR{1}{2}\FR{\dd  \log Z_{1}(k)}{\dd\log\mu} \right), \label{eq:newZ3} \\
\beta^{(\lambda);\{N_i\};\{\sigma_i\}}_{\{k_i\}} 
 &=   \lambda^{\{N_i\}; \{\sigma_i\}}_{\{ k_i \}}
 \left( 
 1 
 -\FR{\dd  \log Z_{1}(0)}{\dd \log \mu} 
 + \FR{\dd  \log Z_{3}(0)}{\dd \log \mu}
 + \frac{1}{2} \sum_i \FR{\dd  \log Z_{1}(k_i)}{\dd \log \mu} 
- \FR{\dd  \log Z^{\{j_i\}}_{\{N_i\};\{\sigma_i\}}(\{ k_j \}) }{\dd \log \mu}
 \right), 
 \label{eq:newZ4} \\
 \beta^{(\td k_F)}  = - \td k_F,   & ~~~~~~~~~
 \beta^{(\td \Lambda)}  = - \td \Lambda,   
\label{eq:newZ5}
\end{align}
and the dynamical critical exponent and the anomalous dimensions as
\bqa
z & = &  1+  \frac{\dd \log Z_1(0)}{\dd\log\mu}
-  \frac{\dd \log Z_3(0)}{\dd\log\mu},  \label{eq:z} \\
\eta_{k}^{(\psi)} &=&
 \frac{1}{2} \frac{\dd \log Z_{1}(k)}{\dd\log\mu}
 - \frac{\dd \log Z_1(0)}{\dd\log\mu}
 +\frac{\dd \log Z_3(0)}{\dd\log\mu}, \label{eq:etapsi4} \\
\eta^{(\Phi)} &=& 
  \frac{\dd \log Z_4(0,0)}{\dd\log\mu}
-  \frac{\dd \log Z_1(0)}{\dd\log\mu}
+ \frac{1}{2} \frac{\dd \log Z_3(0)}{\dd\log\mu} 
- \frac{1}{2} \frac{\dd \log Z_2(0)}{\dd\log\mu}.
\label{eq:etaphi4}
\eqa

The beta functionals 
and the anomalous dimensions 
are obtained by replacing $ \FR{\dd  Z_{i}(k)}{\dd\log\mu} $ with 
\bqa 
 \FR{\dd  Z_{i}(k)}{\dd\log\mu} 
&=&
\FR{\partial  A_{i}(k)}{\partial \log\mu} 
+ \int dp \FR{\delta  A_{i}(k)}{\delta v_p} \beta_p^{(v)}
+ \int dp \FR{\delta  A_{i}(k)}{\delta V_{F,p}}  \beta_p^{(V_F)}
+ \int dp' dp \FR{\delta  A_{i}(k)}{\delta g_{p',p}}   \beta_{p',p}^{(g)} \nn
&&
+ \int dp_1 dp_2 dp_3 
\left.
\FR{1}{\mu}
\FR{\delta  A_{i}(k)}{\delta 
	\bar \lambda^{\{N_i\};\{\sigma_i\} }_{\{ p_i\}} }
	\right|_{\bar \lambda= \lambda \mu^{-1} }
\left(
 - \lambda^{\{N_i\};\{\sigma_i\}}_{\{p_i\}} 
+
\beta^{(\lambda); \{N_i\};\{\sigma_i\}}_{\{k_i\}} 
\right)
\label{eq:dZdmu}
\eqa
in Eqs. (\ref{eq:newZ1}) - 
(\ref{eq:etaphi4}),
and solve the resulting integro-differential equations
for the beta functionals
and the anomalous dimensions.

The beta functionals 
in Eqs. 
(\ref{eq:newZ1})-(\ref{eq:newZ5})
describe the flow of the momentum-dependent coupling functions 
with increasing energy scale $\mu$ at fixed external momenta.  
In Eq. (\ref{eq:RGEq2Point}), the beta functionals appear along with the momentum dilation. 
This is due to the fact that the scale transformation rescales momentum in all directions.
The momentum along the Fermi surface needs to be scaled 
together with the momentum perpendicular to the Fermi surface
because change of momentum along the Fermi surface in non-forward scatterings is proportional to the momentum of the boson that carries a non-zero dimension.
The momentum along the Fermi surface plays a dual role\cite{SHOUVIK2}. 
On the one hand, it labels gapless modes on the Fermi surface,
and the momentum-dependent coupling functions encode how low-energy vertex functions vary along the Fermi surface. 
On the other hand, the momentum acts as a scale 
and is rescaled under the scale transformation.
$z$ represents the dynamical critical exponent that determines
how the frequency is scaled relative to spatial momentum
to keep \eq{eq:VF0equalto1}.

Eq. (\ref{eq:RGEq2Point}) 
relates the vertex function of a theory at one set of frequencies and momenta
with the vertex function of another theory with generally different couplings 
at rescaled frequencies and momenta as 
\begin{align}
\label{eq:SolRGEq2Point}
\begin{split}
&
\varGamma^{(2m,n);  \{N_j\} }(\{ {\bf k}_i\};[\whv,\whg,\whV_{\mathrm{F}}];\td k_F, \td \Lambda
)
=
\exp\left\{\int\limits^{ \ell }_{0}\dd\ell'~
\left[  
(2m+n-1)z(\ell')  -2 +n\eta^{(\Phi)}(\ell') +\sum^{2m}_{j=1} \hat \eta^{(\psi,N_j)}_{k_{N_{j}}(\ell') }(\ell') 
\right]
\right\}\\ &\hspace{1.5cm}\times
\varGamma^{(2m,n);  \{N_j\} }\left(
\{ k_{i,0}(\ell),\vec{k}_i(\ell) \} ;
\left[\whv(\ell), \whg(\ell), \whV_{\mathrm{F}}(\ell), \wh \lambda^{\{N_i\}; \{\sigma_i\}}(\ell)\right];
\td k_F(\ell),
\td \Lambda(\ell)
\right).
	\end{split}
	\end{align}
Here,	
\begin{align}\label{eq:ScalingTransfo}
k_0(\ell)\equiv e^{\int\limits^{\ell}_{0} z(\ell') \dd\ell'  } k_0,
\qquad  
\vec{k}(\ell) \equiv e^{\ell} \vec{k}.
\end{align} 
$\ell$ is the logarithmic length scale.
The scale-dependent coupling functions obey
	\begin{align}
	\begin{split}\label{eq:DifEqv}
	\frac{\partial \whv_{K}(\ell)}{\partial \ell } &= 
	- \beta^{(v)}_{K}(\ell)-K\frac{\partial \whv_{K}(\ell)}{\partial K},
	\end{split}\\
	\begin{split}\label{eq:DifEqV}
	\frac{\partial \whV_{\mathrm{F},K}(\ell)}{\partial \ell } &= -
	\beta^{(V_{\mathrm{F}})}_{K}(\ell) - K\frac{\partial \whV_{\mathrm{F},K}(\ell)}{\partial K},
	\end{split}\\
	\begin{split}\label{eq:DifEqg}
	\hspace{-0.8cm}\frac{\partial \whg_{K',K}(\ell)}{\partial \ell} &=-
	\beta^{(g)}_{K',K}(\ell) - K'\frac{\partial \whg_{K',K}(\ell)}{\partial K'} - K\frac{\partial \whg_{K',K}(\ell)}{\partial K},
	\end{split}\\
	\begin{split}\label{eq:DifElambda}
	\hspace{-0.8cm}\frac{\partial \whl^{\{N_i\};\{\sigma_i\}}_{\{K_i\}}(\ell)}{\partial \ell} &=
	-\beta^{(\lambda);\{N_i\};\{\sigma_i\}}_{\{K_i\}}(\ell) 
	- \sum^{4}_{j=1}K_{j}\frac{\partial \whl^{\{N_i\};\{\sigma_i\}}_{\{K_i\}}(\ell) }{\partial K_{j}},
	\end{split}\\
	\begin{split}\label{eq:DifEqKf}
	\hspace{0cm}	
	\frac{\partial \td k_F(\ell)}{\partial \ell} = \td k_F(\ell), & ~~~~~~~
	\frac{\partial \td \Lambda(\ell)}{\partial \ell} = \td \Lambda(\ell)
	\end{split}
	\end{align}
with the initial conditions, 
$\whv_{K}(0) =\whv_{K}$, 
$\whV_{\mathrm{F},K}(0) =\whV_{\mathrm{F},K}$, 
$\whg_{K',K}(0) = \whg_{K',K}$, 
$\whl^{\{N_i\};\{\sigma_i\}}_{\{K_i\}}(0)=
\whl^{\{N_i\};\{\sigma_i\}}_{\{K_i\}}$,
$\td k_F(0) = \td k_F$
 and 
$\td \Lambda(0)=\td \Lambda$.  
$\hat \eta^{(\psi,N)}_{k \eell}(\ell)$
denotes the anomalous dimension of the fermion measured
at momentum $k \eell$
and energy scale $\Lambda$
in the theory with coupling functions,
$\{ \whv(\ell),$
$\whg(\ell),$
$\whV_{\mathrm{F}}(\ell),$
$\hat \lambda^{\{N_i\}; \{\sigma_i\}}(\ell) \}$.
%
%
%
\eq{eq:SolRGEq2Point} relates
a physical observable measured at $\{ k_{i,0}, \vec k_i \}$ 
in the theory with 
Fermi surface size $\mu \td k_F$,
UV cutoff $\mu \td \Lambda$
and couplings 
$\{ \whv,$ $ \whg,$  $\whV_{\mathrm{F}},$ $\lambda^{\{N_i\}; \{\sigma_i\}} \}$ 
to the  observable measured at  
$\{  e^{\int\limits^{\ell}_{0} z(\ell') \dd\ell'  }k_{i,0}  , \eell \vec k_i \}$ 
in the theory 
with Fermi surface size $ \eell  \mu \td k_F $,
UV cutoff $\eell \mu \td \Lambda$
and couplings
$\{ \whv(\ell),$
$\whg(\ell),$
$\whV_{\mathrm{F}}(\ell),$
$\hat \lambda^{\{N_i\}; \{\sigma_i\}}(\ell) \}$.
It is noted that the RG equation relates observables
in two theories not just with different UV cutoffs 
but with different sizes of Fermi surface, that is, with different numbers of IR degrees of freedom.
Fixed points are characterized by coupling functions
$\{ \whv^*,$ $ \whg^*$  $\whV_{\mathrm{F}}^*,$ $\hat \lambda^{\{N_i\}; \{\sigma_i\} *} \}$
at which the beta functionals in  Eqs. (\ref{eq:DifEqv})-(\ref{eq:DifElambda}) 
vanish, and $\td \Lambda^*=\infty$, $\td k_F^*=\infty$.
%
If the physical observables can be expressed
as regular functions of the renormalized couplings
in the large $\Lambda$ limit,
we can set $\Lambda^* = \infty$ at the fixed point
as is usually done in the continuum limit.
On the other hand, $k_F$ is an IR parameter of the theory
that encapsulates the number of gapless modes in the system.
Since there is no guarantee that  all low-energy physical observables
are well defined in the large $k_F$ limit,
we can not simply ignore the dependence on $k_F$ as we do for $\Lambda$.
This has an obvious consequence:
{\it 
theories with Fermi surfaces
do not have the usual sense of scale invariance
that relates observables defined 
at different scales within one theory 
even in the continuum limit.}
Consequently, \eq{eq:SolRGEq2Point} does not
fully determine how physical observables actually scale 
with energy and momentum in a theory with a fixed $k_F$.
Only for those vertex functions 
that are regular in the large $k_F$ limit,
$\left[  (2m+n-1)z^*    -2 +n \eta^{(\Phi)^*} +
\sum^{2m}_{j=1} \hat \eta^{(\psi)*}_{k_{N_{j}}} \right]$ in \eq{eq:SolRGEq2Point}
determines the actual dependence on energy and momentum.
%
For those observables that are 
singular in the large $k_F$ limit,
the scaling behaviour is modified from what is expected from the predicted scaling dimensions\footnote{
For example,  the response functions to  spatially uniform  thermal/electromagnetic perturbations are sensitive to $k_F$.
}.
There is even no guarantee 
that general low-energy observables depend on energy and momentum in power laws 
at a fixed point\cite{IPSITA,Mandal:2016vh}.  
Without knowing how general observables depend on $k_F$ in priori,
one has to keep $k_F$ as a running coupling  within the theory.

It is noted that the four-fermion couplings has dimension $-1$ at the tree-level. 
This causes the \ffc to decrease in amplitude under the RG flow.
On the other hand, the momentum along the Fermi surface is rescaled,
and the size of the Fermi surface measured in the unit of the
running energy scale increases.
This effectively promotes the four-fermion couplings  
to marginal couplings in the channels in which 
quantum corrections become proportional to the phase space.
For example, this enhancement  of the effective scaling dimension occurs in the pairing channel in which
the phase space for low-energy Cooper pairs
 is extensive.

In keeping track of the momentum dependent coupling functions along the Fermi surface,
it is sometimes convenient to define coupling functions at a fixed physical location on the Fermi surface as
\begin{align}
& 
v_{k}(\ell)  \equiv \hat{v}_{k(\ell)}(\ell), \quad
V_{\mathrm{F},k}(\ell) \equiv \hat{V}_{\mathrm{F}, k(\ell)}(\ell), \quad
\eta^{(\psi)}_{k}(\ell) \equiv \hat{\eta}^{(\psi)}_{k(\ell)}(\ell), \\
& 
g_{k',k}(\ell)  \equiv \hat g_{k'(\ell), k(\ell)}(\ell), \quad
\lambda^{\{N_i\};\{\sigma_i\}}_{\{k_i\}}(\ell) \equiv \hat{\lambda}^{\{N_i\};\{\sigma_i\}}_{ \left\{ k_i(\ell) \right\}} (\ell).
 \label{eq:LambdaAux}
\end{align}
$v_{k}(\ell),V_{\mathrm{F},k}(\ell)$, $g_{k',k}(\ell) $ and 
${\lambda}^{\{N_i\};\{\sigma_i\}}_{\{k_i\}}(\ell)$ satisfy the beta functionals that do not have the momentum dilatation,
\begin{align}
\frac{\partial}{\partial \ell} v_{k}(\ell)&= - \beta^{(v)}_{k}(\ell)
,\label{eq:Dif1}\\
\frac{\partial}{\partial \ell}V_{\mathrm{F},k}(\ell)&= -
\beta^{(V_F)}_{k}(\ell)
,\label{eq:Dif2}\\
\frac{\partial}{\partial \ell} g_{k',k}(\ell) &= -
\beta^{(g)}_{k',k}(\ell)
,\label{eq:Dif3}\\
\frac{\partial }{\partial \ell} {\lambda}^{\{N_i\};\{\sigma_i\}}_{\{k_i\}}(\ell) &=-
\beta^{(\lambda);\{N_i\};\{\sigma_i\}}_{\{k_i\}}(\ell).
\label{eq:Dif4}
\end{align}
%
%
Eqs. (\ref{eq:Dif1}) to (\ref{eq:Dif4}) track the renormalization of the coupling functions with increasing logarithmic length scale $\ell$  at fixed momenta along the Fermi surface. 
It will be useful to go back and forth between
Eqs. (\ref{eq:DifEqv})-(\ref{eq:DifElambda})
and
Eqs. (\ref{eq:Dif1})- (\ref{eq:Dif4})
for different purposes.
The ultimate fate of the system in the low-energy limit
is determined by the RG flow of the full coupling functions, 
$\Bigl\{ v_{k} (\ell)$, $ V_{\mathrm{F},k}(\ell)$, $ g_{k',k}(\ell)$, $\lambda^{\{N_i\}; \{\sigma_i\}}_{ \{k_i\} } (\ell) \Bigr\}$.


\subsection{Quantum effective action}
\label{sec:QEA}

The full quantum effective action is written as 
\begin{align}
\label{eq:QEA}
%
\begin{split}
\boldsymbol{\Gamma}
& = \sum^{8}_{N=1}\sum^{N_c}_{\sigma =1}\sum^{N_f}_{j=1}\int\dd {\bf k}~\psi^{\dagger}_{N,\sigma,j}({\bf k})
\left\{
ik_0+V^{(N)}_{\mathrm{F},k_N}e_{N}[\vec{k};v^{(N)}_{k_N}]
+{\bf \Sigma}_{N}({\bf k}) 
\right\}\psi_{N,\sigma,j}({\bf k})\\
	& + \frac{1}{4} \int \dd {\bf k}~ D^{-1}({\bf k}) \tr{ \Phi({\bf k}) \Phi(-{\bf k}) } \\
&+\sum^{8}_{N=1}\sum^{N_c}_{\sigma\sigma'=1}\sum^{N_f}_{j=1}
\int \dd {\bf k}'  \dd {\bf k} ~ 
\left\{
\frac{1}{\sqrt{N}_f} g^{(N)}_{k_N', k_N}
+ \delta {\bf \Gamma}^{(2,1)}_{N}({\bf k}',{\bf k}) 
\right\}
\psi^\dagger_{N,\sigma',j}({\bf k}')
\Phi_{\sigma' \sigma}({\bf k}'-{\bf k})
\psi_{\bar{N},\sigma,j}({\bf k})\\
&+\sum^{8}_{\{N_i=1\}}\sum^{N_c}_{\{\sigma_i=1\}}\sum^{N_f}_{\{j_i=1\}}
\int \prod^{4}_{i=1} 
\dd {\bf k}_i~~
\delta_{1+2,3+4}
\left\{
\frac{1}{4 \mu} 
\lambda^{\spmqty{N_1 & N_2 \\ N_4 & N_3};\spmqty{\sigma_1 & \sigma_2 \\ \sigma_4 & \sigma_3}}_{\spmqty{k_{1;N_1} & k_{2;N_2} \\ k_{4,N_4} & k_{3;N_3}}}
+ \delta {\bf \Gamma}^{\spmqty{N_1 & N_2 \\ N_4 & N_3};\spmqty{\sigma_1 & \sigma_2 \\ \sigma_4 & \sigma_3}}( \{ {\bf k}_{i; N_i} \} ) 
\right\}  \times \\
& \hspace{5cm}
\psi^\dagger_{N_1,\sigma_1,j_1}({\bf k}_1)\psi^\dagger_{N_2,\sigma_2,j_2}({\bf k}_2)\psi_{N_3,\sigma_3,j_2}({\bf k}_3)\psi_{N_4,\sigma_4,j_1}({\bf k}_4) + ...,
\end{split}
\end{align}
where $...$ represents higher order terms in the  fields.
Here $D^{-1}({\bf k})$ is the boson self-energy.
Since the bare kinetic term of the boson is irrelevant, 
the self-energy determines the entire boson propagator at low energies.  
 ${\bf \Sigma}_{N}({\bf k}) $ is the fermion self-energy.
$\delta {\bf \Gamma}^{(2,1)}_{N}({\bf k}',{\bf k})$ is the quantum correction
to the Yukawa vertex.
Finally,  
$\delta {\bf \Gamma}^{\spmqty{N_1 & N_2 \\ N_4 & N_3};\spmqty{\sigma_1 & \sigma_2 \\ \sigma_4 & \sigma_3}}( \{ {\bf k}_{i; N_i} \} )$
represents the quantum correction to the four-fermion coupling function.

In the present theory, 
the computation of the quantum effective action 
is organized in terms of the small parameter
$v \sim g^2$.
%
At the zeroth order in $v$, 
only $D^{-1}({\bf k})$ is important
among all quantum corrections.
The infinite set of diagrams that contribute to the boson self-energy to the leading order in $v$
can be summed through the Schwinger-Dyson equation in \fig{fig:SDEq}\cite{SCHLIEF}.
Other quantum corrections are at most order of $\frac{g^2}{c} \log(1/v)  \sim \sqrt{v \log(1/v)}$ in the small $v$ limit,
and can be computed perturbatively as functionals of the coupling functions.
The \ffc that is generated 
from the Yukawa coupling through \fig{fig:4f0}
is order of $g^4/c$,
where the phase space enhancement factor of $1/c$ arises because
the largest speed of particles in the loop
is $c$ in one momentum direction.
The interaction energy of two particles, 
given by the eigenvalues of the quartic vertex as in 
\eq{eq:eigenGamma},
goes as $g^4/c^2$, 
where there is an additional factor of $1/c$ 
because the typical momentum transfer of particle goes as $\mu/c$ at energy scale $\mu$\footnote{
For example, the eigenvalue of the four-fermion coupling function in the BCS channel involves an integral over relative momentum of a Cooper pair,
and it is enhanced by $1/c$ due to the slow decay of the coupling function at large momentum in the small $c$ limit.
This will be shown through explicit calculations in Sec. \ref{app:QC2}.
}.
Since $g^4/c^2 \sim v/\log(1/v)$ in the small $v$ limit,
the electron-electron interaction is dominated 
by the interaction 
mediated by the gapless spin fluctuations
whose strength is order of $g^2/c \sim \sqrt{v/\log(1/v)}$.
%
We will see that there exists a large window of energy scale 
in which the four-fermion coupling remains smaller than $g^2/c$ 
before it becomes dominant due to superconducting instabilities at low energies.
Within this window,
the feedback of the four-fermion coupling
to the self-energies and the cubic vertex can be ignored.
Our goal is to extract the universal normal state properties and the evolution of superconducting fluctuations that emerge within this range of energy scale.

\subsubsection{Two faces of the four-fermion coupling}
\label{sec:TwoFaces}
Under the interaction driven scaling,
the Yukawa coupling is marginal,
and quantum corrections 
that include the Yukawa coupling 
are logarithmically divergent, as expected.
On the contrary, the four-fermion coupling
 has dimension $-1$.
Naively, one would not expect infrared divergences
 associated with the four-fermion coupling.
However, the actual degrees of IR divergence
 vary in different diagrams.
In most diagrams that include the four-fermion coupling,
there is indeed no IR divergence
as expected from power-counting.
Some diagrams, however, exhibit logarithmic IR divergences,
defying the expectation based on the scaling dimension.

This disagreement arises because the momentum along the Fermi surface
plays different roles 
in different scattering processes.
In diagrams that involve patches of Fermi surface that are not nested,
a loop momentum that is parallel to the Fermi surface in one patch 
is not parallel to the Fermi surface in another patch.
Consequently, all components of momentum need to be small
in order for virtual fermions to stay close to the Fermi surface.
In this case, all components of momenta  act as scale in the loop, and
the power counting correctly captures the absence of IR divergence.
On the contrary, in diagrams that include only those patches that are nested,
the component of momentum parallel to the patches act as a continuous flavour.
For example, the four-fermion interaction in the pairing channel involves opposite sides of Fermi surface which are perfectly nested in the particle-particle channel.
Because all virtual particles can stay close to the Fermi surface irrespective of
the momentum along the Fermi surface,
actual IR divergences are controlled by the one-dimensional scaling
under which the four-fermion coupling is marginal.
In other words, the integration of the momentum along the nested patches
gives rise to a volume of the low-energy phase space,
and the scale associated with the volume of the phase space effectively
promotes the four-fermion couplings to marginal couplings,
resulting in logarithmic IR divergences.

\subsubsection{Quantum corrections}
\label{sec:QuantumCorrections}

We start with the propagator of the collective mode. 
To the leading order in $v$, 
the dressed boson propagator $D({\bf q})$
satisfies the Schwinger-Dyson equation with the momentum dependent coupling functions,
\bqa
&& D^{-1}({\bf k}) = m_{CT}
+ 2 \sum_{N=1}^8
   \int d {\bf q} ~
   g^{(N)}_{q,q+k} g^{(\bar N)}_{q+k,q} G_{N}({\bf q})  G_{\bar N}({\bf q}+{\bf k}) \nn
 &&  
- \frac{4}{N_c N_f}
\sum^{8}_{N=1}
\int \dd {\bf q} \dd {\bf p}~
g^{(\bar N)}_{p,p+k}
g^{(N)}_{p+k,p+q+k}
g^{(\bar N)}_{p+q+k,p+q}
g^{(N)}_{p+q,p}
G_N({\bf p}+{\bf q}) 
G_{\bar N}({\bf p}) 
G_N({\bf p}+{\bf k}) 
G_{\bar N}({\bf p}+{\bf q}+{\bf k}) D({\bf q}),
\label{eq:fullDyson}
\eqa
where 
$m_{CT}$ is the mass counter term 
that tunes the renormalized mass of the boson to zero.
$G_N({\bf k})$ is the bare fermionic Green's function for hot spot $N$,
\begin{align}\label{eq:G0}
G_{N}({\bf k}) = \frac{1}{ik_0+V^{(N)}_{F,k}e_{N}[\vec{k};v^{(N)}_k]}.
\end{align}
It is noted that $G_N({\bf k})$ depends on momentum not only
through the explicit momentum dependence in the dispersion
but also through $V^{(N)}_{F,k}$ and $v^{(N)}_k$ that
depend on the momentum along the Fermi surface.
Here and henceforth, we drop the patch index $N$ in $k_N$ 
for the momentum along the Fermi surface
when there is no danger of confusion.
For $\lambda \ll g^2/c$, 
the contribution of the short-ranged four-fermion coupling 
to the boson self-energy 
is sub-leading
in the small $v$ limit.
Since the bare kinetic term of the boson is irrelevant under the interaction driven scaling,
we don't need to add any counter term for the boson  except for the mass counter term.
In other diagrams,
we use the non-perturbatively dressed boson propagator.

\begin{figure}[htbp]
	\centering
	\begin{subfigure}[b]{0.30\linewidth}
		\centering
		\includegraphics[scale=0.30]{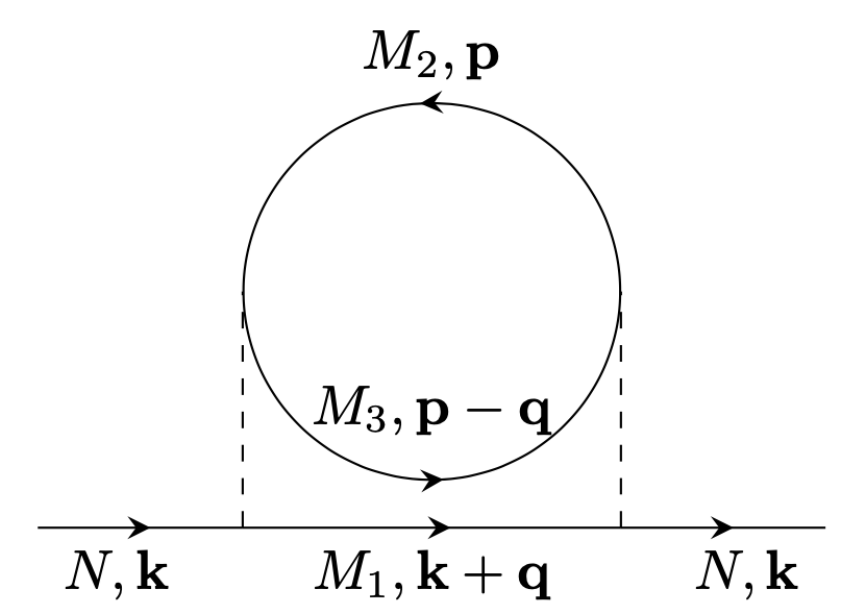}
		\caption{\label{fig:1LFSE4f}}
	\end{subfigure}
	\begin{subfigure}[b]{0.30\linewidth}
		\centering
		\includegraphics[scale=0.30]{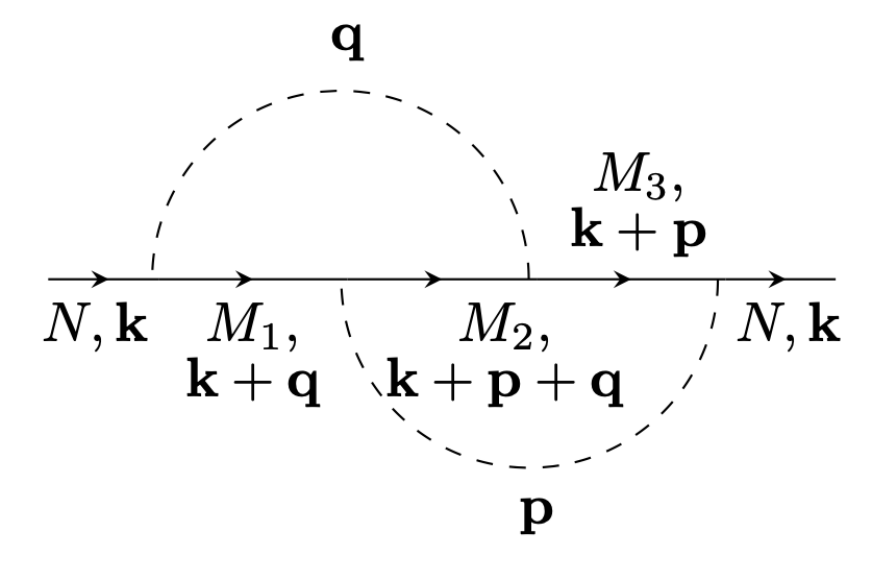}
		\caption{\label{fig:2LSE4f}}
	\end{subfigure}
		\begin{subfigure}[b]{0.30\linewidth}
		\centering
		\includegraphics[scale=0.30]{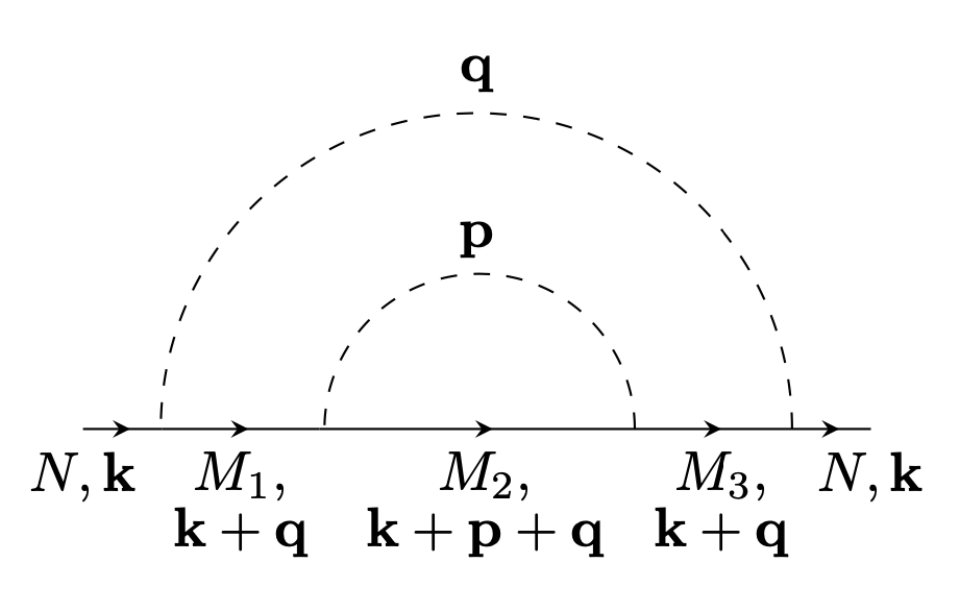}
		\caption{\label{fig:2LSE4ftwo}}
	\end{subfigure}
	\caption{Contributions of the four-fermion coupling to the fermion self-energy.}
	\label{fig:F2LSE}
\end{figure}

To the leading order in $v$,
the fermion self-energy is written as the sum of two terms as
${\bf \Sigma}_N({\bf k}) =
{\bf \Sigma}^{\mathrm{1L}}_{N}({\bf k}) 
+{\bf \Sigma}^{\mathrm{2L}}_{N}({\bf k})$
where
\bqa
{\bf \Sigma}^{\mathrm{1L}}_{N}({\bf k}) 
	&=& - \frac{2(N^2_c-1)}{N_cN_f}\int\dd {\bf q}~ 
	g^{(N)}_{k,k+q} 
	g^{(\bar N)}_{k+q,k}
	G_{\bar{N}}({\bf k}+{\bf q})D({\bf q}), 
	\label{eq:1LoopFSE_full} \\
{\bf \Sigma}^{\mathrm{2L}}_{N}({\bf k}) 
		&=&
\frac{4(N_c^2-1)}{N_c^2N_f^2}
\int\dd {\bf q} \int \dd {\bf p} ~
g^{(N)}_{k,k+q}
g^{(\bar N)}_{k+q,k+p+q}
 g^{(N)}_{k+p+q,k+p} 
g^{(\bar N)}_{k+p,k}  
	D({\bf p})D({\bf q}) 
	G_{\bar{N}}({\bf k}+{\bf p})
	G_{N}({\bf k}+{\bf q}+{\bf p})
	G_{\bar{N}}({\bf k}+{\bf q}). \nn
	\label{eq:2LoopFSE_full} 
\eqa
\eq{eq:1LoopFSE_full} and \eq{eq:2LoopFSE_full}
represent the one-loop and two-loop fermion self-energies
generated from the Yukawa couplings 
as is shown in \fig{fig:1LFSE} and \fig{fig:2LSE}
respectively.
To the leading order in $v$, these are the only diagrams that are important\footnote{
The contribution of the four-fermion coupling 
to the fermion-self energy from \fig{fig:F2LSE}
can be written as
\bqa
{\bf \Sigma}^{' \mathrm{2L}}_{N}({\bf k}) 
		&=&
\frac{1}{4 \mu^2}
\int\dd {\bf q} \int \dd {\bf p} ~ \left[
N_f \lambda^{\spmqty{M_1 & M_3 \\  N &  M_2};\spmqty{\alpha_1 & \alpha_3 \\ \sigma & \alpha_2}}_{\spmqty{k+q  & p-q \\ k & p }}
\lambda^{\spmqty{N & M_2 \\  M_1 &  M_3};\spmqty{\sigma & \alpha_2 \\ \alpha_1 & \alpha_3}}_{\spmqty{k  & p \\ k+q & p-q }}
	G_{M_1}({\bf k}+{\bf q})
	G_{M_2}({\bf p})
	G_{M_3}({\bf p}+{\bf q}) 
	\right. \nn
&& \hspace{2.5cm}
- 
\lambda^{\spmqty{M_1 & M_3 \\  N &  M_2};\spmqty{\alpha_1 & \alpha_3 \\ \sigma & \alpha_2}}_{\spmqty{k+q  & k+p \\ k & k+q+p }}
\lambda^{\spmqty{M_2 & N \\  M_1 &  M_3};\spmqty{\alpha_2 & \sigma \\ \alpha_1 & \alpha_3}}_{\spmqty{k+q+p  & k \\ k+q & k+p }}
	G_{M_1}({\bf k}+{\bf q})
	G_{M_2}({\bf k}+{\bf q}+{\bf p})
	G_{M_3}({\bf k}+{\bf p}) \nn
	&& \hspace{2.5cm}
\left.
-	
\lambda^{\spmqty{M_1 & N \\  N &  M_3};\spmqty{\alpha_1 & \sigma \\ \sigma & \alpha_3}}_{\spmqty{k+q  & k \\ k & k+q }}
\lambda^{\spmqty{M_2 & M_3 \\  M_1 &  M_2};\spmqty{\alpha_2 & \alpha_3 \\ \alpha_1 & \alpha_2}}_{\spmqty{k+q+p  & k+q \\ k+q & k+p+q }}
	G_{M_1}({\bf k}+{\bf q})
	G_{M_2}({\bf k}+{\bf q}+{\bf p})
	G_{M_3}({\bf k}+{\bf q}) 
	\right].	
	\label{eq:2LoopFSEprime_full}
\eqa
For $\lambda \ll g^2/c$,
this is sub-leading compared to 
${\bf \Sigma}^{\mathrm{1L}}_{N}({\bf k})$
and 
${\bf \Sigma}^{\mathrm{2L}}_{N}({\bf k})$.
}.
The one-loop vertex correction 
shown in Fig. \ref{fig:yukawa}
is given by
\begin{align}
	\label{eq:1LY_full}
&&	
\delta {\bf \Gamma}^{(2,1)}_{1}({\bf k}',{\bf k}) = -
\frac{2 }{N_cN^\frac{3}{2}_{f}}
\int \dd {\bf q}~
g^{(N)}_{k',k'+q}
g^{(\bar N)}_{k'+q,k+q}
g^{(N)}_{k+q,k}  ~
D({\bf q}) 
G_{\bar{N}}({\bf k}'+{\bf q})
G_{N}({\bf k}+{\bf q}).
\end{align}
The contribution of the short-range \ffc
to  
$\delta {\bf \Gamma}^{(2,1)}({\bf k}',{\bf k})$
is sub-leading compared to \eq{eq:1LY_full}.

\begin{figure*}
	\centering
	\begin{subfigure}[b]{0.34\linewidth}
		\centering
		\includegraphics[scale=0.925]{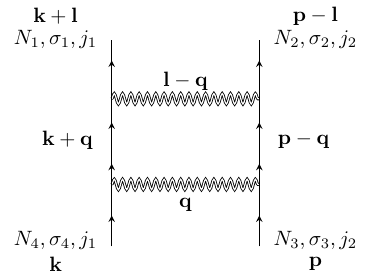}
		\caption{\label{fig:s1}}
	\end{subfigure}
	\begin{subfigure}[b]{0.34\linewidth}
		\centering
		\includegraphics[scale=0.925]{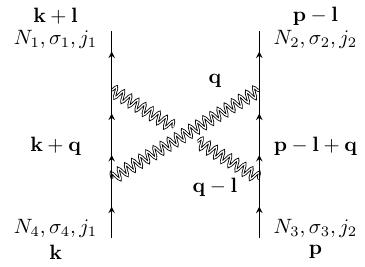}
		\caption{\label{fig:s2}}
	\end{subfigure}
	\caption{
Quantum corrections to the four-fermion couplings that are independent of $\lambda$.
}
	\label{fig:4f0}
\end{figure*}

The one-loop vertex correction to the four-fermion coupling is written as
$
\delta  {\bf \Gamma}^{\{ N_i \}; \{ \sigma_i\} }
=
\sum_{a=0}^2
{\bf \Gamma}_{(a)}^{ \{ N_i \}; \{ \sigma_i\} }$,
where
${\bf \Gamma}_{(a)}^{ \{ N_i \}; \{ \sigma_i\} }$
denotes the vertex correction that is of the $a$-th power of $\lambda$.
The four-fermion vertex that is independent of $\lambda$ is further divided into two parts as
$
{\bf \Gamma}_{(0)}^{ \{ N_i \}; \{ \sigma_i\} }
=
{\bf \Gamma}_{(0)PP}^{ \{ N_i \}; \{ \sigma_i\} }
+
{\bf \Gamma}_{(0) PH}^{\{ N_i \}; \{ \sigma_i\} }$,
where
\bqa
&& {\bf \Gamma}_{(0)PP}^{
\spmqty{N_1 & N_2 \\ N_4 & N_3};\spmqty{\sigma_1 & \sigma_2 \\ \sigma_4 & \sigma_3}
}
\spmqty{{\bf k}+{\bf l} & {\bf p}- {\bf l} \\  {\bf k} &  {\bf p}}
= -\frac{1}{2 N_f^2} \int \dd {\bf q} ~
g^{(N_1)}_{k+l,k+q}
g^{(\bar N_4)}_{k+q,k}
g^{(N_2)}_{p-l,p-q}
g^{(\bar N_3)}_{p-q,p} 
\times \nn && \hspace{1cm}
D({\bf q}) D({\bf l}- {\bf q})
G_{\bar{N}_1}({\bf k}+{\bf q})
G_{\bar{N}_2}({\bf p}- {\bf q})
\delta_{N_1 N_4} \delta_{N_2 N_3}
\mathsf{T}^{\sigma_1 \sigma_2}_{\alpha \beta} 
\mathsf{T}^{\alpha \beta}_{\sigma_4 \sigma_3} 
\label{eq:Gamma_lambda0_PP}
\eqa
is the four-fermion vertex generated from the Yukawa coupling
in the particle-particle channel
(Fig. \ref{fig:s1}),
and
\begin{equation}
\begin{aligned}
& {\bf \Gamma}_{(0)PH}^{\spmqty{N_1 & N_2 \\ N_4 & N_3};\spmqty{\sigma_1 & \sigma_2 \\ \sigma_4 & \sigma_3}}
\spmqty{{\bf k}+{\bf l} & {\bf p}- {\bf l} \\  {\bf k} &  {\bf p}}
= -\frac{1}{2 N_f^2} \int \dd {\bf q} ~
g^{(N_1)}_{k+l,k+q}
g^{(\bar N_4)}_{k+q,k}
g^{(N_2)}_{p-l,p-l+q}
g^{(\bar N_3)}_{p-l+q,p} 
\times \\
& \hspace{1cm}
D({\bf q}) D({\bf l}- {\bf q})
G_{\bar{N}_1}({\bf k}+{\bf q})
G_{\bar{N}_2}({\bf p}- {\bf l}+{\bf q})
\delta_{N_1 N_4} \delta_{N_2 N_3}
 \mathsf{T}^{\beta \sigma_2}_{\sigma_4 \alpha} 
 \mathsf{T}^{\sigma_1 \alpha}_{\beta \sigma_3} 
\end{aligned}
\label{eq:Gamma_lambda0_PH}
\end{equation}
is the four-fermion vertex generated from the Yukawa coupling
in the particle-hole channel
(Fig. \ref{fig:s2}).
$ \mathsf{T}^{\sigma_1  \sigma_2}_{\sigma_4  \sigma_3 }$
is the spin structure factor for the interaction
mediated by the critical spin fluctuations
between incoming electrons with spin $\sigma_4$, $\sigma_3$
and outgoing electrons with spin $\sigma_1$, $\sigma_2$,
\begin{align}
 \mathsf{T}^{\sigma_1  \sigma_2}_{\sigma_4  \sigma_3 } =&
\sum_{a=1}^{N_c^2-1} 
\tau^a_{\sigma_1 \sigma_4}
\tau^a_{\sigma_2 \sigma_3} 
= 
 2\left( \delta_{\sigma_1 \sigma_3}\delta_{\sigma_2 \sigma_4} - \frac{1}{N_c}\delta_{\sigma_1 \sigma_4}\delta_{\sigma_2 \sigma_3} \right).
 \label{eq:Tabcd}
 \end{align}

\begin{figure*}
	\centering
			\begin{subfigure}[b]{0.49\linewidth}
		\centering
		\includegraphics[scale=0.925]{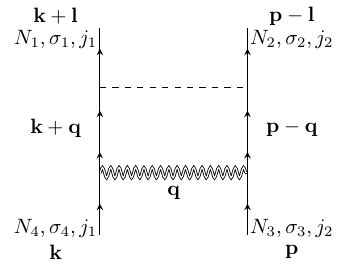}
		\caption{\label{fig:pp1}}
	\end{subfigure}
	\begin{subfigure}[b]{0.49\linewidth}
		\centering
		\includegraphics[scale=0.925]{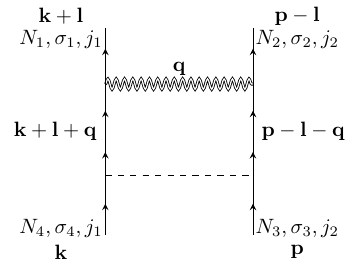}
		\caption{\label{fig:pp2}}
	\end{subfigure} \\
	\begin{subfigure}[b]{0.49\linewidth}
		\centering
		\includegraphics[scale=0.925]{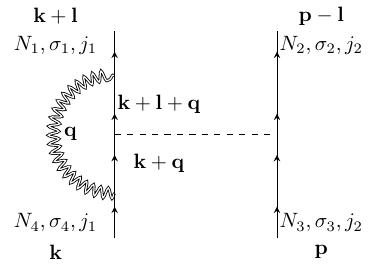}
		\caption{\label{fig:ph1}}
	\end{subfigure}
	\begin{subfigure}[b]{0.49\linewidth}
		\centering
		\includegraphics[scale=0.925]{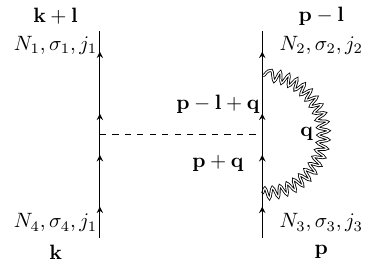}
		\caption{\label{fig:ph2}}
	\end{subfigure}
	\begin{subfigure}[b]{0.49\linewidth}
		\centering
		\includegraphics[scale=0.925]{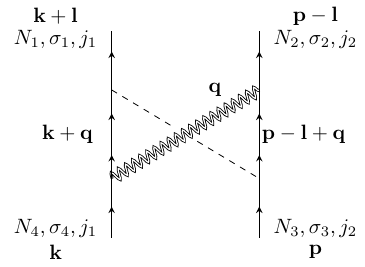}
		\caption{\label{fig:ph3}}
	\end{subfigure}
	\begin{subfigure}[b]{0.49\linewidth}
		\centering
		\includegraphics[scale=0.925]{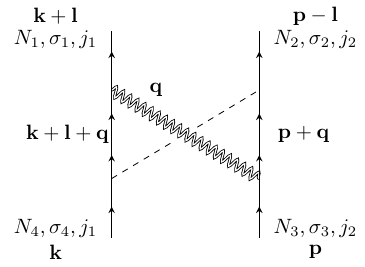}
		\caption{\label{fig:ph4}}
	\end{subfigure}
	\caption{
Quantum corrections to the four-fermion couplings linear in $\lambda$.
}
	\label{fig:4f1}
\end{figure*}

The four-fermion vertex that is linear in $\lambda$ 
can be also divided into the one in which the vertex correction
is in the particle-particle (PP) channel
and the one in which the vertex correction is in the particle-hole (PH) channel, 
$
{\bf \Gamma}_{(1)}^{\spmqty{N_1 & N_2 \\ N_4 & N_3};\spmqty{\sigma_1 & \sigma_2 \\ \sigma_4 & \sigma_3}}
=
{\bf \Gamma}_{(1)PP}^{\spmqty{N_1 & N_2 \\ N_4 & N_3};\spmqty{\sigma_1 & \sigma_2 \\ \sigma_4 & \sigma_3}}
+
{\bf \Gamma}_{(1)PH}^{\spmqty{N_1 & N_2 \\ N_4 & N_3};\spmqty{\sigma_1 & \sigma_2 \\ \sigma_4 & \sigma_3}}$.
The vertex correction in the PP  channel 
(Figs. \ref{fig:pp1} and \ref{fig:pp2}) is 
\bqa
&& {\bf \Gamma}_{(1)PP}^{\spmqty{N_1 & N_2 \\ N_4 & N_3};\spmqty{\sigma_1 & \sigma_2 \\ \sigma_4 & \sigma_3}}
\spmqty{{\bf k}+{\bf l} & {\bf p}- {\bf l} \\  {\bf k} &  {\bf p}}
=  
\frac{1}{4 \mu N_f} \int d {\bf q} ~
g^{(\bar N_4)}_{k+q,k}
g^{(\bar N_3)}_{p-q,p} 
D({\bf q}) 
G_{\bar{N}_4}({\bf k}+{\bf q})
G_{\bar{N}_3}({\bf p}- {\bf q})
\lambda^{\spmqty{N_1 & N_2 \\ \bar N_4 & \bar N_3};\spmqty{\sigma_1 & \sigma_2 \\ \alpha & \beta}}_{\spmqty{k+l  & p-l \\ k+q & p-q }}
 \mathsf{T}^{\alpha \beta}_{\sigma_4 \sigma_3} 
\nn
&& +
\frac{1}{4 \mu N_f} \int d {\bf q} ~
g^{(N_1)}_{k+l,k+l+q}
g^{(N_2)}_{p-l,p-l-q} 
D({\bf q}) 
G_{\bar{N}_1}({\bf k}+ {\bf l}+{\bf q})
G_{\bar{N}_2}({\bf p}- {\bf l}-{\bf q})
    \mathsf{T}^{\sigma_1 \sigma_2}_{\alpha \beta} 
 \lambda^{\spmqty{\bar N_1 & \bar N_2 \\  N_4 &  N_3};\spmqty{\alpha & \beta \\ \sigma_4 & \sigma_3}}_{\spmqty{k+l+q  & p-l-q \\ k & p }}, 
\label{eq:Gamma_lambda1_PP0}
\eqa
and the vertex correction in the PH channel (Figs. \ref{fig:ph1} - \ref{fig:ph4}) is 
\begin{equation}
\begin{aligned}
 {\bf \Gamma}_{(1)PH}^{\spmqty{N_1 & N_2 \\ N_4 & N_3};\spmqty{\sigma_1 & \sigma_2 \\ \sigma_4 & \sigma_3}}
\spmqty{{\bf k}+{\bf l} & {\bf p}- {\bf l} \\  {\bf k} &  {\bf p}}
&=
\frac{1}{4 \mu N_f} \int \dd {\bf q} ~
g^{(\bar N_4)}_{k+q,k}
g^{(N_1)}_{k+l,k+l+q} 
D({\bf q}) 
G_{\bar{N}_1}({\bf k}+ {\bf l}+{\bf q})
G_{\bar{N}_4}({\bf k}+{\bf q})
    \mathsf{T}^{\alpha \sigma_1}_{\sigma_4 \beta} 
	  \lambda^{\spmqty{\bar N_1 & N_2 \\  \bar N_4 &  N_3};\spmqty{\beta & \sigma_2 \\ \alpha & \sigma_3}}_{\spmqty{k+l+q  & p-l \\ k+q & p }}
\\
&
+ \frac{1}{4 \mu N_f} \int \dd {\bf q} ~
g^{(\bar N_3)}_{p+q,p}
g^{(N_2)}_{p-l,p-l+q} 
D({\bf q}) 
G_{\bar{N}_3}({\bf p} + {\bf q})
G_{\bar{N}_2}({\bf p}- {\bf l}+{\bf q})
    \mathsf{T}^{\alpha \sigma_2}_{\sigma_3 \beta} 
	 	 \lambda^{\spmqty{N_1 & \bar N_2 \\  N_4 &  \bar N_3};\spmqty{\sigma_1 & \beta \\ \sigma_4 & \alpha}}_{\spmqty{k+l  & p-l+q \\ k & p+q }}
\\ &
+ \frac{1}{4 \mu N_f} \int \dd {\bf q} ~
g^{(\bar N_4)}_{k+q,k}
g^{(N_2)}_{p-l,p-l+q} 
D({\bf q}) 
G_{\bar{N}_4}({\bf k}+{\bf q})
G_{\bar{N}_2}({\bf p}- {\bf l}+{\bf q})
    \mathsf{T}^{\alpha \sigma_2}_{\sigma_4 \beta} 
 \lambda^{\spmqty{N_1 & \bar N_2 \\  \bar N_4 &   N_3};\spmqty{\sigma_1 & \beta \\ \alpha & \sigma_3}}_{\spmqty{k+l  & p-l+q \\ k+q & p }}
\\ &
+ \frac{1}{4 \mu N_f} \int \dd {\bf q} ~
g^{(\bar N_3)}_{p+q,p}
g^{(N_1)}_{k+l,k+l+q} 
D({\bf q}) 
G_{\bar{N}_1}({\bf k}+ {\bf l}+{\bf q})
G_{\bar{N}_3}({\bf p} + {\bf q})
    \mathsf{T}^{\alpha \sigma_1}_{\sigma_3 \beta} 
 \lambda^{\spmqty{\bar N_1 &  N_2 \\  N_4 &   \bar N_3};\spmqty{\beta & \sigma_2 \\ \sigma_4 & \alpha}}_{\spmqty{k+l+q  & p-l \\ k & p+q }}.
\end{aligned}
\label{eq:Gamma_lambda1_PH0}
\end{equation}

 \begin{figure*}
  	\centering
  	\begin{subfigure}[b]{0.45\linewidth}
  		\centering
  		\includegraphics[scale=1]{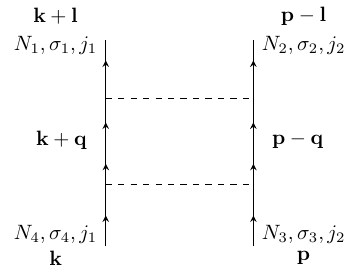}
  		\caption{\label{fig:ppBCS}}
  		\end{subfigure}\\
  		\begin{subfigure}[b]{\linewidth}
  		\centering
  		\includegraphics[scale=0.925]{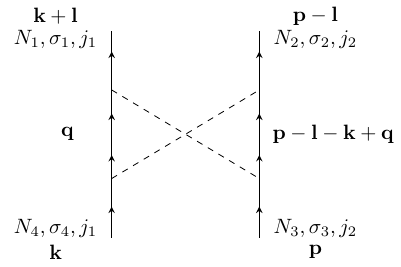}
  		\hspace{1.5cm}\includegraphics[scale=0.925]{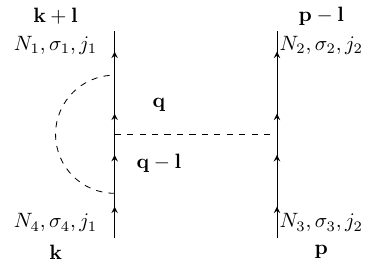}\vspace{1em}
  		\includegraphics[scale=0.925]{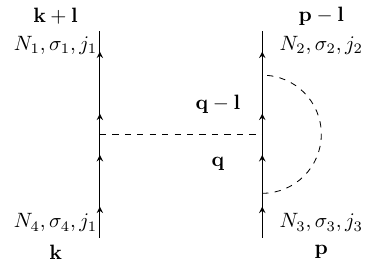}
  		\hspace{1cm}\includegraphics[scale=0.925]{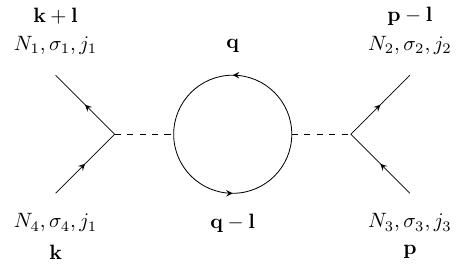}
   		\caption{\label{fig:phfer}}
  	\end{subfigure}
  	\caption{
	Quantum corrections to the four-fermion couplings quadratic in $\lambda$.
}
	\label{fig:4f2}
  	\end{figure*}

The four-fermion vertex that is quadratic in $\lambda$ 
is given by
$
{\bf \Gamma}_{(2)}^{\spmqty{N_1 & N_2 \\ N_4 & N_3};\spmqty{\sigma_1 & \sigma_2 \\ \sigma_4 & \sigma_3}}
=
{\bf \Gamma}_{(2)PP}^{\spmqty{N_1 & N_2 \\ N_4 & N_3};\spmqty{\sigma_1 & \sigma_2 \\ \sigma_4 & \sigma_3}}
+
{\bf \Gamma}_{(2)PH}^{\spmqty{N_1 & N_2 \\ N_4 & N_3};\spmqty{\sigma_1 & \sigma_2 \\ \sigma_4 & \sigma_3}}$,
where the vertex correction in the PP channel (Fig. \ref{fig:ppBCS}) 
is 
\begin{equation}
{\bf \Gamma}_{(2)PP}^{\spmqty{N_1 & N_2 \\ N_4 & N_3};\spmqty{\sigma_1 & \sigma_2 \\ \sigma_4 & \sigma_3}}
\spmqty{{\bf k}+{\bf l} & {\bf p}- {\bf l} \\  {\bf k} &  {\bf p}}
=  
-\frac{1}{8 \mu^2 }
 \int d {\bf q} ~
G_{M_1}({\bf k}+{\bf q})
G_{M_2}({\bf p}- {\bf q})
 \lambda^{\spmqty{N_1 &  N_2 \\  M_1 &   M_2};\spmqty{\sigma_1 & \sigma_2 \\ \beta & \alpha}}_{\spmqty{k+l  & p-l \\ k+q & p-q }}
 \lambda^{\spmqty{M_1 &  M_2 \\  N_4 &   N_3};\spmqty{\beta & \alpha \\ \sigma_4 & \sigma_3}}_{\spmqty{k+q  & p-q \\ k & p }}
 \label{eq:Gamma_lambda2_PP0}
\end{equation}
and  the vertex correction in the PH channel (Fig. \ref{fig:phfer}) is
\begin{equation}
\begin{aligned}
 {\bf \Gamma}_{(2)PH}^{\spmqty{N_1 & N_2 \\ N_4 & N_3};\spmqty{\sigma_1 & \sigma_2 \\ \sigma_4 & \sigma_3}}
\spmqty{{\bf k}+{\bf l} & {\bf p}- {\bf l} \\  {\bf k} &  {\bf p}}
=& 
-\frac{1}{8 \mu^2 }
 \int d {\bf q} ~
\left[
G_{M_1}({\bf q}-{\bf l}) G_{M_2}({\bf q}) 
\Biggl(
- N_f 
  \lambda^{\spmqty{N_1 &  M_1 \\  N_4 &   M_2};\spmqty{\sigma_1 & \alpha \\ \sigma_4 & \beta}}_{\spmqty{k+l  & q-l \\ k & q }}
 \lambda^{\spmqty{M_2 & N_2 \\  M_1 &   N_3};\spmqty{\beta & \sigma_2 \\ \alpha & \sigma_3}}_{\spmqty{q  & p-l \\ q-l & p }}
 \right.
 \\ &  
 + 
   \lambda^{\spmqty{N_1 &  M_1 \\  N_4 &   M_2};\spmqty{\sigma_1 & \alpha \\ \sigma_4 & \beta}}_{\spmqty{k+l  & q-l \\ k & q }}
 \lambda^{\spmqty{M_2 & N_2 \\  N_3 &   M_1};\spmqty{\beta & \sigma_2 \\ \sigma_3 & \alpha}}_{\spmqty{q  & p-l \\ p & q-l }}
 + 
   \lambda^{\spmqty{N_1 &  M_1 \\  M_2 &   N_4};\spmqty{\sigma_1 & \alpha \\ \beta & \sigma_4}}_{\spmqty{k+l  & q-l \\ q & k }}
 \lambda^{\spmqty{M_2 & N_2 \\  M_1 &   N_3};\spmqty{\beta & \sigma_2 \\ \alpha & \sigma_3}}_{\spmqty{q  & p-l \\ q-l & p }}
 \Biggr)
\\ &
 + 
 \left.
 G_{M_1}({\bf q}-{\bf l}+{\bf p}-{\bf k}) G_{M_2}({\bf q}) 
    \lambda^{\spmqty{N_1 &  M_1 \\  M_2 &   N_3};\spmqty{\sigma_1 & \alpha \\ \beta & \sigma_3}}_{\spmqty{k+l  & q-l+p-k \\ q & p }}
 \lambda^{\spmqty{M_2 & N_2 \\  N_4 &   M_1};\spmqty{\beta & \sigma_2 \\ \sigma_4 & \alpha}}_{\spmqty{q  & p-l \\ k & q-l+p-k }}
 \right].
\end{aligned}
\label{eq:Gamma_lambda2_PH0}
\end{equation}

\subsection{Space of IR singularity}
\label{sec:spaceofIRsingularity}

\begin{figure}[h]
	\centering
	\includegraphics[scale=1]{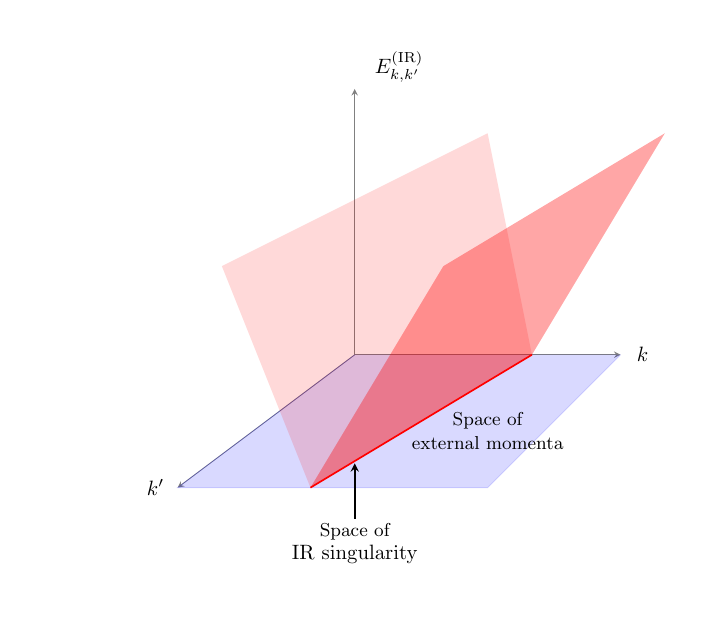}
	\caption{
Schematic diagram of a momentum dependent crossover scale defined in the space of external momenta of a diagram.
The set of external momenta at which the crossover scale vanishes forms the space of IR singularity for the diagram.
Away from the space of IR singularity, the crossover scale becomes non-zero, 
causing a crossover 
from the high-energy region in which the quantum correction is significant 
to the low-energy region in which the quantum correction turns off.
}
\label{fig:space_of_IR_singularity}
\end{figure}

The renormalization of the coupling functions
is determined from quantum corrections
that depend on momenta along the Fermi surface.
In each diagram, 
the momentum dependence of the quantum correction
is controlled by two crucial pieces of information.
The first is {\it the space of IR singularity},
the set of external momenta at which
a diagram is singular in the zero energy limit.  
The second is {\it the momentum dependent crossover scale} 
that cuts off the IR singularity 
when external momenta are away from the space of IR singularity.

Suppose there is a diagram that contributes 
to the vertex function,
where all external electrons are on the Fermi surface.
In the space of external momenta allowed by the momentum conservation,
there exists a subset of external momenta
in which the diagram exhibits an IR singularity
in the limit that all external frequencies become zero.
This subset is referred to as the space of IR singularity for the diagram,
and its dimension is denoted as $d_s$. 
If external momenta lie within the space of IR singularity,
the quantum correction exhibits IR divergence with a strength
determined from the kinematics of virtual particles created within the loop\footnote{
In an $L$-loop diagram, the space of internal three-momenta  
is $3L$-dimensional. 
If there exists a sub-manifold of co-dimension $y$
in which $x$ internal particles  can have zero energy simultaneously, 
the diagram exhibits an IR singularity with degree $y-x$.
If $y=x$ ($x>y$), a logarithmic (power-law) IR divergence can arise.
If $x<y$, there is no IR singularity.}.
If external momenta are outside of the space of IR singularity,
a non-zero energy scale cuts off the IR divergence.
The crossover energy scale is determined 
from the minimum energy that 
internal particles have to carry 
for given external momenta.
At energies above the crossover energy scale,
the quantum correction renormalizes 
the coupling functions 
even if the external momenta are 
outside the space of IR singularity.
At energies below the crossover scale,  
the quantum correction becomes essentially independent of the energy scale,
and the coupling functions 
stop receiving renormalization from the diagram.
This is illustrated in \fig{fig:space_of_IR_singularity}.
In this subsection, we  discuss the space of IR singularity for 
each quantum correction in more detail.

\subsubsection{Fermion self-energy}

An electron has zero energy 
anywhere on the one-dimensional Fermi surface.
The space of IR singularity for a fermion self-energy diagram 
is a subset of the one-dimensional Fermi surface
in which the diagram is singular in the low-energy limit.
To the leading order in $v$, only the diagrams 
in Figs. \ref{fig:1LFSE} and \ref{fig:2LSE} are important.
To be concrete, we consider the self-energy of electron at hot spot $1$.
With the frequency of the external electron set to be zero,
we would like to figure out the set of $k_x$ at which
$ \left.  \frac{ \partial {\bf \Sigma}_1({\bf k}) }{ \partial k_\rho} \right|_{{\bf k}=(0, k_x, -v_{k_x} k_x)}$ 
is IR divergent.
Let us first consider the one-loop self-energy in  \eq{eq:1LoopFSE_full}. 
If the external electron is at the hot spot,
the electron can be scattered right onto the hot spot $4$ by emitting a boson with zero energy.
Because all internal particles can have zero energy at a loop momentum, 
a logarithmic singularity arises.
If the external electron is away from the hot spots, 
there is no choice of loop momentum
at which both electron and boson 
have zero energy in the loop,
which removes the logarithmic singularity.
The same conclusion holds 
for the two-loop self-energy diagrams.
Therefore, the space of IR singularity 
for the fermion self-energy
is the set of hot spots 
with dimension $d_s=0$.
Away from the hot spots, 
the IR singularity is cut off by a momentum dependent scale
that is proportional to $k_x$\footnote{
We will derive the crossover scale in  
Sec. \ref{sec:beta_functionals}
and Appendix 
\ref{app:QC}.
}.
This  means that electrons away from the hot spots
are eventually decoupled from spin fluctuations at sufficiently low energies.

\subsubsection{Electron-boson vertex correction}

Next, let us consider the cubic vertex where an electron
at momentum $(k_x, -v_{k_x} k_x)$ in hot spot $1$
is scattered to  $(k_x', v_{k_x'} k_x')$ in hot spot $4$
(\fig{fig:yukawa}).
In \eq{eq:1LY_full}, the energies of 
the boson and two internal electrons in the loop are given by
 $c \left( |q_x|+|q_y| \right)$,
 $V_{F,k_x+q_x} \left( v_{k_x+q_x}(k_x + q_x) - q_y + v_{k_x} k_x \right)$, 
  and
$V_{F,k_x'+q_x} \left( v_{k_x'+q_x}(k_x' + q_x) + q_y + v_{k_x'} k_x' \right)$, respectively.
As is the case for the self-energy,
all internal particles can have zero energy at $\vec q=0$ if $k_x = k_x'=0$.
This gives rise to the logarithmic IR singularity. \footnote{
This is because the product of the three propagators have IR singularity 
with dimension $-3$ at ${\bf  q}=0$,
which is a subspace of co-dimension $3$ in the space of internal energy-momentum.}
For non-zero $k_x, k_x'$, 
it is impossible to put all internal particles at zero energy,
and the logarithmic singularity disappears. 
The expressions for the IR energy cutoff scales are derived in  
Sec. \ref{sec:beta_functionals}
and Appendix \ref{app:QC}.
Since the vertex correction is singular at zero energy
only when both the incoming and outgoing electrons are at the hot spots,
we have $d_s=0$.
Away from the hot spots,
the quantum corrections `turn off' 
below the crossover energy scale.

\subsubsection{ Four-fermion vertex correction}

\begin{figure}[t]
\centering
\includegraphics[scale=1]{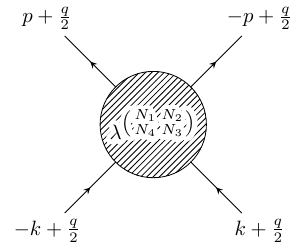}
\caption{
The momentum dependent four-fermion coupling function. 
$q$ denotes the center of momentum in the particle-particle (PP) channel.
$p-k$ and $p+k$ denote the center of mass momenta in two particle-hole (PH) channels.
The gapless spin fluctuations generate singular four-fermion couplings 
in the PP plane with $q=0$ and the PH plane with $p-k=0$.
}
\label{fig:otherlabel}
\end{figure}

At the one-loop order,
there are three types of quantum corrections
that contribute to the beta functional of the four-fermion couplings.
The first is the quantum correction
generated from the Yukawa coupling 
as is shown in \fig{fig:4f0}.
These give rise to contributions that are independent of $\lambda$ in the beta functional,
and act as the sources for the four-fermion coupling.
The second, shown in \fig{fig:4f1}, is the one that 
describes mixing among four-fermion couplings in different channels.
Finally, diagrams in \fig{fig:4f2}
describe the BCS-like scatterings.
Here, we examine the spaces of IR singularity in these quantum corrections.

\paragraph{ The primary couplings }

Through Eqs.  (\ref{eq:Gamma_lambda0_PP})
and (\ref{eq:Gamma_lambda0_PH}),
the primary four-fermion couplings are generated in channel
$\spmqty{N & M \\ N & M}$ for any $N$ and $M$.
Without loss of generality, we can focus on the case with $N=1$
because all other channels are related to the one with $N=1$ 
through the $C_4$ symmetry.
Let us start with the primary couplings with
$(N,M)=(1,1)$ and $(1, 5)$
in Table \ref{table:PrimarySecondary}.
In these channels,
the Fermi velocities of the two internal fermions 
of \fig{fig:4f0} are parallel or anti-parallel to each other at the hot spots.
As a result, there exist channels in which 
a pair of fermions within the loop 
can be far away from the hot spots 
while staying arbitrarily close to the Fermi surface.
Contributions from states far away from the hot spots 
are only suppressed by the energy cost of the boson. 
Since the speed of the boson is $c$,
the phase space of the low-energy states  
becomes proportional to $1 / c$.
Consequently, the quantum correction is enhanced from $g^4$ to $g^4/c$.
For $(N,M)=(1,4)$ and $(1,8)$, 
the patches of Fermi surface are not perfectly nested,
and the virtual electronic excitations in the loop 
can not stay on the Fermi surface at large momentum.
Nonetheless, the main energy penalty for creating virtual excitations 
far away from the hot spots still comes from the boson
because the nesting angle $v$ is smaller than $c$ in the small $v$ limit as is shown in \eq{eq:CofV}.
Since the phase space of low-energy states is still controlled by the speed of the boson,
the diagrams are still enhanced by $1/c$.
For the primary couplings in group 3
in Table \ref{table:PrimarySecondary},
the Fermi velocities of the two internal fermions 
are almost perpendicular to each other.
In this case, the Fermi velocity,
which is order of $1$,
controls the phase space of virtual electronic excitations,
and the quantum corrections are simply order of $g^4$. 
Therefore, we can ignore the couplings in group 3
to the leading order in $v$.
The couplings in group $4$ 
are not generated from spin fluctuations.

When all external fermions are at the hot spots in \fig{fig:4f0},
there exists a choice of the loop momentum 
at which all four internal particles (two fermions and two bosons) have zero energy.
Because four internal particles have zero three-momentum
at the origin (the manifold of co-dimension three)
in the space of loop energy-momentum, 
it exhibits an IR singularity with degree $-1$.
However, this power-law IR divergence is cut off 
as soon as
any of the external momenta becomes non-zero.
If an external electron is away from the hot spots,
the electron has to absorb or emit a boson with non-zero momentum 
to scatter onto the Fermi surface inside the loop.
Alternatively absorbing or emitting a boson with zero momentum (relative to $\vec Q_{AF}$), a virtual electron has to be away 
from the Fermi surface.
Since the power-law divergence is removed 
for any external momentum away from the hot spots,
the space of IR singularity with degree $-1$ is only zero-dimensional ($d_s=0$).
Under the extended minimal subtraction scheme,
no counter term is needed for 
the IR divergence with degree $-1$
localized within a zero-dimensional manifold 
of external momenta
because \eq{eq:stringentF5} remains finite 
in the small $\mu$ limit.
Although the IR singularity with degree $-1$ is localized 
within the zero-dimensional manifold of external momenta,
IR singularities with degree $0$ (logarithmic divergence)
can arise in an extended space.
These are the quantum corrections
for which counter terms are needed.
Below, we identify the channels 
in groups 1 and 2
for which logarithmic IR singularities arise 
in spaces with $d_s \geq 1$.

In the $\spmqty{1 & 4 \\ 1 & 4}$ channel,
there is no space of IR singularity with $d_s > 0$.
A logarithmic IR divergence can arise only 
if there exists a choice of internal momentum at which
both electrons and at least one boson in the loop have zero energy
in \fig{fig:4f0}\footnote{
The case in which two bosons and only one electron
have zero energy does not 
give rise to an IR singularity 
because the electron propagator is odd under ${\bf k} \rightarrow -{\bf k}$.
}.
This forces a pair of external momenta in hot spots $1$ and $4$ to be zero.
The fact that the patches at hot spots $1$ and $4$ are not nested with each other\footnote{
It is important to consider a small but non-zero $v$.}, 
combined with the constraint that external electrons are on the Fermi surface,
further forces the other two external momenta to be zero as well.
This shows that there is no extended space of IR singularity
in the $\spmqty{1 & 4 \\ 1 & 4}$ channel.
The exact same argument applies to the  $\spmqty{1 & 8 \\ 1 & 8}$ channel.

\begin{figure*}
	\centering
		\begin{subfigure}[b]{0.4\linewidth}
		\centering
		\includegraphics[scale=0.75]{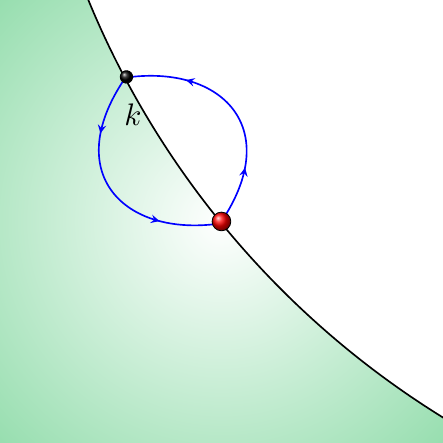}
		\caption{\label{fig:4f0_1111_b}}
	\end{subfigure}
	\begin{subfigure}[b]{0.4\linewidth}
		\centering
		\includegraphics[scale=0.75]{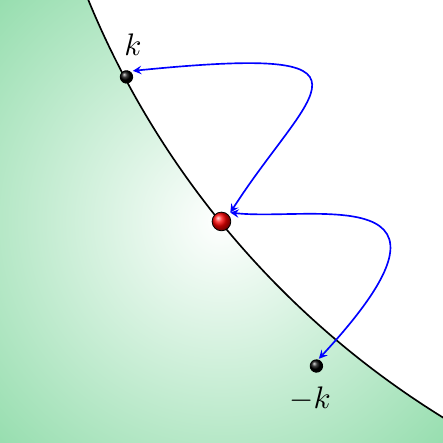}
		\caption{\label{fig:4f0_1111_a}} \end{subfigure} \caption{ \textcolor{blue}{(a)}
	The forward scattering between  an electron at a hot spot and an electron away from the hot spot. Irrespective of the shape of the Fermi surface, the forward scattering receives singular quantum corrections. \textcolor{blue}{(b)} The Fulde–Ferrell–Larkin–Ovchinnikov (FFLO) scattering where a pair of electrons are scattered in and out of a hot spot. The FFLO scattering remains singular at low energies only if the Fermi surface has the particle-hole symmetry, for example,  if the Fermi surface is straight. 
	In general, two electrons from a hot spot can not scatter onto the Fermi surface due to the curvature of Fermi surface. 
	} \label{fig:4f0_1111} 
\end{figure*}

In contrast, the quartic coupling 
in the $\spmqty{1 & 1 \\ 1 & 1}$ channel
supports an extended space of IR singularity.
\fig{fig:s2} is logarithmically divergent  
in the $\spmqty{1 & 1 \\ 1 & 1}$ channel
as far as a pair of external electron and hole
are at the hot spots
even if the other pair are away from the hot spots.
Because there exists a choice of loop momentum at which
two fermions and one boson in the loop have zero energy
when one external electron-hole pair are at the hot spot,
it gives rise to the logarithmic singularity.
The space of IR singularity is one-dimensional ($d_s=1$)
because the other external momenta are arbitrary. 
The \ffc in the one-dimensional space of IR singularity
is parameterized as
\bqa
\left\{
\lambda^{\spmqty{1 & 1 \\ 1 & 1}}_{\spmqty{0 & k \\ k & 0}},
\lambda^{\spmqty{1 & 1 \\ 1 & 1}}_{\spmqty{k & 0 \\ 0 & k}}
\right\}.
\label{eq:SofIS1}
\eqa
This coupling describes the forward scattering
between an electron at the hot spot and an electron  at a general momentum on the Fermi surface
(\fig{fig:4f0_1111_b}).
%


\begin{figure*}
	\centering
	\begin{subfigure}[b]{0.24\linewidth}
		\centering
		\includegraphics[scale=0.8]{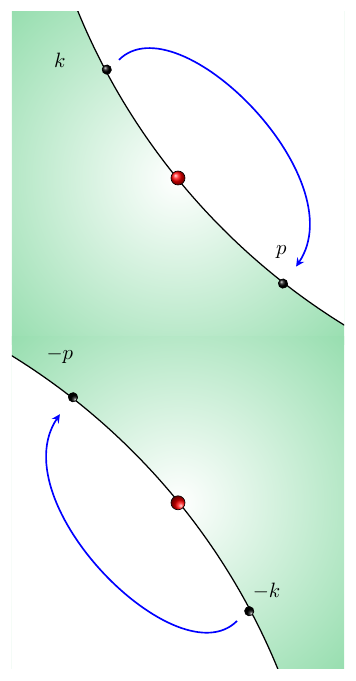}
		\caption{\label{fig:4f0_1515_a}}
	\end{subfigure}
	\hspace{1cm}
	\begin{subfigure}[b]{0.24\linewidth}
		\centering
		\includegraphics[scale=0.8]{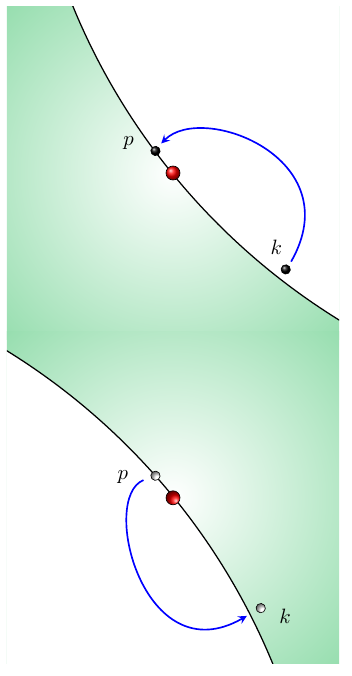}
		\caption{\label{fig:4f0_1515_b}}
	\end{subfigure}
	\caption{
	\textcolor{blue}{(a)}
The Bardeen-Cooper-Schrieffer (BCS) scattering between a pair of electrons with zero center of mass momentum.
Irrespective of the shape of the Fermi surface, the BCS scattering receives  singular quantum corrections.
\textcolor{blue}{(b)}
The $2k_F$ particle-hole scattering associated with a pair of particle and hole on the anti-podal patches of Fermi susrface.
The $2k_F$ particle-hole scattering remains singular at low energies only if the Fermi surface has the particle-hole symmetry.
In general, a non-zero curvature of Fermi surface prevents particle-hole pairs with fixed momentum from staying on the Fermi surface.
}
	\label{fig:4f0_1515}
\end{figure*}

The space of IR singularity is even bigger 
for the $\spmqty{1 & 5 \\ 1 & 5}$ channel in group $2$.
As far as the center of mass momentum of the electron pair is zero
in \fig{fig:s1},
the quantum correction is logarithmically divergent
irrespective of the relative momentum of the 
incoming {\it and} outgoing pairs.
When the total momentum of the electron pair is zero,
two internal electrons in the loop
can stay close to the Fermi surface 
irrespective of the relative momentum along the Fermi surface.
Since the two internal fermions can have zero energy 
within the manifold of internal energy-momentum with co-dimension $2$,
it gives rise to the logarithmic IR singularity.
It is noted that this IR singularity is generated purely from gapless fermions,
and whether the boson is gapless or not does not matter.
Since the relative momentum of incoming and outgoing fermion pairs
can be arbitrary, $d_s=2$.
The \ffc  in the two-dimensional space of IR singularity
is parameterized as
\bqa
\left\{
\lambda^{\spmqty{1 & 5 \\ 1 & 5}}_{\spmqty{p & -p \\ k & -k}}
\right\}.
\label{eq:SofIS2}
\eqa
This coupling describes the BCS pairing interaction for pairs of electrons
with zero total momentum  (\fig{fig:4f0_1515_a}).  

In summary, 
\fig{fig:4f0} generates primary four-fermion couplings 
within extended spaces of external momenta with $d_s >0$
in groups 1 and 2 to the leading order in $v$.
In the following, we consider the couplings that are further 
generated from the primary couplings through operator mixing.

\paragraph{ The secondary couplings  }

\begin{figure*}
	\centering
		\includegraphics[scale=0.75]{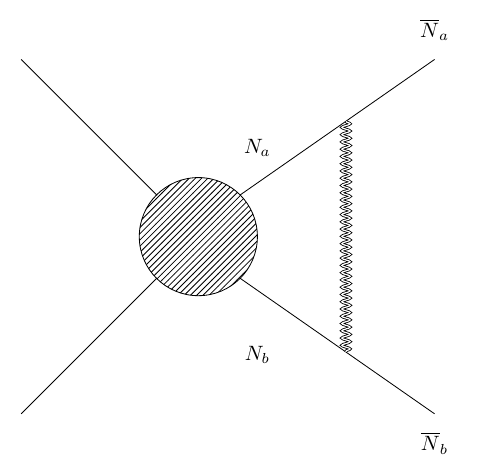}
	\caption{
The quantum correction that linearly mixes four-fermion couplings with different hot spot indices to the lowest order in $v$.
A pair of fermions can change their hot spot indices from $(N_a, N_b)$ 
to  $(\bar N_a, \bar N_b)$ 
by exchanging a boson.
}
	\label{fig:NiNjbar}
\end{figure*}

\begingroup
\renewcommand*\arraystretch{3.75}
\begin{table*}
	\centering
	\begin{tabular}{|c|c|c|c|}
		\hline
&  $d_s$ &  Primary  couplings & Secondary couplings    \\
		\hline
Group 1 & $1$ & 
$\lambda^{\spmqty{1 & 1 \\ 1 & 1}}_{\spmqty{0 & k \\ k & 0}}$,
$\lambda^{\spmqty{1 & 1 \\ 1 & 1}}_{\spmqty{k & 0 \\ 0 & k}}$
&
$\lambda^{\spmqty{4 & 1 \\ 1 & 4}}_{\spmqty{0 & k \\ k & 0}}$,
$\lambda^{\spmqty{1 & 4 \\ 4 & 1}}_{\spmqty{k & 0 \\ 0 & k}}$ 
\\
 		\hline	
Group 2  &  $2$ & 
$\lambda^{\spmqty{1 & 5 \\ 1 & 5}}_{\spmqty{p & -p \\ k & -k}}$, 
$\lambda^{\spmqty{4 & 8 \\ 4 & 8}}_{\spmqty{p & -p \\ k & -k}}$ 
&
$\lambda^{\spmqty{4 & 5 \\ 1 & 8}}_{\spmqty{p & -p \\ k & -k}}$, 
$\lambda^{\spmqty{1 & 8 \\ 4 & 5}}_{\spmqty{p & -p \\ k & -k}}$  \\
\hline
	\end{tabular}
	\caption{
The primary four-fermion couplings generated from the Yukawa coupling
and the secondary couplings further generated 
through the linear mixings 
in the spaces of IR singularity with dimensions $d_s \geq 1$. 
}
		\label{table:mixing1}
\end{table*}
\endgroup

%
%

Once the Yukawa coupling generates the primary four-fermion couplings 
in the $\spmqty{1 & 1 \\ 1 & 1}$ and  $\spmqty{1 & 5 \\ 1 & 5}$ channels,
the vertex corrections in
\fig{fig:4f1}
generate secondary couplings 
by scattering a pair of electrons in the PP 
and PH channels, respectively.
For example, a pair of indices 
in $\lambda^{\spmqty{N_1 & N_2 \\ N_4 & N_3}}_{\spmqty{k_1 & k_2 \\ k_4 & k_3}}$
can change from
$(N_a, N_b)$ 
to 
$(\bar N_a, \bar N_b)$ through the one-loop mixing
as is shown in 	\fig{fig:NiNjbar}.
The resulting operators can generate 
yet another set of operators through mixing. 
This in general creates a network of operators
described by a mixing matrix in the space of 
hot spot indices, spin indices and momentum.
However, we only need to focus on those channels 
in which operator mixings are present 
within the space of IR singularity with $d_s > 0$.
We don't need to add counter terms 
for the mixings that are present 
in zero-dimensional space of IR singularity.
The secondary couplings that are generated 
within the extended spaces of IR singularity 
are summarized  in Table \ref{table:mixing1}.

In the $\spmqty{1 & 1 \\ 1 & 1}$ channel,
the quantum correction is IR singular 
only when the external legs that are associated with the vertex correction
carry zero momenta.
Each of
$\lambda^{\spmqty{1 & 1 \\ 1 & 1}}_{\spmqty{0 & k \\ k & 0}}$ and
$\lambda^{\spmqty{1 & 1 \\ 1 & 1}}_{\spmqty{k & 0 \\ 0 & k}}$
mixes with
$\lambda^{\spmqty{4 & 1 \\ 1 & 4}}_{\spmqty{0 & k \\ k & 0}}$ and
$\lambda^{\spmqty{1 & 4 \\ 4 & 1}}_{\spmqty{k & 0 \\ 0 & k}}$, 
respectively.
The linear mixing is generated from Figs. \ref{fig:ph3} and \ref{fig:ph4}.
In the $\spmqty{1 & 5 \\ 1 & 5}$ channel,
the vertex correction is IR singular within
a two-dimensional manifold   
in which a particle-particle pair 
in hot spots $1$ and $5$ carry net zero momenta.
As a result, 
$\lambda^{\spmqty{1 & 5 \\ 1 & 5}}_{\spmqty{p & -p \\ k & -k}}$
mixes with
$\lambda^{\spmqty{4 & 8 \\ 1 & 5}}_{\spmqty{p' & -p' \\ k & -k}}$,
$\lambda^{\spmqty{1 & 5 \\ 4 & 8}}_{\spmqty{p & -p \\ k' & -k'}}$
through Figs. \ref{fig:pp1} and \ref{fig:pp2}.


At the quadratic order in $\lambda$, 
the standard BCS diagram
give rise to the logarithmic divergence
for the coupling in group 2
within the two-dimensional manifold of external momenta
in which the center of mass momentum of Cooper pair is zero.

\paragraph{Additional channels that become singular 
in the presence of the particle-hole symmetry}

In the presence of the PH symmetry,  $v_k = v_{-k}$.
In this case, the phase space of low-energy scatterings is further enlarged due to an enhanced nesting,
and additional couplings receive singular quantum corrections 
in extended spaces of IR singularity.
Although we focus on the generic case 
in which the PH symmetry is absent,
here we list those additional couplings for completeness. 

With $v_k = v_{-k}$, a pair of electrons with momenta $k$ and $-k$ 
can simultaneously stay on the Fermi surface near one hot spot.
As a result, \fig{fig:s1} is also logarithmically divergent 
in the $\spmqty{1 & 1 \\ 1 & 1}$ channel.
The coupling function that receives IR singular quantum corrections
is parameterized as
\bqa
\left\{
\lambda^{\spmqty{1 & 1 \\ 1 & 1}}_{\spmqty{0 & 0 \\ k & -k}},
\lambda^{\spmqty{1 & 1 \\ 1 & 1}}_{\spmqty{k & -k \\ 0 & 0}}
\right\}.
\label{eq:SofIS1_2}
\eqa
This coupling describes the processes where
a pair of electrons with total momentum $2k_F$
scatter in and out of the hot spots
 in the Fulde–Ferrell–Larkin–Ovchinnikov (FFLO) pairing channel (\fig{fig:4f0_1111_a}).
In group 2,
\fig{fig:s2} is also IR singularity for
\bqa
\left\{
\lambda^{\spmqty{1 & 5 \\ 1 & 5}}_{\spmqty{p & k \\ k & p}}
\right\}.
\label{eq:SofIS2_2}
\eqa
These couplings are defined in the two-dimensional space with zero center of mass momentum
in the PH channel.
\eq{eq:SofIS2_2} describes scatterings 
of particle-hole pairs with momentum 
$2k_F$ (\fig{fig:4f0_1515_b}). 
In the presence of the PH symmetry,
this plane of the IR singularity
intersects with \eq{eq:SofIS2}
at a line with zero center of mass momentum 
both in the PP
and PH channels.

Once those additional primary couplings are generated,
additional secondary couplings are generated.
In group 1,
$\lambda^{\spmqty{1 & 1 \\ 1 & 1}}_{\spmqty{0 & 0 \\ k & -k}}$,
$\lambda^{\spmqty{1 & 1 \\ 1 & 1}}_{\spmqty{k & -k \\ 0 & 0}}$,
mixes with
$\lambda^{\spmqty{4 & 4 \\ 1 & 1}}_{\spmqty{0 & 0 \\ k & -k}}$,
$\lambda^{\spmqty{1 & 1 \\ 4 & 4}}_{\spmqty{k & -k \\ 0 & 0}}$
through Figs. \ref{fig:pp1} and \ref{fig:pp2}.
In group 2,
$\lambda^{\spmqty{1 & 5 \\ 1 & 5}}_{\spmqty{k & p \\ p & k}}$
mixes with
$\lambda^{\spmqty{4 & 5 \\ 1 & 8}}_{\spmqty{k' & p \\ p & k'}}$
and
$\lambda^{\spmqty{1 & 8 \\ 4 & 5}}_{\spmqty{k & p' \\ p' & k}}$
through Figs. \ref{fig:ph3} and \ref{fig:ph4}.


\subsection{Adiabaticity}
\label{sec:adiabatic}

Now we turn our attention to the computation of 
the quantum effective action
expressed as integrations of loop momenta in
Sec. \ref{sec:QEA}.
The salient feature of theories with 
continuously many gapless degrees of freedom 
is that quantum corrections
are functionals of coupling functions.
Even if the coupling functions are independent of momentum at one scale,
they in general acquire non-trivial momentum dependences at low energies unless protected by symmetry.
This makes it necessary to 
compute quantum corrections
in the presence of general momentum dependent coupling functions.
A simplification arises for quantum corrections generated from small angle scatterings 
in which
loop momentum is bounded by
the external energy.
Thanks to the locality in real space, 
the rate at which coupling functions vary in momentum is controlled by the energy scale at which the coupling functions are defined.
If the momentum carried by virtual particles in a loop is limited by a  momentum that is proportional to external frequencies,
quantum corrections can be computed approximately
by treating the coupling functions as constants as far as the coupling functions do not vary significantly over that momentum scale.
This motivates us to 
introduce the notion of {\it adiabaticity}
in the momentum space.
We say that coupling functions are adiabatic in a diagram 
if no coupling functions change appreciably 
within the range of loop momenta
from which IR divergent contributions arise.
Below, we make this precise and examine how adiabaticity is invoked to efficiently compute
 quantum corrections for non-nested diagrams.

We start with the Schwinger-Dyson equation 
for the boson propagator in \eq{eq:fullDyson}. 
At energy scale $\mu$,
we need to know the boson propagator 
up to momentum $\mu/c$,
where $c$ is the speed of the boson.
After the mass renormalization is subtracted,
\eq{eq:fullDyson} is finite
in the limit that $k_F$ and $\Lambda$ are large.
With all momentum cutoffs set to be infinite, the external momentum is the only scale in the integration, and 
it plays the role of a soft UV energy cutoff
for the loop integration.
When the external momentum is $\mu/c$,
the upper bound 
for the soft UV energy cutoff is $V_{F,0} \mu/c$.
This is because 
the singular renormalization of the collective mode arises from electrons near the hot spots,
and $V_{F,0}$ is the largest component of velocity near the hot spots.
This energy cutoff is translated into 
the upper bound for the momentum cutoff,
\bqa
\Pi_\mu = 
\frac{1}{V_{F,0} v_0 }
\frac{V_{F,0}\mu}{c}
=
\frac{\mu}{c v_0},
\eqa
where we use the fact that $V_{F,0} v_0$ is the smallest component of velocity\footnote{For example,
with energy  $V_{F,0} \mu/c$,
electron-hole pairs
can be created near hot spots $1$ and $4$ 
up to momentum $k_x \sim \frac{\mu}{v_0 c}$
away from the hot spots. }.
%
%
We say the coupling functions satisfy adiabaticity at energy scale $\mu$ 
if the relative variation of $v_k, V_{F,k}$ and $g_{k',k}$ 
is small within the range of momentum $\Pi_\mu$, that is,
\begin{align}
\epsilon_\mu 
\equiv \max\limits_{
|k_i'-k_i|< \Pi_\mu
}
\Bigg\{
\left| 
\frac{ V_{F,k_1'} - V_{F,k_1} }{V_{F,k_1}} \right|,
\left| 
\frac{ v_{k_1'} - v_{k_1} }{v_{k_1}} \right|,
\left| 
\frac{ g_{k_1',k_2'} - g_{k_1,k_2} }{g_{k_1,k_2}} \right|
\Bigg\}
\ll 1,
\label{eq:adiabatic_condition}
\end{align}
where all couplings are defined at energy $\mu$.
Here, we don't expect the four-fermion coupling function satisfies adiabaticity 
 at general momenta because the four-fermion coupling function, being irrelevant, has stronger momentum dependence than the marginal coupling functions. 
In Sec. \ref{sec:fixedpoint},
we will show that 
if  $\epsilon_\Lambda \ll 1$\footnote{This is  a reasonable starting point in that bare coupling functions  are smooth functions of momentum},
$\epsilon_\mu \ll 1$ at all $\mu \leq \Lambda$.
For now, we assume that the adiabaticity is satisfied at all energy scales and show how it simplifies the computation of quantum corrections.
With \eq{eq:adiabatic_condition},
one can ignore the variation of the coupling functions within the loop,
and \eq{eq:fullDyson} can be approximated as
\begin{equation}
\begin{aligned}
& D^{-1}({\bf k}) = m_{CT}
+ 2 g_{0,0}^2  \sum_{N=1}^8
   \int \dd {\bf q} ~
   G_{N}({\bf q} | 0)  G_{\bar N}({\bf q} +{\bf k}| 0) \\
 &  
- \frac{4}{N_c N_f}
g_{0,0}^4
\sum^{8}_{N=1}
\int \dd {\bf q} \dd {\bf p}~
G_N({\bf p}+{\bf q}|0) G_{\bar N}({\bf p} | 0) G_N({\bf p} +{\bf k}| 0) G_{\bar N}({\bf p} + {\bf q} + {\bf k}| 0) D({\bf q}),
\end{aligned}
\label{eq:fullDyson_adia}
\end{equation}
where the coupling functions are evaluated at the hot spots, and
\bqa
G_{N}({\bf k} | k') = \frac{1}{ik_0+V^{(N)}_{F,k'}e_{N}[\vec{k};v^{(N)}_{k'}]}
\eqa
is the fermion propagator at ${\bf k}$
in which the coupling functions
inside the propagator
are evaluated at $k'$.
As in \eq{eq:CofV},
the self-consistent equation 
in \eq{eq:fullDyson_adia} gives 
\bqa
c(v_0) = \sqrt{\frac{v_0}{8N_c N_f}\log\left(\frac{1}{v_0}\right)},
\label{eq:CofV2} 
\eqa
where $v_0$ denotes the nesting angle at the hot spots.
Provided  that \eq{eq:adiabatic_condition}  is satisfied at energy scale $\mu$,  
\eq{eq:BosProp} is the  valid expression up to momentum $\mu/c$ with
the boson velocity given by 
\eq{eq:CofV2}.

Now, let us consider other vertex functions.
Unlike the boson propagator,
other diagrams that contribute 
to the counter terms in \eq{eq:CTAction}
are in general UV divergent logarithmically 
in $\Lambda$, $k_F$ or both.
This implies that internal momenta in those diagrams
can be much bigger than $\mu$,
and we can not use adiabaticity to compute
the quantum corrections. 
However, what we need is the derivative of the counter terms with respect to $\log \mu$ not the counter terms themselves.
The derivatives capture the contributions generated 
from the fast modes 
within an infinitesimal window of energy scales,
and determine the beta functionals.
Because $\Lambda$ is the UV energy cutoff,
the derivative of the counter terms
are finite in the large $\Lambda$ limit.
However, they are not necessarily finite in the large $k_F$ limit 
as there are gapless fermionic modes with large momenta.

Therefore, we are led to consider 
two types of UV divergent quantum corrections separately.
The quantum corrections of the first type are those whose derivative with respect to $\log \mu$ are finite 
in the large $k_F$ limit. 
Quantum corrections that are generated by small-angle scatterings belong to this type.
For example, in diagrams that involve only the Yukawa couplings,
spin fluctuations scatter electrons 
between patches that are not nested with each other 
for $v \neq 0$.
Consequently, internal fermions can not be scattered
far away from  external momenta modulo $\vec Q_{AF}$.
In the derivative of quantum corrections,
all components of internal momenta become 
dynamically bounded by the external energy,
and its contribution to the beta functional is finite even in the large $k_F$ limit.
Their contributions 
to the beta functional can be computed by treating the coupling functions inside loops 
 as constants as far as the adiabaticity condition is satisfied.
The quantum corrections of the second type  are the ones whose 
derivative with respect to $\log \mu$ 
are not finite in the large $k_F$ limit 
while being finite in the large $\Lambda$ limit.
The second type of quantum corrections
arise from large-angle scatterings
in the channels in which fermions can stay close
to the Fermi surface over an extended phase space
that is not bounded by energy.
For example, Cooper pairs with zero net momentum 
can be created far away from the hot spots 
without electronic energy penalty,
and the phase space for virtual low-energy excitations is not limited by the external energy.
In those cases, even the derivative of the counter terms remain sensitive to the size of the Fermi surface, and
we can not ignore the momentum profiles 
of the coupling functions within loops.
For the second type, the beta functionals should be expressed 
as integrations over the momenta along 
the Fermi surface. 
We will come back to the second type of quantum corrections when we explicitly compute quantum corrections that include
the four-fermion couplings in the nested channels
in Sec. \ref{sec:fullbetalambda}.
In the rest of the section, 
we discuss how adiabaticity is invoked 
to simplify the computation of quantum corrections
of the first type.

Let us start with 
the contributions of $A^{(i)}(k)$ with $i=1,2,3$ to the beta functionals.
Let ${\bf \Sigma}_1({\bf k})$ be the exact fermion-self energy 
in \eq{eq:1LoopFSE_full} at hot spot $1$.
While 
$\frac{\partial {\bf \Sigma}_1({\bf k}) }{\partial k_\rho}$
is UV divergent logarithmically,
$
\frac{\partial }{\partial \log \mu}
\left.
\frac{\partial {\bf \Sigma}_1({\bf k}) }{\partial k_\rho}
\right|_{{\bf k}=(\mu, k_x, -v_{k_x} k_x)}
$ is UV finite.
Since external fermions are on the Fermi surface with frequencies that are $O(\mu)$,
the soft UV cutoff for the loop momentum is $\mu$
for the UV finite integral.
Since there is no loop that is purely made of fermions in this case,
one necessarily creates bosonic virtual particle in the loop
whose energy increases as $c |q_x|+c|q_y|$, where $\vec q$ is the loop momentum.
This makes the momentum cutoff for the internal loop to be $\mu /c$.
Therefore, the adiabaticity is satisfied
if the variation of the coupling functions can be ignored over $\mu/c$.
If \eq{eq:adiabatic_condition} is satisfied,
this condition is automatically satisfied. 
Because the integrand is peaked at $q=0$,
we can replace
$g_{k+q,k}g_{k,k+q}$, $v_{k+q}$ and $V_{F,k+q}$ 
with those evaluated at $q=0$ in \eq{eq:1LoopFSE_full}
to the leading order in $\epsilon_\mu$.
The derivative of the counter term for the 
fermion kinetic term can be written as
\bqa
\frac{\partial }{\partial \log \mu}
\left.
\frac{\partial {\bf \Sigma}_1({\bf k})}{\partial k_\rho}
\right|_{{\bf k}=(\mu, k_x, -v_{k_x} k_x)}
=
\frac{\partial}{\partial \log \mu}
\left.
\frac{\partial  \Sigma_1({\bf k})}{\partial k_\rho}
\right|_{{\bf k}=(\mu, k_x, -v_{k_x} k_x)} + ...,
\label{eq:Sigma1_adia}
\eqa
where $\Sigma_1({\bf k})$ is the self-energy computed in the adiabatic limit,
\begin{align}
	\label{eq:1LoopFSE_adia}
	\Sigma_{1}({\bf k}) 
= - g_{k,k}^2
	\frac{2(N^2_c-1)}{N_cN_f}\int\dd {\bf q}~ 
	G_{4}({\bf k}+{\bf q} | k)D({\bf q}),
\end{align}
and $...$ represents corrections that are  further suppressed by $\epsilon_\mu$.
In Sec. \ref{sec:fixedpoint},
it will be shown that 
$\epsilon_\mu \sim  \sqrt{v}$.
Therefore, the terms that are ignored in the adiabatic approximation are sub-leading in the small $v$ limit.
In \eq{eq:Sigma1_adia},
the momentum dependent coupling functions are evaluated
at the external momentum, $k_x$
because the integrand is peaked at ${\bf q}=0$
with the soft UV cutoff that is order of $\mu/c$.
It is noted that external fermions can be far away from the hot spots,
and momentum dependent coupling functions can not be
replaced with those evaluated at the hot spots in general.
However, one can still invoke adiabaticity locally in the momentum space
at any point on the Fermi surface.
Even if a coupling function at $k$
is significantly different from its value at the hot spots,
the coupling functions are  still adiabatic 
if their variation is slow locally at that momentum.
Since the derivative of the coupling functions is largest near the hot spots as will be shown in Sec. \ref{sec:fixedpoint},
the adiabatic condition is most stringent near the hot spots.
Namely, 
\eq{eq:adiabatic_condition} is satisfied for general $k_i$'s on the Fermi surface,
if it is satisfied
near the hot spots.

Similarly, the contribution of the vertex correction in \fig{fig:yukawa} 
to the beta functional of the Yukawa coupling is given by  
\begin{align}
\label{eq:GenQC}
&\frac{ \partial A^{(4)}(k_{x}', k_{x}) }{\partial \log \mu}
= \left.		
\frac{2 }{N_cN^\frac{3}{2}_{f}}
\frac{ \partial  }{\partial \log \mu}	
\int \dd {\bf q} ~
g_{k',k'+q}
g_{k'+q,k+q}
g_{k+q,k}  
D({\bf q}) 
G_{4}({\bf k}'+{\bf q} )
G_{1}({\bf k}+{\bf q})
\right|_{
 \scriptsize
  \begin{array}{l}
  {\bf k}' =(2 \mu, k'_{x},-v_{k'_{x}} k'_{x}) \\
  {\bf k}=( \mu, k_{x},v_{k_{x}} k_{x})
 \end{array}.
}
\end{align}
The degree of divergence of $A^{(4)}$ is $0$, 
and its derivative with respect to $\mu$ is UV finite.
Consequently, 
the singular part of \eq{eq:GenQC} can be computed as
\bqa
&\frac{ \partial A^{(4)}(k_{x}', k_{x}) }{\partial \log \mu}
= \left.		
		\frac{2 }{N_cN^\frac{3}{2}_{f}}
		g_{k',k'}
g_{k',k}
g_{k,k}  
\frac{ \partial  }{\partial \log \mu}	
\int \dd {\bf q}~
D({\bf q}) 
G_{4}({\bf k}'+{\bf q}|k')
G_{1}({\bf k}+{\bf q}|k)
\right|_{
 \scriptsize
   \begin{array}{l}
  {\bf k}' =(2 \mu, k'_{x},-v_{k'_{x}} k'_{x}) \\
  {\bf k}=( \mu, k_{x},v_{k_{x}} k_{x})  
 \end{array}
}
+ ... 
\label{eq:dAdmu}
\eqa
to the leading order in $v$.



\section{Beta functionals}
\label{sec:beta_functionals}

\fbox{\begin{minipage}{48em}
{\it
\begin{itemize}
\item
The RG flow of the nesting angle, the Fermi velocity,
 the Yukawa coupling and the forward scattering amplitude 
 are driven by small angle scatterings.
As a result, the beta functionals for those coupling functions at a momentum
depend only on coupling functions evaluated at the same momentum.
\item
The pairing interaction can be 
renormalized by large momentum scatterings, 
and its beta functional
is written as an integration of coupling functions along the Fermi surface.
\item
Theories with general coupling functions suffer from UV/IR mixing
as low-energy operators  with large differences in momentum along the Fermi surface
can mix with each other through high-energy boson.
The UV/IR mixing excludes some one-particle irreducible vertex functions from being universal low-energy observables.
Nonetheless, the low-energy effective theory remains predictive for `legitimate' low-energy observables.
\item
In the space of coupling functions with small nesting angle,
there is a unique interacting fixed point.
At the fixed point, 
the coupling functions are momentum independent,
the nesting angle vanishes,
the dynamical critical exponent is $1$,
and the boson has anomalous dimension $1$.
\end{itemize}
}
\end{minipage}}
\vspace{0.5cm}

In this section, we compute the beta functionals 
for the nesting angle, Fermi velocity, electron-boson coupling
and four-fermion coupling functions.

\subsection{Nesting angle, Fermi velocity and electron-boson coupling}

To the leading order in $v$,
the counter terms 
to the fermion kinetic term
and the cubic coupling
are given by (see Appendix~\ref{app:QC} for derivation)
\begin{equation}
\begin{aligned}
A^{(1)}(k)
&= -
h^{(1)}_k \log\left( \FR{\Lambda}{\mathscr{H}_1( \mu, 2v_kc k)} \right)
+ A^{(1)}_{reg}(k),
\\
A^{(2)}(k) &=   
\frac{2}{\pi} h^{(1)}_k 
\frac{c}{V_{F,k}}
\log\left(\frac{V_{F,k}}{c}\right)
\log\left(\frac{\Lambda}{\mathscr{H}_1(\mu, 2 v_k c k )}\right)
+
3 h^{(2)}_k
\log\left(\frac{\Lambda}{\mathscr{H}_1(\mu,4V_{F,k}v_k k)}\right)
+ A^{(2)}_{reg}(k),
\\
A^{(3)}(k) &= - 
\frac{2}{\pi} h^{(1)}_k 
\frac{c}{V_{F,k}}
\log\left(\frac{V_{F,k}}{c}\right)\log\left(\frac{\Lambda}{\mathscr{H}_1(\mu, 2 v_k c k)}\right)
- h^{(2)}_k
\log\left(\frac{\Lambda}{\mathscr{H}_1(\mu,4V_{F,k}v_k k)}\right)
+ A^{(3)}_{reg}(k),
\\
A^{(4)}(k',k)
&=
-h^{(1)}_{k',k}
 \log
 \left(
 \frac{\Lambda}
{\mathscr{H}_3
( \mu,
2v_{k}c k,
2v_{k'}c k',
\mathscr{R}_{k',k})}
\right) + A^{(4)}_{reg}(k',k),
\end{aligned}
\label{eq:allcounterterms}
\end{equation}
where 
\bqa
h^{(1)}_k &=& \frac{N_c^2-1}{\pi^2 N_cN_f} \FR{g_k^2}{cV_{F,k}}, \label{eq:hk} \\
h^{(1)}_{k',k} &=& \frac{2 g_kg_{k'}}{\pi^2cN_cN_{f}(V_{F,k}+V_{F,k'})} \log\left(\frac{c(V_{F,k}^{-1}+V_{F,k'}^{-1})}{v_k+v_{k'}}\right),
\label{eq:hkk}\\
h^{(2)}_k &=& \frac{N_c^2-1}{2\pi^4 N_c^2N_f^2} 
\frac{g_k^4}{c^2 V_{F,k}^2 } 
\log^2 \PFR{V_{F,k} v_k}{c} 
\label{eq:hkp}
%
\eqa
with 
$g_k\equiv g_{k,k}$
and
$\mathscr{R}_{k',k} = 4(v_{k'}k'+v_{k}k)/(V_{F,k'}^{-1}+V_{F,k}^{-1})$.
 $\mathscr{H}_i(x_1,k_2,..,x_{i+1})$'s 
represent smooth crossover functions that satisfy
\bqa
\mathscr{H}_i(x_1,x_2,..,x_{i+1})
\approx |x_j| ~~~
\mbox{ if $|x_j| \gg |x_1|,..,|x_{j-1}|,|x_{j+1}|,.., |x_{i+1}|$ }.
\label{eq:Hi}
\eqa
The form of the crossover functions depend on the subtraction scheme.
The specific form of  
$\mathscr{H}_i$ is not important for us
as far as the counter term removes 
all IR divergences
in physical observables.
Choosing different $\mathscr{H}_i$ 
amounts to 
modifying the finite parts, $\mathscr{F}_i$  in Eqs. (\ref{eq:RG2})-(\ref{eq:RG5}) 
and  imposing a different set of  renormalization conditions.  
Each counter term is proportional to
\bqa
\log
\left[
\frac{\Lambda}{\mathscr{H}_i( \mu, \Delta_1,..,\Delta_i)} 
\right],
\label{eq:loglhi}
\eqa
where $\mu$ is the energy scale
at which the RG condition is imposed,
and $\Delta_i$ represents crossover energy scales 
that depend on external momenta.
Physically, these  energy scales correspond to 
the energies that virtual particles 
have to carry  within loops 
for given external momenta.
If $\mu$ is much larger than all $\Delta_i$'s,
\eq{eq:loglhi} becomes $\log \left( \frac{\Lambda}{\mu} \right)$,
and the quantum correction gives rise to 
the logarithmic flow of the coupling functions as the energy is lowered.
If $\mu$ becomes much smaller than any of $\Delta_i$,
the quantum correction becomes independent of $\mu$,
and no longer contributes to the flow of coupling functions. 
Roughly speaking, the contribution to the beta function 
turns off below the energy scale
which is given by the largest of $\Delta_i$'s.
Since quantum corrections turn off at different energy scales
at different points on the Fermi surface,
the renormalized coupling functions acquire momentum dependence
at low energies
even if one starts with momentum independent coupling functions
at the UV cutoff scale.
$A^{(i)}_{reg.}$ represent terms
that are regular 
in the limit that 
$\log \frac{\Lambda}{\mathscr{H}_i}$ is large.
In our minimal subtraction scheme,
we can set $A^{(i)}_{reg.}=0$ for $i=1,4$.
One still needs to include 
$A^{(2)}_{reg.}$ and $A^{(2)}_{reg.}$
to enforce the RG condition  in \eq{eq:RG1}
because the finite parts of the fermion self-energy 
affect the shape of the renormalized Fermi surface.
However, the contribution of the regular counter terms 
to the flow of the coupling function diminishes 
in the small $\mu$ limit.

$A^{(i)}(k)$ for $i=1,2,3$ are the counter terms
for the kinetic term of the electron at momentum $k$ away from the hot spot $1$
along the Fermi surface.
$A^{(4)}(k',k)$ is the counter term for the 
cubic vertex that describes the scattering of electron
from momentum $k$ near hot spot $4$
to momentum $k'$ near hot spot $1$.
In general, the quantum corrections are 
functionals of the coupling functions.
However, $h^{(1)}_k$ and $h^{(2)}_k$ depend only on 
the coupling functions at momentum $k$.
Similarly, $A^{(4)}(k',k)$ depends only on 
the coupling functions 
that are defined at $k$ and $k'$ due to adiabaticity.
Through explicit calculations, 
we will show that the coupling functions 
that satisfy the adiabaticity at a UV scale
continue to satisfy the adiabaticity 
below the UV scale.

The counter terms that include $h^{(1)}_k$ are the contributions of the one-loop fermion self-energy in \fig{fig:1LFSE}. Besides the factor of $g_k^2$,
there are factors of $1/V_{F,k}$ and $1/c$ inside $h^{(1)}_k$
because the phase space of the loop momentum
is controlled by the Fermi velocity in one direction
and the boson velocity in another direction.
For $A^{(2)}(k)$ and   $A^{(3)}(k)$, there is an additional factor of $c/V_{F,k}$
because the quantum correction that renormalizes the Fermi velocity is controlled by the boson velocity\footnote{
The one-loop self-energy diagram 
depends on the external momentum only through the combination of $c \vec k$.}.
Inside $\mathscr{H}_1$, 
$2 v_k c k$ denotes the crossover energy scale
determined from the kinematics of the virtual particles in the loop.
For the external fermion at momentum 
$(k_x, -v_{k_x} k_x)$ on the Fermi surface near hot spot $1$,
the energies of the intermediate boson and electron in the loop can be written as
$ c\left( |q_x|+|q_y| \right)$ and 
$V_{F,k_x+q_x} \left( 
v_{k_x+q_x}(k_x + q_x) - q_y + v_{k_x} k_x
\right)$, respectively,
where $\vec q$ is the momentum carried by the internal boson. 
For $\vec k=0$,
the electron can be scattered right onto the hot spot $4$  
by emitting or absorbing a boson with  zero energy.
In this case, 
all virtual particles have zero energy at $\vec q=0$,
and an infrared divergence arises in the low-energy limit.
If the external electron is away from the hot spot ($k \neq 0$), 
it is impossible for the electron to be
scattered onto the Fermi surface with a zero-energy boson.
If it is to be scattered onto the Fermi surface in hot spot $4$,
it must create a boson with energy that is order of $2 c v_{k_x} k_x$.
On the other hand, if the virtual boson carries zero energy,
the internal electron should be created away from the Fermi surface 
with energy that is order of $V_{F, k_x} v_{k_x} k_x$
as is illustrated in \fig{fig:Scatterings}.
Since the boson is slower than fermion ($c \ll V_{F}$) in the small $v$ limit,
it is energetically `cheaper' to create a bosonic excitation while keeping virtual fermions
on the Fermi surface.
The crossover scale is given by the minimum energy that virtual particles have to carry  among all possible choices of internal momentum,
and
the crossover scale for the one-loop self-energy becomes
\begin{align}
                \label{eq:E1Lk}   
        \Eonex     &= 
        2 v_k c  |k|.
        \end{align}
Below the crossover energy scale,
the energy cost for creating virtual excitations becomes bigger than $\mu$,
and the one-loop quantum correction is dynamically turned off.

The counter terms that are proportional to $h^{(2)}_k$ are 
from the two-loop fermion self-energy shown in \fig{fig:2LSE}.
In the two-loop diagram, 
the crossover energy scale is different from that of the one-loop diagram.
The energies of the two internal bosons
and three internal fermions created in the loops can be written as
$E_1= c\left( |q_x|+|q_y| \right)$, 
$E_2 =  c\left( |p_x|+|p_y| \right)$, 
$E_3 =$$ V_{F,k_x+q_x}$ $ \left( v_{k_x+q_x}(k_x + q_x) - q_y + v_{k_x} k_x \right)$,
$E_4 =$ $V_{F,k_x+p_x}$ $ \left( v_{k_x+p_x}(k_x + p_x) - p_y + v_{k_x} k_x\right)$,
$E_5=$ $V_{F,k_x+q_x+p_x}$  
$\left( v_{k_x+p_x+q_x}(k_x+p_x+q_x) -v_{k_x} k_x + p_y + q_y\right)$,
as functions of internal momenta $\vec q$ and $\vec p$.
For $k \neq 0$,
it is kinematically impossible to put all virtual particles at zero energy.
If all internal fermions are to be on the Fermi surface,
at least one boson has to carry energy that is order of $c k_x$.
Alternatively, zero-energy bosons with $\vec q=\vec p=0$ 
put internal fermions away from the Fermi surface 
with energy that is order of $V_F v k$.
Since $c \gg V_F v$ in the small $v$ limit, 
it is energetically favourable to create 
fermions away from the Fermi surface 
while keeping bosons at zero energy.
The crossover energy scale, obtained by minimizing
$\mbox{max} \{ |E_1|, |E_2|, |E_3|, |E_4|, |E_5| \}$
over difference choices of $\vec q$ and $\vec p$,
becomes
\begin{align}
\label{eq:E2Lk}  
\Etwox      &=  4 V_{F,k} v_k| k|.
\end{align}
Since $\Etwox > \Eonex$ for $k \neq 0$, 
electrons away from the hot spots 
disengage with the two-loop quantum correction 
at higher energy scales
than the one-loop correction.

%
%
%


The counter term that is proportional to $h^{(1)}_{k',k}$ 
is from the vertex correction in \fig{fig:yukawa}.
When the electron at momentum $k$ near hot spot $1$
is scattered to momentum $k'$ near hot spot $4$,
the virtual particles are created with energies,
$E_1=c\left( |q_x|+|q_y| \right)$,
$E_2=V_{F,k_x+q_x} \left( v_{k_x+q_x}(k_x + q_x) - q_y + v_{k_x} k_x \right)$ and
$E_3=V_{F,k'_x+q_x} \left( v_{k'_x+q_x}(k'_x + q_x) + q_y + v_{k_x'} k_x' \right)$,
where $\vec q$ is the momentum of the internal boson.
The crossover energy scale is given by minimizing
$\mbox{max} \{ |E_1|, |E_2|, |E_3|\}$  over $\vec q$.
To have a rough estimation of the crossover scale,
one can first set the energy of one of the internal fermions (say $E_2$) to zero 
by choosing $q_y= v_{k_x+q_x}(k_x + q_x)  + v_{k_x} k_x$.
With difference choices of $q_x$, one can tilt the balance
between $E_1$ and $E_3$.
Since $c \gg v$,
it is energetically favourable to minimize the energy
of boson at the expense of the energy of fermion :
with $q_x = 0$,
we have 
$E_1=2 c v_{k_x} k_x$
and
$E_3=2 V_{F,k'_x} \left( v_{k'_x}k'_x + v_{k_x} k_x \right)$.
The bigger between these two determines the crossover scale.
An explicit calculation (Appendix \ref{app:QC} \ref{sec:Yukawa_app_1L})
shows that the crossover scale is symmetric between 
$k$ and $k'$, and can be written as 
$\max \{ \Eonex$, $E^{(1L)}_{k'}$, $\Eonexpx \}$,
where
        \begin{align}
        \label{eq:E1Lkkp}   
        \Eonexpx &=
        \frac{4v_kk+4v_{k'}k'}{V_{F,k}^{-1}+V_{F,k'}^{-1}}.
        \end{align}
%

%
%
%

%


From 
Eqs. (\ref{eq:newZ1}) - (\ref{eq:newZ3}),
one obtains the beta functionals 
for $v_k, V_{F,k}$ and $g_{k',k}$,
and the anomalous dimension of fermion $\eta^{(\psi)}_k$.
To the leading order in $v$, 
solving the beta functionals 
from the quantum corrections 
can be greatly simplified 
as one can use
\bqa
\FR{\dd  Z_{i}(k)}{\dd\log\mu} 
=
\FR{\partial  A_{i}(k)}{\partial \log\mu}
\label{eq:dZdlogmu}
\eqa
in Eqs. (\ref{eq:newZ1}) - (\ref{eq:newZ5})
and Eqs. (\ref{eq:z})-(\ref{eq:etaphi4}).
The rest of the terms in \eq{eq:dZdmu}
are of higher order in $v$.
At low energies, 
one can drop the contributions 
from the finite counter terms.
The resulting beta functionals 
and the anomalous dimension are given by
\begin{align}
    \beta^{(v)}_k 
    &= v_k 
\Biggl[
        \FR{4}{\pi} 
        h^{(1)}_k 
        \frac{c}{V_{F,k}} \log\PFR{V_{F,k}}{c}\theta_1(\mu,\Eonex) 
        + 4 h^{(2)}_k         \theta_1(\mu,\Etwox) 
\Biggr], 
\label{betaVgen1}\\
\beta^{(V_{F})}_k  &= V_{F,k}
\Biggl[ 
-  \FR{2}{\pi} h^{(1)}_k 
\frac{c}{V_{F,k}}
\log\PFR{V_{F,k}}{c}\theta_1(\mu,\Eonex)
+h^{(1)}_k \theta_1(\mu, \Eonex)
 + \frac{2}{\pi} h^{(1)}_0 c \log\PFR{1}{c}
 -h^{(1)}_0 
 +h^{(2)}_0
 -h^{(2)}_k\theta_1(\mu, \Etwox)
\Biggr], 
\label{betaVfgen1}\\
\beta^{(g)}_{k',k} &= g_{k',k}
\Biggl[ 
        h^{(1)}_{0,0} 
      -h^{(1)}_{k',k}    \theta_3(\mu, 
      E^{(1L)}_{k}, 
      E^{(1L)}_{k'}, 
      \Eonexpx)
       -h^{(1)}_0  
       +\frac{ h^{(1)}_k  \theta_1(\mu, \Eonex)  + h^{(1)}_{k'} \theta_1(\mu, E^{(1L)}_{k'})}{2} 
        + 2 h^{(2)}_0  
+\frac{2}{\pi}
        h^{(1)}_0 c \log \frac{1}{c}
\Biggr], \label{mainTextBetaGgen1} \\
\eta^{(\psi)}_{k} &=   
\frac{h^{(1)}_k}{2}  \theta_1(\mu, \Eonex)  
- (z-1),
 \label{eq:main_eta} 
\end{align}
where
$h^{(1)}_k$, $h^{(1)}_{k',k}$ and $h^{(2)}_k$ 
are defined in 
Eqs. (\ref{eq:hk})-(\ref{eq:hkp}),
and
\bqa
\theta_i( \mu, \Delta_1,..,\Delta_i) =  \frac{ \partial \log \mathscr{H}_i( \mu, \Delta_1, .., \Delta_i )}{\partial \log \mu }
\label{eq:theta_i}
\eqa
is the derivative of the crossover function with respect to the energy scale.
It controls whether each term in the beta functional is turned on or off
depending on whether $\mu$ is greater or less than the momentum
dependent the crossover energy scales.

\subsection{Four-fermion coupling}
\label{sec:fullbetalambda}

In this section, we compute the beta functional for the four-fermion coupling.
To the lowest order in $v$,
\eq{eq:newZ4} can be written as
\begin{equation}
\beta^{(\lambda);\{N_i\};\{\sigma_i\}}_{\{k_{i}\}} = \left( 1 + 
3(z-1) + \sum_{j=1}^4 \eta^{(\psi,N_j)}_{k_j} 
\right)\lambda^{\{N_i\};\{\sigma_i\}}_{\{k_{i}\}} 
- 4\mu \frac{\partial \mathbf{\tilde \Gamma}_{CT}^{ \{ N_i \}; \{ \sigma_i \} }(\{ k_{i} \})}{\partial\log\mu},
\label{eq:betalambda2_main}
\end{equation}
where $\mathbf{\tilde \Gamma}_{CT}^{ \{ N_i \}; \{ \sigma_i \} }(\{ k_{i,N_i} \})$ represents
the local counter terms that are needed to remove the singular parts of the vertex correction.
On the right hand side of \eq{eq:betalambda2_main}, 
the first term represents the fact that the four-fermion couplings 
have scaling dimension $-1$ at the tree-level.
The next two terms are the contributions
from the anomalous dimension of frequency $(z-1)$
and the momentum-dependent anomalous dimension
of the fermion field ($\eta^{(\psi,N)}_k$), respectively.
These are common in all channels.
What is channel dependent is the last term
that represents the vertex corrections.
To the leading order, 
there are three distinct 
vertex corrections.
The first is the one in which
four-fermion couplings are generated
from the spin fluctuations.
This gives rise to a term in the beta function
that is independent of the four-fermion couplings.
The diagrams that source the four-fermion coupling 
at the lowest order in $v$ are shown in \fig{fig:4f0}.
The second type of the vertex correction
describes the processes in which spin fluctuations mix
quartic fermion operators in different channels and momenta.
The leading order diagrams 
that describe the  linear mixing 
are shown in  \fig{fig:4f1}.
It generates a term 
that is linear in the four-fermion couplings
but off-diagonal in the space of channel and momentum.
Finally, quantum corrections that are quadratic in the four-fermion couplings 
are shown in \fig{fig:4f2}.
This is the process that drives the pairing instability 
(or particle-hole instability if there is a nesting)
in the presence of attractive interactions
in Fermi liquids.

For general Fermi surface without the PH symmetry,
we only need to consider 
the  forward scattering channel in group 1
and the BCS channel in group 2.
For the derivation of the counter terms that lead to the beta functional through  \eq{eq:betalambda2_main}, see Appendix \ref{app:QC2}.
In Appendix \ref{sec:additional_beta},
we discuss the additional channels 
that need to be considered 
in the presence of the PH symmetry.

\subsubsection{Group 1 : small-angle scatterings }
\label{sec:fullbetalambda_1}

Before we show the result of explicit calculation,
let us first describe the physics that determines
the beta functionals.
In group 1, the critical spin fluctuations generate 
the source term in the $\spmqty{1 & 1 \\ 1 & 1}$ channel.
In the small $v$ limit,
only the ladder diagrams shown in \fig{fig:4f0}
give the leading order contribution.
It is noted that tree-diagrams do not contribute to the 1PI vertex function.
In the ladder diagrams, 
a pair of electron and hole near hot spot $1$
are scattered to intermediate states near hot spot $4$
before they are scattered back to the region near hot spot $1$
by exchanging critical bosons. 
In the absence of the PH symmetry, 
there is no IR singularity for the diagram in the PP channel
due to a lack of nesting.
The coupling in the forward scattering channel
that is potentially IR singular 
can be written as
$\lambda^{\spmqty{1 & 1 \\ 1 & 1}}_{\spmqty{p & k \\ k & p}} $.
On the dimensional ground, 
one expects to encounter a logarithmic divergence
for general $k$ and $p$
because both fermions in the loop of \fig{fig:4f0}
can stay on the Fermi surface
irrespective of the relative momentum along the Fermi surface
\footnote{
Even if both $k$ and $p$ are non-zero, 
a boson that carries a non-zero momentum 
can scatter external fermions 
to virtual states on the Fermi surface. 
One naively expects a logarithmic divergence
as two fermions can have zero three-momentum 
in the space of internal three-momentum 
with co-dimension $2$.
}.
However, 
the actual IR singularity arises only for
\bqa
\lambda^{\spmqty{1 & 1 \\ 1 & 1}}_{\spmqty{0 & k \\ k & 0}}, ~~
\lambda^{\spmqty{1 & 1 \\ 1 & 1}}_{\spmqty{k & 0 \\ 0 & k}} ,
\label{eq:lambda1111kk00}
\eqa
where at least one electron-hole pair are at the hot spot.
This is because
the Pauli exclusion principle suppresses the low-energy phase space 
for particle-hole pairs created on one side of the Fermi surface.
For example, the phase space that is available
for an electron-hole pair 
with total momentum $\vec q$ 
vanishes in the zero $\vec q$ limit.
As a result, the diagram exhibits a logarithmic singularity
only with the help of the critical boson.
When a pair of external electron-hole pair are at the hot spot $1$, 
they can be scattered right onto the hot spot $4$
through the boson with zero energy.
Since two fermions and one boson can simultaneously
have zero three-momentum, 
a logarithmic divergence arises.
Once the spin fluctuations generate the source term
in the $\spmqty{1 & 1 \\ 1 & 1}$ channel,
it spreads to other channels through linear mixing (\fig{fig:4f1}).
The same reasoning that determines the IR singularity of the source term shows that 
the mixing terms exhibit IR singularity 
only for external momenta shown in \eq{eq:lambda1111kk00}.
In the $\spmqty{1 & 1 \\ 1 & 1}$ channel, 
the diagrams in \fig{fig:4f2} do not have IR singularity
because the critical boson is not involved in those processes.
%

\begin{figure*}
	\centering
	\includegraphics[scale=1]{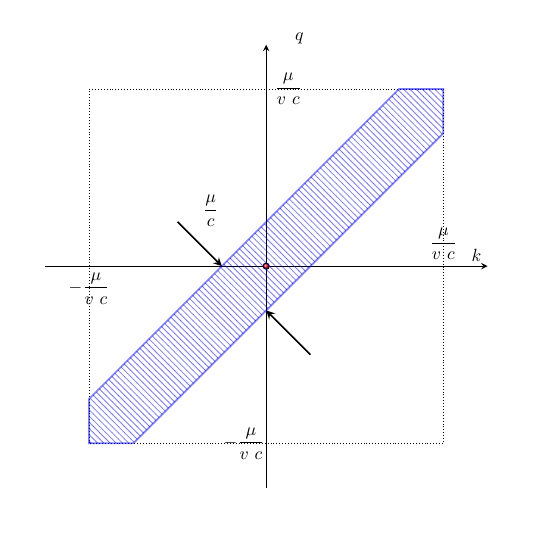}
	\caption{ 
When an electron with momentum $k$ near hot spot $1$ on the Fermi surface 
is scattered to momentum $p$ near 
hot spot $4$ on the Fermi surface 
by exchanging a boson with energy $\mu$,
the interaction is largest in the shaded region in which both $k$ and $p$ are less than $\mu/(vc)$ and their difference is less then $\mu/c$ in magnitude.
Outside the shaded region, the interaction decays in a power-law.
	}
	\label{fig:D_support}
\end{figure*}

The fact that the coupling function 
receives singular vertex correction 
only for \eq{eq:lambda1111kk00}
can be checked from the full expression 
of the beta functional for 
$\lambda^{\spmqty{1 & 1 \\ 1 & 1}}_{\spmqty{p & k \\ k & p}} $ at general $k$ and $p$
\begin{equation}
\begin{aligned}
& \beta^{(\lambda);\spmqty{1 & 1 \\ 1 & 1};\spmqty{\sigma_1 & \sigma_2 \\ \sigma_4 & \sigma_3}}_{\spmqty{p & k \\ k & p}} =
 \left( 1 + 3(z-1) + 2\eta^{(\psi)}_{p} + 2\eta^{(\psi)}_{k}  \right)\lambda^{\spmqty{1 & 1 \\ 1 & 1};\spmqty{\sigma_1 & \sigma_2 \\ \sigma_4 & \sigma_3}}_{\spmqty{p & k \\ k & p}}
\\ & -\int\dd\rho(q)\left\lbrace \frac{\mathsf{D}_{\mu}(k;q)^2}{2\pi N_f g^2_{q,k}}
 \lambda^{\spmqty{1 & 4 \\ 4 & 1}; \spmqty{\sigma_1 & \alpha \\ \beta & \sigma_3}}_{\spmqty{p & q \\ q & p}}
  \mathsf{T}^{\beta \sigma_2}_{\sigma_4 \alpha} 
+ \frac{\mathsf{D}_{\mu}(p;q)^2}{2 \pi N_f g^2_{p,q}}
  \mathsf{T}^{\sigma_1 \alpha}_{\beta \sigma_3} 
\lambda^{\spmqty{4 & 1 \\ 1 & 4}; \spmqty{\beta & \sigma_2 \\ \sigma_4 & \alpha}}_{\spmqty{q & k \\ k & q}} \right. \\
& -\left.\frac{ \mathsf{D}_{\mu}(p;q)\mathsf{D}_{\mu}(q;k)}{\pi N_f^2}
  \mathsf{T}^{\beta \sigma_2}_{\sigma_4 \alpha} 
  \mathsf{T}^{\sigma_1 \alpha}_{\beta \sigma_3} 
\left(\frac{ \mathsf{D}_{\mu}(q;k)}{ g^2_{q,k}} + \frac{\mathsf{D}_{\mu}(p;q)}{ g^2_{p,q}}\right)  \right\rbrace.
\label{eq:beta1111PH1_0}
\end{aligned}
\end{equation}
Here all repeated spin indices are understood to be summed over.
$ \mathsf{T}^{\sigma_1  \sigma_2}_{\sigma_4  \sigma_3 } $
is defined in \eq{eq:Tabcd}.
$\int \dd\rho(q) = \int \frac{\dd q }{2\pi \mu V_{F,q}}$
represents integration along the Fermi surface
with the measure normalized by 
energy scale $\mu$ and the Fermi velocity.
\begin{align}
\mathsf{D}_{\mu}(q;k) = & 
g_{q,k}^2 
\frac{  \mu}{\mu + c\left(\abs{q - k}_\mu + \abs{v_{q}q + v_{k} k}_\mu \right)}
\label{eq:Dqk}
\end{align}
with  $\abs{k}_\mu = \sqrt{ k^2 + \mu^2}$
represents the contribution of the 
gapless spin fluctuations that renormalizes the short-range four-fermion 
coupling at energy scale $\mu$.
This arises from local counter terms that remove IR singularities
of the loop correction in the small $\mu$ limit.
For $|q-k| \gg \mu$, 
$\mathsf{D}_{\mu}(q;k)$ 
coincides with the  
interaction mediated by the collective mode 
with spatial momentum $( q-k, v_q q + v_k k)$
at energy $\mu$,
where
$( q-k, v_q q + v_k k)$ denotes the momentum 
that is needed to scatter an electron
from $(k, -v_k k)$ near hot spot $1$
to $(q, v_q q)$ near hot spot $4$ on the Fermi surface.
In the limit that $q$ and $k$ are small,
$\mathsf{D}_{\mu}(q;k)$ smoothly saturates to a constant
and represents a local interaction in the real space.
The main support of 
$\mathsf{D}_{\mu}(q;k)$  
in the space of $q$ and $k$ is given by
$S_{D} = \{
(q,k) | |q-k|< \mu/c, |q|<\mu/(vc), |k|<\mu/(vc)
\}$ 
as is illustrated in \fig{fig:D_support}.
$\eta^{(\psi)}_k$ is the momentum dependent anomalous dimension of fermion
defined in  \eq{eq:main_eta}. 
To the leading order in $v$, $\eta^{(\psi)}_k$ is given by
\bqa
\eta^{(\psi)}_k
= \frac{N_c^2-1}{2 \pi^2 N_cN_f} \FR{g_k^2}{cV_{F,k}} \frac{\mu}{\mu + 2 c v_k |k|_\mu} 
- (z-1) .
\eqa
The first line in   \eq{eq:beta1111PH1_0}
 is the contribution
of the tree-level scaling dimension
and the anomalous dimensions of frequency and the fermion fields.
The second line represents the mixings between 
$\lambda^{\spmqty{1 & 1 \\ 1 & 1};\spmqty{\sigma_1 & \sigma_2 \\ \sigma_4 & \sigma_3}}_{\spmqty{p & k \\ k & p}}$
and
$\left\{
\lambda^{\spmqty{1 & 4 \\ 4 & 1}; \spmqty{\sigma_1 & \alpha \\ \beta & \sigma_3}}_{\spmqty{p & q \\ q & p}}\right.$,
$\left. \lambda^{\spmqty{4 & 1 \\ 1 & 4}; \spmqty{\beta & \sigma_2 \\ \sigma_4 & \alpha}}_{\spmqty{q & k \\ k & q}}  \right\}$.
Since the momentum along the Fermi surface acts as a continuous flavour,
the mixing with coupling functions at different momenta 
is represented as the integration over $q$.
The last line is the contribution of the ladder diagrams
that source the four-fermion coupling.
It is essentially the convolution of two boson propagators
that are needed to scatter an electron from momentum $k$ to the intermediate momentum $q$ 
and finally to momentum $p$.
The additional factor 
$\left(\frac{\mathsf{D}_{\mu}(q;k)}{ g^2_{k,q}} + \frac{\mathsf{D}_{\mu}(q;p)}{ g^2_{p,q}}\right)$
arises from the derivative of $\mathsf{D}_{\mu}(q;k)$ with respect to $\log \mu$
to the leading order in $v$ limit\footnote{
Because the beta functional is given by the derivative of the quantum correction
with respect to $\log \mu$,
and the gapless fermions alone do not exhibit IR singularity in the small $\mu$ limit,
the derivative only acts on the boson propagator. }.
It is noted that the  full vertex correction is written as integrations over the momentum along the Fermi surface  in \eq{eq:beta1111PH1_0}.

%
If both $p$ and $k$ are away from the hot spots, 
the vertex corrections vanish in the small $\mu$ limit.
On the other hand, the phase space in which both $k$ and $p$ are near the hot spot is negligible.
Therefore, we focus on the forward scattering between an electron near the hot spot 
and an electron away from the hot spot.
For $k \approx 0$  and $|p| \gg \mu/(vc)$,
$\lambda_{\spmqty{p & q \\ q & p}}$
and
$\mathsf{D}_{\mu}(q;p) $
as functions of $q$
vary much more slowly 
compared to
$\mathsf{D}_{\mu}(k;q)^2$
which is sharply peaked at $q=k$\footnote{
The profile of $\lambda$ will be confirmed from the solution of the beta functional in Sec.  \ref{sec:fixedpoint}.}.
Then, we can use
$
\int dq   ~ 
\mathsf{D}_{\mu}(k;q)^2
f(q;p) \approx
f(k;p) 
\int dq   ~
\mathsf{D}_{\mu}(k;q)^2$  
for
$ f(q;p)=
 \lambda_{\spmqty{p & q \\ q & p}}$
 or 
 $ \mathsf{D}_{\mu}(q;p) $
to simplify the beta functional as\footnote{
We can also use
$
\int dq   ~ 
\mathsf{D}_{\mu}(p;q)^2
f(q;k) \approx
f(p;k) 
\int dq   ~
\mathsf{D}_{\mu}(p;q)^2$  
because both sides vanish in the small $\mu$ limit
for $k \approx 0$  and $|p| \gg \mu/(vc)$.
}
%
\begin{equation}
\begin{aligned}
& \beta^{(\lambda);\spmqty{1 & 1 \\ 1 & 1};\spmqty{\sigma_1 & \sigma_2 \\ \sigma_4 & \sigma_3}}_{\spmqty{p & k \\ k & p}} =
 \left( 1 + 3(z-1) + 2\eta^{(\psi)}_{p} + 2\eta^{(\psi)}_{k}  \right)\lambda^{\spmqty{1 & 1 \\ 1 & 1};\spmqty{\sigma_1 & \sigma_2 \\ \sigma_4 & \sigma_3}}_{\spmqty{p & k \\ k & p}}
\\ & - \frac{g^2_{k,k}}{2 \pi^2 c N_f V_{F,k}}\frac{\mu}{\mu + 2v_kc|k|_\mu}
\lambda^{\spmqty{1 & 4 \\ 4 & 1}; \spmqty{\sigma_1 & \alpha \\ \beta & \sigma_3}}_{\spmqty{p & k \\ k & p}}
 \mathsf{T}^{\beta \sigma_2}_{\sigma_4 \alpha} 
 - \frac{g^2_{p,p}}{2\pi^2 c N_f V_{F,p}}\frac{\mu}{\mu + 2v_pc|p|_\mu}
   \mathsf{T}^{\sigma_1 \alpha}_{\beta \sigma_3} 
\lambda^{\spmqty{4 & 1 \\ 1 & 4}; \spmqty{\beta & \sigma_2 \\ \sigma_4 & \alpha}}_{\spmqty{p & k \\ k & p}}
\\ & +
 \mathsf{T}^{\beta \sigma_2}_{\sigma_4 \alpha} 
   \mathsf{T}^{\sigma_1 \alpha}_{\beta \sigma_3} 
\frac{\mathsf{D}_{\mu}(p;k)}{\pi^2 c N_f^2}\left[\frac{\mu g^2_{k,k}}{V_{F,k}(\mu + 2v_kc|k|_\mu)}+\frac{\mu g^2_{p,p}}{V_{F,p}(\mu + 2v_pc|p|_\mu)}\right]
 \label{eq:beta1111PH1}
\end{aligned}
\end{equation}
in the small $\mu$ limit.
Here, $\frac{g^2_{k,k}}{2 \pi^2 c N_f V_{F,k}}\frac{\mu}{\mu + 2vc|k|_\mu}$
corresponds to the momentum dependent mixing matrix element.
For  $k=0$ and $p \neq 0$
($k\neq 0$ and $p = 0$),
$\lambda^{\spmqty{1 & 1 \\ 1 & 1};\spmqty{\sigma_1 & \sigma_2 \\ \sigma_4 & \sigma_3}}_{\spmqty{p & k \\ k & p}}$
mixes only with
$\lambda^{\spmqty{1 & 4 \\ 4 & 1}; \spmqty{\sigma_1 & \alpha \\ \beta & \sigma_3}}_{\spmqty{p & k \\ k & p}}$
$\left(\lambda^{\spmqty{4 & 1 \\ 1 & 4}; \spmqty{\beta & \sigma_2 \\ \sigma_4 & \alpha}}_{\spmqty{p & k \\ k & p}}\right)$
in the small $\mu$ limit.
This implies that the forward scattering amplitude 
is mainly renormalized through 
the electrons at the hot spots
as is indicated in Table. \ref{table:mixing1}.

Since the beta functional for 
$\lambda^{\spmqty{1 & 1 \\ 1 & 1}}$ 
depend on
$\lambda^{\spmqty{1 & 4 \\ 4 & 1}}$,
$\lambda^{\spmqty{4 & 1 \\ 1 & 4}}$,
and their beta functionals depend on 
$\lambda^{\spmqty{4 & 4 \\ 4 & 4}}$,
we need to compute the beta functionals for those couplings 
to have a closed set of beta functionals.
The beta functionals for the other couplings are obtained to be
\begin{equation}
\begin{aligned}
& \beta^{(\lambda);\spmqty{1 & 4 \\ 4 & 1};\spmqty{\sigma_1 & \sigma_2 \\ \sigma_4 & \sigma_3}}_{\spmqty{p & k \\ k & p}} =
 \left( 1 + 3(z-1) + 2\eta^{(\psi)}_{p} + 2\eta^{(\psi)}_{k}  \right)\lambda^{\spmqty{1 & 4 \\ 4 & 1};\spmqty{\sigma_1 & \sigma_2 \\ \sigma_4 & \sigma_3}}_{\spmqty{p & k \\ k & p}}
\\ & - \frac{g^2_{k,k}}{2 \pi^2 c N_f V_{F,k}}\frac{\mu}{\mu + 2v_kc|k|_\mu}
\lambda^{\spmqty{1 & 1 \\ 1 & 1}; \spmqty{\sigma_1 & \alpha \\ \beta & \sigma_3}}_{\spmqty{p & k \\ k & p}}
 \mathsf{T}^{\beta \sigma_2}_{\sigma_4 \alpha} 
 - \frac{g^2_{p,p}}{2\pi^2 c N_f V_{F,p}}\frac{\mu}{\mu + 2v_pc|p|_\mu}
   \mathsf{T}^{\sigma_1 \alpha}_{\beta \sigma_3} 
\lambda^{\spmqty{4 & 4 \\ 4 & 4}; \spmqty{\beta & \sigma_2 \\ \sigma_4 & \alpha}}_{\spmqty{p & k \\ k & p}},
 \label{eq:beta1111PH2}
\end{aligned}
\end{equation}
\begin{equation}
\begin{aligned}
& \beta^{(\lambda);\spmqty{4 & 1 \\ 1 & 4};\spmqty{\sigma_1 & \sigma_2 \\ \sigma_4 & \sigma_3}}_{\spmqty{p & k \\ k & p}} =
 \left( 1 + 3(z-1) + 2\eta^{(\psi)}_{p} + 2\eta^{(\psi)}_{k}  \right)\lambda^{\spmqty{4 & 1 \\ 1 & 4};\spmqty{\sigma_1 & \sigma_2 \\ \sigma_4 & \sigma_3}}_{\spmqty{p & k \\ k & p}}
\\ & - \frac{g^2_{k,k}}{2 \pi^2 c N_f V_{F,k}}\frac{\mu}{\mu + 2v_kc|k|_\mu}
\lambda^{\spmqty{4 & 4 \\ 4 & 4}; \spmqty{\sigma_1 & \alpha \\ \beta & \sigma_3}}_{\spmqty{p & k \\ k & p}}
 \mathsf{T}^{\beta \sigma_2}_{\sigma_4 \alpha} 
 - \frac{g^2_{p,p}}{2\pi^2 c N_f V_{F,p}}\frac{\mu}{\mu + 2v_pc|p|_\mu}
   \mathsf{T}^{\sigma_1 \alpha}_{\beta \sigma_3} 
\lambda^{\spmqty{1 & 1\\ 1 & 1}; \spmqty{\beta & \sigma_2 \\ \sigma_4 & \alpha}}_{\spmqty{p & k \\ k & p}},
 \label{eq:beta1111PH3}
\end{aligned}
\end{equation}
\begin{equation}
\begin{aligned}
& \beta^{(\lambda);\spmqty{4 & 4 \\ 4 & 4};\spmqty{\sigma_1 & \sigma_2 \\ \sigma_4 & \sigma_3}}_{\spmqty{p & k \\ k & p}} =
 \left( 1 + 3(z-1) + 2\eta^{(\psi)}_{p} + 2\eta^{(\psi)}_{k}  \right)\lambda^{\spmqty{4 & 4 \\ 4 & 4};\spmqty{\sigma_1 & \sigma_2 \\ \sigma_4 & \sigma_3}}_{\spmqty{p & k \\ k & p}}
\\ & - \frac{g^2_{k,k}}{2 \pi^2 c N_f V_{F,k}}\frac{\mu}{\mu + 2v_kc|k|_\mu}
\lambda^{\spmqty{4 & 1 \\ 1 & 4}; \spmqty{\sigma_1 & \alpha \\ \beta & \sigma_3}}_{\spmqty{p & k \\ k & p}}
 \mathsf{T}^{\beta \sigma_2}_{\sigma_4 \alpha} 
 - \frac{g^2_{p,p}}{2\pi^2 c N_f V_{F,p}}\frac{\mu}{\mu + 2v_pc|p|_\mu}
   \mathsf{T}^{\sigma_1 \alpha}_{\beta \sigma_3} 
\lambda^{\spmqty{1 & 4 \\ 4 & 1}; \spmqty{\beta & \sigma_2 \\ \sigma_4 & \alpha}}_{\spmqty{p & k \\ k & p}}
\\ & +
 \mathsf{T}^{\beta \sigma_2}_{\sigma_4 \alpha} 
   \mathsf{T}^{\sigma_1 \alpha}_{\beta \sigma_3} 
\frac{\mathsf{D}_{\mu}(p;k)}{\pi^2 c N_f^2}\left[\frac{\mu g^2_{k,k}}{V_{F,k}(\mu + 2v_kc|k|_\mu)}+\frac{\mu g^2_{p,p}}{V_{F,p}(\mu + 2v_pc|p|_\mu)}\right].
 \label{eq:beta1111PH4}
\end{aligned}
\end{equation}
The beta functional for
$\lambda^{\spmqty{4 & 4 \\ 4 & 4};\spmqty{\sigma_1 & \sigma_2 \\ \sigma_4 & \sigma_3}}_{\spmqty{p & k \\ k & p}}$
takes the same form as
that of 
$\lambda^{\spmqty{1 & 1 \\ 1 & 1};\spmqty{\sigma_1 & \sigma_2 \\ \sigma_4 & \sigma_3}}_{\spmqty{p & k \\ k & p}}$
because those two are related to each other through the $C_4$ symmetry.
On the other hand,
there is no source term for
$\lambda^{\spmqty{1 & 4 \\ 4 & 1};\spmqty{\sigma_1 & \sigma_2 \\ \sigma_4 & \sigma_3}}_{\spmqty{p & k \\ k & p}}$ and
$\lambda^{\spmqty{4 & 1 \\ 1 & 4};\spmqty{\sigma_1 & \sigma_2 \\ \sigma_4 & \sigma_3}}_{\spmqty{p & k \\ k & p}}$
to the leading order in $v$
as is shown  in Eqs. (\ref{eq:beta1111PH1}) and (\ref{eq:beta1111PH4}).

\subsubsection{Group 2 : BCS pairing }
\label{sec:fullbetalambda_2}


In group $2$, the ladder diagrams in \fig{fig:4f0}
generate $\lambda^{\spmqty{1 & 5 \\ 1 & 5}}$.
It then mixes with
$\lambda^{\spmqty{4 & 8 \\ 1 & 5}}$,
$\lambda^{\spmqty{1 & 5 \\ 4 & 8}}$
through diagrams in  \fig{fig:4f1}.
In terms of how couplings are mixed in the space of hot spots,
the structure of the beta functionals
is similar to 
the $\spmqty{1 & 1 \\ 1 & 1}$ channel
except that the mixing occurs in the particle-particle channel for general $k$ and $p$ in group 2.
There are more important differences 
in how operators with different momenta mix
in this channel compared to the 
$\spmqty{1 & 1 \\ 1 & 1}$ channel.
The difference arises from the fact that 
the couplings in the $\spmqty{1 & 5 \\ 1 & 5}$ channel
describe scatterings of fermions on the opposite sides of the Fermi surface,
and the phase space of a pair of fermions in 
antipodal patches is not suppressed by the Pauli exclusion principle.
Namely, a Cooper pair with zero center of mass momentum
can be placed anywhere above but arbitrarily close
to the Fermi surface.
As a result,  IR divergences arise within a two-dimensional 
space of external momenta 
irrespective of the relative momenta 
of electrons 
within incoming and outgoing Cooper pairs.

The beta functional for
$\lambda^{\spmqty{1 & 5 \\ 1 & 5};\spmqty{\sigma_1 & \sigma_2 \\ \sigma_4 & \sigma_3}}_{\spmqty{p & -p \\ k & -k}}$ at generic $k$ and $p$ is given by
\begin{equation}
\begin{aligned}
 \beta^{(\lambda);\spmqty{1 & 5 \\ 1 & 5};\spmqty{\sigma_1 & \sigma_2 \\ \sigma_4 & \sigma_3}}_{\spmqty{p & -p \\ k & -k}} &=  \left( 1 + 3(z-1) + 2 \eta^{(\psi)}_{p} + 2 \eta^{(\psi)}_{k}  \right)
\lambda^{\spmqty{1 & 5 \\ 1 & 5};\spmqty{\sigma_1 & \sigma_2 \\ \sigma_4 & \sigma_3}}_{\spmqty{p & -p \\ k & -k}}
\\ & +\int \dd\rho(q) \left\lbrace -
\frac{ \mathsf{D}_{\mu}(p;q)}{2 \pi N_f}  
 \mathsf{T}^{\sigma_1 \sigma_2}_{\alpha \beta} 
 \lambda^{\spmqty{4 & 8 \\ 1 & 5}; \spmqty{\alpha & \beta \\ \sigma_4 & \sigma_3}}_{\spmqty{q & -q \\ k & -k}}
  -\frac{\mathsf{D}_{\mu}(q;k)}{2\pi N_f} 
 \lambda^{\spmqty{1 & 5 \\ 4 & 8}; \spmqty{\sigma_1 & \sigma_2 \\ \alpha & \beta}}_{\spmqty{p & -p \\ q & -q}}
  \mathsf{T}^{\alpha \beta}_{\sigma_4 \sigma_3} \right.
\\ & +\frac{1}{\pi N_f^2}
 \mathsf{T}^{\sigma_1 \sigma_2}_{\alpha \beta} 
 \mathsf{T}^{\alpha \beta}_{\sigma_4 \sigma_3} 
\mathsf{D}_{\mu}(p;q)
\mathsf{D}_{\mu}(q;k)
\\ & \left. + \frac{1}{4\pi}
\left(\lambda^{\spmqty{1 & 5 \\ 1 & 5}; \spmqty{\sigma_1 & \sigma_2 \\ \beta & \alpha}}_{\spmqty{p & -p \\ q & -q}}\lambda^{\spmqty{1 & 5 \\ 1 & 5}; \spmqty{\beta & \alpha \\ \sigma_4 & \sigma_3}}_{\spmqty{q & -q \\ k & -k}} + \lambda^{\spmqty{1 & 5 \\ 4 & 8}; \spmqty{\sigma_1 & \sigma_2 \\ \beta & \alpha}}_{\spmqty{p & -p \\ q & -q}}\lambda^{\spmqty{4 & 8 \\ 1 & 5}; \spmqty{\beta & \alpha \\ \sigma_4 & \sigma_3}}_{\spmqty{q & -q \\ k & -k}} \right)\right\rbrace.
\label{eq:beta1515PP1}
\end{aligned}
\end{equation}
$\lambda^{\spmqty{1 & 5 \\ 1 & 5};\spmqty{\sigma_1 & \sigma_2 \\ \sigma_4 & \sigma_3}}_{\spmqty{p & -p \\ k & -k}}$ describes the interaction 
in which a pair of electrons with
zero center of mass momentum are scattered
in antipodal hot patches.
The first line in \eq{eq:beta1515PP1} is the contribution
from the tree-level scaling dimension and the anomalous dimensions.   
The second line represents the linear mixing between 
$\lambda^{\spmqty{1 & 5 \\ 1 & 5};\spmqty{\sigma_1 & \sigma_2 \\ \sigma_4 & \sigma_3}}_{\spmqty{p & -p \\ k & -k}}$
and
$\left\{ \lambda^{\spmqty{4 & 8 \\ 1 & 5}; \spmqty{\alpha & \beta \\ \sigma_4 & \sigma_3}}_{\spmqty{q & -q \\ k & -k}},
 \lambda^{\spmqty{1 & 5 \\ 4 & 8}; \spmqty{\sigma_1 & \sigma_2 \\ \alpha & \beta}}_{\spmqty{p & -p \\ q & -q}} \right\}$.
Unlike in  \eq{eq:beta1111PH1_0}, the mixing between coupling functions at different momenta 
decays only in a single power of $\mathsf{D}_\mu$,
and there is no additional suppression even when both $p$ and $k$ are way from the hot spots.
This is because the vertex correction is IR divergent for any $p$ and $k$.
For example, the first term in the second line describes the process in which 
a pair of electrons at momenta $k$ and $-k$ near hot spots $1$ and $5$
are scattered to momenta $q$ and $-q$ near hot spots $4$ and $8$
through the short-range four-fermion interaction
and then to momenta $p$ and $-p$ near hot spots $1$ and $5$
by exchanging a boson. 
The next term can be understood similarly.
In this process, the internal fermions can simultaneously stay outside  
but close to the Fermi surface for any $q$.
This gives rise to a logarithmic IR singularity 
that is not tied to the criticality of the boson.
In the beta functional that
is given by the integration over $q$,
the amplitude of mixing is simply controlled by the 
interaction mediated by the boson that carries the momentum
needed to scatter the pair of electrons within the Fermi surface.
Since the mixing amplitude scales as 
$\mathsf{D}_{\mu}(q,k) \sim g_{q,k}^2/|q-k|$ at large momenta,
the contributions from large-angle scatterings are not strongly suppressed.

If the coupling functions are weakly dependent on momentum,
the slow decay of the mixing matrix gives rise to a logarithmic divergence $\log \Lambda'/\mu$,
where $\Lambda'$ is a scale at which the large momentum is cut off\footnote{
For example, the scale associated with the  irrelevant kinetic term of the boson 
can act as a large momentum cutoff. 
}.
The explicit dependence of the beta functional on a UV scale 
is a manifestation of the fact 
that the quantum correction
itself exhibits a $\log^2$ divergence,
where one logarithm is from the BCS scatterings of the gapless fermions 
and the other logarithm is from the criticality of the boson.
In the functional RG formalism,
the physical origins of these two logarithms are naturally resolved,
and they manifest themselves in different ways :
$\mu$ in the fermionic $\log$ 
acts as an IR energy cutoff that 
controls the `distance' away from the Fermi surface 
and
$\mu$ in the bosonic $\log$ 
acts as an IR cutoff 
that regulates how operators 
with different momentum along the Fermi surface mix.
Interestingly, the mixing between operators with momenta $q$ and $k$ along the Fermi surface is controlled by $\mathsf{D}_{\mu}(q,k)$.
The fact that the mixing between low-energy operators with large $q-k$ is determined from the dynamics of the high-energy boson implies that
%
%
%
%
%
the four-fermion coupling function
is not a low-energy observable 
that can be predicted within the low-energy effective field theory.
While the 1PI quartic vertex function,
represented by the four-fermion coupling function,
can not be predicted within the  low-energy effective field theory,
the theory is still predictive for 
a different low-energy observable
that captures the strength of two-body interaction.
In Sec. \ref{sec:fixedpoint},
this will be discussed in full details.
For now, let us set this issue aside and complete the rest of the beta functionals.
Once the full beta functionals are completed,
we will be in the better position to address the issue of UV/IR mixing more systematically.
The term in the third line is the source that is generated from the spin fluctuations.
Because there exists the fermionic logarithmic singularity associated
with the virtual fermions that are on the antipodal patches of the Fermi surface,
the contribution to the beta function is simply given by a convolution of two 
boson propagators  without an additional suppression as in \eq{eq:beta1111PP}.
Finally, the last line is the usual term that drives the
BCS instability in the presence of attractive four-fermion couplings in Fermi liquids.
Its contribution is expressed as a convolution of two four-fermion coupling functions.\footnote{
Here, we focus on the four-fermion couplings
that are generated from the critical spin fluctuations to the leading order in $v$.
In the presence of a bare four-fermion coupling,
more channels 
such as 
$\lambda^{\spmqty{1 & 5 \\ 2 & 6}}
\lambda^{\spmqty{2 & 6 \\ 1 & 5}}$
and
$\lambda^{\spmqty{1 & 5 \\ 3 & 7}}
\lambda^{\spmqty{3 & 7 \\ 1 & 5}}$
should be included in the BCS term.
}

Together with the beta functional for
$ \lambda^{\spmqty{1 & 5 \\ 1 & 5}; \spmqty{\sigma_1 & \sigma_2 \\ \sigma_4 & \sigma_3}}_{\spmqty{p & -p \\ k & -k}}$,
the beta functionals for
$ \lambda^{\spmqty{4 & 8 \\ 1 & 5}; \spmqty{\sigma_1 & \sigma_2 \\ \sigma_4 & \sigma_3}}_{\spmqty{p & -p \\ k & -k}}$,
$ \lambda^{\spmqty{1 & 5 \\ 4 & 8}; \spmqty{\sigma_1 & \sigma_2 \\ \sigma_4 & \sigma_3}}_{\spmqty{p & -p \\ k & -k}}$,
$ \lambda^{\spmqty{4 & 8 \\ 4 & 8}; \spmqty{\sigma_1 & \sigma_2 \\ \sigma_4 & \sigma_3}}_{\spmqty{p & -p \\ k & -k}}$ 
form a closed set of flow equations at generic momenta 
in the two-dimensional plane of IR singularity.
The beta functionals for the remaining coupling functions are written as
\bqa
 \beta^{(\lambda);\spmqty{4 & 8 \\ 1 & 5}; \spmqty{\sigma_1 & \sigma_2 \\ \sigma_4 & \sigma_3}}_{\spmqty{p & -p \\ k & -k}} &=&
 \left( 1 + 3(z-1) + 2 \eta^{(\psi)}_{p} + 2 \eta^{(\psi)}_{k} \right)
 \lambda^{\spmqty{4 & 8 \\ 1 & 5}; \spmqty{\sigma_1 & \sigma_2 \\ \sigma_4 & \sigma_3}}_{\spmqty{p & -p \\ k & -k}}
\nn
&& + \int \dd\rho(q) \left\lbrace -\frac{1}{2\pi N_f}\left[
\mathsf{D}_{\mu}(p;q) 
 \mathsf{T}^{\sigma_1 \sigma_2}_{\alpha \beta} 
\lambda^{\spmqty{1 & 5 \\ 1 & 5} ; \spmqty{\alpha & \beta \\ \sigma_4 & \sigma_3}}_{\spmqty{q & -q \\ k & -k}} 
+ 
\mathsf{D}_{\mu}(q;k) 
\lambda^{\spmqty{4 & 8 \\ 4 & 8} ; \spmqty{\sigma_1 & \sigma_2 \\ \alpha & \beta}}_{\spmqty{p & -p \\ q & -q}} 
 \mathsf{T}^{\alpha \beta}_{\sigma_4 \sigma_3} 
\right] \right.
\nn
&&
\left. 
+ \frac{1}{4\pi}
\left(\lambda^{\spmqty{4 & 8 \\ 1 & 5}; \spmqty{\sigma_1 & \sigma_2 \\ \beta & \alpha}}_{\spmqty{p & -p \\ q & -q}}\lambda^{\spmqty{1 & 5 \\ 1 & 5}; \spmqty{\beta & \alpha \\ \sigma_4 & \sigma_3}}_{\spmqty{q & -q \\ k & -k}} + \lambda^{\spmqty{4 & 8 \\ 4 & 8}; \spmqty{\sigma_1 & \sigma_2 \\ \beta & \alpha}}_{\spmqty{p & -p \\ q & -q}}\lambda^{\spmqty{4 & 8 \\ 1 & 5}; \spmqty{\beta & \alpha \\ \sigma_4 & \sigma_3}}_{\spmqty{q & -q \\ k & -k}} \right)\right\rbrace,  
\label{eq:beta1515PP2} \\
%
 \beta^{(\lambda);\spmqty{1 & 5 \\ 4 & 8}; \spmqty{\sigma_1 & \sigma_2 \\ \sigma_4 & \sigma_3}}_{\spmqty{p & -p \\ k & -k}} 
 &=&
 \left( 1 + 3(z-1) +  2 \eta^{(\psi)}_{p} + 2 \eta^{(\psi)}_{k}  \right)\lambda^{\spmqty{1 & 5 \\ 4 & 8}; \spmqty{\sigma_1 & \sigma_2 \\ \sigma_4 & \sigma_3}}_{\spmqty{p & -p \\ k & -k}}
\nn
&& + \int \dd\rho(q) \left\lbrace -\frac{1}{2\pi N_f}\left[
\mathsf{D}_{\mu}(p;q) 
 \mathsf{T}^{\sigma_1 \sigma_2}_{\alpha \beta} 
\lambda^{\spmqty{4 & 8 \\ 4 & 8} ; \spmqty{\alpha & \beta \\ \sigma_4 & \sigma_3}}_{\spmqty{q & -q \\ k & -k}} 
+ 
\mathsf{D}_{\mu}(q;k) 
\lambda^{\spmqty{1 & 5 \\ 1 & 5} ; \spmqty{\sigma_1 & \sigma_2 \\ \alpha & \beta}}_{\spmqty{p & -p \\ q & -q}} 
 \mathsf{T}^{\alpha \beta}_{\sigma_4 \sigma_3} 
\right] \right.
\nn 
&&\left. 
+ \frac{1}{4\pi}
\left(\lambda^{\spmqty{1 & 5 \\ 1 & 5}; \spmqty{\sigma_1 & \sigma_2 \\ \beta & \alpha}}_{\spmqty{p & -p \\ q & -q}}\lambda^{\spmqty{1 & 5 \\ 4 & 8}; \spmqty{\beta & \alpha \\ \sigma_4 & \sigma_3}}_{\spmqty{q & -q \\ k & -k}} + \lambda^{\spmqty{1 & 5 \\ 4 & 8}; \spmqty{\sigma_1 & \sigma_2 \\ \beta & \alpha}}_{\spmqty{p & -p \\ q & -q}}\lambda^{\spmqty{4 & 8 \\ 4 & 8}; \spmqty{\beta & \alpha \\ \sigma_4 & \sigma_3}}_{\spmqty{q & -q \\ k & -k}} \right)\right\rbrace, 
\label{eq:beta1515PP3} \\
%
 \beta^{(\lambda);\spmqty{4 & 8 \\ 4 & 8};\spmqty{\sigma_1 & \sigma_2 \\ \sigma_4 & \sigma_3}}_{\spmqty{p & -p \\ k & -k}} &=&  \left( 1 + 3(z-1) +  2 \eta^{(\psi)}_{p} + 2 \eta^{(\psi)}_{k} \right)
\lambda^{\spmqty{4 & 8 \\ 4 & 8};\spmqty{\sigma_1 & \sigma_2 \\ \sigma_4 & \sigma_3}}_{\spmqty{p & -p \\ k & -k}}
\nn
&& +\int \dd\rho(q) \left\lbrace -
\frac{ \mathsf{D}_{\mu}(p;q)}{2 \pi N_f}  
 \mathsf{T}^{\sigma_1 \sigma_2}_{\alpha \beta} 
 \lambda^{\spmqty{1 & 5 \\ 4 & 8}; \spmqty{\alpha & \beta \\ \sigma_4 & \sigma_3}}_{\spmqty{q & -q \\ k & -k}}
  -\frac{\mathsf{D}_{\mu}(q;k)}{2\pi N_f} 
 \lambda^{\spmqty{4 & 8 \\ 1 & 5}; \spmqty{\sigma_1 & \sigma_2 \\ \alpha & \beta}}_{\spmqty{p & -p \\ q & -q}}
  \mathsf{T}^{\alpha \beta}_{\sigma_4 \sigma_3} \right.
\nn
&& +\frac{1}{\pi N_f^2}
 \mathsf{T}^{\sigma_1 \sigma_2}_{\alpha \beta} 
 \mathsf{T}^{\alpha \beta}_{\sigma_4 \sigma_3} 
%
%
\mathsf{D}_{\mu}(p;q)
\mathsf{D}_{\mu}(q;k)
\nn 
&& \left. + \frac{1}{4\pi}
\left(\lambda^{\spmqty{4 & 8 \\ 1 & 5}; \spmqty{\sigma_1 & \sigma_2 \\ \beta & \alpha}}_{\spmqty{p & -p \\ q & -q}}\lambda^{\spmqty{1 & 5 \\ 4 & 8}; \spmqty{\beta & \alpha \\ \sigma_4 & \sigma_3}}_{\spmqty{q & -q \\ k & -k}} + \lambda^{\spmqty{4 & 8 \\ 4 & 8}; \spmqty{\sigma_1 & \sigma_2 \\ \beta & \alpha}}_{\spmqty{p & -p \\ q & -q}}\lambda^{\spmqty{4 & 8 \\ 4 & 8}; \spmqty{\beta & \alpha \\ \sigma_4 & \sigma_3}}_{\spmqty{q & -q \\ k & -k}} \right)\right\rbrace.
\label{eq:beta1515PP4}
\eqa
\eq{eq:beta1515PP4}  is related to 
\eq{eq:beta1515PP1} through the $C_4$ symmetry.
Similarly,
\eq{eq:beta1515PP3}  is related to 
\eq{eq:beta1515PP2} through the symmetry.
It is noted that there is no $\lambda$-independent source term for \eq{eq:beta1515PP2} and \eq{eq:beta1515PP3}.

\subsection{The true fixed point}

The full beta functionals 
that describe the flow of the coupling functions 
under the lowering of the energy scale 
and the dilatation of momentum along the Fermi surface
are Eqs. (\ref{eq:DifEqv})-(\ref{eq:DifElambda}),
where 
$\beta^{(v)}_k$,
$\beta^{(V_F)}_k$,
$\beta^{(g)}_{k'k}$ and
$\beta^{(\lambda)}_{\{k_i\}}$
are given by 
Eqs. 
(\ref{betaVgen1})-
(\ref{eq:main_eta}), 
Eqs. 
(\ref{eq:beta1111PH1})
-(\ref{eq:beta1111PH4}),
Eqs. 
(\ref{eq:beta1515PP1})
-(\ref{eq:beta1515PP4})
\footnote{
It is reminded that 
$\beta^{(v)}_k$,
$\beta^{(V_F)}_k$,
$\beta^{(g)}_{k'k}$ and
$\beta^{(\lambda)}_{\{k_i\}}$
describe
the flow of the coupling functions with increasing energy 
at {\it fixed} momenta without momentum dilatation.}.
In the space of the coupling functions,
a fixed point arises at
\bqa
 v_k=0, ~~   V_{F,k}=1, ~~ 
g_{k' k} = 0, ~~ 
\lambda^{\spmqty{N_1 & N_2 \\ N_3 & N_4};\spmqty{\sigma_1 & \sigma_2 \\ \sigma_4 & \sigma_3}}_{\spmqty{k+l & p-l \\ k & p}} =0,
\label{eq:true_fp}
\eqa
with $\frac{ g_{k' k}^2}{v_k}= \frac{\pi}{2}$.
Since the coupling functions in \eq{eq:true_fp} are independent of momentum,
$\beta^{(v)}_k$,
$\beta^{(V_F)}_k$,
$\beta^{(g)}_{k'k}$ and
$\beta^{(\lambda)}_{\{k_i\}}$
also vanish at the fixed point.
In the space of coupling functions,
\eq{eq:true_fp} is a singular point, and 
the theory with $g_{k'k}=0$ and $v_k=0$ 
is well defined 
only after the ratio 
between the Yukawa coupling and the nesting angle
is specified as the singular point is approached.
Since the leading-order quantum corrections 
for the renormalized boson propagator
are proportional to $g^2/v$,
the ratio determines 
the anomalous dimension of the boson.
For the non-interacting Gaussian theory, 
$\frac{ g_{k' k}^2}{v_{k''}}=0$.
With $\frac{ g_{k' k}^2}{v_{k''}}=\frac{\pi}{2}$,
there are infinitely many diagrams
that renormalize the boson propagator non-perturbatively.
These are included in Schwinger-Dyson equation in \eq{eq:BosProp},
which gives rise to anomalous dimension 
$1$ for the boson.
On the other hand,  
the anomalous dimension of the fermion 
is proportional to $g^2/c$
(\eq{eq:main_eta}).
At the interacting fixed point with
$\frac{ g_{k' k}^2}{v_{k''}}=\frac{\pi}{2}$,
$g_{k'k}^2/c$ vanishes
because $v_k$ and $c$ are related to each other through \eq{eq:CofV}.
Consequently, the fermion has no anomalous dimension,
and the dynamical critical exponent is $1$ at the fixed point.

It turns out that \eq{eq:true_fp} is an unstable fixed point as a generic perturbation added to the fixed point drives the theory toward a superconducting state at low energies.
We establish this 
by solving the full beta functionals 
of theories that are tuned away from the fixed point.
To be concrete, we consider a UV theory with 
\begin{align}
{v}_{k}=v_0\ll 1, ~~~
{V}_{\mathrm{F},k} =1, \nn
{g}_{k',k}  =\sqrt{\frac{\pi }{2}v_0}, ~~~
{\lambda}^{ \{N_i\}; \{ \sigma_i \} }_{ \{ k_i \} }=0
\label{eq:allIncond}
\end{align}
at a UV energy scale $\Lambda$.
Here, we consider a small but non-zero nesting angle.
The UV theory has momentum independent nesting angle,  Fermi velocity and 
Yukawa coupling with zero four-fermion coupling.
Considering this particular UV theory is  not a strong constraint for the following reasons.
First, although the bare four-fermion coupling is zero at a UV scale in \eq{eq:allIncond},
non-zero four-fermion couplings are generated 
from the spin fluctuations at low energies.
Conclusions drawn for the UV theory in \eq{eq:allIncond} also apply to theories 
in which the bare four-fermion coupling 
is weaker than the four-fermion coupling generated 
from the spin fluctuations.
Second, even if the bare coupling functions are chosen to be momentum-independent in \eq{eq:allIncond},
renormalized coupling functions acquire non-trivial momentum dependence at low energies.
We will see that 
the universal momentum profiles 
of the coupling functions
become singular near the hot spots
in the low energy limit\footnote{
Below the superconducting transition temperature scale,
the singularity is cut off.
}.
If a UV theory has momentum dependent bare coupling functions,
which must be smooth as functions of 
momentum due to locality,
the renormalized coupling functions acquire the
same universal singularities near the hot spots
on top of the smooth profile of coupling functions.
Third, the value of the Yukawa coupling is merely a convention.
It can be chosen to be any $O(1)$ number 
in the absence of the bare boson kinetic term 
that can be dropped at low energies
in the vicinity of the interacting fixed point.
Finally, the simple momentum independent  coupling functions chosen in \eq{eq:allIncond}  happens to possess the PH symmetry.
In this case, 
the four-fermion couplings also receive 
singular quantum corrections
in the $2k_F$ scattering channels 
in which the external momenta take values shown in \eq{eq:SofIS1_2} and \eq{eq:SofIS2_2}.
The full beta functionals  that include the $2k_F$ scattering channels are derived in Appendix \ref{sec:additional_beta}.
However, at low energies,
one can ignore the contributions from those additional couplings to the  flow of the couplings in the  BCS pairing channel that drives the superconducting instability.
This is because the volume of the phase space in which the $2k_F$ scattering channels overlap with the BCS pairing channels vanishes in the low energy limit.
For details, see  Appendix \ref{sec:additional_beta}.

As it is the case for the hot spot theory,
we organize quantum corrections in terms of $v \sim g^2$
to understand the full functional RG flow in the vicinity
of the fixed point. 
$g^2/c \sim \sqrt{ v/\log(1/v)}$ controls the strength of the
quantum corrections associated with the Yukawa coupling 
beyond the non-perturbative quantum corrections 
that is included through \eq{eq:BosProp}.
The four-fermion coupling 
generated from the critical spin fluctuations
is order of $g^4/c$.
Since the four-fermion coupling generated from the loop corrections is smaller 
than the tree-level interaction 
mediated by the spin fluctuations,
one can understand 
the RG flow of $v$, $V_F$ and $g$
without including the feedback of 
the four-fermion coupling 
until the four-fermion coupling become large 
due to superconducting instability
in the low-energy limit.
We will see that 
in the small $v$ limit
there is a large window of 
energy scale in which the feedback 
of the four-fermion coupling can be ignored.
In the low-energy limit in which the four-fermion coupling becomes stronger than the 
interaction mediated by the spin fluctuations,
the four-fermion coupling dominates the physics and the spin fluctuations become largely unimportant.

\section{Quasi-fixed points}
\label{sec:fixedpoint}

\fbox{\begin{minipage}{48em}
{\it
\begin{itemize}
\item
A theory that is tuned away from the fixed point with a non-zero nesting angle develops non-trivial momentum profiles for the coupling functions  at low energies even if the couplings are momentum independent at the UV scale.

\item
If the bare nesting angle is small and momentum independent,
the full RG flow stays close to the subspace  with a fixed nesting angle 
over a large window of length scale
due to the slow flow of the nesting angle.
The rest of the coupling functions run significantly faster,
and the RG flow within a subspace 
with a fixed nesting angle is characterized by quasi-fixed points which are non-Hermitian 
at non-zero nesting angles.
The true Hermitian fixed point arises at the end point of one parameter families of the quasi-fixed points.

\item
In the limit that the nesting angle is small,
the proximity of the quasi-fixed points to 
the space of Hermitian theories creates
a bottleneck region with constricted RG flow of Hermitian theories. 
\item

With a non-zero bare nesting angle,
the four-fermion coupling in the pairing channel
inevitably flows to the strong coupling regime 
in the low-energy limit, 
signifying a superconducting instability. 
\item
The functional renormalization group flow projected to a finite dimensional subspace of the four-fermion coupling functions exhibits qualitatively different flow depending on the relative weight between the hot and cold electrons.
In channels with too much weight on hot or cold electrons, superconducting instabilities are suppressed 
by the strong pair breaking effect caused by spin fluctuation-induced incoherence
or the lack of pairing glue, respectively.
However, there always exist channels  that include lukewarm electrons that are unstable.

 \end{itemize}
}
\end{minipage}}
\vspace{0.5cm}

In this section, we study the RG flow of the UV theory 
that are tuned away from the fixed point as in \eq{eq:allIncond}. 
Because the feedback of the four-fermion coupling to 
 the nesting angle, Fermi velocity and the Yukawa coupling 
 is negligible over a large window of length scale,
we can first focus on the flow equations of $v, V_F$ and $g$.
Then, we discuss the beta functional 
for the four-fermion coupling.

\subsection{Fermi velocity and electron-boson coupling}

From 
Eqs. 
(\ref{betaVgen1})-
(\ref{eq:main_eta}), 
the beta functionals for $v_k, V_{F,k}$ and $g_{k',k}$ 
are written as
\begin{align}
    &\FR{\di v_k(\ell)}{\di\ell}    
= v_k(\ell) 
\Biggl[ 
        -\FR{4(N_c^2-1)}{\pi^3N_cN_f}\FR{g_k(\ell)^2}{V_{F,k}(\ell)^2}\log\PFR{V_{F,k}(\ell)}{c(\ell)}\Theta_1(\ell,\elltwokf) 
\nn &\hspace{1.0cm}
        - \FR{2(N_c^2-1)}{\pi^4N_c^2N_f^2}
        \FR{g_k(\ell)^4}{c(\ell)^2V_{F,k}(\ell)^2}
            \log^2\PFR{V_{F,k}(\ell)v_k(\ell)}{c(\ell)}
        \Theta_1(\ell,\ellonekf) 
\Biggr], 
\label{betaVgen3}\\
&\FR{\di V_{F,k}(\ell)}{\di\ell}  = V_{F,k}(\ell)
\Biggl[ 
        \FR{2(N_c^2-1)}{\pi^3N_cN_f}\FR{g_k(\ell)^2}{V_{F,k}(\ell)^2}\log\PFR{V_{F,k}(\ell)}{c(\ell)}\Theta_1(\ell,\elltwokf)
        -\FR{N_c^2-1}{\pi^2N_cN_f}\FR{g_k(\ell)^2}{c(\ell)V_{F,k}(\ell)}\Theta_1(\ell, \elltwokf)
\nn &\hspace{1.0cm}
        -\frac{3}{2} \FR{N_c^2-1}{\pi^2N_cN_f}v_0(\ell)\log\PFR{1}{c(\ell)}
        +\FR{N_c^2-1}{2\pi N_cN_F}w_0(\ell)
\nn &\hspace{1.0cm}
        + \FR{(N_c^2-1)}{2\pi^4N_c^2N_f^2}
        \FR{g_k(\ell)^4}{c(\ell)^2V_{F,k}(\ell)^2}
            \log^2\PFR{V_{F,k}(\ell)v_k(\ell)}{c(\ell)}
        \Theta_1(\ell,\ellonekf) 
\Biggr], 
\label{betaVfgen3}\\
&\FR{\di g_{k',k}(\ell)}{\di\ell} = g_{k',k}(\ell)
\Biggl[ 
        -\FR{1}{2\pi N_cN_f}w_0(\ell)\log\PFR{1}{w_0(\ell)}
        +\FR{N_c^2-1}{2\pi N_cN_f}w_0(\ell)
        -\FR{(N_c^2-1)}{\pi^2N_cN_f}v_0(\ell)\log\PFR{1}{v_0(\ell)} 
\nn &\hspace{1.0cm}
        -\FR{(N_c^2-1)g_k(\ell)^2  \Theta_1(\ell, \elltwokf) }{2\pi^2N_cN_fc(\ell)V_{F,k}(\ell) }
        -\FR{(N_c^2-1)g_{k'}(\ell)^2 \Theta_1(\ell, \elltwokpf) }{2\pi^2N_cN_fc(\ell)V_{F,k'}(\ell)}
\nn &\hspace{1.0cm}
        +\frac{2 g_k(\ell)g_{k'}(\ell)}{\pi^2c(\ell)N_cN_{f}(V_{F,k}(\ell)+V_{F,k'}(\ell))} \log\left(\frac{c(V_{F,k}(\ell)^{-1}+V_{F,k'}(\ell)^{-1})}{v_k(\ell)+v_{k'}(\ell)}\right)
        \Theta_2(\ell, 
        \elltwokf,
        \elltwokpf,
        \ellthreekkpf)
\Biggr].
\label{mainTextBetaGgen3}
%
%
\end{align}
Here  $\ell$ is the logarithmic length scale, 
$\ell = \log \Lambda /\mu$.
$
\Theta_i( \ell, \ell_1,..,\ell_i) = 
\theta_i( \Lambda \emell, \Lambda e^{-\ell_1},..,\Lambda e^{-\ell_i})
$
are crossover functions 
written in terms of
the logarithmic length scale,
where $\theta_i$ is defined in \eq{eq:theta_i}.
At short  (long) distance scales with $\ell \ll \min_i\{\ell_i\}$ ($\ell \gg \min_i\{\ell_i\}$),
$\Theta_i( \ell_1,..,\ell_i, \ell) \approx 1$
($\Theta_i( \ell_1,..,\ell_i, \ell) \approx 0$),
and the corresponding term in the beta function is turned on (off).
$\ellonekpf$, $\elltwokpf$ and $\ellthreekkpf$ 
are the logarithmic
length scales that mark three crossovers, 
\bqa
\ellonekf     = \log \left( \frac{\Lambda  }{\Etwox} \right), ~~
\elltwokf     = \log \left( \frac{\Lambda  }{\Eonex } \right),~~
\ellthreekkpf = \log \left( \frac{\Lambda  }{\Eonexpx } \right),
\label{eq:defLs}
\eqa
where 
$\ellonekf$,
$\elltwokf$ and
$\ellthreekkpf$
are implicit functions of $\ell$
through 
$v_k(\ell)$, 
$V_{F,k}(\ell)$ 
and $c(\ell)$
that enter in Eqs.
\eqref{eq:E1Lk}-
\eqref{eq:E1Lkkp}.
Crossovers occur when
$\ell$ crosses one or more of these length scales 
defined by the self-consistent equations\footnote{
It is noted that $\Eonex $, $\Eonexpx$ and $\Etwox$ are functions of $\ell$ because  the coupling functions in Eqs. (\ref{eq:E1Lk})-(\ref{eq:E1Lkkp})  depend on $\ell$.
}, 
\begin{equation}
\begin{aligned}
 \ellonek  = & L^{(2L)}(k; \ellonek), \\
 \elltwok   = &  L^{(1L)}(k; \elltwok), \\
 \ellthreekkp  = & 
 \min
 \Bigl\{ 
L^{(1L)}(k; \ellthreekkp),
L^{(1L)}(k'; \ellthreekkp),  
L^{(1L)}(k',k; \ellthreekkp) 
\Bigr\}.
\end{aligned}
\label{SelfConsistentLengthScales}
\end{equation}
The solutions to 
\eq{SelfConsistentLengthScales} are given by 
(see Appendices
\ref{sec:FirstCrossoverScale})
\begin{equation}
\begin{aligned}
\ellTwoLoopX & \approx  \log\PFR{\Lambda}{4v_0(0)k} 
+\log\left( 1 + \FR{1}{\ell_0}  \log\PFR{\Lambda}{4v_0(0)k}\right), \\
\ellOneLoopX & \approx \ellTwoLoopX +   \log\FR{2}{c(\ellonek)}, \\
\ellthreekkp & \approx \min(\ell^{(2L)}_{(k+k')/2},\ell^{(1L)}_k,\ell^{(1L)}_{k'}).
\end{aligned}
\label{eq:crossoverscales}
\end{equation}
Here $\ell_0$ is the length scale that parameterizes the bare nesting angle at $\ell=0$ as defined in \eq{eq:L0}.

\begin{figure*}
	\centering
	\includegraphics[scale=1]{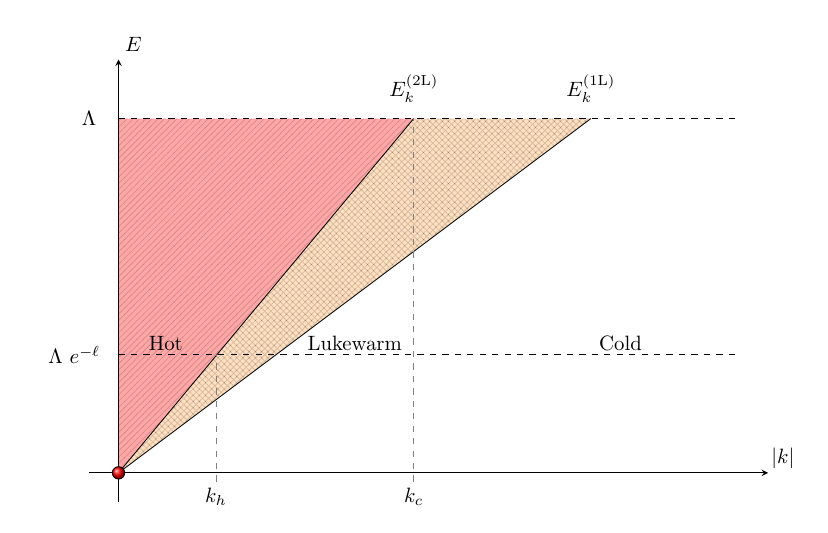}
	\caption{
Crossover energy scales associated 
with the two-loop fermion self-energy ($E^{(2L)}_k$)
and the one-loop fermion self-energy ($E^{(1L)}_k$),
respectively.
Since the flow of 
the couplings 
between 
$E^{(2L)}_k$ and  $E^{(1L)}_k$ is negligible,
one can consider only
one crossover, 
say  $E^{(2L)}_k$.
At energy scale 
$\Lambda \emell$,
the momentum space is divided into three regions.
In the hot region ($k<k_h$), electrons receive quantum correction from the UV scale all the way down to the current energy scale 
$\Lambda \emell$.
In the lukewarm region ($k_h< k < k_c$), electrons received some quantum correction at high energies but are decoupled from spin fluctuations at the current energy scale.
In the cold region ($k>k_c$), electrons do not receive any quantum correction.
}
	\label{fig:crossover}
\end{figure*}

Since the RG flow equations for 
$v_k, V_{F,k}, g_k$ 
do not depend on the off-diagonal 
Yukawa coupling $g_{k',k}$, 
we first solve the beta functions
for  $v_k(\ell), V_{F,k}(\ell), g_k(\ell)\equiv g_{kk}(\ell)$. 
For $k'=k$, $\ellthreekkp = \ell^{(2L)}_{k}$. 
The beta functionals in \eq{mainTextBetaGgen1}
take different forms in each of the three windows of length scale :
(i) $\ell < \ellTwoLoopX$,
(ii) $\ellTwoLoopX < \ell < \ellOneLoopX$,
and (iii) $\ellOneLoopX < \ell$. 
These regions are depicted in \fig{fig:crossover}.
In region i), electrons receive renormalization 
from all quantum corrections generated 
by spin fluctuations to the leading order in $v$.
In region ii), the two-loop correction 
is turned off, but the one-loop quantum corrections are still  present.
In region iii), electrons are decoupled from spin fluctuations.
To have the global solution,
we solve the beta functions to obtain
 $J_k^{(i)}(\ell)$, $J_k^{(ii)}(\ell)$, $J_k^{(iii)}(\ell)$ for $J_k=v_k,V_{F,k},g_k$ 
 in each energy window,
 and the coupling functions are matched at the boundary to ensure continuity.

In the small $v$ limit, 
$h^{(1)}_k$,
$h^{(2)}_k$,
$h^{(1)}_{k'k}$
scale as $w=v/c \sim \sqrt{v/\log(1/v)}$.
As a result, the beta functionals for $J_k=v_k,V_{F,k},g_k$ are bounded by
$ \frac{1}{J} \frac{\partial J}{\partial \ell} \sim w \ll 1$.
Since $\ellOneLoopX - \ellTwoLoopX \sim \log 1/c$,
the net change of the couplings that occur 
in $\ellTwoLoopX < \ell < \ellOneLoopX$
is only 
$\frac{ | J(\ellOneLoopX) - J(\ellTwoLoopX)|}{J(\ellTwoLoopX)} \sim w \log(1/c) \ll 1$ in the small $v$ limit. 
Since the flow of couplings can be ignored in region (ii),
we can use $J^{(iii)}_k$ 
in both regions (ii) and (iii). 
In this case, 
we only need to consider one crossover scale
$E^{(2)}_k$ besides the UV cutoff scale $\Lambda$.
At a given length scale $\ell$,
the Fermi surface can be 
divided into hot, lukewarm and cold regions
separated by two momentum scales 
$k_h(\ell)$ and $k_c(\ell)$
defined by
\begin{align}
\label{defnX1X2}
\ell^{(2L)}_{k_h(\ell)} = \ell, \quad
\ell^{(2L)}_{k_c(\ell)} = 0.
\end{align}
Below, we summarize the 
solution to the RG equation in each region.
The derivation can be found in Appendix \ref{appendixA_SingleParticle}.
\begin{itemize}
\item
Hot region :
In  $0 \le |k| \le k_h(\ell)$,
the momentum  dependent IR cutoffs 
are smaller than the energy scale ($\Lambda e^{-\ell}$),
and electrons remains strongly coupled with spin fluctuations.
Accordingly,
$v_k$, $V_{F,k}$ and $g_k$ 
flow in the same way as in the hot spots : 
$v_k(\ell)  =	
\FR{\pi^2 N_c N_f}{2(N_c^2-1)}  \FR{1}{(\ell+\ell_0)\log(\ell+\ell_0)}$,
$V_{F,k}(\ell)=1$
and
$g_k(\ell)=\sqrt{\frac{\pi v_0(\ell) }{2}}$ as in  \eq{eq:ZeroMomParam}.
Here $V_{F,k}(\ell)$ at the hot spot is chosen to be $1$ 
as  the reference speed with respect to which
all other speeds are measured.
\item
Lukewarm region :
In $k_h(\ell)  \le |k| \le  k_c(\ell)$,
the momentum dependent IR cutoff 
is larger than the floating energy scale, $\mu = \Lambda \emell$,
but smaller than the UV cut off, $\Lambda$.
Electrons in this range of momentum receive quantum corrections 
within a window of energy scale given by $  \Lambda e^{ -\ellonek } < E < \Lambda$
before they decouple below $ \Lambda e^{ -\ellonek }$.
Once electrons are decoupled,
the coupling with spin fluctuations  
decreases with increasing $\ell$
while $v_k(\ell)$ freezes out.
Since spin fluctuations no longer slow electrons down,
$V_{F,k}(\ell)$ increases relative to $V_{F,0}(\ell)$.
The momentum profiles of the coupling functions
are determined from the momentum dependent IR cutoff.
Since electrons farther away from the hot spots 
decouple from spin fluctuations at higher energies,
$g_k(\ell)$ decays while
$v_k(\ell)$ and $V_{F,k}(\ell)$ 
increases with increasing momentum.
It is noted that the division of 
the hot and lukewarm regions depends
on the energy scale.
As the energy is lowered,
the hot region shrinks
while the lukewarm region grows
as more electrons on the Fermi surface
are decoupled from spin fluctuations.
\item
Cold region :
In $|k| \ge  k_c(\ell)$,
electrons are too far away from the hot spots
to receive any significant renormalization
from spin fluctuations 
at energies below $\Lambda$.
In this region, 
$v_k(\ell)$ does not run,
while $g_k(\ell)$ ($V_{F,k}(\ell)$) constantly 
decreases (increases) with increasing $\ell$.
\end{itemize}

\begin{figure}[ht!]
        \centering
        \begin{subfigure}[t]{0.49\textwidth}
        \includegraphics[width=\linewidth]{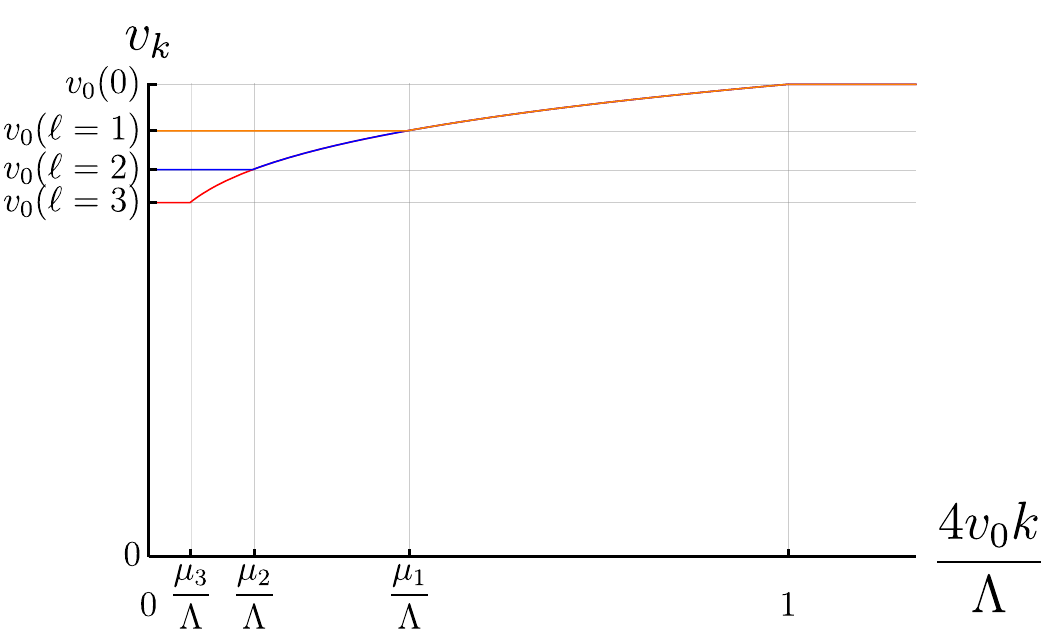}
        \caption{}
        \label{figVPLOT1}
        \end{subfigure}
        %
        \begin{subfigure}[t]{0.49\textwidth}
        \includegraphics[width=\linewidth ]{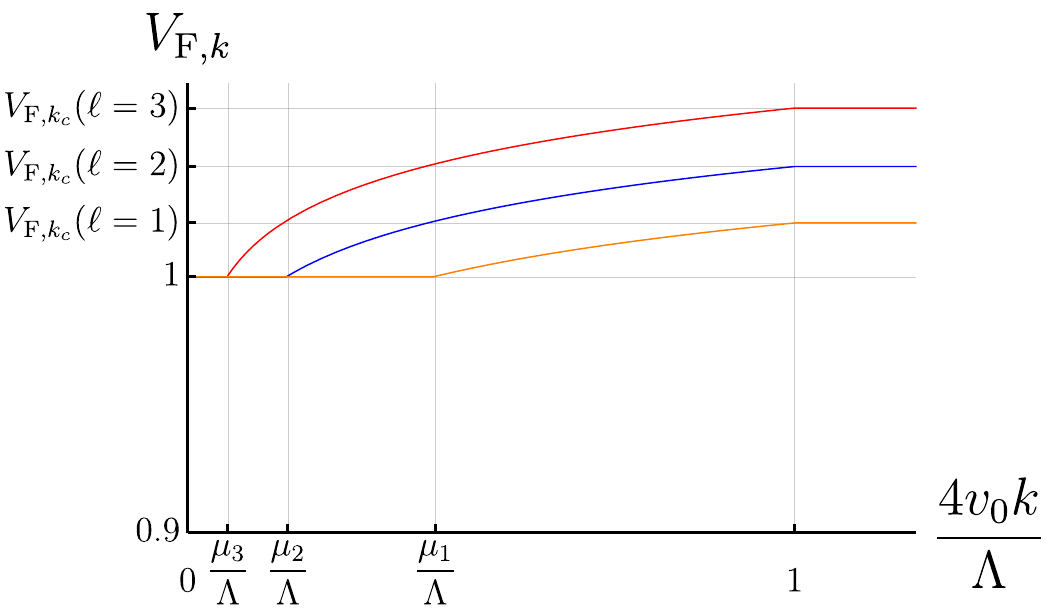}
        \caption{}
        \label{figVFPLOT2}
        \end{subfigure}
        \\
        \begin{subfigure}[t]{0.49\textwidth}
        \includegraphics[width=\linewidth]{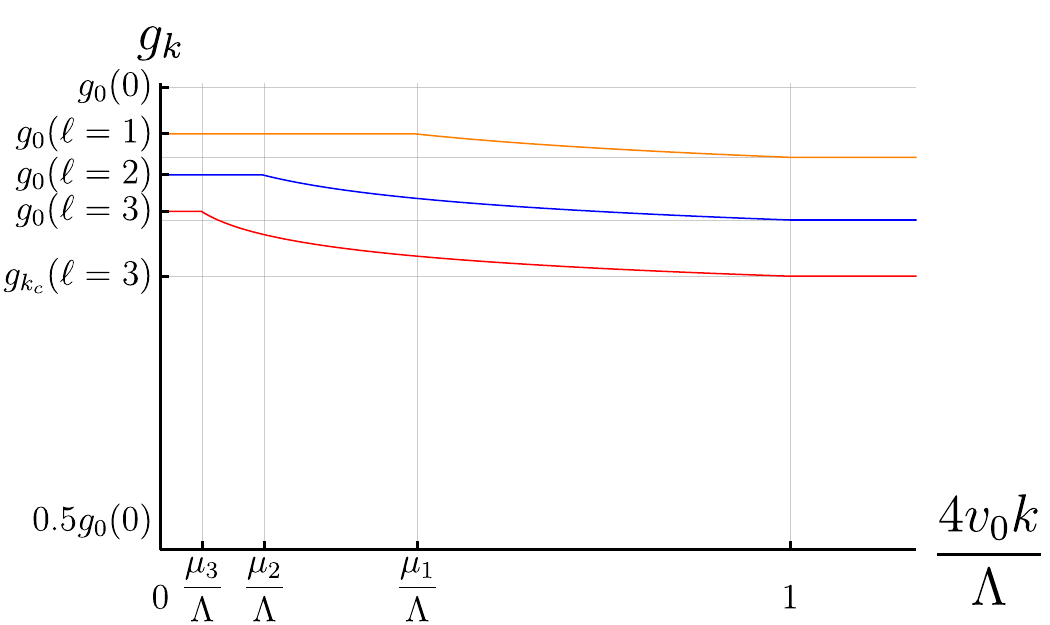}
        \caption{}   
        \label{figGPLOT1}
        \end{subfigure}
         %
        \begin{subfigure}[t]{0.49\textwidth}
        \includegraphics[width=\linewidth]{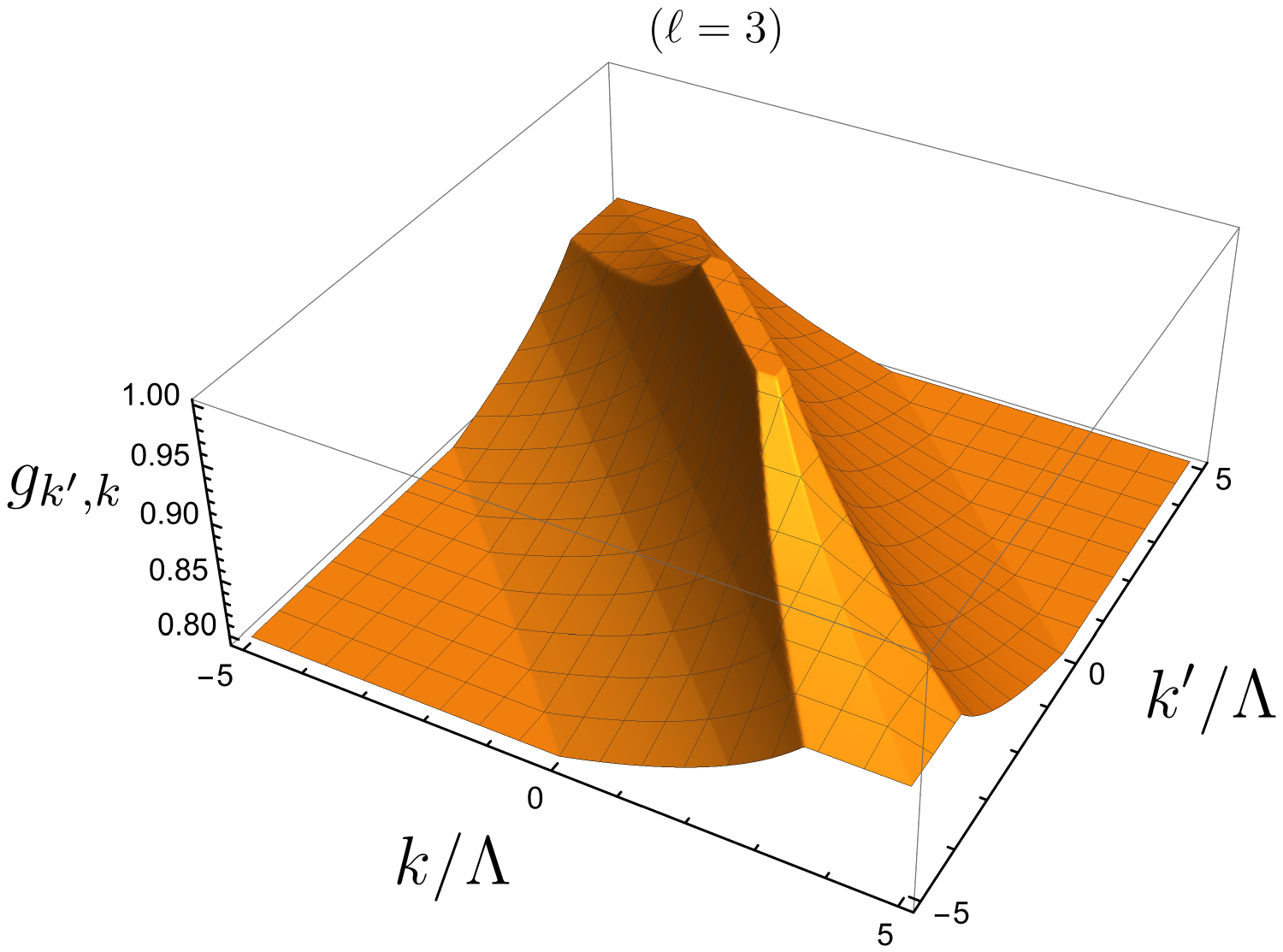}
        \caption{}
        \label{figGKKP1}
        \end{subfigure}
        \caption{
(a) $v_k(\ell)$,
(b) $V_{F,k}(\ell)$,
(c) $g_k(\ell)\equiv g_{kk}(\ell)$
and
(d) $g_{k'k}(\ell)$ 
plotted as functions of momentum along the Fermi surface for the theory with a constant bare nesting angle, $v_0(0) = 0.1$.
For plots in (a)-(c), the coupling functions are shown for $\ell=1,2,3$,
where 
$\mu_\ell = \Lambda e^{-\ell}$
denotes the energy scale associated with $\ell=1,2,3$. 
In the hot region near $k=0$, the coupling functions are essentially independent of momentum, 
taking the values of the  coupling constants of the hot spot theory. 
The coupling functions are also constants in the cold region as electrons far away from the hot spots are not normalized.
In the lukewarm region that interpolates the hot and cold regions, the coupling functions acquire non-trivial momentum dependence.
As the energy is lowered, the size of the hot region near the hot spot decreases  as more electrons become decoupled from spin fluctuations.
At a fixed energy,
$v_k$ and $V_{F,k}$ decreases with decreasing $k$ because spin fluctuations remain coupled with electrons down to lower energies
($V_{F,k}$ is measured in the unit of the hot spot velocity which is set to be $1$).
On the contrary, the Yukawa coupling increases with decreasing $k$.
The kinks in the plots are  the artifact of not keeping the precise crossover functions between the regions.
For the off-diagonal Yukawa coupling in (d), $\ell=3$ is chosen.
}
\label{fig:vVFg}
\end{figure}

At a given scale $\ell$,
the momentum dependent coupling functions are obtained to be
\begin{align}
\label{vMPsimplified}
v_k(\ell) &= 
\begin{cases}
v_0(\ell)                           & k < k_h(\ell)\\
v_0(\ellonek)                       & k_h(\ell) < k < k_c(\ell)\\
v_0(0)                              &  k_c(\ell) < k
\end{cases}
\\
\label{gxxMPsimplified}
g_k(\ell) &= 
\begin{cases}
\sqrt{\pi v_0(\ell)/2}                             & k < k_h(\ell)\\
\sqrt{\FR{\pi}{2} v_0(\ellonek)} ~ \mc E_0(\ell;\ellonek) & k_h(\ell) < k < k_c(\ell)\\
\sqrt{\pi v_0(0)/2}  ~
\mc E_0(\ell;0)               &  k_c(\ell) < k
\end{cases},\\
\label{vfMPsimplified}
V_{F,k}(\ell) &= 
\begin{cases}
1     & k < k_h(\ell)\\
\mc E_1(\ell;\ellonek)              & k_h(\ell) < k < k_c(\ell)\\
\mc E_1(\ell;0)                     &  k_c(\ell) < k
\end{cases},
\end{align}
where
\begin{align}
v_0(\ell)  &=	\FR{\pi^2 N_c N_f}{2(N_c^2-1)}  \FR{1}{(\ell+\ell_0)\log(\ell+\ell_0)}, \\ 
\mc E_0(X,Y) &\equiv \exp\left(-\FR{\sqrt{X+\ell_0}-\sqrt{Y+\ell_0}}{\sqrt{N_c^2-1}}\right),\\
\mc E_1(X,Y) &\equiv
\exp\left(\sqrt{N_c^2-1}\left(\mathrm{Ei}(\log\sqrt{X+\ell_0})-\mathrm{Ei}(\log\sqrt{Y+\ell_0})\right)\right).
\end{align}
For the general Yukawa coupling,
the solution of the beta functional is given by
\begin{align}
\label{eq:gofffinal}
g_{k,k'}(\ell) &= 
\begin{cases}
\sqrt{\FR{\pi}{2}v_0(\ell)}                                   & \ell \le \ellthreekkp\\
\sqrt{\FR{\pi}{2}v_0(\ellthreekkp)}\mc E_0(\ell,\ellthreekkp) & \ell \ge \ellthreekkp.
\end{cases}
\end{align}
%
\eq{eq:gofffinal}
can be understood in the following way.
The RG flow of $g_{k',k}$ is  
controlled by the vertex correction 
and the self-energy corrections
to the electrons at momenta $k$ and $k'$.
In the small $v$ limit, the vertex correction is the dominant factor, 
and the momentum dependence of the renormalized Yukawa coupling 
is largely determined by the crossover scale at which the vertex correction turns off.
At the hot spots, the vertex correction is present at all energy scales.
The vertex correction tends to enhances 
the Yukawa coupling at low energies
due to the anti-screening effect 
associated with the non-Abelian nature 
of the $SU(N_c)$ group\cite{SHOUVIK3}.
Once this vertex correction is absorbed by
the field renormalization of the boson,
the Yukawa coupling between 
the boson and electrons near the hot spots 
is kept to be the order of $\sqrt{v_0}$ 
at all energy scales, 
while the boson is endowed with 
the large anomalous dimension.
For electrons away from hot spots,
the vertex correction turns off for $\ell > \ellthreekkp$.
At low energies, the Yukawa coupling decays 
because the boson that remains strongly renormalized by electrons near the hot spots is too `heavy' 
to stay coupled with electrons away from hot spots 
without the help of the anti-screening vertex correction :
the large scaling dimension of the boson caused by hot electrons makes the boson to decouple from lukewarm 
electrons at low energies.
The momentum dependent coupling functions are shown in \fig{fig:vVFg}. 

It is noted that $v_0(\ell)$ flows to zero in the large $\ell$ limit.
Therefore, only $v_0=0$ is a true fixed point.
However, for $\ell_0 \gg 1$
there exists a large window of scale $0< \ell \ll \ell_0$
within which $v_k(\ell)$ does not change appreciably as a function of $k$ and $\ell$.
Within this window of length scale, 
$v_k(\ell)$ is well approximated by $v_0(0)$,
and physical observables obey 
scaling relations controlled
by exponents that depend on $v_0(0)$.
Therefore, we can consider an one-parameter family of {\it quasi-fixed points} labeled by $v_0$.
To characterize those quasi-fixed points,
let us first extract the scaling behaviour of the coupling functions.
Eqs. (\ref{vfMPsimplified})-(\ref{eq:gofffinal}) describes
how the coupling functions at a fixed physical momentum
evolve as the energy scale is lowered.
The coupling functions will exhibit scale invariance
if the momentum along the Fermi surface is simultaneously scaled  
 as the energy scale is lowered.
To find a scale invariant fixed point,
we consider the coupling functions
defined in \eq{eq:LambdaAux},
\bqa
\hat v_K = v_{k}, \quad
\hat V_{F,K} = V_{F,k}, \quad
\hat g_{K,K'} = g_{k, k'},
\label{eq:vGghat}
\eqa 
where $K=k \eell$, $K'=k' \eell$ are the rescaled momenta.
For fixed $K$ and $K'$,  
the hatted coupling functions in \eq{eq:vGghat}
become independent of $\ell$ to the leading order in $\ell/\ell_0$,
\bqa
\hat v_K &= & v_0(0), \label{hatvMPsimplified} \\
\hat V_{F,K} &= &
\begin{cases}
1     &  |K|  < K_h \\
\left( \frac{|K|}{K_h}\right)^{\alpha_1(0) }                 &K_h  < |K| <  K_c   \\
\left( \frac{K_c}{K_h}\right)^{\alpha_1(0) }                 &   K_c < |K|  \\
\end{cases},
\label{hatvfMPsimplified} \\
\hat g_K &= &
\begin{cases}
\sqrt{\FR{\pi}{2} v_0(0)}                              &|K|< K_h \\
\sqrt{\FR{\pi}{2} v_0(0)} ~ 
\left( \frac{K_h}{|K|}\right)^{\alpha_0(0)  }     &K_h  < |K| <  K_c   \\
\left( \frac{K_h}{K_c}\right)^{\alpha_0(0) }                 &   K_c < |K|  \\
\end{cases},
\label{hatgxxMPsimplified} \\
\hat g_{K,K'} &= &
\begin{cases}
\sqrt{\FR{\pi}{2}v_0(0)}                                   &
\max\{ \frac{| K+ K'|}{2},  \frac{c |K|}{2},   \frac{c |K'|}{2} \} 
< K_h\\
\sqrt{\FR{\pi}{2}v_0(0)}     \left( \frac{K_h}{
\max\{ \frac{| K+ K'|}{2},  \frac{c |K|}{2},   \frac{c |K'|}{2} \} 
}\right)^{\alpha_0(0)  }                               &  K_h < 
\max\{ \frac{| K+ K'|}{2},  \frac{c |K|}{2},   \frac{c |K'|}{2} \} 
< K_c\\
\sqrt{\FR{\pi}{2}v_0(0)}     \left( \frac{K_h}{K_c} \right)^{\alpha_0(0)  }                               
&K_c <
\max\{ \frac{| K+ K'|}{2},  \frac{c |K|}{2},   \frac{c |K'|}{2} \} 
\end{cases}, 
\label{hatgxxpMPsimplified}
\eqa
where
$K_h = \eell k_h(\ell) = \frac{ \Lambda}{4 v_0(0)}$ and
$K_c = \eell k_c(\ell) = \frac{ \Lambda \eell}{4 v_0(0)}$
represent the rescaled crossover momenta 
from hot to lukewarm,
and from lukewarm to cold region, 
respectively, and
\begin{equation}
\begin{aligned}
\alpha_1(\ell) =& \frac{\sqrt{N_c^2-1}}{\sqrt{\ell_0+\ell} \log (\ell_0+\ell)}, \\
\alpha_0(\ell) =&\frac{1}{2 \sqrt{N_c^2-1} \sqrt{\ell_0+\ell}  }
\end{aligned}
\label{eq:alpha01}
\end{equation}
are the critical exponents that govern the universal power-law decays
of the coupling functions in the momentum space.
The exponents only depend on the nesting angle
within the line of quasi-fixed points.
It is noted that 
$\hat V_{F,K}$ and $\hat g_{K',K}$ depend on momentum
more strongly than  $\hat v_K$.
As a result, the momentum dependence of the former two
can not be ignored even if the latter is regarded
as constant to the leading order in $\ell/\ell_0$.
Across the lukewarm region,
each of the coupling functions changes by
$\frac{\hat v_{K_c}}{\hat v_{K_h}} \approx
  1 + \frac{\ell}{\ell_0} $,
$\frac{\hat V_{F,K_c}}{\hat V_{F,K_h}} \approx
e^{ \sqrt{N_c^2-1} \frac{\ell}{\sqrt{\ell_0} \log \ell_0 } }$,
and
$\frac{\hat g_{K_c}}{\hat g_{K_h}} \approx 
e^{-\frac{1}{2 \sqrt{N_c^2-1}} \frac{\ell}{\sqrt{\ell_0}} }$. 
For $\ell/\ell_0 \ll 1$,
$\frac{\hat v_{K_c}}{\hat v_{K_h}} \approx 1$.
However, the variations of 
the Fermi velocity and the Yukawa coupling are
not negligible if $\ell/\sqrt{\ell_0} \geq 1$.

Although the coupling functions acquire
non-trivial momentum profiles 
at low energies,
they 
do not vary much over momentum scales that are proportional to
the energy scale $\mu$.
From the momentum profiles of the coupling functions
in  Eqs.  \eqref{hatvMPsimplified}- \eqref{hatgxxpMPsimplified}, 
 we can now check the validity of the adiabaticity in \eq{eq:adiabatic_condition}.
The relative variation of the couplings within the range of $\Pi_\mu = \frac{\mu}{vc}$ is given by
\bqa
\epsilon_\mu
\sim
\alpha_0 \log\frac{1}{vc}
\eqa
in the small $v$ limit.
Up to a logarithmic correction, 
$\epsilon_\mu \sim \sqrt{v} \ll 1$
and
\eq{eq:adiabatic_condition} is satisfied at all energy scales.

$\hat g_{K',K}$ in \eq{hatgxxpMPsimplified} decays
in a power law as a function of momenta
in all directions in the space of $K'$ and $K$.
The power-law decay of the off-diagonal elements of $\hat g_{K',K}$
signifies the importance of the non-forward scatterings.
Consequently, the number of electrons at each 
momentum is not conserved.
%
%
There is still a sense in which
a large symmetry emerges at low energies without the four-fermion coupling\cite{PhysRevX.11.021005}.
In the space of rescaled momentum,
$k_F$ runs to infinity in the low-energy limit.
Because the range of the off-diagonal elements are fixed
and the size of Fermi surface blows up under the RG flow,
the number of electrons within a finite fraction
of the total system becomes better
conserved as the energy is lowered.
This can be also understood by examining
the coupling function in the space
of physical momenta,
$k=K \emell$, $k'=K' \emell$. 
If the large $\ell$ limit is taken for fixed $k$ and $k'$,
the Yukawa coupling function vanishes
for $k$ and $k'$ away from the hot spots.
The four-fermion coupling functions generated from the Yukawa coupling give rise to stronger large-angle scatterings 
that invalidate the patch description more severely. 
In the following  two sections, we examine the RG flow of the four-fermion coupling function projected to the space of a fixed nesting angle.

\subsection{Four-fermion coupling in group 1 : the singular Landau function}
\label{sec:four_fermion_fp1}

We now turn our attention to the four-fermion coupling functions.
The couplings in group 1 describes the forward scattering
whose beta functional is written as
\begin{equation}
\begin{aligned}
\frac{\partial}{\partial\ell}
{\lambda}_{\spmqty{p & k \\  k &  p}}^{\spmqty{N_1 & N_2 \\ N_4 & N_3};\spmqty{\sigma_1 & \sigma_2 \\ \sigma_4 & \sigma_3}}
 & = -\left(1+3(z-1)+2 \eta^{(\psi)}_k+2 \eta^{(\psi)}_p\right)
{\lambda}_{\spmqty{p & k \\  k &  p}}^{\spmqty{N_1 & N_2 \\ N_4 & N_3};\spmqty{\sigma_1 & \sigma_2 \\ \sigma_4 & \sigma_3}}
 -\frac{S_{k,p}}{N_f^2}
  \mathsf{T}^{\beta \sigma_2}_{\sigma_4 \alpha} 
  \mathsf{T}^{\sigma_1 \alpha}_{\beta \sigma_3} 
 \delta^{N_1}_{N_4}
 \delta^{N_2}_{N_3}
\\
& +\frac{B_k}{2N_f} 
\lambda^{\spmqty{N_1 & \bar N_2 \\ \bar N_4 &  N_3};\spmqty{\sigma_1 & \alpha \\ \beta & \sigma_3}}_{\spmqty{p  & k \\ k & p }}
 \mathsf{T}^{\beta \sigma_2}_{\sigma_4 \alpha} 
 + \frac{B_p}{2N_f} 
  \mathsf{T}^{\sigma_1 \beta}_{\alpha \sigma_3} 
 \hat\lambda^{\spmqty{\bar N_1 & N_2 \\ N_4 & \bar N_3};\spmqty{\alpha & \sigma_2 \\ \sigma_4 & \beta}}_{\spmqty{p  & k \\ k & p }}.
 \label{eq:lambda1111all}
\end{aligned}
\end{equation}
Here
$\spmqty{N_1 & N_2 \\ N_4 & N_3}$
is one of the elements in the set of
\bqa
H_{1111}^{PH}=\Big\{ 
\spmqty{1 & 1 \\ 1 & 1},
\spmqty{1 & 4 \\ 4 & 1},
\spmqty{4 & 1 \\ 1 & 4},
\spmqty{4 & 4 \\ 4 & 4}
\Bigr\},
\eqa
and
\bqa
B_{k} = \frac{g^2_{k}}{\pi^2 c V_{F,k}}\frac{\mu }{\mu  + 2v_kc|k|_\mu}~, ~~~~
S_{k,p} = \mathsf{D}_{\mu}(k;p)(B_k+ B_p)
\label{eq:VXSKP0}
\eqa
represent the momentum dependent vertex correction
and the source term generated from the spin fluctuations, respectively.
Since the coupling measured in the unit of the Fermi velocity represents the physical strength of interaction\footnote{
For example, the perturbation series in the four-fermion coupling is organized in terms of the ratio between the four-fermion coupling and the Fermi velocity.}, we define
\bqa
{\lambda}^{V; \spmqty{N_1 & N_2 \\ N_4 & N_3}; \spmqty{\alpha & \beta \\ \gamma & \delta}}_{1PH \spmqty{p & k \\ k & p}} 
=
\frac{1}{\sqrt{ V_{F,p} V_{F,k}}}
{\lambda}^{\spmqty{N_1 & N_2 \\ N_4 & N_3}; \spmqty{\alpha & \beta \\ \gamma & \delta}}_{1PH \spmqty{p & k \\ k & p}},
\label{eq:lambdaV11}
\eqa
where the coupling is divided by 
$\sqrt{ V_{F,p} V_{F,k}}$
to keep $\lambda^V$ symmetric.
Its beta functional reads
\begin{equation}
\begin{aligned}
\frac{\partial}{\partial\ell}
{\lambda}_{1PH \spmqty{p & k \\  k &  p}}^{V; \spmqty{N_1 & N_2 \\ N_4 & N_3};\spmqty{\sigma_1 & \sigma_2 \\ \sigma_4 & \sigma_3}}
 & = -\left(1 + \eta_k+ \eta_p\right)
{\lambda}_{1PH \spmqty{p & k \\  k &  p}}^{V; \spmqty{N_1 & N_2 \\ N_4 & N_3};\spmqty{\sigma_1 & \sigma_2 \\ \sigma_4 & \sigma_3}}
 -\frac{S^V_{k,p}}{N_f^2}
  \mathsf{T}^{\beta \sigma_2}_{\sigma_4 \alpha} 
  \mathsf{T}^{\sigma_1 \alpha}_{\beta \sigma_3} 
 \delta^{N_1}_{N_4}
 \delta^{N_2}_{N_3}
\\
& +\frac{B_k}{2N_f} 
\lambda^{V; \spmqty{N_1 & \bar N_2 \\ \bar N_4 &  N_3};\spmqty{\sigma_1 & \alpha \\ \beta & \sigma_3}}_{1PH \spmqty{p  & k \\ k & p }}
 \mathsf{T}^{\beta \sigma_2}_{\sigma_4 \alpha} 
 + \frac{B_p}{2N_f} 
  \mathsf{T}^{\sigma_1 \beta}_{\alpha \sigma_3} 
 \hat\lambda^{V; \spmqty{\bar N_1 & N_2 \\ N_4 & \bar N_3};\spmqty{\alpha & \sigma_2 \\ \sigma_4 & \beta}}_{1PH \spmqty{p  & k \\ k & p }},
 \end{aligned}
 \label{eq:lambda1111all_2}
\end{equation}
where
\bqa
\eta_{k} &=&
 \frac{(N_c^2 -1) {g}_{k}^2}{ 2 \pi^2  N_c N_f c V_{F,k}}\frac{\mu}{\mu + 2 c  v_{k} \abs{k}_\mu },
\label{eq:etak0} \\
S^V_{k,p} & = &
\frac{1}{\sqrt{ V_{F,p} V_{F,k}}}
\mathsf{D}_{\mu}(k;p)(B_k+ B_p).
\label{eq:VXSKP1}
\eqa

Just as the Fermi velocity and the Yukawa coupling functions acquire momentum profiles that are solely determined from the nesting angle at the quasi-fixed point, the forward scattering amplitude acquires a singular momentum profile near the hot spots.
To extract the universal momentum profile of the forward scattering amplitude 
that is scale invariant at the quasi-fixed point, 
we consider the beta functional for the four-fermion coupling functions 
defined in the space of rescaled momentum,
${\hat \lambda}_{1PH; \spmqty{K_1 & K_2 \\  K_4 &  K_3}}^{\spmqty{N_1 & N_2 \\ N_3 & N_4};\spmqty{\sigma_1 & \sigma_2 \\ \sigma_4 & \sigma_3}} =
{\lambda}_{1PH; \spmqty{k_1 & k_2 \\  k_4 &  k_3}}^{V; \spmqty{N_1 & N_2 \\ N_3 & N_4};\spmqty{\sigma_1 & \sigma_2 \\ \sigma_4 & \sigma_3}}$,
where $K_i = \eell k_i$.
The beta functionals for the four-fermion coupling functions defined in the space of rescaled momentum
is written as
\begin{equation}
\begin{aligned}
\left[\frac{\partial}{\partial\ell}+K\frac{\partial}{\partial K} + P\frac{\partial}{\partial P}\right]{\hat \lambda}_{1PH; \spmqty{P & K \\  K &  P}}^{\spmqty{N_1 & N_2 \\ N_4 & N_3};\spmqty{\sigma_1 & \sigma_2 \\ \sigma_4 & \sigma_3}}
 = -\left(1+\hat\eta_K+\hat \eta_P\right)
{\hat \lambda}_{1PH; \spmqty{P & K \\  K &  P}}^{\spmqty{N_1 & N_2 \\ N_4 & N_3};\spmqty{\sigma_1 & \sigma_2 \\ \sigma_4 & \sigma_3}}
\\
+\frac{\hat B_K}{2N_f} 
\hat \lambda^{\spmqty{N_1 & \bar N_2 \\ \bar N_4 &  N_3};\spmqty{\sigma_1 & \alpha \\ \beta & \sigma_3}}_{1PH; \spmqty{P  & K \\ K & P }}
 \mathsf{T}^{\beta \sigma_2}_{\sigma_4 \alpha} 
 + \frac{\hat B_P}{2N_f} 
  \mathsf{T}^{\sigma_1 \beta}_{\alpha \sigma_3} 
 \hat \lambda^{\spmqty{\bar N_1 & N_2 \\ N_4 & \bar N_3};\spmqty{\alpha & \sigma_2 \\ \sigma_4 & \beta}}_{1PH; \spmqty{P  & K \\ K & P }} 
 -\frac{\hat S_{K,P}}{N_f^2}
  \mathsf{T}^{\beta \sigma_2}_{\sigma_4 \alpha} 
  \mathsf{T}^{\sigma_1 \alpha}_{\beta \sigma_3} 
 \delta^{N_1}_{N_4}
 \delta^{N_2}_{N_3},
\end{aligned}
\end{equation}
where
$\hat \eta_{K} = \eta_{K \emell}$, 
$\hat B_{K} = B_{K \emell}$ and
$\hat S_{K,P} = S^V_{K \emell, P \emell}$.
%
%
We combine the four coupling functions into a matrix as
\begin{equation}
 \hat {\lambda}^{ \spmqty{\sigma_1 & \sigma_2 \\ \sigma_4 & \sigma_3}}_{1PH\spmqty{P & K \\  K & P}} = 
\pmqty{
\hat{\lambda}^{\spmqty{1 & 1 \\ 1 & 1};\spmqty{\sigma_1 & \sigma_2 \\ \sigma_4 & \sigma_3}  }_{\spmqty{P & K \\ K & P}} & 
\hat{\lambda}^{\spmqty{1 & 4 \\ 4 & 1} ;\spmqty{\sigma_1 & \sigma_2 \\ \sigma_4 & \sigma_3}}_{\spmqty{P & K \\ K & P}} \\ 
\hat{\lambda}^{\spmqty{4 & 1 \\ 1 & 4};\spmqty{\sigma_1 & \sigma_2 \\ \sigma_4 & \sigma_3}}_{\spmqty{P & K \\ K & P}} & 
\hat{\lambda}^{\spmqty{4 & 4 \\ 4 & 4};\spmqty{\sigma_1 & \sigma_2 \\ \sigma_4 & \sigma_3}}_{\spmqty{P & K \\ K & P}}} 
\end{equation}
to write the set of beta functionals as
a matrix differential equation,
\begin{equation}
\begin{aligned}
& \left[\frac{\partial}{\partial\ell}+K\frac{\partial}{\partial K} + P\frac{\partial}{\partial P}\right]
 \hat {\lambda}^{ \spmqty{\sigma_1 & \sigma_2 \\ \sigma_4 & \sigma_3}}_{1PH\spmqty{P & K \\  K & P}} 
 =  -\left( 1 + \hat\eta_K + \hat\eta_P \right)
 \hat {\lambda}^{ \spmqty{\sigma_1 & \sigma_2 \\ \sigma_4 & \sigma_3}}_{1PH\spmqty{P & K \\  K & P}} 
 \\&
+
 \frac{\hat B_{K}}{2N_f} 
 \hat {\lambda}^{ \spmqty{\sigma_1 & \alpha \\ \beta & \sigma_3}}_{1PH\spmqty{P & K \\  K & P}} 
 \mathsf{T}^{\beta \sigma_2}_{\sigma_4 \alpha} 
 \pmqty{0& 1 \\ 1 & 0}
+
\frac{\hat B_P}{2N_f}
   \mathsf{T}^{\sigma_1 \beta}_{\alpha \sigma_3} 
 \pmqty{0& 1 \\ 1 & 0}
 \hat {\lambda}^{ \spmqty{\alpha & \sigma_2 \\ \sigma_4 & \beta}}_{1PH\spmqty{P & K \\  K & P}} 
 - \frac{\hat S_{K,P}}{N_f^2}
 \mathsf{T}^{\beta \sigma_2}_{\sigma_4 \alpha} 
   \mathsf{T}^{\sigma_1 \alpha}_{\beta \sigma_3} 
 \pmqty{1& 0 \\ 0 & 1}.
\end{aligned}
\label{eq:betabigmatrix_10_5PPH}
\end{equation}
In the PH channel,
the spin tensor of the interaction can be decomposed as
\bqa
 \mathsf{T}^{\alpha  \beta}_{  \gamma  \delta } =
Y_{PH}^{(t)}  \mathsf{I}^{\alpha \beta}_{\gamma \delta}
+ Y_{PH}^{(a)}  \mathsf{\chi}^{\alpha \beta}_{\gamma \delta}.
\label{eq:TinPH}
\eqa
Here
\bqa
\mathsf{I}^{\sigma_1 \sigma_2}_{\sigma_4 \sigma_3} = \frac{1}{N_c}  \delta_{\sigma_1 \sigma_3}\delta_{\sigma_2 \sigma_4}, 
&~~~~&
\mathsf{\chi}^{\sigma_1 \sigma_2}_{\sigma_4 \sigma_3} = \left(
\delta_{\sigma_1 \sigma_4}\delta_{\sigma_2 \sigma_3} -\frac{1}{N_c} \delta_{\sigma_1 \sigma_3}\delta_{\sigma_2 \sigma_4} \right)
\label{eq:IChi}
\eqa
project a spin state of a pair of particle and hole into the
trivial representation 
and the adjoint representation of the $SU(N_c)$ group, respectively,
and
\bqa
Y_{PH}^{(t)} =  2 \left( N_c - \frac{1}{N_c} \right), 
~~
Y_{PH}^{(a)} =   -  \frac{2}{N_c}
\label{eq:Yta}
\eqa
denotes the eigenvalue of 
$\mathsf{T}^{\alpha  \beta}_{  \gamma  \delta }$
in each channel.
Similarly, the matrix that controls the mixing in the space of hot spots
can be decomposed as
\bqa
\pmqty{0 & 1 \\ 1 & 0} =  1^s \mathscr{P}_s + 1^d \mathscr{P}_d, ~~~
\eqa
where
\bqa
\mathscr{P}_s= \frac{1}{2} \pmqty{1 & 1 \\ 1 & 1}, ~~~ \mathscr{P}_d = \frac{1}{2} \pmqty{1 & -1 \\ -1 & 1} 
\label{eq:PsPd}
\eqa
project the four-fermion couplings into the s-wave and d-wave channels, respectively,
and
\bqa 
1^s=1, ~~~ 1^d=-1
 \label{eq:1sd}
\eqa 
are the associated eigenvalues.
Naturally, the coupling function is decomposed into four different channels as
 \begin{equation}
\hat {\lambda}^{\spmqty{\sigma_1 & \sigma_2 \\ \sigma_4 & \sigma_3}}_{1PH \{K_i\}} = 
\hat {\lambda}^{(t)(s)}_{1PH \{K_i\}}\mathsf{I}^{\sigma_1 \sigma_2}_{\sigma_4 \sigma_3} 
\mathscr{P}_s
+ \hat {\lambda}^{(t)(d)}_{1PH \{K_i\}}\mathsf{I}^{\sigma_1 \sigma_2}_{\sigma_4 \sigma_3} 
\mathscr{P}_d
+ \hat {\lambda}^{(a)(s)}_{1PH \{K_i\}}\mathsf{\chi}^{\sigma_1 \sigma_2}_{\sigma_4 \sigma_3}
\mathscr{P}_s
+ \hat {\lambda}^{ (a)(d)}_{1PH \{K_i\}}\mathsf{\chi}^{\sigma_1 \sigma_2}_{\sigma_4 \sigma_3}
\mathscr{P}_d.
\end{equation}
The beta functional for the coupling function in each channel becomes
\begin{equation}
 \left[
\frac{\partial}{\partial\ell}+K\frac{\partial}{\partial K} + P\frac{\partial}{\partial P}
\right]
\hat\lambda^{\spmqty{t \\ a} , \spmqty{s \\ d}}_{1PH\spmqty{P & K \\ K & P}}
  =  -\left( 
  1+\hat\eta_K+\hat\eta_P 
  - \frac{1^{\spmqty{s \\ d}} Y_{PH}^{\spmqty{t \\ a}}}{ 2 N_f} 
  \left[
  \hat B_{K} + \hat B_{P}
  \right] \right)
  \hat\lambda^{\spmqty{t \\ a} , \spmqty{s \\ d}}_{1PH\spmqty{P & K \\ K & P}}  
   -  \left(Y_{PH}^{\spmqty{t \\ a}}\right)^2
\frac{\hat S_{K,P}}{N_f^2}.
\label{eq:beta_1ph_matrix}
\end{equation}
%
%
%
%
%
%

Due to the momentum dilatation,
the coupling functions at different momenta mix under the RG flow.
However, only the overall magnitude of external momenta is rescaled,
and the relative magnitudes do not change.
To integrate the beta functional,
it is convenient to introduce a polar coordinate 
for the space of external momenta,
\bqa
(K,P) = X \hat \Omega,
\eqa
where $X = \sqrt{K^2+P^2}$ 
represents the overall magnitude,
and $\hat \Omega = (  \Omega_K,  \Omega_P)$ 
represents the unit vector that specifies direction 
in the space of external momenta.
At the fixed point, the coupling function satisfies
\bqa
X\frac{\partial}{\partial X} 
\hat\lambda^{*\spmqty{t \\ a} , \spmqty{s \\ d}}_{1PH, X, \hat \Omega}
 +\left( 
  1+\hat\eta_{X \Omega_K}+\hat\eta_{X \Omega_P} 
  - \frac{1^{\spmqty{s \\ d}} Y_{PH}^{\spmqty{t \\ a}}}{ 2 N_f} 
  \left[
 \hat B_{X \Omega_K}
+\hat B_{X \Omega_P}
  \right] 
\right)
  \hat\lambda^{*\spmqty{t \\ a} , \spmqty{s \\ d}}_{1PH,X, \hat \Omega}  
     + \frac{1}{N_f^2} 
   \left(Y_{PH}^{\spmqty{t \\ a}}\right)^2 
\hat S_{X, \hat \Omega} = 0. 
\label{eq:beta_s_11all_fixed}
\eqa
The fixed point solution is readily obtained to be
\begin{align}
\hat \lambda^{*\spmqty{t \\ a} , \spmqty{s \\ d}}_{1PH, X, \hat \Omega} = 
\psi^{\spmqty{s \\ d}}(X)
\left[
\frac{ 
\hat \lambda^{\spmqty{t \\ a} , \spmqty{s \\ d}}_{1PH, \Lambda_0, \hat \Omega} 
}{\psi^{\spmqty{s \\ d}}(\Lambda_0)}
-
\frac{1}{N_f^2}
   \left.Y_{PH}^{\spmqty{t \\ a}}\right.^2
\int_{\Lambda_0}^X \frac{dX'}{X'} \frac{1}{ \psi^{\spmqty{s \\ d}}(X')}  \hat S_{X', \hat \Omega}
\right],
\label{eq:fxptsoln110}
\end{align}
where 
\begin{align}
\psi^{\spmqty{s \\ d}}(X) = {\rm Exp}\left[
-\int_{\Lambda}^X
\frac{dX'}{X'}\left(1 + \hat{\eta}_{X' \Omega_K}+ \hat{\eta}_{X' \Omega_P}
- \frac{1^{\spmqty{s \\ d}} Y_{PH}^{\spmqty{t \\ a} }}{2 N_f} 
(\hat B_{X' \Omega_K}
+\hat B_{X' \Omega_P}
)
\right)\right]
\end{align}
and 
$\Lambda_0$ is a reference scale at which the boundary condition is imposed for
$\hat \lambda^{*\spmqty{t \\ a} , \spmqty{s \\ d}}_{1PH, X, \hat \Omega}$.
Since $\psi^{\spmqty{s \\ d}}(X)$
diverges in the small $X$ limit,
the solution that is regular at the hot spot is obtained by choosing $\Lambda_0=0$\footnote{
The coupling functions are regular at all momenta including the hot spots at any non-zero energy scale.
},
\begin{align}
\hat \lambda^{*\spmqty{t \\ a} , \spmqty{s \\ d}}_{1PH, X, \hat \Omega} = 
-
\frac{1}{N_f^2}
   \left.Y_{PH}^{\spmqty{t \\ a}}\right.^2
\psi^{\spmqty{s \\ d}}(X)
\int_{0}^X \frac{dX'}{X'} \frac{1}{ \psi^{\spmqty{s \\ d}}(X')}   \hat S_{X', \hat \Omega}.
\label{eq:fxptsoln11}
\end{align}

\begin{figure*}
	\centering
	\includegraphics[scale=0.3]{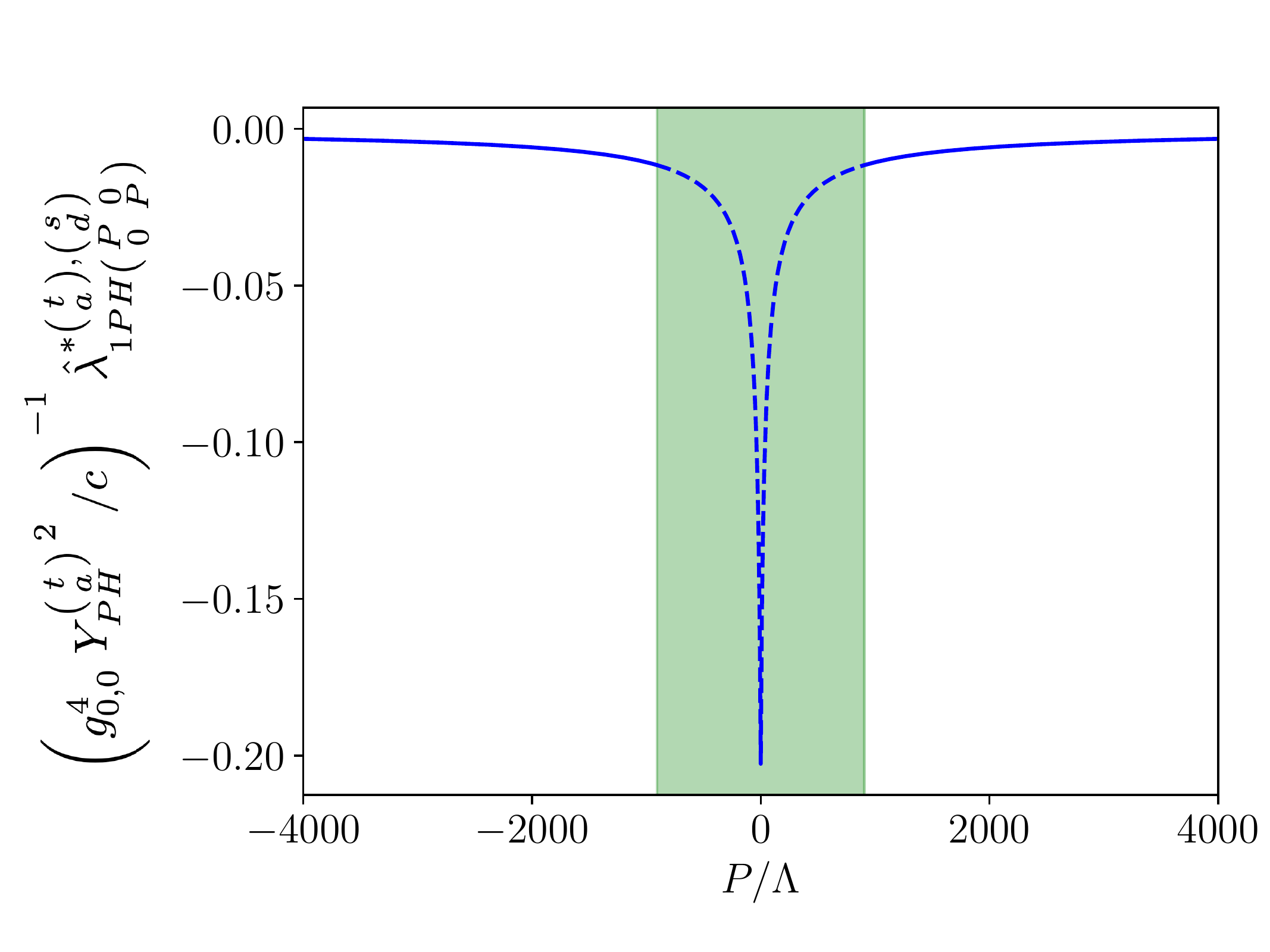}
\caption{
The forward scattering amplitude 
$\hat \lambda^{*\spmqty{t \\ a} , \spmqty{s \\ d}}_{1PH\spmqty{P & 0 \\ 0 & P}}$
in Eq.~\eqref{eq:lambda_fp_1PH_1}
plotted in units of
$ \left( g^4_{0,0} \left.Y_{PH}^{\spmqty{t \\ a}}\right.^2/c \right)$
as a function of $P/\Lambda$ at $K = 0$ for
$\ell_0 = 50$ 
(equivalently,  $1/(v_0c)\approx 907$). 
The width of the shaded region is $2/(v_0c)$. 
}
\label{fig:4f0_1111_plot}
\end{figure*}


In the small $v$ limit, 
$\psi^{\spmqty{s \\ d}}(X)$ is well approximated by $\Lambda/X$, and 
the anomalous dimension and the $\hat B$ terms are sub-leading. 
At the fixed point, the coupling function takes the  scale invariant form given by
\begin{equation}
\begin{aligned}
 \hat \lambda^{*\spmqty{t \\ a} , \spmqty{s \\ d}}_{1PH\spmqty{P & K \\ K & P}}
= 
-\frac{\hat g^2_{P,K}   \left.Y_{PH}^{\spmqty{t \\ a}}\right.^2
 }{\pi^2c N_f^2\sqrt{\hat V_{F,K} \hat V_{F,P}}}
 & \left[\frac{\hat g^2_{K,K}}{\hat V_{F,K}}\frac{\Lambda  \log \left(\frac{c |\hat v_K K + \hat v_P P|_\Lambda +c |K-P|_\Lambda +\Lambda }{2 \hat v_Kc|K|_\Lambda +\Lambda }\right)}{c (|\hat v_K K + \hat v_P P|_\Lambda +|K-P|_\Lambda -2 \hat v_K| K|_\Lambda )} \right.
\\
& \left. + \frac{\hat g^2_{P,P}}{\hat V_{F,P}}\frac{\Lambda  \log \left(\frac{c |\hat v_K K + \hat v_P P|_\Lambda +c |K-P|_\Lambda +\Lambda }{2\hat v_Pc|P|_\Lambda +\Lambda }\right)}{c (|\hat v_K K + \hat v_P P|_\Lambda +|K-P|_\Lambda -2 \hat v_P| P|_\Lambda )}\right]
\end{aligned}
 \label{eq:lambda_fp_1PH_1}
\end{equation}
to the leading order in $v$\footnote{
In the small $v$ limit, 
we ignored the momentum dependence of 
the coupling functions
and use the approximation
$\Lambda + c | K |_\Lambda \approx \Lambda + c|K|$
in $\hat S_{K,P}$.
}.
Just as other coupling functions discussed in the previous section,
the forward scattering amplitude becomes independent of $\ell$ 
when the rescaled momenta $K$ and $P$ are fixed.
This implies that 
the scale invariance emerges 
when the forward scattering amplitude is probed at momenta increasingly closer to the hot spots 
as the energy is lowered.
%
If one of the electron is in the hot region and the other is at momentum $p$ 
 far away from the hot region,
the forward scattering amplitude decays as $1/p$ with a logarithmic correction as
\bqa
\frac{1}{4\mu} 
\lambda^{V;
  \spmqty{t \\ a} , \spmqty{s \\ d}}_{1PH\spmqty{p & 0 \\ 0 & p}}
=
-\frac{ g^2_{p,0} g^2_0
\left.Y_{PH}^{\spmqty{t \\ a}}\right.^2}{4 \pi^2c^2 N_f^2\sqrt{V_{F,p}}}
 \frac{1}{|p|}
\log \left(\frac{c |p| }{\mu }\right),
\eqa
where $p=P \emell$ represents the physical momentum.
The momentum profile of the forward scattering amplitude
is shown in \fig{fig:4f0_1111_plot}.


\subsection{Four-fermion coupling in group 2}
\label{sec:four_fermion_fp2}


%

\subsubsection{UV/IR mixing }
\label{subsubsection:uvirmixing}

In group 2, 
$  {\lambda}^{\spmqty{1 & 5 \\ 1 & 5};\spmqty{\sigma_1 & \sigma_2 \\ \sigma_4 & \sigma_3}}_{\spmqty{p & -p \\ k & -k}} $,
$  {\lambda}^{\spmqty{4 & 8 \\ 4 & 8};\spmqty{\sigma_1 & \sigma_2 \\ \sigma_4 & \sigma_3}}_{\spmqty{p & -p \\ k & -k}} $,
$  {\lambda}^{\spmqty{1 & 5 \\ 4 & 8};\spmqty{\sigma_1 & \sigma_2 \\ \sigma_4 & \sigma_3}}_{\spmqty{p & -p \\ k & -k}} $,
$  {\lambda}^{\spmqty{4 & 8 \\ 1 & 5};\spmqty{\sigma_1 & \sigma_2 \\ \sigma_4 & \sigma_3}}_{\spmqty{p & -p \\ k & -k}} $ 
form a closed set of beta functionals
given by
Eqs.
(\ref{eq:beta1515PP1})-
(\ref{eq:beta1515PP4}).
The beta functionals are expressed as 
integrations over $q$ because of 
the significant mixing between operators 
that carry different momenta
along the Fermi surface.
To simplify the system of beta functionals,
we combine the four coupling functions into one matrix,
%
\begin{align}
 {\lambda}^{\spmqty{\sigma_1 & \sigma_2 \\ \sigma_4 & \sigma_3}}_{2PP \spmqty{p &- p \\  k & -k}} = 
& \pmqty{{\lambda}^{\spmqty{1 & 5 \\ 1 & 5}; \spmqty{\sigma_1 & \sigma_2 \\ \sigma_4 & \sigma_3}}_{\spmqty{p & -p \\ k & -k}} ~&~ {\lambda}^{\spmqty{1 & 5 \\ 4 & 8}; \spmqty{\sigma_1 & \sigma_2 \\ \sigma_4 & \sigma_3}}_{\spmqty{p & -p \\ k & -k }} \\
\\ {\lambda}^{\spmqty{4 & 8 \\ 1 & 5}; \spmqty{\sigma_1 & \sigma_2 \\ \sigma_4 & \sigma_3}}_{\spmqty{p & -p \\ k & -k }} ~&~ {\lambda}^{\spmqty{4 & 8 \\ 4 & 8}; \spmqty{\sigma_1 & \sigma_2 \\ \sigma_4 & \sigma_3}}_{\spmqty{p & -p \\ k & -k }}}
\label{eq:propmatrix0}
\end{align}
to rewrite 
Eqs.
(\ref{eq:beta1515PP1})-
(\ref{eq:beta1515PP4})
as
\begin{align}
& \frac{\partial}{\partial\ell}{\lambda}^{\spmqty{\sigma_1 & \sigma_2 \\ \sigma_4 & \sigma_3}}_{2PP \spmqty{p & -p \\ k & -k}} = 
 -\left( 1 + 3(z-1) + 2 {\eta}^{(\psi)}_k+ 2 {\eta}^{(\psi)}_p \right)
 {\lambda}^{\spmqty{\sigma_1 & \sigma_2 \\ \sigma_4 & \sigma_3}}_{2PP \spmqty{p & -p \\ k & -k}}
\nn
& - \frac{1}{4\pi}\int \frac{\dd q}{2 \pi \mu V_{F,q}}  
\left[ 
{\lambda}^{\spmqty{\sigma_1 & \sigma_2 \\ \beta & \alpha}}_{2PP \spmqty{p & -p \\ q & -q}} 
-\frac{ 2 \mathsf{T}^{\sigma_1  \sigma_2}_{  \beta  \alpha }}{N_f}
\mathsf{D}_{\mu}(p; q)  \pmqty{0 & 1 \\ 1 & 0} \right]
\left[ 
{\lambda}^{\spmqty{\beta & \alpha \\ \sigma_4 & \sigma_3}}_{2PP \spmqty{q & -q \\ k & -k}} 
-\frac{2 \mathsf{T}^{ \beta  \alpha}_{  \sigma_4  \sigma_3 }}{N_f}\mathsf{D}_{\mu}(q; k)  \pmqty{0 & 1 \\ 1 & 0} \right].
\label{eq:betabigmatrixsimple_main0}
\end{align}
As in \eq{eq:lambdaV11},
we symmetrically normalize the four-fermion coupling
in the unit of the Fermi velocity
by defining
\bqa
{\lambda}^{V; \spmqty{\alpha & \beta \\ \gamma & \delta}}_{2PP \spmqty{p & -p \\ k & -k}} 
=
\frac{1}{\sqrt{ V_{F,p} V_{F,k}}}
{\lambda}^{\spmqty{\alpha & \beta \\ \gamma & \delta}}_{2PP \spmqty{p & -p \\ k & -k}}.
\label{eq:lambdaV}
\eqa
Its beta functional is given by
\begin{align}
& \frac{\partial}{\partial\ell}
{\lambda}^{V; \spmqty{\sigma_1 & \sigma_2 \\ \sigma_4 & \sigma_3}}_{2PP \spmqty{p & -p \\ k & -k}} 
=
 -\left( 1 +  \eta_k+ \eta_p \right)
 {\lambda}^{V; \spmqty{\sigma_1 & \sigma_2 \\ \sigma_4 & \sigma_3}}_{2PP \spmqty{p & -p \\ k & -k}} \nn
 & 
 - \frac{1}{4\pi}
\int \frac{\dd q}{2\pi\mu}
\left[
 {\lambda}^{V; \spmqty{\sigma_1 & \sigma_2 \\ \beta & \alpha}}_{2PP \spmqty{p & -p \\ q & -q}} 
- \frac{2  \mathsf{T}^{\sigma_1  \sigma_2}_{  \beta \alpha }}{N_f } 
\frac{\mathsf{D}_{\mu}(p; q)}{\sqrt{ V_{F,p} V_{F,q}}}
  \pmqty{0 & 1 \\ 1 & 0}
\right]
\left[
 {\lambda}^{V; \spmqty{\beta & \alpha \\ \sigma_4 & \sigma_3}}_{2PP \spmqty{q & -q \\ k & -k}} 
- \frac{2  \mathsf{T}^{\beta  \alpha}_{  \sigma_4  \sigma_3 }}{N_f } 
\frac{\mathsf{D}_{\mu}(q; k)}{\sqrt{ V_{F,q} V_{F,k}}}
  \pmqty{0 & 1 \\ 1 & 0}
\right],
%
\label{eq:betabigmatrixsimple_main01}
\end{align}
where $\eta_{k}$  is defined in
\eq{eq:etak0}.

In the space of spin wavefunctions for two electrons,
 the four-fermion coupling can be decomposed into 
 the symmetric and anti-symmetric channels of $SU(N_c)$.
For $N_c=2$, the symmetric and anti-symmetric representations
 correspond to the spin triplet and spin singlet representations, respectively.
 Combined with the wavefunction defined in the space of hot spot indices, 
 the four-fermion coupling can be decomposed into
 spin-symmetric s-wave ($+,s$),
 spin-symmetric d-wave ($+,d$),
 spin-anti-symmetric s-wave ($-,s$)
 and
 spin-anti-symmetric d-wave ($-,d$)
 channels as
 \begin{equation}
{\lambda}^{V; \spmqty{\sigma_1 & \sigma_2 \\ \sigma_4 & \sigma_3}}_{2PP \{k_i\}} = 
{\lambda}^{V; (+)(s)}_{2PP \{k_i\}}\mathsf{S}^{\sigma_1 \sigma_2}_{\sigma_4 \sigma_3} 
\mathscr{P}_s
+ {\lambda}^{V; (+)(d)}_{2PP \{k_i\}}\mathsf{S}^{\sigma_1 \sigma_2}_{\sigma_4 \sigma_3} 
\mathscr{P}_d
+ {\lambda}^{V; (-)(s)}_{2PP \{k_i\}}\mathsf{A}^{\sigma_1 \sigma_2}_{\sigma_4 \sigma_3}
\mathscr{P}_s
+ {\lambda}^{V; (-)(d)}_{2PP \{k_i\}}\mathsf{A}^{\sigma_1 \sigma_2}_{\sigma_4 \sigma_3}
\mathscr{P}_d.
\label{eq:SPdecomposition}
\end{equation}
Here
$P_s$ and $P_d$ are defined in \eq{eq:PsPd}.
\bqa
\mathsf{S}^{\sigma_1 \sigma_2}_{\sigma_4 \sigma_3} =
\frac{1}{2} \left(
\delta_{\sigma_1 \sigma_4}\delta_{\sigma_2 \sigma_3} + \delta_{\sigma_1 \sigma_3}\delta_{\sigma_2 \sigma_4}
\right), &~~~~&
\mathsf{A}^{\sigma_1 \sigma_2}_{\sigma_4 \sigma_3} =
\frac{1}{2} \left(
\delta_{\sigma_1 \sigma_4}\delta_{\sigma_2 \sigma_3} - \delta_{\sigma_1 \sigma_3}\delta_{\sigma_2 \sigma_4}
\right)
\label{eq:spinSA}
\eqa
project a spin state of two particles into the $SU(N_c)$ symmetric
and anti-symmetric representations.
The spin dependence in the interaction mediated by the spin fluctuations
can be resolved into
\bqa
  \mathsf{T}^{\alpha  \beta}_{  \gamma  \delta } =
Y_{PP}^{(+)}  \mathsf{S}^{\alpha \beta}_{\gamma \delta}
+ Y_{PP}^{(-)}  \mathsf{A}^{\alpha \beta}_{\gamma \delta}
\eqa
with
\bqa
Y_{PP}^{(+)} = 2 \left( 1-  \frac{1}{N_c}  \right), ~~Y_{PP}^{(-)} = -2 \left(  1+ \frac{1}{N_c} \right).
 \label{eq:Ypm}
\eqa
The beta functional for the coupling function in each channel is written as
\begin{equation}
\begin{aligned}
& \frac{\partial}{\partial\ell}
{\lambda}^{V; (\pm),\spmqty{s \\ d}}_{2PP \spmqty{p & -p \\ k & -k}} 
=
 -\left( 1 +  \eta_k+ \eta_p \right)
 {\lambda}^{V; (\pm),\spmqty{s \\ d}}_{2PP \spmqty{p & -p \\ k & -k}} \\
 &
 - 
\frac{1}{4\pi}
\int \frac{\dd q}{2\pi\mu}
\left[
 \lambda^{V; (\pm),\spmqty{s \\ d}}_{2PP \spmqty{p & -p \\ q & -q}} 
 - \frac{2}{N_f} 
 Y_{PP}^{(\pm)} 
1^{\spmqty{s \\ d}}
\frac{ \mathsf{D}_{\mu}(p; q)}{ \sqrt{ V_{F,p} V_{F,q} } }
\right]
 \left[
 \lambda^{V; (\pm),\spmqty{s \\ d}}_{2PP \spmqty{q & -q \\ k & -k}} 
 - \frac{2}{N_f} 
 Y_{PP}^{(\pm)} 
1^{\spmqty{s \\ d}}
\frac{ \mathsf{D}_{\mu}(q; k)}{ \sqrt{ V_{F,q} V_{F,k} } }
\right],
\end{aligned}
\label{eq:betabigmatrixsimple_main01V}
\end{equation}
where $1^{\spmqty{s \\ d}}$ is defined in \eq{eq:1sd}.

What is interesting for the beta functionals in group 2 
 is the fact that 
 the strength of mixing between 
 two low-energy operators
 defined near the Fermi surface 
 but with a large difference in momentum 
is controlled by high-energy bosons.
Since $D_\mu(q;k)$ decays as $\frac{g_{qk}^2}{c|q-k|}$ at large $|q-k|$,
the contribution from $q$ far away from $k$ 
can be important.
The potential UV divergence associated with the 
$q$ integration in 
\eq{eq:betabigmatrixsimple_main01V}  
is cut off by either the momentum profile of the Yukawa coupling included in $D_\mu(q;k)$ at large $|q-k|$ in \eq{eq:Dqk}
or the irrelevant boson kinetic term\footnote{
For instance, the scale associated with the crossover from high-energy `Gaussian physics'  to low-energy `critical physics'  can act as a cutoff momentum for $q$ integration.}.
The trouble is that whichever cuts off the UV divergence is related to the dynamics of boson at large energy/momentum, which is not a part of the universal low-energy data.
This implies that the four-fermion coupling can not be determined without including the high-energy physics.
{\it Namely, the one-particle irreducible (1PI) quartic vertex function is not an observable that can be determined solely in terms of other low-energy observables.}
What is then the low-energy observable that measures the strength of the two-body interaction?
To identify the right low-energy observable, 
we note that the four-point functions,
which determine physical susceptibilities,
are determined by the sum of 
the 1PI four-point function 
and the tree-diagram that involves the 1PI three-point functions and the boson propagator (see	\fig{fig:4pt3pts}).
Therefore, we consider the net two-body interaction that combines the contributions of the four-fermion coupling and the  interaction mediated by the spin fluctuations,
\begin{align}
{\lambda}^{' (\pm),\spmqty{s \\ d} }_{2PP \spmqty{p & -p \\ k & -k}} = 
 \lambda^{V; (\pm),\spmqty{s \\ d}}_{2PP \spmqty{p & -p \\ k & -k}} 
 - \frac{2}{N_f} 
 Y_{PP}^{(\pm)} 
1^{\spmqty{s \\ d}}
\frac{ \mathsf{D}_{\mu}(p; k)}{ \sqrt{ V_{F,p} V_{F,k} } }.
\label{eq:lambdaprime0}
\end{align}
The net two-body interaction is 
what  determines the pairing interaction 
as is shown in the complete square 
 term 
 in \eq{eq:betabigmatrixsimple_main01V}\footnote{
The Wilsonian RG scheme is also naturally formulated in terms of the net two-body interaction\cite{PhysRevB.91.115111}.}.
The RG flow equation for the net two-body interaction reads
\begin{equation}
\begin{aligned} 
 \frac{\partial }{\partial\ell}
 {\lambda}^{' (\pm),\spmqty{s \\ d}}_{2PP \spmqty{p & -p \\ k & -k}} 
 = &
 -\left(1 + 
   \eta_k + \eta_p 
\right)
 {\lambda}^{' (\pm),\spmqty{s \\ d}}_{2PP \spmqty{p & -p \\ k & -k}}
 -\frac{1}{4\pi} \int \frac{\dd q}{2\pi\mu}
 {\lambda}^{' (\pm ),\spmqty{s \\ d}}_{2PP \spmqty{p & -p \\ q & -q}} 
 {\lambda}^{' (\pm),\spmqty{s \\ d}}_{2PP \spmqty{q & -q \\ k & -k}} 
  - \frac{2}{N_f}
Y_{PP}^{(\pm)} 
1^{\spmqty{s \\ d}}
{\mathsf{r}}(p, k ),
\label{eq:betaallwaves_main00}
\end{aligned}
\end{equation}
where
\bqa
{\mathsf{r}}(k, p) 
&=& 
\left( -\mu \frac{\partial}{\partial \mu} 
+ 1 + \eta_k + \eta_p \right) 
\frac{ \mathsf{D}_{\mu}(p; k)}{ \sqrt{ V_{F,p} V_{F,k} } }
\label{eq:RetaKK00}
\eqa
corresponds to the contribution of the spin fluctuations that arises between energy $\mu$  and $\mu-d\mu$.
The spin fluctuations generate attractive interactions
in the spin anti-symmetric d-wave channel
and the spin symmetric s-wave channel.
Its magnitude is strongest near the hot spots.
To the leading order in $v$, 
${\mathsf{r}}(k, p) =
\frac{  {g}_{k, p}^2}{\sqrt{ V_{F,k} V_{F,p}}} \frac{\mu^2}{\left(\mu + c \abs{k -p} + c \abs{ v_{k}k + v_{p}p}\right)^2}$.
Because  ${\mathsf{r}}(k, p)$  is the low-energy contribution to the net two-body interaction, it decays as $1/|k-p|^2$ at large momentum, and its contribution becomes negligible 
at momentum much larger than $\mu/(vc)$
Due to the fast decay of $\mathsf{r}(k,p)$ at large momenta,
the flow of the net two-body interaction is no longer sensitive to UV  scales.
Once the net two-body interaction is known at a scale, 
within a power-law accuracy 
its value at a lower energy scale can be determined 
from \eq{eq:betaallwaves_main00}  
without having to resort to UV physics.

\subsubsection{Absence of Hermitian quasi-fixed point at non-zero nesting angle}

To find a fixed point of the beta functional,
we have to examine the flow equation for the four-fermion coupling function defined in the space of rescaled momentum,
\begin{align}
\tilde {\lambda}^{ (\pm),\spmqty{s \\ d} }_{2PP \spmqty{P & -P \\ K & -K}} = 
{\lambda}^{' (\pm),\spmqty{s \\ d} }_{2PP \spmqty{p & -p \\ k & -k}},
\label{eq:tildelambda0}
\end{align}
where $K = \eell k$, $P= \eell p$.
The beta functionals for $\tilde \lambda$ becomes 
\begin{equation}
\begin{aligned} 
 \frac{\partial }{\partial\ell}
 \tilde{\lambda}^{(\pm),\spmqty{s \\ d}}_{2PP \spmqty{P & -P \\ K & -K}} 
 = &
 -\left(1 + K\frac{\partial}{\partial K} + P\frac{\partial}{\partial P} 
  + \hat \eta_K + \hat \eta_P 
\right)
 \tilde{\lambda}^{ (\pm),\spmqty{s \\ d}}_{2PP \spmqty{P & -P \\ K & -K}}
 -\frac{1}{4\pi} \int \frac{\dd Q}{2\pi\Lambda}
 \tilde{\lambda}^{ (\pm ),\spmqty{s \\ d}}_{2PP \spmqty{P & -P \\ Q & -Q}} 
 \tilde{\lambda}^{  (\pm),\spmqty{s \\ d}}_{2PP \spmqty{Q & -Q \\ K & -K}} 
 \\ & 
  - \frac{2}{N_f}
Y_{PP}^{(\pm)} 
1^{\spmqty{s \\ d}}
\hat{\mathsf{R}}(P, K ),
\label{eq:betaallwaves_main0}
\end{aligned}
\end{equation}
where
\bqa
\hat{ \mathsf{R}}(K, P) = 
\frac{  \hat{g}_{K, P}^2}{\sqrt{\hat V_{F,K} \hat V_{F,P}}} \frac{\Lambda^2}{\left(\Lambda + c \abs{K -P} + c \abs{\hat v_{K}K + \hat v_{P}P}\right)^2}, &~~&
\hat{\eta}_K  = 
  \frac{(N_c^2 -1) \hat{g}_{K}^2}{2\pi^2 c N_c N_f \hat V_{F,K}}\frac{\Lambda}{\Lambda + 2 c \hat v_{K} \abs{K}}.
\label{eq:RetaKK}
\eqa
Here, the attractive interaction 
$\hat{ \mathsf{R}}(K, P)$
generated from the spin fluctuations tends to drive the system to a superconducting state.
On the other hand, $\hat \eta_K$,
that represents the momentum dependent anomalous dimension, tends to suppress growth of the four-fermion coupling by making electrons incoherent.
The fate of the theory is determined by the competition between the attractive interaction 
that favours superconductivity and 
the pair-breaking effect caused by incoherence.
If the pairing effect dominates, 
the theory flows to a superconducting state, and
quasi-fixed points arise 
only outside the space of Hermitian theories.
On the other hand, if the pair-breaking effect dominates, 
there can be Hermitian quasi-fixed points. 
Once the theory is attracted to the quasi-fixed point with a non-zero $v$,
the theory would gradually flow to the true fixed point at $v=0$
under the full RG flow.
In the latter case, a stable non-Fermi liquid state would be realized at zero temperature.
This scenario is realized near three space dimensions where the co-dimension of the Fermi surface is close to $2$\cite{LUNTS,SCHLIEF2}.
Our goal is to understand the fate of the system that results as an outcome of this competition in two space dimensions.

In principle, there can be multiple quasi-fixed points.
Here we focus on the one-parameter families of quasi-fixed points 
that are continuously connected to the true fixed point at $v=0$.
It is difficult to write down the momentum dependent coupling function explicitly
at the quasi-fixed points.
However, we can understand the asymptotic form 
of $\tilde \lambda$ at large momenta.
To the leading order in $v$,
the last term in \eq{eq:betaallwaves_main0}
decays in a power-law as
$ \hat{ \mathsf{R}}(K, P) 
\sim \frac{\Lambda^2}{\left( c \abs{K -P} + c \abs{\hat v_{K}K + \hat v_{P}P}\right)^2}$
at large $|K|$ and $|P|$.
Since
 $ \left(1 + K\frac{\partial}{\partial K} + P\frac{\partial}{\partial P} \right)
\hat{\mathsf{R}}(K, P)
\approx - \hat{\mathsf{R}}(K, P)$,
at large momenta
the coupling function at the quasi fixed-point 
can be determined from balancing the first and the last terms of the beta functional as
\bqa
\tilde{\lambda}^{* \spmqty{\sigma_1 & \sigma_2 \\ \sigma_4 & \sigma_3}}_{2PP \spmqty{P & -P \\ K & -K}} 
\approx
\frac{2}{N_f} \hat{\mathsf{R}}(K, P) 
  \mathsf{T}^{\sigma_1 \sigma_2}_{\sigma_4 \sigma_3}
(\mathscr{P}_s - \mathscr{P}_d ).
\label{eq:tildelambda_fixedpt_largekp}
\eqa 
Here $\tilde \lambda^2$ term and $\hat \eta \tilde \lambda$ term
can be ignored in the small $v$ limit.
This is a special solution of the fixed point equation
which can be augmented with a homogeneous solution
of $(1+K \partial_K+P \partial_P)\tilde \lambda = 0$.
The homogeneous solution also vanishes as $1/K$ or $1/P$ at large momenta.
This shows that  
$ \tilde{\lambda}^{(\pm),\spmqty{s \\ d}}_{2PP \spmqty{P & -P \\ K & -K}}$ 
has to vanish in the limit that either $K$ or $P$, or both are large
at the quasi-fixed points.

We now prove that the quasi-fixed points have to be
non-Hermitian at $v \neq 0$.
To show this,  
we rewrite \eq{eq:betaallwaves_main0} as a matrix equation,
\bqa
 && \frac{\partial }{\partial\ell}
 \tilde{\lambda}^{(n)}
 =
 - \tilde{\lambda}^{(n)} 
 - L \tilde{\lambda}^{(n)} 
 - \tilde{\lambda}^{(n)}  L^\dagger
 - H \tilde{\lambda}^{(n)} 
 -\tilde{\lambda}^{(n)}  H
 - \frac{1}{4 \pi} \tilde{\lambda}^{(n)}   \tilde{\lambda}^{(n)}
 + \alpha^{(n)} R. 
 \label{eq:betaallwaves_matrix}
 \eqa
Here, 
$\tilde{\lambda}^{(n)}_{QK}  = \tilde{\lambda}^{(n)}_{2PP \spmqty{Q & -Q \\ K & -K}}$
with $n=(+,s), (+,d), (-,s), (-,d)$
represent the four-fermion coupling functions written as matrices in the space of momentum.
$L_{PK} = 2 \pi \Lambda
P \partial_P \delta(P-K)$,
$H_{PK}= 2 \pi  \Lambda \hat \eta_P \delta(P-K)$,
$R_{PK}=\hat{\mathsf{R}}(P,K)$
are also viewed as matrices,
where the multiplication of matrices is defined as
$(AB)_{PK}=
\int \frac{\dd Q}{2\pi\Lambda} A_{PQ} B_{QK}$.
$\alpha^{(n)}$ denotes the parameter 
that determines the sign and the strength of interaction in each channel,
\bqa
\alpha^{(+,s)} = -\frac{4}{N_f}\left( 1-\frac{1}{N_c} \right),~ 
\alpha^{(+,d)} = \frac{4}{N_f}\left( 1-\frac{1}{N_c} \right), ~
\alpha^{(-,s)} = \frac{4}{N_f}\left( 1+\frac{1}{N_c} \right), ~
\alpha^{(-,d)} = -\frac{4}{N_f}\left( 1+\frac{1}{N_c} \right).
\label{eq:alphas}
\eqa
The beta functionals  can be written in the complete square form as
\bqa
 && \frac{\partial }{\partial\ell}
 \tilde{\lambda}^{(n)}
 =
  - \frac{1}{4 \pi} 
  \left[
  \tilde{\lambda}^{(n)}   
  + 4 \pi \left( \frac{ I}{2} + L + H \right)
  \right] 
  \left[
  \tilde{\lambda}^{(n)}
    + 4 \pi \left( \frac{ I}{2} + L^\dagger + H \right)
  \right]
  +{\cal D}^{(n)},
\eqa
where ${\cal D}^{(n)}$ is the discriminant matrix given by
 \bqa
 {\cal D}^{(n)} =
  \alpha^{(n)} R + 
  4 \pi
  \left(
    \frac{I}{ 2}  + L + H
  \right)
    \left(
   \frac{I}{2}  + L^\dagger + H
  \right).
\eqa
$I$ is the identity matrix with $I_{PK} = 2 \pi \Lambda \delta(P-K)$.
At the fixed point,
the four-fermion coupling function should satisfy
\bqa
  \left[
  \tilde{\lambda}^{(n)}   
  + 4 \pi \left( \frac{ I}{2} + L + H \right)
  \right] 
  \left[
  \tilde{\lambda}^{(n)}
    + 4 \pi \left( \frac{ I}{2} + L^\dagger + H \right)
  \right]
  - 4 \pi{\cal D}^{(n)} = 0.
   \label{eq:lambda_fp_eq}
\eqa
It is noted that  
${\cal D}^{(n)}$
and 
$H$ are Hermitian matrices,
but $L$ is not.
 Even if 
 $ \left[
  \tilde{\lambda}^{(n)}   
  + 4 \pi \left( \frac{ I}{2} + L^\dagger + H \right)
  \right] $  is not Hermitian, 
  the polar decomposition theorem 
  guarantees that
  there exists a unitary matrix $U$
that makes $U \left[
  \tilde{\lambda}^{(n)}   
  + 4 \pi \left( \frac{ I}{2} + L^\dagger + H \right)
  \right] $ Hermitian.
The solution 
to  \eq{eq:lambda_fp_eq}
is then written as
  \bqa
 \tilde{\lambda}^{(n)} = 
- 4 \pi \left( \frac{I}{2}  + L^\dagger + H \right)
  +
( 4 \pi )^{1/2} U^\dagger   {\cal E}^{(n)},
  \eqa
  where  $  {\cal E}^{(n)}$ represents
  a matrix that satisfies
   $[{\cal E}^{(n)}]^2 = {\cal D}^{(n)}$\footnote{ 
For an $N \times N$ Hermitian matrix ${\cal D}^{(n)}$,
there are $2^{(N-N_0)}$
distinct solution for $[{\cal E}^{(n)}]^{2} = {\cal D}^{(n)}$,
where $N_0$ is the number of zero eigenvalues. }.

To show that $\tilde{\lambda}^{(n)}$
is non-Hermitian at the quasi-fixed point
with $v \neq 0$, we consider a vector of the form,
\bqa
f_K = \left( \frac{ \Lambda}{|K|} \right)^{1/2} e^{
\int_{\Lambda'}^{|K|} \frac{dK'}{K'} \hat \epsilon_{K'}
}
\eqa
with a real function $\hat \epsilon_K$
and a scale $\Lambda'$.
For $f_K$ to be square integrable, 
$\hat \epsilon_K$ has to be positive (negative) in the limit that $|K|$ is small (large).
The expectation value of the both sides of \eq{eq:lambda_fp_eq} for $f_K$ 
is written as
\bqa
\int \frac{d K}{2 \pi \Lambda} \tilde f_K^{'*} \tilde f_K
   - 4 \pi \alpha^{(n)} 
\int \frac{d K dP}{(2 \pi \Lambda)^2}  f_K^*  R_{KP} f_{P} 
 - ( 4 \pi )^2 \int  \frac{d K}{2 \pi \Lambda} (\hat \eta_K - \hat \epsilon_K)^2 | f_K |^2
=0,
   \label{eq:lambda_exp}
\eqa
where
$\tilde f_K =  
\int \frac{d P}{2 \pi \Lambda} 
  \tilde{\lambda}^{(n)}_{K P}
  ~ f_{P}
  + 4 \pi  (\hat \eta_K - \hat \epsilon_K) f_K$
  and
$\tilde f'_K =  
\int \frac{d P}{2 \pi \Lambda} 
  \tilde{\lambda}^{(n) \dagger}_{K P}
  ~ f_{P}
    + 4 \pi  (\hat \eta_K - \hat \epsilon_K) f_K$.
If $\tilde \lambda$ is Hermitian,
$\tilde f'_K = \tilde f_K$,
and the first term in \eq{eq:lambda_exp}
is non-negative.
The second term 
is strictly positive
for $n=(-,d)$ and $(+,s)$ because 
$\alpha^{(n)}$ is negative for these channels
(see \eq{eq:alphas}),
and $R_{KP}$, $f_K$, $f_P$ are positive for all $K$ and $P$.
The third term is negative, 
but we can make it arbitrarily small
by tuning   $\hat \epsilon_K$ and $\Lambda'$ as far as 
$\hat \eta_K$ goes to zero in the large $K$ limit.
Let us choose $\hat \epsilon_K$ to be
\bqa
\hat \epsilon_K =
\left\{  \begin{array}{ccc}
\hat \eta_K & \mbox{for} & 
|K| < \Lambda', \\
 - \delta & \mbox{for} & 
 |K| > \Lambda'
\end{array}
\right.
\eqa
with 
$\delta > 0$.
Since $\hat \eta_K$ approaches zero in the large $K$ limit,
for any non-zero $\delta$, there exists a sufficiently large $\Lambda'$
such that $|\hat \eta_K| \ll \delta$ 
for $|K| > \Lambda'$.
In this case, 
\bqa
 \int \frac{dK}{2\pi \Lambda} (\hat \eta_K - \hat \epsilon_K)^2 | f_K |^2
 \approx
2  \delta^2 \int_{\Lambda'}^\infty  
\frac{dK}{2\pi K}  
\left( \frac{\Lambda'}{K} \right)^{2 \delta}
=
\frac{\delta}{2\pi}.
\eqa 
This can be made arbitrarily small by choosing $\delta$ that is nonzero but small enough.
On the other hand, the second term in \eq{eq:lambda_exp}
remains strictly positive even in the limit in which
$\delta$ is small and $\Lambda'$ is large.
This implies that there exist normalizable vectors
for which the left hand side of \eq{eq:lambda_exp} is positive definite 
if $\tilde \lambda$ was Hermitian.
This proves that $\tilde \lambda$ can not be Hermitian,
and {\it the quasi-fixed point must be non-Hermitian 
for  $v \neq 0$}\cite{Gorbenko:2018vt}.

Because the beta functionals have real coefficients,
non-Hermitian quasi-fixed points
arise in pairs that are related to each other 
through the Hermitian conjugation.
On the other hand, the true Hermitian fixed point in \eq{eq:true_fp}
is at $\tilde \lambda=0$ in the $v \rightarrow 0$ limit.
As $v$ approaches zero,
a pair of non-Hermitian quasi-fixed points
should merge into the true Hermitian-fixed point
due to continuity.
This implies that at least one pair of non-Hermitian fixed points 
are close to the space of Hermitian theories for a small $v$. 
This is illustrated in        
\fig{fig:non_Hermitian_fp}.

\subsubsection{Collision of projected quasi-fixed points}

\begin{figure}[h]
\centering
\includegraphics[scale=1]{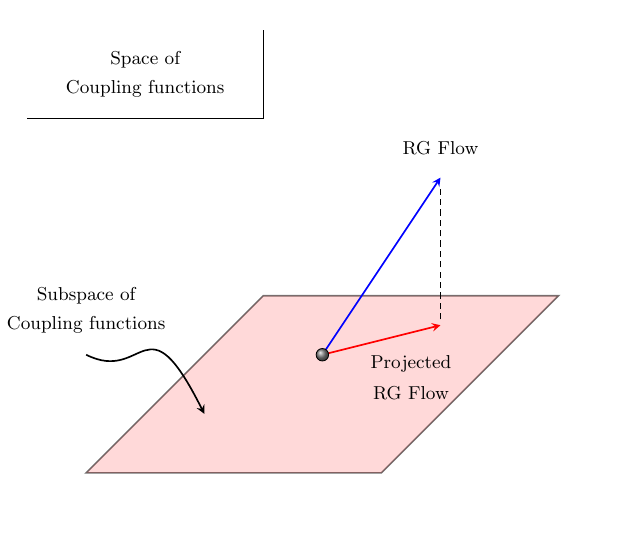}
\caption{
The full RG flow defined 
in the infinite dimensional space of coupling functions can be projected onto 
a finite dimensional subspace 
by restricting the flow vector 
to the tangent space of the subspace.
}
\label{fig:projected_RG_flow}
\end{figure}

Given that the RG flow is defined in the 
infinite dimensional space of coupling functions,
it is not easy to visualize it.
However, one can have a glimpse of the RG flow
by projecting it onto 
a finite dimensional subspace of coupling functions.
For this, we view the space of coupling functions as a vector space
and decompose the four-fermion coupling function
as an infinite sum of orthonormal basis,
\begin{equation}
\tilde{\lambda}^{(n)}_{QK}(\ell) = \sum_{j = 0}^{\infty}c_{n,j}(\ell)
\tilde{\lambda}^{[j]}_{QK}.
\end{equation}
Here $\tilde{\lambda}^{[j]}_{QK}$ represents 
the $j$-th basis of the coupling function 
that obeys the orthonormality condition,
$\int\frac{\dd K \dd P}{(2\pi\Lambda)^2}\tilde{\lambda}^{[i]}_{KP}\tilde{\lambda}^{[j]}_{PK} = \delta_{ij}$.
$c_{n,j}(\ell)$ denotes the strength of the coupling function
projected to the $j$-th basis function in channel $n$.
The space of coupling function is
viewed as the infinite dimensional space of 
the coupling constants, $\{ c_{n,j}(\ell) \}$,
and
the full beta functional can be written as 
a coupled differential equations for $c_{n,j}(\ell)$.
Within the infinite dimensional space of coupling constants,
let us consider a subspace 
spanned by one basis coupling,
\begin{equation}
\tilde{\lambda}^{[0]}_{KP}(\ell) = f_K f_P^*,
\label{eq:lambdatf}
\end{equation}
where $f_K$ denotes a wavefunction 
defined in the space of 
relative momentum of a Cooper pair 
with zero center of mass momentum.
The corresponding coupling, 
denoted as $c_{n,0}(\ell)= t(\ell) $,
measures the strength  of the interaction 
in that specific pairing channel.
For Hermitian theories, $t(\ell)$ is real,
but here we allow it to be complex to  accommodate non-Hermitian quasi-fixed points.
We choose the Cooper pair wavefunction to be of the form, 
\begin{equation}
f_K = A \sqrt{\frac{\Lambda}{\abs{K}}}\Theta(\abs{K}-\varepsilon)\Theta(\Delta -\abs{K}),
\label{eq:basis0}
\end{equation}
where
$A = \left[ \frac{1}{\pi} \log\left(\frac{\Delta}{\varepsilon}\right) \right]^{-\frac{1}{2}}$
is the normalization constant
\footnote
{
The rest of the basis coupling functions can be chosen to be orthogonal to 
\eq{eq:basis0}.
}.
Here $\varepsilon$ and $\Delta$
correspond to the small and large 
momentum cutoffs
for the wavefunction, respectively.
Since the norm of the wavefunction diverges both in the small $\varepsilon$
and the large $\Delta$ limits, 
one has to consider 
finite $\Delta / \varepsilon$.
With decreasing $\varepsilon$, 
the wavefunction has more weight 
for electrons near the hot spots.
In contrast, a larger $\Delta$
puts more weight on cold electrons away from the hot spots.
The nature of the projected RG flow 
depends on the relative weight between hot and cold electrons.

Even if one starts with a theory within the subspace 
in \eq{eq:lambdatf}, 
the theory in general flows out of the subspace because
the beta functions for other coupling constants are not generally zero within the subspace.
Here, we consider the RG flow that is projected
onto the subspace (see \fig{fig:projected_RG_flow}).
The projected beta function\cite{Lee:2014uf} 
is defined as
\begin{equation}
\frac{\partial t(\ell)}{\partial\ell} =
\int\frac{\dd K \dd P}{(2\pi\Lambda)^2}
\left[ 
\frac{\partial\tilde{\lambda}^{(n)}_{KP}(\ell)}{\partial\ell}
\right]_{ 
c_{n,0}=t,~
c_{n,j\neq0}=0 
}
\tilde{\lambda}^{[0]}_{PK}.
\label{eq:tproj}
\end{equation}
From Eqs. (\ref{eq:betaallwaves_matrix}) and (\ref{eq:lambdatf}),
the projected beta function can be written as 
\bqa
\frac{\partial t(\ell)}{\partial\ell} 
&=&
-
\left\{
  \int\frac{\dd P \dd K}{(2\pi\Lambda)^2}f_P^*
  \left(1 + K\frac{\partial}{\partial K} + P\frac{\partial}{\partial P} 
  + \hat \eta_K + \hat \eta_P 
\right)
 \tilde{\lambda}^{(n)}_{PK}f_K \right. \nn
 & &
\hspace{1cm} \left. 
+   \frac{1}{4\pi} \int \frac{\dd P \dd Q \dd K}{(2\pi\Lambda)^3}
f_P^* \tilde{\lambda}^{(n)}_{PQ} 
 \tilde{\lambda}^{(n)}_{QK} f_K
 - \alpha^{(n)}
\int\frac{\dd P \dd K}{(2\pi\Lambda)^2}f_P^* R(P, K ) f_K
\right\},
\label{eq:betafullf}
\eqa
where the momentum dilatation on the first line only act on $\tilde{\lambda}^{(n)}_{PK}$.
The resulting projected beta function
is written as a differential equation for $t$ only,
\begin{equation}
\begin{aligned} 
\frac{\partial t(\ell)}{\partial\ell} 
= & - \left[
 \frac{1}{4\pi} t(\ell)^2 + 2 \langle f | \hat{\eta} | f \rangle  t(\ell) 
 -
 \alpha^{(n)}
 \langle f | R | f \rangle
 \right],
\end{aligned}
\label{eq:betatpolynomial}
\end{equation}
where
$\langle f | T | f \rangle  
\equiv
\int \frac{dP dK}{(2\pi\Lambda)^2}
f_K^* T_{KP} f_P$.
The fixed points of the projected beta function
arise at 
\begin{equation}
t_{\star} = -4\pi\langle f | \eta | f \rangle
 \pm 
2\pi
\sqrt{
    d_{\star} 
},
\label{eq:quasifixedpoints}
\end{equation}
where 
$d_{\star}  \equiv  
4 \langle f | \eta | f \rangle^2  + \frac{ \alpha^{(n)}}{\pi}  \langle f |R| f \rangle$
is the discriminant.
$\langle f | \eta | f \rangle$ 
is the contribution from the anomalous dimension of electrons, 
and $\langle f |R| f \rangle$ is from the interaction  generated from the spin fluctuations. 
The anomalous dimension makes electrons incoherent and tends to suppress pairing instability.
On the other hand,  
the attractive interaction promotes superconductivity in the channels with $\alpha^{(n)} < 0$.
While  $\langle f | \eta | f \rangle$  contributes  to the discriminant with the  higher power than  $\langle f |R| f \rangle$, for a non-zero $w$,  the discriminant can be dominated by either one of the two depending on the choice of $\varepsilon$ and $\Delta$.
In channels in which the pair breaking effect (attractive interaction) dominates over the other, $d_\star > 0$ ($d_\star <0$) and
the quasi-fixed points are Hermitian (non-Hermitian).

 \begin{figure}[t]
\centering
\includegraphics[width=0.32\linewidth]{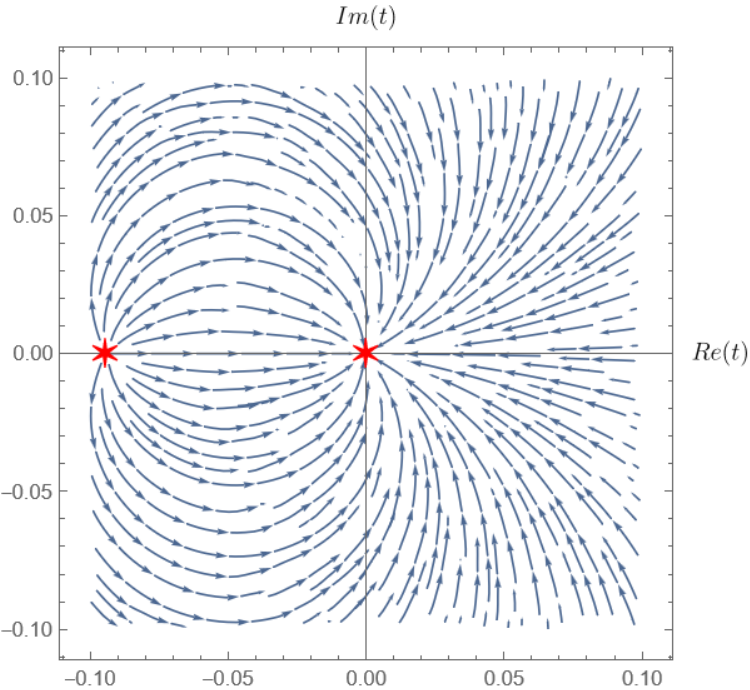}
\includegraphics[width=0.32\linewidth]{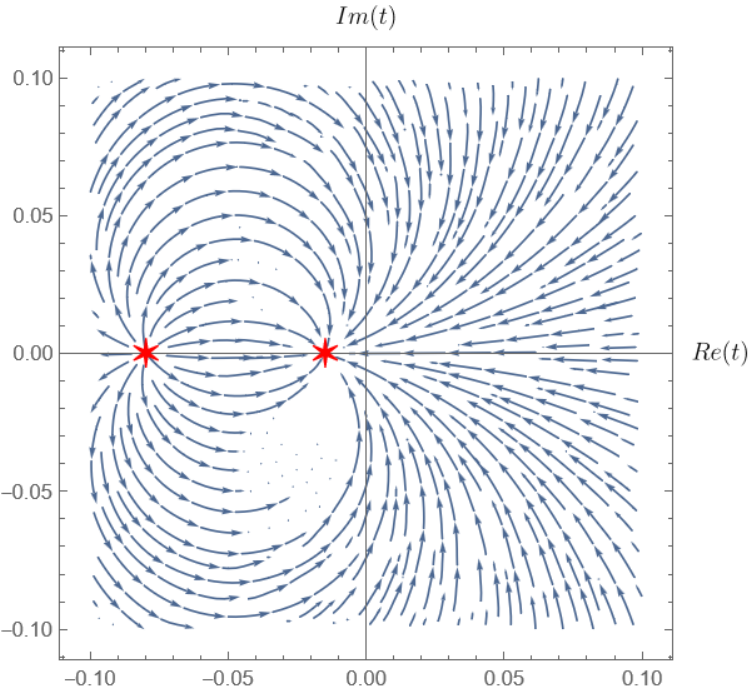}
\includegraphics[width=0.32\linewidth]{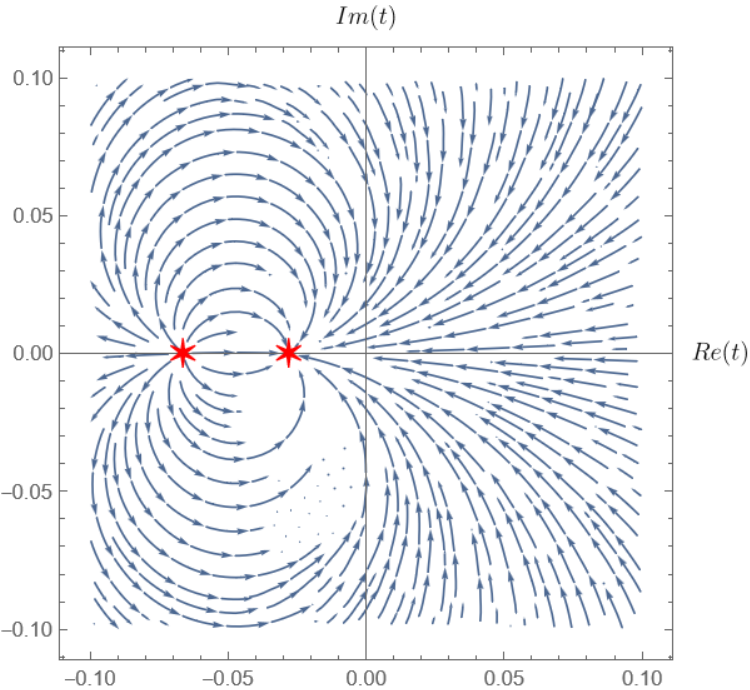}
\\ \includegraphics[width=0.32\linewidth]{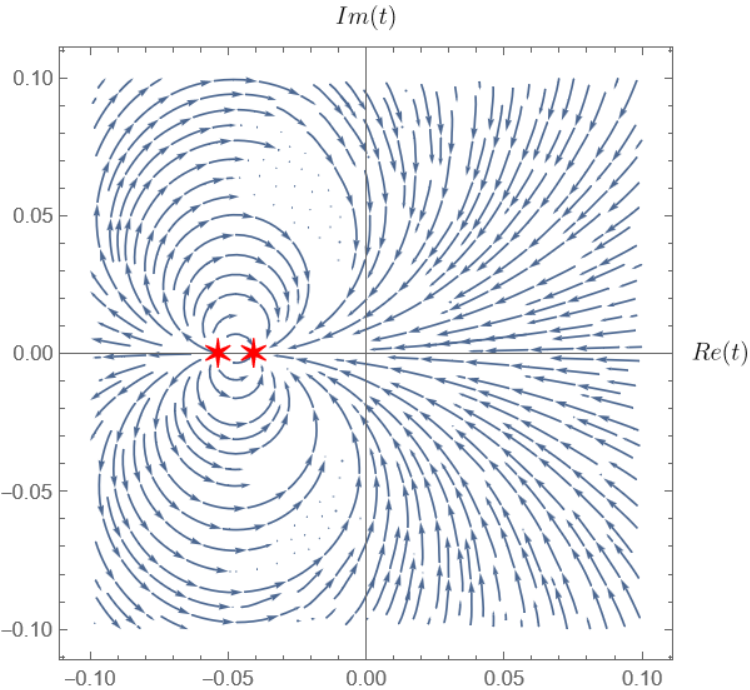}
\includegraphics[width=0.32\linewidth]{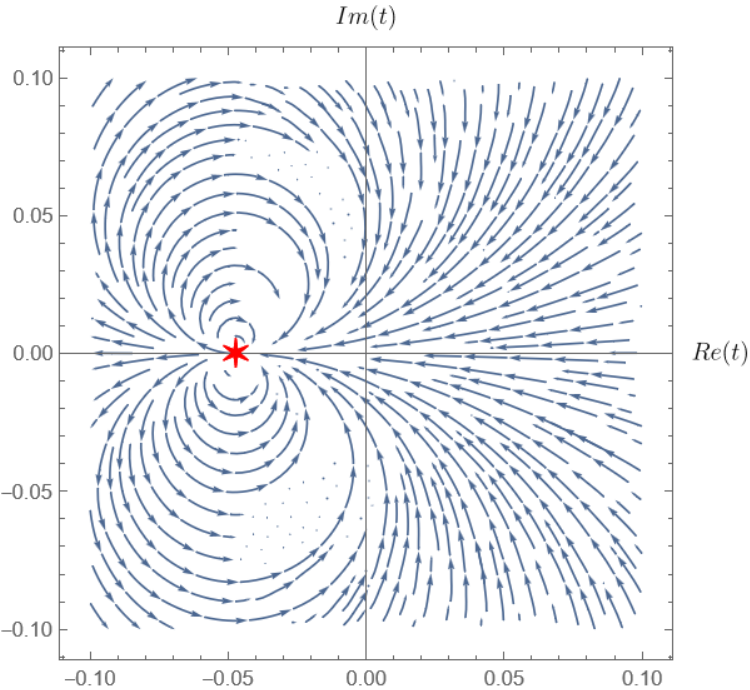}
\includegraphics[width=0.32\linewidth]{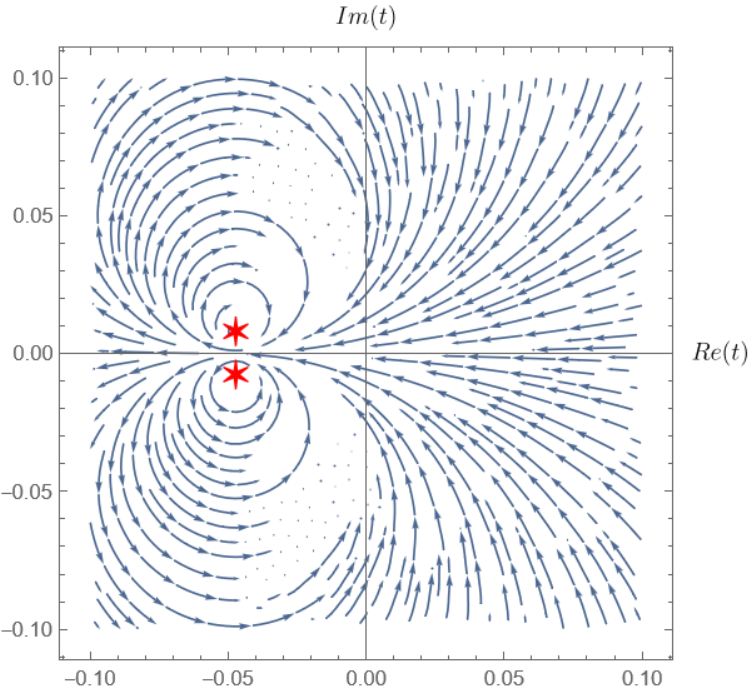}
\\ \includegraphics[width=0.32\linewidth]{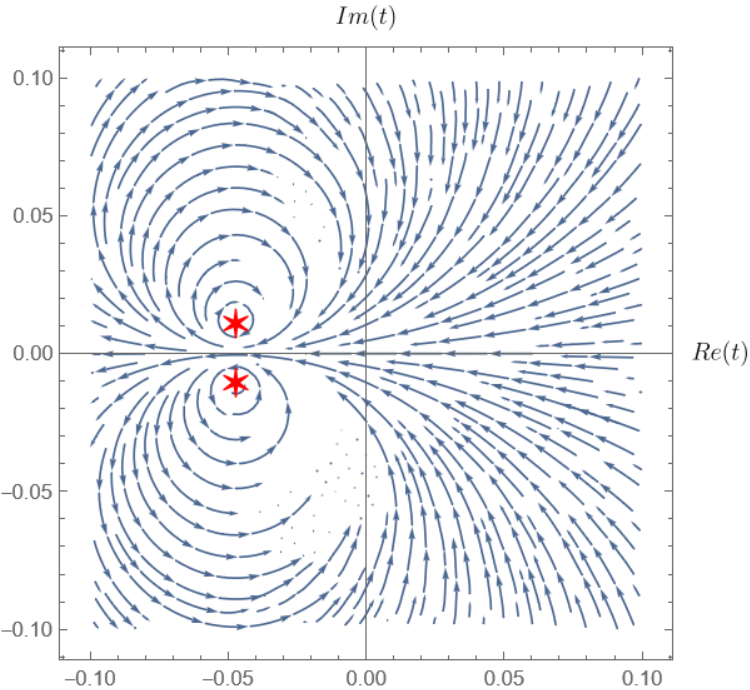}
\includegraphics[width=0.32\linewidth]{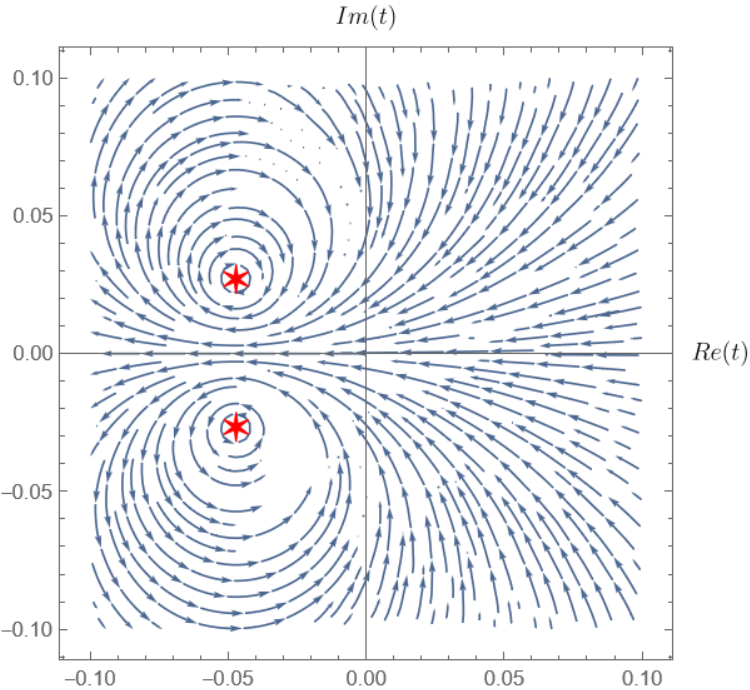}
\includegraphics[width=0.32\linewidth]{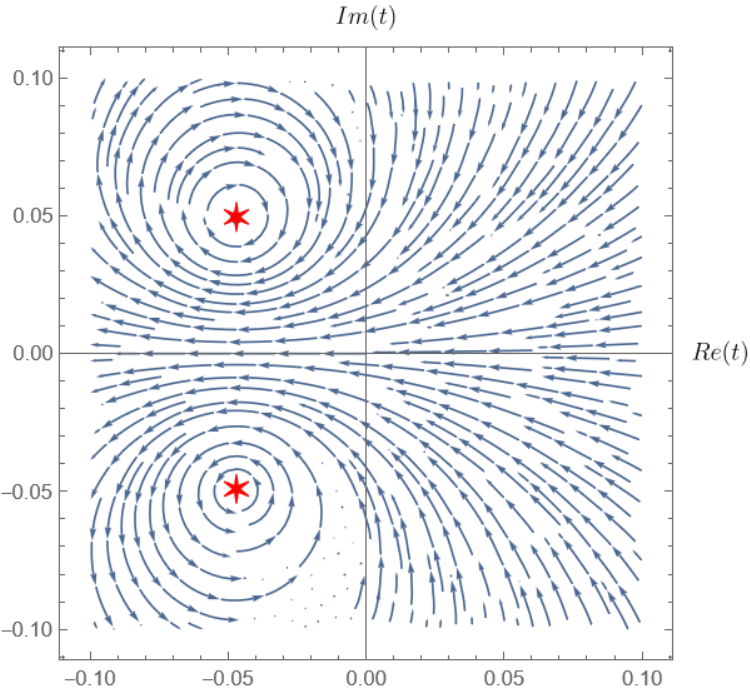}
\caption{
The RG flow projected onto the subspace of one complex four-fermion coupling 
in the spin anti-symmetric d-wave 
pairing channel with 
Cooper-pair wave function given in \eq{eq:basis0}
for 
$v = 0.000476257$, 
$N_c = 2$ and $N_f = 1$.
Here $\Delta/\Lambda = 1000 /(v c) \approx 1.39\times10^8$  
and $\varepsilon /\Lambda$ is chosen to be  $\exp{(-10^a)}$ 
with 
$a=6$, $ 4$,  $3.8$,
$ 3.73$, 
$ 3.721826341$,
$3.71$, 
$3.7$, $3.6$ and $3.4$ 
from the top left panel 
to the bottom right.
The quasi-fixed points are marked as (red) stars. 
For small values of  $\varepsilon /\Lambda$, 
the pair breaking effect for hot electrons dominates, 
resulting in the quasi-fixed points 
on the real axis.
At the (approximate) critical value $\varepsilon/\Lambda = \exp(-10^{3.721826341})$,
the stable and unstable quasi-fixed points collide.
For larger values of $\varepsilon/\Lambda$,
the pair forming effect dominates,
and the quasi-fixed points
move away from the real axis,
resulting in a runaway flow 
for Hermitian theories on the real axis.
}
\label{fig:streamflowepsilon}
\end{figure}

Let us examine the discriminant in the limit
that $\varepsilon/\Lambda \ll 1/c$  and $\Delta / \Lambda \gg 1/(vc)$.
In this case, 
$\langle f | \eta | f \rangle$
and $\langle f |R| f \rangle$
can be computed analytically,
\begin{align}
\langle f | \hat{\eta} | f \rangle = 
 \frac{(N_c^2 - 1)}{4\pi N_c N_f }
 w
\frac{\log\left(\frac{\Lambda}{2 v c \varepsilon}\right)}{
\log\left(\frac{\Delta}{\varepsilon}\right) 
}, 
&~~~~
\langle f | R | f \rangle = 
 \frac{1}{2} w
\log\left(\frac{2}{v}\right)
\frac{1}{\log\left(\frac{\Delta}{\varepsilon}\right) }.
\label{eq:Revf}
\end{align}
Here, 
$K\partial_K f_K = -\frac{1}{2}f_K + 
A \sqrt{\varepsilon\Lambda}\sgn(K)\delta(\abs{K}-\varepsilon) 
- A \sqrt{\Delta\Lambda}\sgn(K)\delta(\Delta - \abs{K})$ is used and 
$w \equiv v/c$ with
$v=v_0$ and $c = c(v_0)$.
In the spin anti-symmetric d-wave channel, the discriminant  becomes
\begin{equation}
    d_{\star} 
=    
\frac{1}{\log\left(\frac{\Delta}{\varepsilon}\right) }
\left[
\frac{(N_c^2 - 1)^2}{4\pi^2 N_c^2 N_f^2 }
 w^2
\frac{\log^2\left( \frac{\Lambda}{2 v c \varepsilon} \right)}{ \log\left(\frac{\Delta}{\varepsilon}\right) }
- \frac{2}{\pi N_f}\left(1+\frac{1}{N_c}\right)w
\log\left(\frac{2}{v}\right)
\right].
    \label{eq:explicitdiscriminant}
\end{equation}
The relative magnitude between the two terms in \eq{eq:explicitdiscriminant}
is controlled by
$\frac{\log^2\left( \frac{\Lambda}{2 v c \varepsilon} \right)}{ \log\left(\frac{\Delta}{\varepsilon}\right) }$.
If we take the small $\varepsilon/\Lambda$ limit for a fixed $\Delta/\Lambda$, 
the wavefunction has a large weight for incoherent electrons close to the hot spots.
While the attractive interaction is also strong near the hot spots, the interaction is not singular enough to overcome the pair breaking effect.
In those channels with small $\varepsilon/\Lambda$, 
the discriminant is positive
and the projected RG flow supports two fixed points on the real axis of the coupling.
One is a stable fixed point and the other is an unstable fixed point.
Near the stable fixed point, the pairing interaction does not grow due to the pair breaking effect.
Alternatively, if we take the large $\Delta/\Lambda$ limit for a fixed $\varepsilon/\Lambda$,
the wavefunction has a large weight for cold electrons.
Since electrons away from the hot spots are largely coherent, they are more susceptible to pairing instability.
While the attractive interaction is also weak away from the hot spots,
the pairing effect prevails over the pair breaking effect in these channels
because
$\langle f | \eta | f \rangle^2$
goes to zero faster than
$\langle f |R| f \rangle$
with increasing $\Delta/\Lambda$.
As a result,
the discriminant is negative
in the channels with large $\Delta/\Lambda$\footnote{
However,  $d_\star$ eventually  approaches zero in the $\Delta /  \Lambda \rightarrow \infty$ limit because the attractive interaction becomes vanishingly small for  electrons that are very far from the hot spots. 
Therefore, there exists an optimal   choice of $\varepsilon$ and $\Delta$ at which the discriminant is most negative.
},
and the fixed points arise away from the real axis.
This corresponds to a non-Hermitian   fixed point for the projected RG flow.
On the real axis, the couplings 
in those channels
exhibit run-away flows toward 
the strong coupling regime 
with large attractive interactions, 
signifying a superconducting instability.
These two different behaviours are separated by critical wavefunctions at which 
the discriminant vanishes
and two fixed points collide on the real axis\cite{PhysRevD.80.125005}.
Here, the collision of fixed points arises within one theory as the plane onto which the RG flow is projected is rotated in the space of coupling functions.
The evolution of the projected RG flow with different values of $\varepsilon/\Lambda$ and $\Delta/\Lambda$ is shown in  \fig{fig:streamflowepsilon}.

A small perturbation around the projected quasi-fixed point
evolves under the linearized beta function given by
\begin{equation}
\frac{\partial\delta t(\ell)}{\partial\ell} =  
\mp 
\sqrt{d_\star}
\ \delta t(\ell),
\end{equation}
where
$\delta t(\ell) = t(\ell) - t_{\star} $.
The eigenvalues
$\mp \sqrt{d_\star}$
are real at the Hermitian fixed points
and purely imaginary at the non-Hermitian quasi-fixed points.
The RG flow in the vicinity of the real quasi-fixed points exhibits the usual converging or diverging behaviour depending on whether the fixed point is stable or unstable, respectively.
Near the non-Hermitian fixed points,
the RG flow is rotational, exhibiting limit cycles\cite{
VEYTSMAN1993315,
PhysRevLett.89.230401,
PhysRevB.69.020505,
PhysRevD.75.025005,
PhysRevLett.108.131601}.

\begin{figure}[t]
\centering
\includegraphics[width=0.45\linewidth]{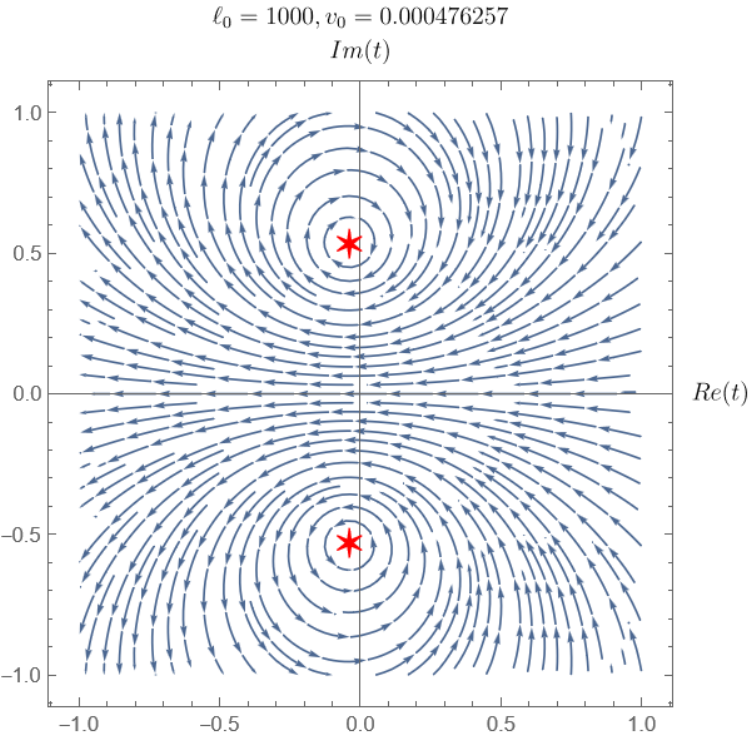}
\includegraphics[width=0.45\linewidth]{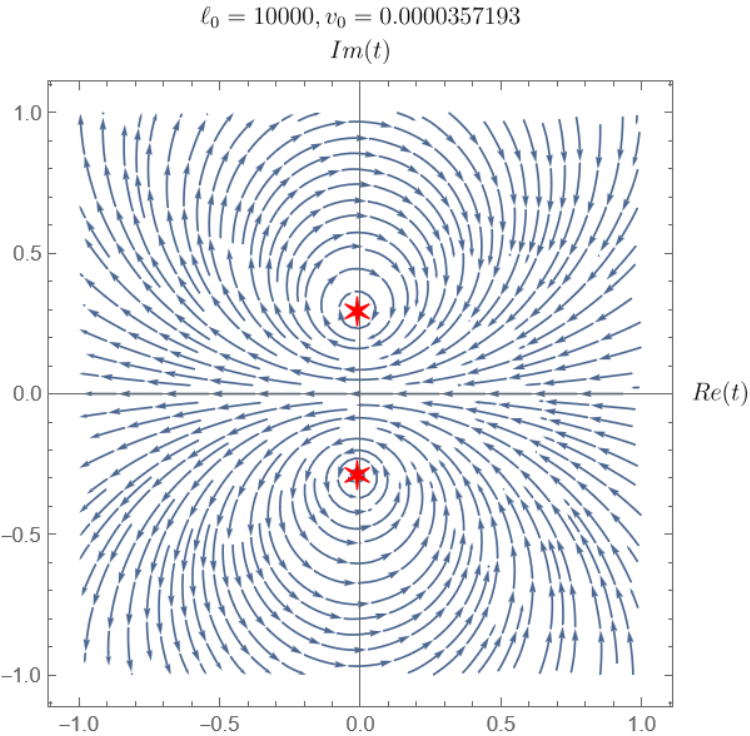}\\
\includegraphics[width=0.45\linewidth]{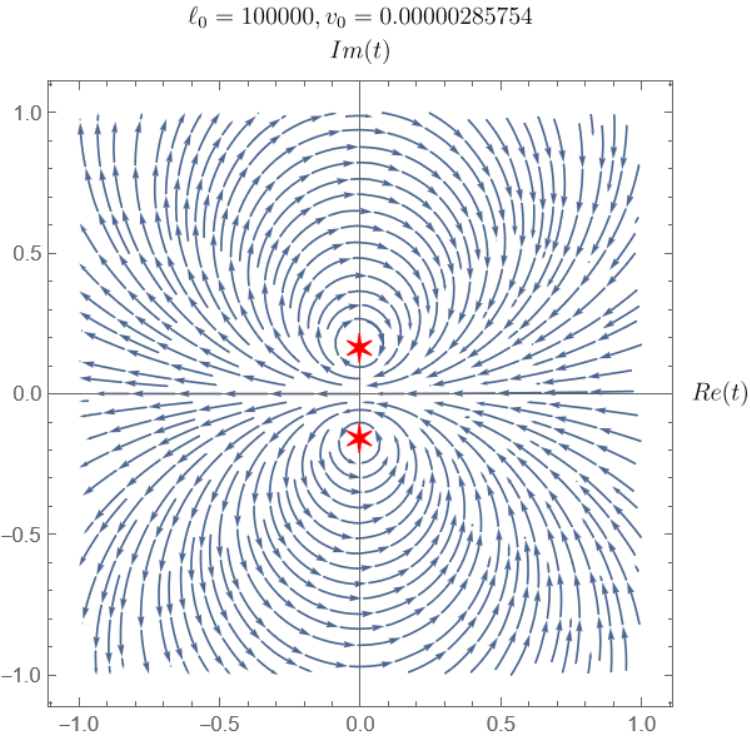}
\includegraphics[width=0.45\linewidth]{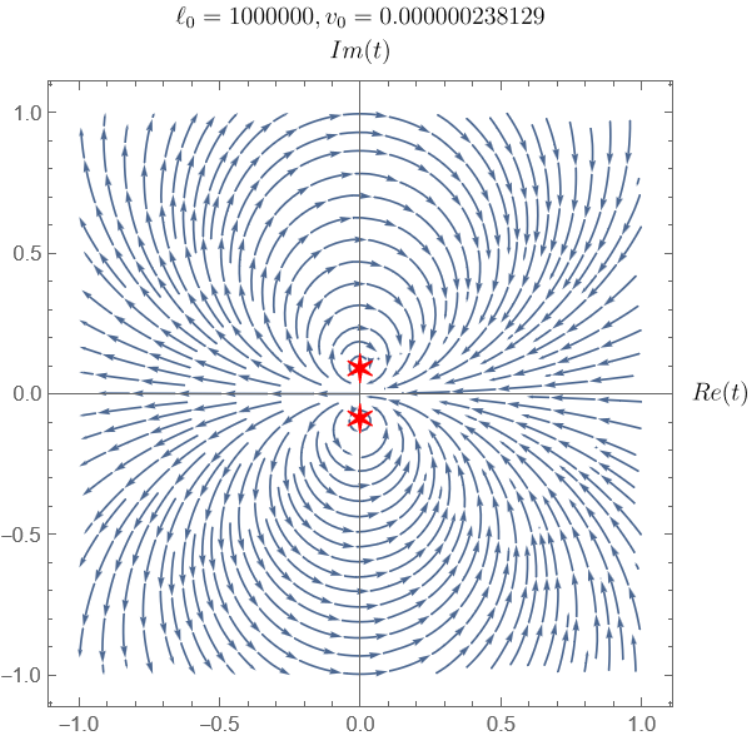}
\caption{
The RG flow projected onto the subspace of one complex four-fermion coupling 
in the spin anti-symmetric d-wave 
pairing channel with 
Cooper-pair wave function given in \eq{eq:basis0}
for 
$N_c = 2$,
$N_f = 1$,
$\Delta / \Lambda = 10^{8}$ and
$\varepsilon / \Lambda =10^{-10}$
with
$\ell_0 =$
$10^3, 10^4, 10^5, 10^6$
from top left to bottom right.
With increasing $\ell_0$ (decreasing $v_0(0)$),
the pair of complex quasi-fixed points approach to the real axis, creating a bottleneck region in the real axis.
}
\label{fig:streamflow}
\end{figure}

The existence of channels in which the quasi-fixed points arise away from the real axis suggests that a Hermitian theory eventually flows toward a superconducting fixed point at low energies.
When the nesting angle is small,
we expect a weaker superconducting instability as the pairing interaction generated from the spin fluctuations is weak.
Indeed, the non-Hermitian quasi-fixed points get closer to the space of Hermitian theories as $v$ decreases.
In Fig. \ref{fig:streamflow}, 
we plot the projected RG flow around the non-Hermitian quasi fixed point
in the spin anti-symmetric d-wave channel
for different values of $v$.
With decreasing $v$,  
the quasi-fixed points approach the real axis,
creating a bottleneck in the RG flow of Hermitian theories.
The bottleneck creates a large window of length scale in which the theory exhibits an approximate scale invariance before the system eventually becomes superconducting in the low energy limit.

Our next task is to understand 
 how Hermitian theories undergo superconducting instabilities at low energies.
We can not use \eq{eq:betatpolynomial} to describe the actual superconducting instability
because the full RG flow does not stay within the one-dimensional subspace spanned by \eq{eq:lambdatf}.
The RG flow that is projected to a fixed channel does not capture how the Cooper pair wavefunction evolves under the RG flow.
In particular, the dilatation term in the beta functional continuously push electrons from the hot region to the cold region under the RG flow, which broadens the width of the pair wavefunction relative to the width of the hot spot region\footnote{
This reflects the fact that more and more electrons are decoupled from spin fluctuations as the low-energy limit is taken.}.
Therefore, we go back to the full beta functional to address superconducting instability.

\section{Superconducting instability}
\label{sec:super}

\fbox{\begin{minipage}{48em}
{\it
\begin{itemize}
\item
Theories whose bare four-fermion interactions are attractive 
and stronger than the interaction mediated by the critical spin fluctuations 
develop superconducting instabilities 
via the conventional BCS mechanism in which
the nearby non-Fermi liquid fixed point plays little role.
\item
Theories whose bare four-fermion interactions are not strongly attractive in any channel
necessarily flow through the bottleneck region controlled by the non-Fermi liquid fixed point.
It creates a window of energy scale for quasi-universal scaling,
where the energy window increases with decreasing bare nesting angle.
\item
The superconducting transition temperature and the pairing wavefunction
of those theories that go through the bottleneck region
are determined from the bare nesting angle that sets the distance of the bare theory from the non-Fermi liquid fixed point.
 \end{itemize}
}
\end{minipage}}
\vspace{0.5cm}

In this section, we study superconducting instability by solving the beta functional 
for the four-fermion coupling function 
in the pairing channel with a fixed nesting angle.
This is justified because 
the theory undergoes a superconducting instability before the nesting angle changes appreciably in the small $v$ limit,
as will be shown later.
Since we already know that there is no Hermitian fixed point with $v \neq 0$,
for the purpose of understanding superconductivity,
it is simpler to use the beta functional for the coupling function 
defined in the space of physical momentum.
The way superconductivity emerges 
at low energies can be understood from
the solution of the beta functional for the four-fermion coupling function.
Since 
${\lambda}^{V;  (\pm),\spmqty{s \\ d}}_{2PP \spmqty{p & -p \\ k & -k}}$  
is sensitive to the boson propagator 
at large momenta,
only the net two-body interaction function defined in \eq{eq:lambdaprime0}
can be determined within 
the low-energy effective field theory.
Nonetheless, we can still understand superconducting instabilities 
by solving \eq{eq:betabigmatrixsimple_main01V}
for 
${\lambda}^{V;  (\pm),\spmqty{s \\ d}}_{2PP \spmqty{p & -p \\ k & -k}}$,
assuming that the boson propagator takes the form of \eq{eq:Dqk} at all momenta.
Although 
$\mathsf{D}_{\mu}(q;k)$ in \eq{eq:Dqk} and
${\lambda}^{V;  (\pm),\spmqty{s \\ d}}_{2PP \spmqty{p & -p \\ k & -k}}$ obtained from
\eq{eq:betabigmatrixsimple_main01V}
are not individually reliable at large momenta,
the combination in \eq{eq:lambdaprime0}
is insensitive to the unknown UV physics.
In particular, 
if the net two-body interaction
diverges due to a superconducting instability,
so does  ${\lambda}^{V;  (\pm),\spmqty{s \\ d}}_{2PP \spmqty{p & -p \\ k & -k}}$
because \eq{eq:Dqk} is regular.
Therefore, we directly solve the beta functional
in \eq{eq:betabigmatrixsimple_main01V}.
The manner superconductivity arises crucially depends 
on whether the bare coupling function
has any channel with attractive interaction 
which is stronger than the interaction mediated by the spin fluctuations or not \cite{PhysRevLett.15.524}.
Therefore, discussion on superconductivity is divided into two parts.

\subsection{Attractive bare interaction}

If the bare interaction is attractive in any channel with its strength greater than $w=v/c$,
\eq{eq:betabigmatrixsimple_main01V} is dominated by the BCS term that is quadratic
in the four-fermion coupling.
In this case,  \eq{eq:betabigmatrixsimple_main01V} is well approximated by
\begin{align}
& \frac{\partial}{\partial\ell}
{\lambda}^{V;  (\pm),\spmqty{s \\ d}}_{2PP \spmqty{p & -p \\ k & -k}} 
=
 -  {\lambda}^{V;  (\pm),\spmqty{s \\ d}}_{2PP \spmqty{p & -p \\ k & -k}}
 - 
\frac{1}{4\pi}
\int \frac{\dd q}{2\pi\mu}
 \lambda^{V; (\pm),\spmqty{s \\ d}}_{2PP \spmqty{p & -p \\ q & -q}} 
 \lambda^{V;  (\pm),\spmqty{s \\ d}}_{2PP \spmqty{q & -q \\ k & -k}}. 
\label{eq:betabigmatrixsimple_main02}
\end{align}
The first term on the right hand side reflects the fact that 
the four-fermion coupling is irrelevant by power-counting
under the scaling in which all components of momentum 
are scaled. 
The flip side of the scaling is the scale dependent measure
 $\frac{\dd q}{2\pi\mu}$
 in the second term.
It describes the BCS process in which 
Cooper pairs are scattered to intermediate states on the Fermi surface.
The volume of the phase space for virtual Cooper pairs measured in the unit of the running energy scale $\mu= \Lambda \emell$ 
increases with decreasing energy. 
The enhancement from the phase space volume compensates 
the suppression from the power-counting, effectively promoting
the four-fermion coupling to a marginal coupling as expected.  
This can be easily seen by absorbing 
a factor of $\eell$ into the coupling to 
write the beta functional as 
\begin{align}
& \frac{\partial}{\partial\ell}
\left( \eell
{\lambda}^{V;  (\pm),\spmqty{s \\ d}}_{2PP \spmqty{p & -p \\ k & -k}} 
\right)
=
 - 
\frac{1}{4\pi}
\int \frac{\dd q}{2\pi\Lambda}
\left( \eell \lambda^{V;  (\pm),\spmqty{s \\ d} }_{2PP \spmqty{p & -p \\ q & -q}} \right)
\left( \eell
 \lambda^{V;  (\pm),\spmqty{s \\ d} }_{2PP \spmqty{q & -q \\ k & -k}} \right). 
\label{eq:betabigmatrixsimple_main03}
\end{align}
Its solution is given by
\bqa
{\lambda}^{V; (\pm),\spmqty{s \\ d}}(\ell) 
=
\emell
{\lambda}^{V; (\pm),\spmqty{s \\ d}}(0) 
\left[
1 + \frac{\ell }{4\pi}
{\lambda}^{V; (\pm),\spmqty{s \\ d}}(0) 
\right]^{-1}.
\eqa
Here ${\lambda}^{V; (\pm),\spmqty{s \\ d}}$ is viewed as a matrix defined in the space of momentum. 
In this expression, the matrix multiplications are defined with the measure
$(A B)_{p,k} = \int \frac{\dd q}{2\pi\Lambda} A_{pq} B_{qk}$.
If ${\lambda}^{V}(0)$ has any channel with negative eigenvalue,
the four-fermion coupling blows up around scale
$\ell_c \sim \frac{4 \pi}{|E|}$,
where $E<0$ is the most negative eigenvalue of the bare coupling. 
For $|E| > w$, 
$\ell_c < 1/w \ll \ell_0$,  
and the flow of $v$ is negligible between $\ell=0$ and $\ell_c$.
When the bare interaction is attractive and stronger than  $w$,
the superconducting transition temperature
and the pairing wavefunction are 
sensitive to the bare four-fermion coupling.
In this case, 
gapless spin fluctuations have little effect on superconductivity,
and the manner in which superconductivity emerges is not universal.

\subsection{Repulsive bare interaction}

Theories in which the bare coupling is not strongly attractive in any channel 
are more interesting
in that the emergence of superconductivity 
is governed by the universal physics associated with the nearby non-Fermi liquid fixed point.
This is because those theories necessarily 
flow through the bottleneck region 
where the RG flow is constricted.
To see this, 
we view  $\lambda^V_{2PP}$ 
in \eq{eq:lambdaV} as a matrix and
decompose it as
${\lambda}^{V}_{2PP}
=  \sum_i  {\lambda}_i 
 | i  \rangle
\langle  i |$.
Here, the channel indices ($(\pm),\spmqty{s \\ d}$) are dropped to avoid clutter in notation.
$| i \rangle$'s ($\langle i |$'s) represent normalized column (row) eigenvectors 
that diagonalize the four fermion coupling function, and
${\lambda}_i$'s represent the eigenvalues.
Eigenvalues and eigenvectors obey the flow equations given by
\begin{equation}
\begin{aligned}
\frac{\partial}{\partial \ell}  {\lambda}_i
= &  -  {\lambda}_i
-  2 \langle i | \eta  | i \rangle_\mu {\lambda}_i
- \frac{1}{4 \pi} 
\langle i| ( \lambda' )^2 | i \rangle_\mu,
%
\\
\frac{\partial | i \rangle }{\partial \ell} = &
- \sum_{j \neq i} 
\frac{ (  \lambda_i +  \lambda_j ) \left(\langle j | \eta | i \rangle_\mu - \frac{\alpha^{(n)}}{4\pi}\langle j | \mathsf{D}^V | i \rangle_\mu \right) + \frac{(\alpha^{(n)})^2}{4\pi} \langle j | \left( \mathsf{D}^V \right)^2 | i \rangle_\mu}{ \lambda_i- \lambda_j} |j \rangle.
\end{aligned}
\end{equation}
Here, 
$ \langle i | C  | j \rangle_\mu 
\equiv \int \frac{\dd k \dd p}{(2\pi\mu)^2}
f_{i,k}^* C_{kp} f_{j,p}$
with $f_{i,k}$ representing the 
$i$-th eigenvector written in the momentum space.
$\eta$ is a diagonal matrix with 
$\eta_{p,k}=
2 \pi \mu 
\delta(p-k) \eta_p$.
$\lambda^{' (\pm),\spmqty{s \\ d}}_{\spmqty{p & -p \\ q & -q}} $ is defined in \eq{eq:lambdaprime0}
and
$D^V_{p,k} \equiv \frac{\mathsf{D}_\mu(p;k)}{\sqrt{V_{F,p}V_{F,k}}}$.
Since $\eta$ and $(\lambda')^2$ are non-negative matrices,
$\frac{d {\lambda}_i}{d \ell} \leq 0$
for $\lambda_i \geq 0$.
%
%
%
This means that theories with repulsive couplings
flow toward $\lambda_i=0$ at low energies.
In the small $v$ limit, 
theories with bare repulsive couplings 
flow to the fixed point at $\lambda_i=0$.
For $v \neq 0$, $\lambda_i=0$ is no longer a fixed point.
Since $\frac{\partial {\lambda}_i}{\partial \ell} < 0$ for $\lambda_i  \geq 0$,
all theories with $v \neq 0$ develop at least one channel 
with attractive interactions 
at sufficiently low energies\footnote{
This is expected from the absence of Hermitian quasi-fixed points with $v \neq 0$.}.
Although $\lambda = 0$ is not a fixed point,
it still acts as an approximate focal point in the space of theories
because
at $\lambda=0$ 
the beta functional is proportional to $\mathsf{D}_\mu^2$ 
whose eigenvalues are order of $w^2$.
In the small $v$ limit,
the slow RG speed near $\lambda=0$ creates a bottleneck region
in which a theory
spends a long RG `time'.
Consequently, 
bare theories with $O(1)$ repulsive interactions  
are naturally attracted to the region with $|\lambda_i| \leq w$ 
at scale $\ell^* \sim w^{-1}$ before they become negative.
For $v \neq 0$, there is no perfect focusing of the RG flow.
Nonetheless, theories spend longer RG time 
in the bottleneck region as $v$ decreases.
This makes the theory within the bottleneck 
an approximate attractor of UV theories\cite{PhysRevLett.71.2421,PhysRevX.7.031051}.
Once theories are attracted to the bottleneck region,
the superconducting transition temperature is determined by
the RG time that is needed for theories to pass through it.
In this section, 
we examine how superconductivity emerges
in a theory 
that is within the bottleneck  region with $\lambda \approx 0$ 
at a scale $\ell^*$.

To remove the explicit scale dependence in the measure 
of the momentum integration of the beta functional, we consider
\bqa
\bar{\lambda}^{V; (\pm),\spmqty{s \\ d}}_{2PP\spmqty{p & -p \\ k & -k}}
=
\eell
 \lambda^{V; (\pm),\spmqty{s \\ d}}_{2PP \spmqty{p & -p \\ k & -k}}. 
\label{eq:barlambda_def}
\eqa 
Its beta functional  reads
\bqa
&& \frac{\partial}{\partial\ell}
\bar {\lambda}^{V; (\pm),\spmqty{s \\ d}}_{2PP \spmqty{p & -p \\ k & -k}} 
=
 -\left( \eta_k+ \eta_p \right)
 \bar {\lambda}^{V; (\pm),\spmqty{s \\ d}}_{2PP \spmqty{p & -p \\ k & -k}} \nn
 &&
 - 
\frac{1}{4\pi}
\int \frac{\dd q}{2\pi\Lambda}
\left[
\bar \lambda^{V; (\pm),\spmqty{s \\ d}}_{2PP \spmqty{p & -p \\ q & -q}} 
 - \frac{2}{N_f} 
 Y_{PP}^{(\pm)} 
1^{\spmqty{s \\ d}}
\frac{ \eell  {\mathsf{D}}_{\Lambda \emell} (p; q)}{ \sqrt{ V_{F,p} V_{F,q} } }
\right]
 \left[
\bar \lambda^{V; (\pm),\spmqty{s \\ d}}_{2PP \spmqty{q & -q \\ k & -k}} 
 - \frac{2}{N_f} 
 Y_{PP}^{(\pm)} 
1^{\spmqty{s \\ d}}
\frac{ \eell {\mathsf{D}}_{\Lambda \emell} (q; k)}{ \sqrt{ V_{F,q} V_{F,k} } }
\right].
  \label{eq:betaallwaves_main}
\eqa
Here, $ {\mathsf{D}}_{\mu} (q; k)$ is defined in \eq{eq:Dqk}.
A theory that is at the bottleneck point at scale $\ell^*$ corresponds 
to the `initial' condition
\bqa
\bar {\lambda}^{V; (\pm),\spmqty{s \\ d}}_{2PP \spmqty{p & -p \\ k & -k}}(\ell^*) =0.
\label{eq:tilde_lambda_initial0}
\eqa
In the following, we focus on the d-wave and spin anti-symmetric sector
in which the attractive interaction 
is strongest\cite{SCALAPINO,PhysRevB.81.224505}.
At energy scales that are not too smaller than $\Lambda^*=\Lambda e^{-\ell^*}$,
we can ignore $\bar \lambda_{2PP}^V$ 
on the right hand side of \eq{eq:betaallwaves_main}.
As the energy scale is lowered, 
the spin fluctuations generate
attractive interaction
which, in turn, accelerates the flow of $\bar \lambda_{2PP}^V$\footnote{
This follows from 
the fact that the largest eigenvalue  of ${\mathsf{D}}_\mu$ is positive
and
$ Y_{PP}^{(-)} 1^{\spmqty{ d}} = 2 \left(  1+ \frac{1}{N_c} \right) > 0$.
}.
At sufficiently low energies,
the magnitude of $\bar \lambda_{2PP}^V$ 
surpasses that of 
$
 - \frac{2}{N_f} 
 Y_{PP}^{(\pm)} 
1^{\spmqty{s \\ d}}
\frac{ \eell  {\mathsf{D}}_{\Lambda \emell} (p; q)}{ \sqrt{ V_{F,p} V_{F,q} } }$
in \eq{eq:betaallwaves_main}.
As the four-fermion coupling  becomes stronger than the attractive interaction generated by spin fluctuations at low energies, the further growth of the four-fermion coupling is dominated by the BCS process.
We denote this crossover scale as $\ell_1$.
Since the beta function is dominated by different terms 
below and above the crossover scales,
we write the approximate solution of the beta functional as 
\bqa
\bar{\lambda}_{2PP}^V = 
\left\{
\begin{array}{cl}
\bar{\lambda}_{I} & ~~~\mbox{for $\ell < \ell_1$} \\
\bar{\lambda}_{II} & ~~~\mbox{for $\ell > \ell_1$}
\end{array}
\right..
\label{eq:lambdaIandII}
\eqa
For $\ell < \ell_1$,
the RG flow is approximated by
\bqa
&& \frac{\partial}{\partial\ell}
\bar{\lambda}^{(-),\spmqty{d}}_{I, \spmqty{p & -p \\ k & -k}} 
\approx
 - 
\frac{1}{4\pi}
 \left( \frac{2 Y_{PP}^{(-)} }{N_f}  \right)^2
\int \frac{\dd q}{2\pi\Lambda}
\frac{ \eell  {\mathsf{D}}_{\Lambda \emell} (p; q)}{ \sqrt{ V_{F,p} V_{F,q} } }
\frac{ \eell {\mathsf{D}}_{\Lambda \emell} (q; k)}{ \sqrt{ V_{F,q} V_{F,k} } }.
  \label{eq:betaallwaves_main_2}
\eqa
It describes the process in which the four-fermion coupling
is generated from gapless spin fluctuations.
The contribution of the anomalous dimension can be also ignored because $\lambda_{2PP}^V$ is small.
The solution of \eq{eq:betaallwaves_main_2} is written as
\bqa
\bar{\lambda}^{(-),\spmqty{d}}_{I, \spmqty{p & -p \\ k & -k}} (\ell)
&=&
 - 
\frac{1}{4\pi}
 \left( \frac{2 Y_{PP}^{(-)} }{N_f}  \right)^2 
  \int_{\ell^*}^{\ell} d \ell'
 \frac{1}{
\sqrt{ V_{F,p} V_{F,k} } 
}
\times \nn
 &&
\int \frac{d q}{2\pi V_{F,q}}
\frac{ g_{p,q}^2  g_{q,k}^2  \Lambda }{
\left[ \mu' + c \abs{p - q}_{\mu'} + c\abs{v_{p}p + v_{q} q}_{\mu'}  \right]
\left[ \mu'+ c \abs{q - k}_{\mu'} + c \abs{v_{q}q + v_{k} k}_{\mu'} \right]},
  \label{eq:betaallwaves_main_sol1}
\eqa
where $\mu'=\Lambda e^{-\ell'}$,
and all coupling functions on the right hand side of the equation
are evaluated at scale $\ell'$.
Let us denote the most negative eigenvalue 
and the associated eigenvector 
of $\bar \lambda_I(\ell)$ as 
$E_0(\ell)$ and $F_k(\ell)$, respectively. 
The crossover scale $\ell_1$ 
is determined from the condition that $E_0(\ell)$ 
becomes comparable 
to the spin fluctuation-induced interaction projected onto $F_k(\ell)$, 
\bqa
E_0(\ell_1) \sim 
  \frac{2}{N_f} 
 Y_{PP}^{(-)} 
 \left\langle 
 \frac{ e^{\ell_1} {\mathsf{D}}_{\Lambda e^{-\ell_1}} (p; k)}{ \sqrt{ V_{F,p} V_{F,k} } }
 \right\rangle_{F},
\label{eq:crossover_ell1}
\eqa
where
$ \left\langle 
 \frac{ e^{\ell_1} {\mathsf{D}}_{\Lambda e^{-\ell_1}} (p; k)}{ \sqrt{ V_{F,p} V_{F,k} } }
 \right\rangle_{F}
= 
\int \frac{dp dk}{(2 \pi \Lambda)^2}
\frac{ e^{\ell_1} {\mathsf{D}}_{\Lambda e^{-\ell_1}} (p; k)}{ \sqrt{ V_{F,p} V_{F,k} } }
F_p^{*}(\ell_1) 
F_k(\ell_1)$.
%
%
%
%
To the leading order in $v$,
the four-fermion coupling function generated by the Yukawa coupling 
in \eq{eq:betaallwaves_main_sol1}
decays as $1/|p-k|$ 
at large momenta
up to a logarithmic correction.
The slow decay of the large-angle scatterings give rise to inter-patch couplings that
invalidates the patch description. 

Since it is difficult to analytically compute the eigenvector of  
$\bar{\lambda}^{(-),\spmqty{d}}_{I, \spmqty{p & -p \\ k & -k}} (\ell)$, 
we first estimate the crossover scale 
using a simple Ansatz. 
At energy scale $\Lambda^*$,
electrons within the range of 
momentum $|k| < \Lambda^*/(vc)$
are strongly coupled with spin fluctuations. 
Therefore, we consider 
a Cooper pair wavefunction
whose width is order of $\Lambda^*/(vc)$
in the momentum space, 
\begin{equation}
f_k= 
\left(
\frac{2\pi vc}{\Lambda^*}
\right)^{1/2}
\Theta \left( \frac{1}{2} - \frac{ v c\abs{k}}{\Lambda^*} \right).
\label{eq:boxAnsatz}
\end{equation}
The expectation value of 
$\frac{e^{\ell}\mathsf{D}_{\mu}}{\sqrt{V_F}\sqrt{V_F}}$ 
for this Ansatz is written as
\begin{equation}
\begin{aligned}
 \left\langle \frac{e^{\ell}\mathsf{D}_{\mu}}{\sqrt{V_F}\sqrt{V_F}} \right\rangle 
 & = 
\int\frac{\dd k \dd p}{(2\pi\Lambda)^2}
\frac{g_{kp}(\ell)^2}{
 \sqrt{ V_{F,k}(\ell) V_{F,p}(\ell) }
}
\frac{\Lambda}{\Lambda \emell +c\abs{k-p} + c v\abs{k+p}}f^{*}_{k}(\ell)f_{p}(\ell),
\end{aligned}
\label{eq:devintegral}
\end{equation}
where 
\begin{equation}
g_{k,p}(\ell)  =  g_{k,p}(0)
\left(
\frac{\mu}{ \text{ max} \{\mu, 2 v \abs{k+p}, 2 v c \abs{k}, 2 v c\abs{p}\} }
\right)^{ \alpha_{0}(\ell_0) }, 
~~
V_{F,k}(\ell)  = 
\left(
\frac{\mu}{ \text{max}\{ \mu, 4v|k| \} }
\right)^{\alpha_1(\ell_0)/2}
\end{equation}
with $\mu=\Lambda \emell$
and
$\alpha_0$, $\alpha_1$ defined in \eq{eq:alpha01}.
To simplify the computation of 
\eq{eq:devintegral},
we first use a few assumptions,
and later justify them from the solution.
First, we assume that the Yukawa coupling function does not change significantly as a function of $\ell$ for $\ell^*< \ell< \ell_1$ :
\begin{equation}
\frac{g_{k,p}(\ell_1)}{g_{k,p}(\ell^*)} 
\sim 1,
\label{eq:yukawanochange}
\end{equation}
where 
$\ell^*$ is the scale at which the theory is in the bottleneck region 
and
$\ell_1$ is the crossover scale.
If \eq{eq:yukawanochange} is satisfied,
the scale dependence of $V_{F,k}(\ell)$ can be also ignored in $\ell^*< \ell < \ell_1$ because $\alpha_1 \ll \alpha_0$ in the small $v$ limit.
Second, we assume that the coupling functions are almost constant within the support of $f_k$ :
\begin{equation}
V_{F,k}(\ell) \approx 1, \quad 
g_{k,p}(\ell) \approx g_{0}(\ell)
\label{eq:couplingsnomom}
\end{equation}
for $|k|<\Lambda^*/(vc)$ and $\ell^* < \ell < \ell_1$. 
Finally, we assume that the crossover occurs at an energy that is much smaller than $\Lambda^*$,
\begin{equation}
\ell_1 \gg \ell^*.
\label{eq:largemomlimit}
\end{equation}
These assumptions allow us to approximate  \eqref{eq:devintegral} as
\begin{equation}
\left\langle \frac{e^{\ell}\mathsf{D}_{\mu}}{\sqrt{V_F}\sqrt{V_F}} \right\rangle 
\approx  g_{0}(\ell^*)^2\int\frac{\dd k \dd p}{(2\pi\Lambda)^2}\frac{\Lambda}{c\abs{k-p} + c v\abs{k+p}}f^{*}_{k}(\ell)f_{p}(\ell)
\label{eq:devintegral2}
\end{equation}
for $\ell^* < \ell < \ell_1$. 
The direct integrations over the momenta gives
\begin{equation}
\left\langle \frac{e^{\ell}\mathsf{D}_{\mu}}{\sqrt{V_F}\sqrt{V_F}} \right\rangle 
\approx
\frac{g_0(\ell^*)^2}{\pi c}
\log\left(\frac{1}{v}\right) =  \frac{w_0(\ell^*)}{2}\log\left(\frac{1}{v}\right).
 \label{eq:estimationofD}
\end{equation}
\eq{eq:estimationofD} is largely independent of $\ell$\footnote{
While 
 $\frac{ e^{\ell} {\mathsf{D}}_{\Lambda e^{-\ell}} (p; k)}{ \sqrt{ V_{F,p} V_{F,k} } }$ 
at the hot spot ($p=k=0$)
increases without a bound 
with increasing $\ell$,
its eigenvalues remain bounded at all $\ell$.
In particular, 
$
\int_{-k_F}^{k_F}\frac{\dd p}{2\pi\Lambda} \int_{-k_F}^{k_F}\frac{\dd k}{2\pi\Lambda} 
f_p^*~
\frac{\Lambda}{ c \abs{k-p} + c v_0 \abs{k+p}} ~f_k
$
is finite for all square integrable functions $f_k$.
}.
On the other hand,
$E_0(\ell)$ grows linearly in $\ell$ as
\begin{equation}
\begin{aligned}
E_0(\ell) \sim
 -\frac{1}{4\pi N_f^2}\left(Y_{PP}^{(-)}\right)^2 
(\ell - \ell^*) 
\left\langle \frac{e^{\ell}\mathsf{D}_{\mu}}{\sqrt{V_F}\sqrt{V_F}} \right\rangle_f^2 
\sim -\frac{1}{4\pi N_f^2} \left(Y_{PP}^{(-)}\right)^2  (\ell-\ell^*)w^2 \log^2\left(\frac{1}{v}\right).
\label{eq:E0_estimation}
\end{aligned}
\end{equation}
Inserting Eqs. 
\eqref{eq:estimationofD}-\eqref{eq:E0_estimation}
to \eq{eq:crossover_ell1}, 
we obtain the crossover scale to be
\bqa
\ell_1 \sim \ell^* + 
\frac{2\pi N_f}{\left(1+\frac{1}{N_c}\right)}\frac{1}{w \log(1/v)},
\label{eq:ell1}
\eqa
and the strength of the four-fermion coupling at the crossover scale is
\begin{equation}
E_0(\ell_1)
\sim
-\frac{2}{ N_f}\left(1+\frac{1}{N_c}\right) w \log\left(\frac{1}{v}\right).
\label{eq:E0atell1}
\end{equation}
%
The consistency of 
 the assumptions used in Eqs.  \eqref{eq:yukawanochange} -\eqref{eq:largemomlimit}
 can be checked\footnote{
Eq. \eqref{eq:largemomlimit} directly follows from \eq{eq:ell1}.
With $\alpha_0$ and $c$ expressed 
in terms of $v$ as 
$\alpha_0(v) = \frac{\sqrt{v\log(1/v)}}{\sqrt{2}\pi \sqrt{N_c N_f}}$, 
$c(v) = \sqrt{\frac{v}{8 N_c N_f}\log\left(\frac{1}{v}\right)}$,
\eq{eq:yukawanochange} becomes
$\frac{g_{k,p}(\ell_1)}{g_{k,p}(\ell^*)} 
\sim  e^{-\alpha_0(\ell_0)(\ell_1-\ell^*)} 
\sim
e^{-\frac{1}{2(N_c+1)}} \sim 1$.
For $N_c = 2, N_f = 1$,
$g(\ell_1)/g(\ell^*) \approx 0.846482$.
Eq. \eqref{eq:couplingsnomom} for the diagonal Yukawa coupling
can be checked from
$ \frac{g_{\frac{\Lambda^*}{vc}}(\ell_1)}{g_{0}(\ell_1)} \sim \left(\frac{\Lambda e^{-\ell_1}}{\Lambda^*}\right)^{\alpha_0} = e^{-\alpha_0(\ell_1-\ell^*)} \sim 1$.
Similarly, Eq. \eqref{eq:couplingsnomom} 
for the off-diagonal Yukawa coupling 
and $V_{F,k}$ follow.
}.

\begin{figure}[t]
	\centering
	\begin{subfigure}[b]{0.49\linewidth}
		\centering
		\includegraphics[scale=0.5]{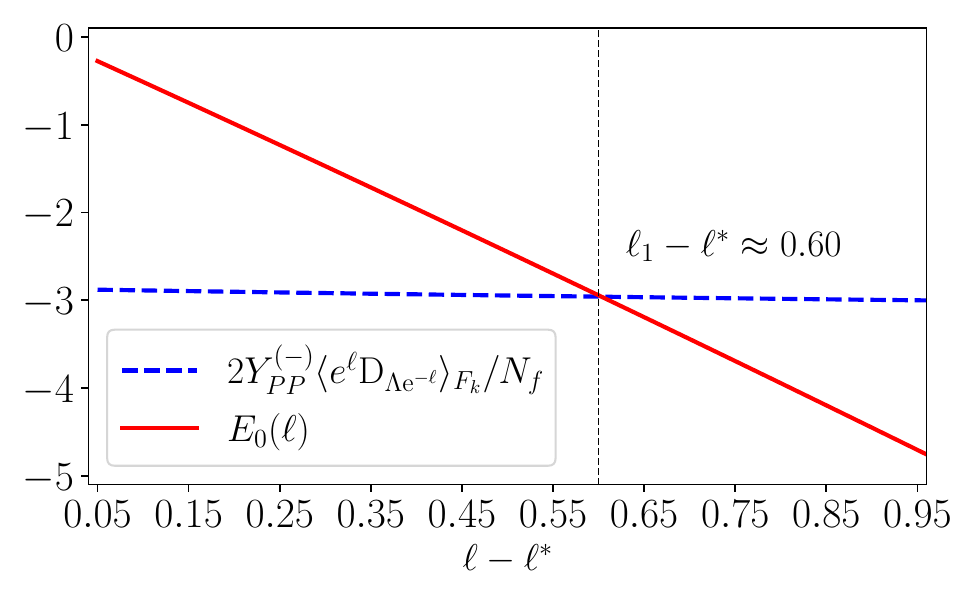}
		\caption{\label{fig:crossover_l0=100}}
	\end{subfigure}
	\begin{subfigure}[b]{0.49\linewidth}
		\centering
		\includegraphics[scale=0.51]{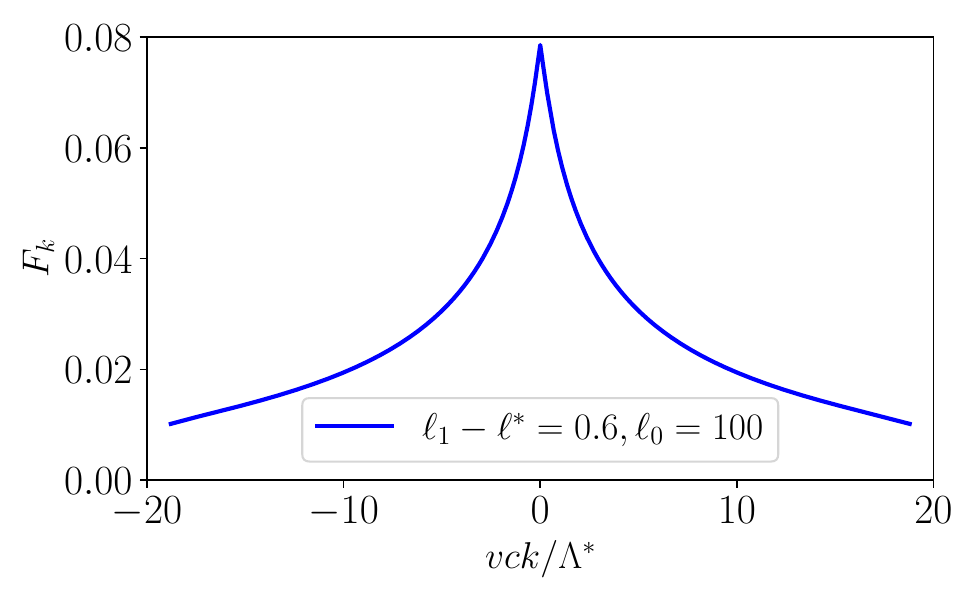}
		\caption{\label{fig:crossover_eigvec}}
	\end{subfigure}
		\begin{subfigure}[b]{0.49\linewidth}
		\centering
		\includegraphics[scale=0.5]{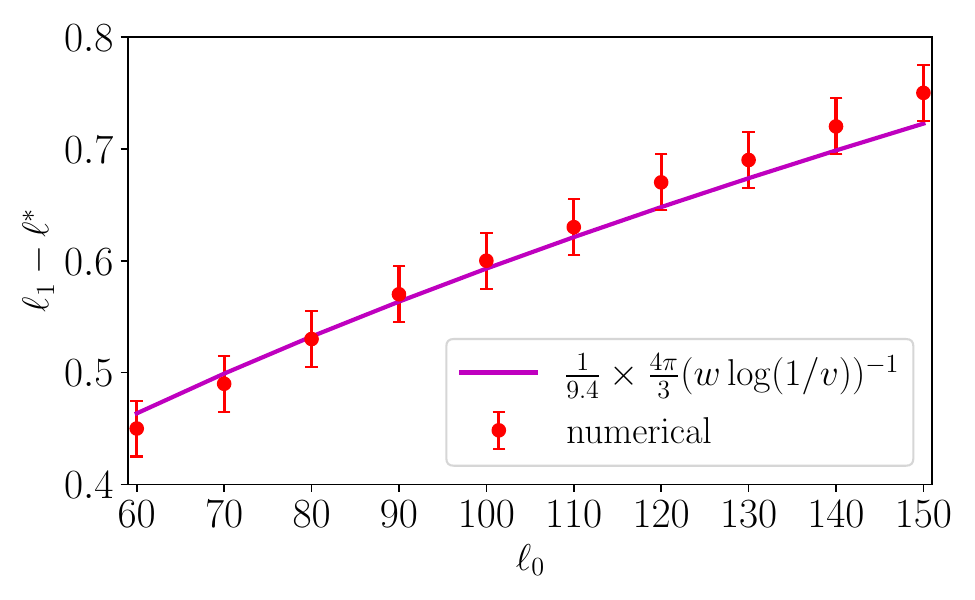}
		\caption{\label{fig:crossover_l0}}
	\end{subfigure}
	\caption{\label{fig:source_crossover}
Numerical results for the crossover scale and the Cooper pair wavefunction.
(a)  The solid and dashed lines denote the most negative eigenvalue
of  \eq{eq:betaallwaves_main_sol1}
($E_0(\ell)$) and the expectation value of ${\mathsf{D}}_{\mu}(k,p)$ 
in \eq{eq:devintegral}
for the associated eigenvector, respectively, plotted as functions of $\ell-\ell^*$ 
for $\ell_0 = 100$, $N_c = 2$ and $N_f = 1.$ 
The eigenvalue and the expectation value cross at scale $\ell \approx \ell^*+0.6$.
(b) The normalized eigenvector  associated with the  most negative eigenvalue of  \eq{eq:betaallwaves_main_sol1}
at the crossover scale.
(c) The $\ell_0$ dependence of the crossover scale.
The solid line represents the analytic estimation for the crossover scale obtained in \eq{eq:ell1} with a multiplicative factor determined from
 a fit of the numerical crossover scales denoted as dots.
The uncertainty in the numerical data is due to the grid size of $\ell-\ell^*$, which is taken to be $0.05$.}	
\end{figure}

We confirm this estimation of the crossover scale by  numerically diagonalizing  
\eq{eq:betaallwaves_main_sol1}.
\fig{fig:crossover_l0=100} shows the numerical results for
the most negative eigenvalue of 
$\bar{\lambda}^{(-),\spmqty{d}}_{I, \spmqty{p & -p \\ k & -k}} (\ell)$
and the expectation value of 
$\frac{ e^{\ell_1} {\mathsf{D}}_{\Lambda e^{-\ell_1}} (p; k)}{ \sqrt{ V_{F,p} V_{F,k} } }$ 
evaluated for the corresponding eigenvector.
While the magnitude of the former increases more or less linearly in $\ell-\ell^*$,
the expectation value of the latter is largely constant, as expected.
This results in the crossover 
at a scale $\ell_1$.
At the crossover scale, the eigenvector 
shown in 
\fig{fig:crossover_eigvec}
is peaked at the hot spot 
but its support is extended to 
$\Lambda^*/(vc)$.
As is expected, the crossover scale increases 
with decreasing $v$ (or increasing $\ell_0$) 
as is shown in \fig{fig:crossover_l0}.

For $\ell > \ell_1$,
the magnitude of $\bar \lambda^V$ exceeds
the contribution of 
$\frac{2}{N_f} 
 Y_{PP}^{(\pm)} 
1^{\spmqty{s \\ d}}
\frac{ \eell {\mathsf{D}}_{\Lambda \emell} (q; k)}{ \sqrt{ V_{F,q} V_{F,k} } }$
in \eq{eq:betaallwaves_main} 
at least in the channel with the most negative eigenvalue.
The growth of the eigenvalue in that channel is then mostly driven by 
$\bar \lambda^V$ itself.
This marks the start of the second stage.
In the small $v$ limit, 
$E_0(\ell_1) \gg
 \int \frac{dk}{2 \pi \Lambda}
 \eta_k
|f_k(\ell_1)|^2$,
and we can further ignore the contribution of the anomalous dimension
in \eq{eq:betaallwaves_main}
to write the beta functional as
\begin{equation}
\frac{\partial }{\partial\ell}\bar{\lambda}^{(-),\spmqty{d}}_{II,  \spmqty{p & -p \\ k & -k}} (\ell) \approx
-\frac{1}{4\pi}
\int \frac{\dd q}{2\pi\Lambda}
  \bar{\lambda}^{(-),\spmqty{d}}_{II,  \spmqty{p & -p \\ q & -q}} 
  \bar{\lambda}^{(-),\spmqty{d}}_{II,  \spmqty{q & -q \\ k & -k}}
\label{eq:BCSdom_main}
\end{equation}
with the matching condition,
$\bar{\lambda}^{(-),\spmqty{d}}_{II,  \spmqty{p & -p \\ k & -k}} (\ell_1) 
=\bar{\lambda}^{(-),\spmqty{d}}_{I, \spmqty{p & -p \\ k & -k}} (\ell_1)
$.
%
The solution to \eq{eq:BCSdom_main} is given by 
$ \bar{\lambda}^{(-),\spmqty{ d}}_{II}(\ell)   =
 \frac{
\bar{\lambda}^{(-),\spmqty{ d}}_I(\ell_1)  
 }{ 
1 +
 \frac{1}{4\pi}(\ell - \ell_1)
 \bar{\lambda}^{(-),\spmqty{d}}_I(\ell_1)
 }$,
and the renormalized four-fermion coupling 
blows up around the scale,
$\ell_c \sim
\ell_1
+ \frac{4\pi }{ |E_0(\ell_1)| } = \ell^* + \frac{4\pi N_f}{\left(1+\frac{1}{N_c}\right)w\log\left(\frac{1}{v}\right)}$.
The superconducting transition temperature is given by
\begin{equation}
T_c \sim
\Lambda^*
e^{- \frac{4\pi N_f}{\left(1+\frac{1}{N_c}\right)w\log\left(\frac{1}{v}\right)}}.
\label{eq:ellc}
\end{equation}
In the second stage of the RG flow, 
the eigenvector with the most negative eigenvalue is more or less frozen,
and
the eigenvectors of
$\bar{\lambda}^{(-),\spmqty{d}}_{I, \spmqty{p & -p \\ k & -k}} (\ell_1)$ (\fig{fig:crossover_eigvec}) 
determines the channel that becomes superconducting.
It is noted that the pairing wavefunction is extended far beyond the hot spot region defined at scale $\ell_c$.
This shows the importance of large-angle scatterings beyond the hot spot region.
Therefore, the hot spot theory can not capture the superconducting instability properly.
Since  $\ell_c - \ell^* \sim w \log(1/v) \ll 1/v$ in the small $v$ limit, 
the nesting angle does not change much between the scale where the theory is in the bottleneck 
and the scale at which the superconductivity sets in.
This justifies our analysis in which the flow of the nesting angle is ignored.

For the superconductivity,
the physics of non-Fermi liquid and the Fermi liquid play
important roles in different length scales.  
When the theory is within the bottleneck region say at energy scale $\Lambda^*$,
the gapless spin fluctuations generate an attractive interaction for electrons 
within the range of momentum 
$|k|< \frac{\Lambda^*}{vc}$ 
from the hot spots.
The spin fluctuations also
 make those same electrons incoherent,
 causing a `pair-breaking' effect.
At lower energy scale, 
the momentum region 
where the spin fluctuations generate
attractive interaction  
becomes increasingly localized near the hot spots
as more and more electrons are decoupled from spin fluctuations.
At the same time, the pair breaking effect 
caused by the spin fluctuations subsides 
 except for the small region near the hot spots.
At low energies,
what remains away from the hot spots
is the heavy but largely coherent electrons that are 
subject to the attractive  four-fermion interaction
that has been accumulated 
at high energies.
The spin fluctuations continue to add 
more attractive interaction near the hot spots.
But, once the accumulated four-fermion coupling becomes comparable to 
the interaction mediated by the spin fluctuations,
the further growth of the four-fermion coupling 
is mainly driven by those more abundant cold electrons 
through the BCS scatterings\footnote{
This is in contrast to the cases in which
the Fermi surface is coupled with 
a critical bosonic mode
centered at zero momentum.
In those cases, 
the entire Fermi surface remains strongly renormalized
down to the zero energy, 
and the pairing must arise out of hot fermions.}.
The RG time that is needed to  reach the crossover scale ($\ell_1-\ell^*$) is comparable to the RG time that is further needed for the BCS process to finally drive the instability from the crossover ($\ell_c-\ell_1$).
Interestingly, the residual attractive interaction that is left for the coherent electrons at low energies is only dependent on the bare nesting angle
and so is the superconducting transition temperature.

\section{Hot spots as critical points in momentum space}
\label{sec:normal}

\fbox{\begin{minipage}{48em}
{\it
\begin{itemize}
\item
In the normal state, 
electrons exhibit non-Fermi liquid behaviours near the hot spots, 
while the Fermi surface far away from the hot spots
support well-defined quasiparticles.
\item
The hot spots act as critical points in the momentum space where the non-Fermi liquid behaviour persists down to the low energy limit.
In the space of energy and momentum along the Fermi surface, 
the quasiparticle decay rate exhibits a critical fan centered at the hot spots. 
\item
As the hot-spot momentum is approached on the Fermi surface,
the quasiparticle weight,
the deformed shape of Fermi surface
and the Fermi velocity decay following 
universal functions controlled by the bare nesting angle. 
 \end{itemize}
}
\end{minipage}}
\vspace{0.5cm}

A theory with $v \neq 0$
always flows to a superconducting state 
in the low energy limit.
However, the theories 
whose bare interactions 
are not strongly attractive 
compared to the interaction 
mediated by the spin fluctuations
necessarily stay in the bottleneck 
region for an RG scale that is order of 
$\Delta \ell \sim \frac{1}{w \log (1/v)}$
before superconducting instabilities arise.
This window of length scale  becomes large 
as the bare nesting angle decreases.
In this section, 
we examine universal scaling behaviours 
that the single electron spectral function exhibits in the normal state.

The spectral function 
$\mc A (\omega, \vec k)$ 
describes the probability that  
an electron with momentum $\vec k$ 
has energy $\omega$.
The spectral function can be obtained from 
the RG equation for the two-point function in \eq{eq:SolRGEq2Point}
under the assumption that
the two-point function takes 
the form of the free fermion 
with $V_{F,k}=1$ and $v_k=v_0$
at the UV cutoff energy scale $\Lambda$\footnote{
$\Lambda$ should be
small enough for the effective field theory description to be valid at energies below $\Lambda$.
Electrons at energy scale $\Lambda$ are already dressed by quantum fluctuations between the lattice scale and $\Lambda$.
Nonetheless, those high-energy quantum corrections only introduce 
smooth variations in electronic properties as a function of energy and momentum.
The high-energy renormalization does not affect the singular part that emerges in the low-energy limit.
Therefore, it is okay to use the free propagator at energy $\Lambda$ 
for the purpose of extracting 
the  universal singularity that arises at low energies.
}.
Since the biggest quantum correction to the electron spectral function arises from the one-loop quantum correction
which exhibits a crossover at  $\ellOneLoopX$, 
we divide the length scale into three regions :
the short-distance region ($\ell<\ellTwoLoopX$) that is far less than the crossover scale,
the intermediate region ($\ellTwoLoopX < \ell < \ellThree$)
that includes the crossover scale
and
the long-distance region ($\ellThree < \ell$) 
that is far bigger than the crossover scale,
where
$\ellTwoLoopX = \ellOneLoopX - \log \frac{2}{c(\ellOneLoopX)}$ 
and 
$\ellThree = \ellOneLoopX + \log \frac{2}{c(\ellOneLoopX)}$.
The intermediate region is chosen to be big enough to capture the  smooth crossover
($\ellTwoLoopX \ll \ellOneLoopX \ll \ellThree$)
yet it is small enough that we can simplify the calculation by ignoring the running of the coupling functions within the window\footnote{
For example,
$
c(\ellOneLoopX)
\approx 
c(\ellTwoLoopX)
\approx 
c(\ellThree)
$.}.
We set $\ell$ 
such that $k_0(\ell) = \Lambda$
in \eq{eq:ScalingTransfo}
and perform the analytic continuation
to obtain the spectral function near hot spot $1$ as 
(See Appendix \ref{chap:2PFELE} for   the details)
\begin{align}
\mc A(\omega,\vec k) &= 
-   \frac1\pi\frac{\mathrm{Im}(G_{\mathrm{R}}(\omega,\vec{k})^{-1})}{\mathrm{Re}(G_{\mathrm{R}}(\omega,\vec{k})^{-1})^2+\mathrm{Im}(G_{\mathrm{R}}(\omega,\vec{k})^{-1})^2},
   \label{eq:A} 
\end{align}
where
\begin{align}
    \mathrm{Re}( G^{-1}_{\mathrm{R}}(\omega,\vec{k}) )
    &= -\omega\mathrm{Re}(\mathcal{F}_z(\omega,k)) + e[\vec k,v_k(\lor)],
\label{D28}
    \\
    \mathrm{Im}( G^{-1}_{\mathrm{R}}(\omega,\vec{k}) )
    &= - \omega
    \mathrm{Im}(\mathcal{F}_z(\omega,k))
\label{D29}
\end{align}
are the real and imaginary parts of the inverse of the retarded Green's function.
$\mathcal{F}_z(\omega,k)$
is the leading contribution to the fermion self-energy given by
\begin{equation}
\begin{aligned}
&   \mathcal{F}_z(\omega,k) = \\
&
    \begin{cases}
    \mc E_1(\lor,0)\left(  1+\FR{i\pi \alpha_1(\lor)}{2}
    \right) 
    &
    \lor < \ellTwoLoopX,\\
    \mc E_1(\ellTwoLoopX,0)
    \PFR{  
        1+e^{\ellOneLoopX-\ellTwoLoopX}
    }{   
        \sqrt{ 1 +e^{2\ellOneLoopX-2\lor} }
    }^{\alpha_1(\ellTwoLoopX)}
    \Bigl(
        1 + i \alpha_1(\ellTwoLoopX) \arctan(e^{\ellOneLoopX-\lor})
    \Bigr)
    &
    \ellTwoLoopX < \lor < \ellThree,\\
    \begin{aligned}
    &\mc E_1(\ellTwoLoopX,0) 
     \PFR{1+e^{\ellOneLoopX-\ellTwoLoopX}}
    {1+e^{\ellOneLoopX-\ellThree}}^{\alpha_1(\ellTwoLoopX)}
    e^{\alpha_3(\ellTwoLoopX)}
\left[ 
        1 +
        i \alpha_3(\ellTwoLoopX)
        e^{-\lor+\ellThree}
        \mc{E}_0(\lor,\ellThree)^2
    \right]
    \end{aligned}
    &
    \ellThree <  \lor 
    \\
    \end{cases}
\end{aligned}
\end{equation}
where
\begin{align}
\label{eq:ellomega}
\lor
= 
 \log\FR{\Lambda}{\omega} - \FR{ \sqrt{N_c^2-1}\sqrt{ \ell_0 + \log\FR{\Lambda}{\omega} } }{ \log\sqrt{\ell_0 + \log\FR{\Lambda}{\omega}} } + \FR{ \sqrt{N_c^2-1}\sqrt{ \ell_0 } }{ \log\sqrt{\ell_0} }
\end{align}
represents the logarithmic length scale associated with energy $\omega$,
$\alpha_1(\ell)$ is defined in 
\eq{eq:alpha01}
and 
$\alpha_3(\ell)  =
\FR{\pi}{4}\FR{1}{(\ell+\ell_0)\log(\ell+\ell_0)}$.

 \begin{figure*}
	\centering 
	\includegraphics[width=0.49\linewidth]{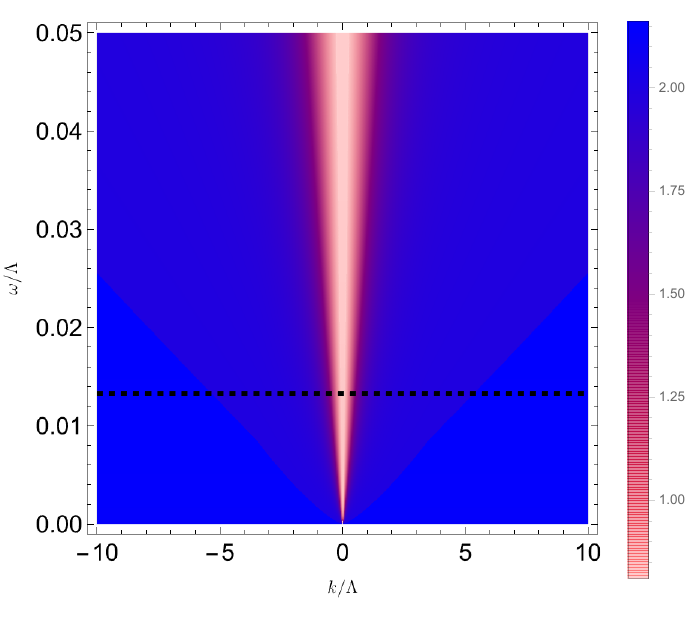}
	\includegraphics[width=0.49\linewidth]{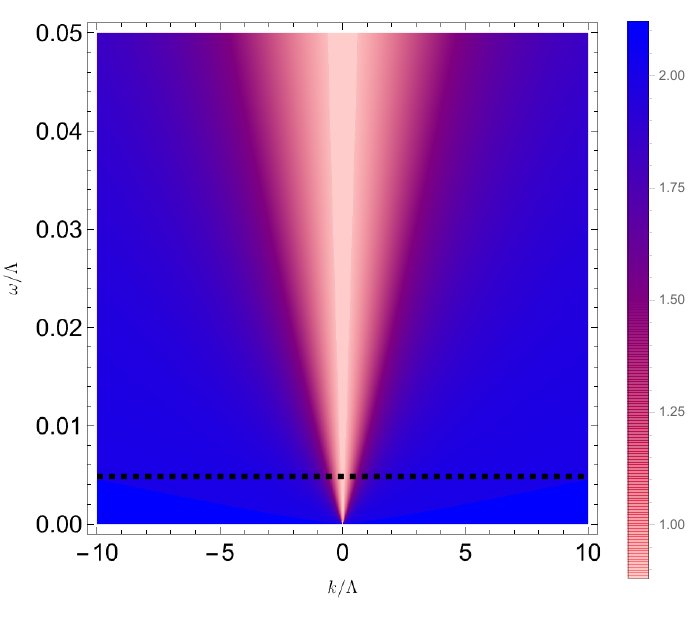}
	\includegraphics[width=0.49\linewidth]{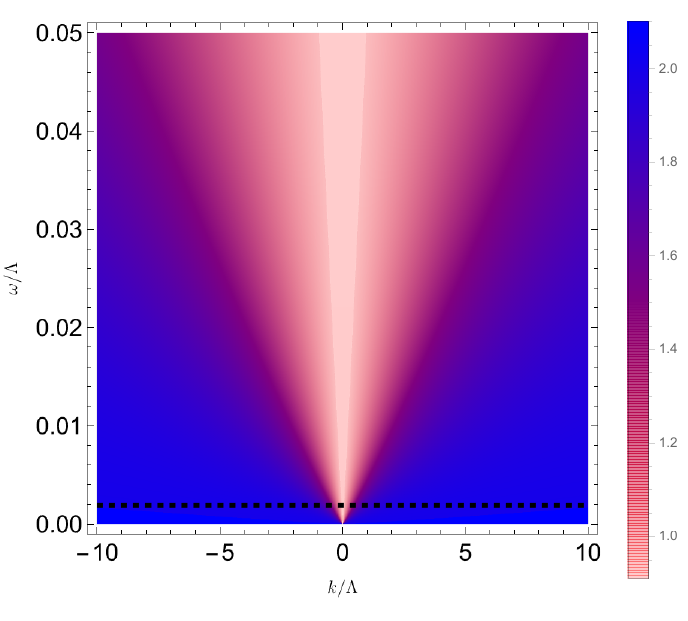}
	\includegraphics[width=0.49\linewidth]{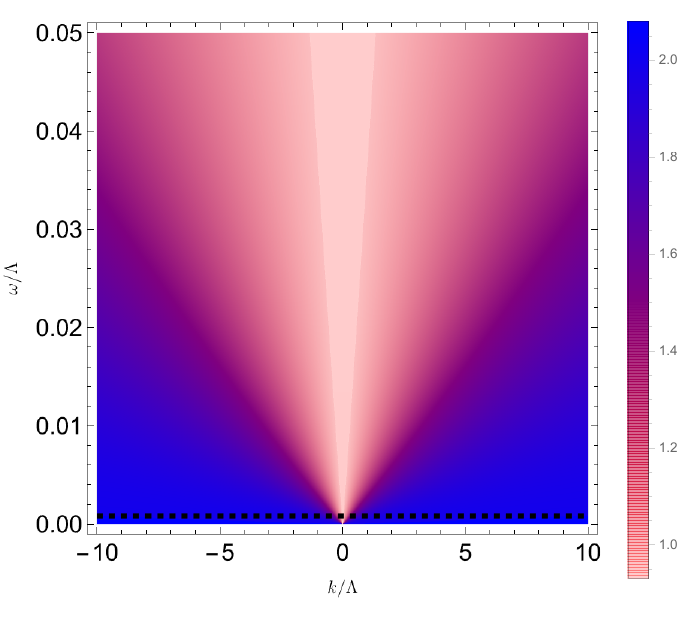}
	\caption{
$\FR{ \di\log\ImG}{\di\log\omega}$ plotted
as a function of momentum along the Fermi surface and  energy for different values of $\ell_0$ :
$\ell_0=10,20,30,40$
from top left to bottom right
with the corresponding values of the nesting angle 
$v(0,\ell_0) \approx 0.14,  0.05, 0.03, 0.02$, respectively.
The horizontal dashed lines represent   the superconducting transition temperatures estimated from \eq{eq:ellc}.
As $\ell_0$ increases ($v$ decreases),
the superconducting transition temperature is suppressed, opening up a larger range of energy scales for the normal state that exhibits non-Fermi liquid scaling near the hot spots.
In the last panel, 
the superconducting transition temperature is too small to be visible in the scale.
The crossover scale $\omega_0$ is not shown as it is far below the superconducting transition temperature for all choices of $\ell_0$.
} 
\label{fig:dlogim}
\end{figure*}

At general $\vec k$, the spectral function does not support a sharp peak, 
but we can still define the `dispersion relation', 
$\omega_{\vec k}$ through
$-\omega_{\vec k} \mathrm{Re}(\mathcal{F}_z(\omega_{\vec k},k)) + e[\vec k,v_k(\ell^{Re}_{\omega_{\vec k}})]=0$.
This approximately corresponds to the energy at which the spectral function is peaked.
Around the peak, 
$\mathrm{Im}(G_{\mathrm{R}}(\omega_{\vec k},\vec{k})^{-1})$ 
controls the width of the spectral function. 
To characterize 
how the imaginary part of the self-energy 
scales with the frequency,
we consider the local exponent with which the imaginary part of the self-energy 
scales with frequency,
\begin{align}
\FR{ \di\log\ImG}{\di\log\omega}
= 
\begin{cases}
1 -  \alpha_1(\lor)
&\lor < \ellTwoLoopX
\\
1
+
\FR{e^{\ellOneLoopX-\lor}}
{1+e^{2\ellOneLoopX-2\lor}}
\left(
\FR{1}{\arctan(e^{\ellOneLoopX-\lor})}
-\alpha_1(\ellTwoLoopX) e^{\ellOneLoopX-\lor}
\right)
&\ellTwoLoopX <  \lor <  \ellThree
\\
2+2\alpha_0(\lor)
&\ellThree < \lor
\end{cases},
\label{d22}
\end{align}
where 
$\alpha_0(\ell)$ and
$\alpha_1(\ell)$ 
are defined in 
\eq{eq:alpha01}.
\eq{d22} is a function of 
energy and momentum along the Fermi surface
\footnote{
There is one-to-one correspondence between $(k, \omega_{\vec k})$ and $(k_x, k_y)$.
}.
In \fig{fig:dlogim},  the exponent is plotted as a function of the momentum along the Fermi surface and energy for different choices of $\ell_0$ ($v$).
At the hot spot ($k=0$), 
the exponent is less than $1$
at all energy scales above zero,
exhibiting the non-Fermi liquid behaviour.
Since the imaginary part of the self-energy is bigger than the peak energy, 
there is no well-defined quasiparticle.
At energy scales far above 
$\omega_0 = \Lambda e^{- (1+ \alpha_1(0)) \ell_0}$\footnote{$\omega_0=\Lambda e^{- z(0) \ell_0}$ to the leading order in (small) $v_0(0)$. See Eqs. \eqref{eq:zhot} and \eqref{eq:SpectralHotspots}.},
the flow of the nesting angle can be ignored, 
and the spectral function scales in frequency with exponent $\FR{ \di\log\ImG}{\di\log\omega}
  = 1 - \alpha_1(0)$ 
to the leading order in the small $v$ limit.
Below energy $\omega_0$,
the logarithmic flow of the nesting angle makes the transient exponent runs toward $1$.
In the low-energy limit,
the spectral function scales linearly in $\omega$ with super-logarithmic correction 
as is shown in 
\eq{eq:SpectralHotspots}.
However, this ultra-low energy scaling is not accessible because the superconducting instability kicks in  at the energy scale higher than $\omega_0$.
For $k \neq 0$, 
electrons decouple from spin fluctuations 
and the exponent approaches 
$2$ in the low-energy limit,
exhibiting the Fermi liquid behaviour\footnote{
At sufficiently low energies, 
the $\omega^2 \log \omega$ contribution
from the short-range four-fermion coupling should be also included.
}.
The high-energy non-Fermi liquid behaviour 
and  the low-energy Fermi liquid behaviour 
is divided by a crossover region
that interpolates the exponent smoothly.
The crossover from the non-Fermi liquid to the Fermi liquid behaviours
arises around the momentum-dependent energy scale 
$\ell_{\omega}= \ellOneLoopX$.
In $\omega > \omega_0$,
the crossover occurs at
$\omega \sim k^{1+\alpha_1(0)}$.
The crossover takes the from of a critical fan in the space of momentum and energy, 
where the momentum along the Fermi surface plays the role of a parameter 
that is tuned to reach the `critical point', that is, the hot spot momentum.


\begin{figure*}
	\centering 
	\begin{subfigure}[b]{0.49\linewidth}
		\centering
	\includegraphics[width=0.9\linewidth]{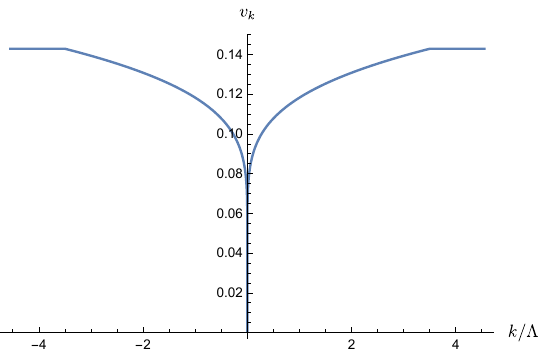}~~
		\caption{\label{fig:vk_linear}}
	\end{subfigure}
		\begin{subfigure}[b]{0.49\linewidth}
		\centering
	\includegraphics[width=\linewidth]{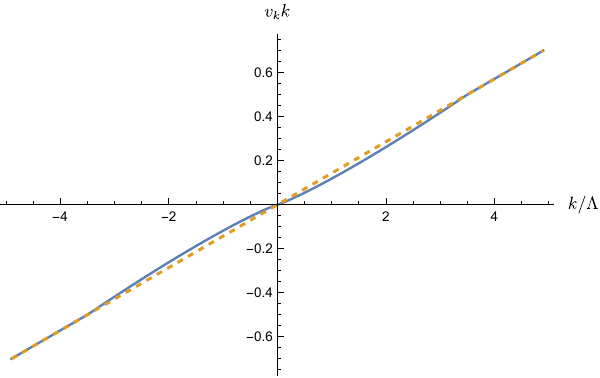}
		\caption{\label{fig:vk_log}}
	\end{subfigure}
	\caption{
(a) The nesting angle for the fully renormalized Fermi surface at $\ell=\infty$ plotted as a function of momentum along the Fermi surface.
(b) The fully renormalized Fermi surface (solid line) that arises from a straight bare
Fermi surface (dashed line) with $\ell_0=10$.
It is noted that the renormalized nesting angle vanishes only at the hot spots and the slope of the Fermi surface remains non-zero away from the hot spots.
} 
\label{figRENORMFERMISURFACE}
\end{figure*}

\begin{figure*}
	\centering 
	\includegraphics[scale=0.7]{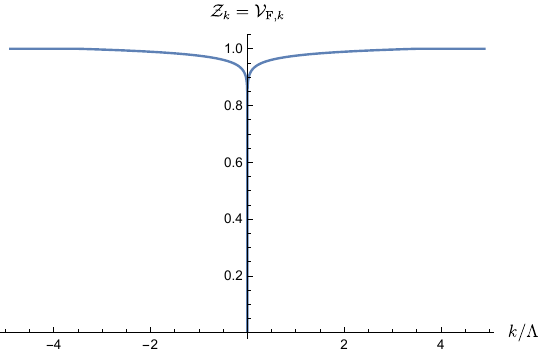}
	\caption{
The quasiparticle weight (and the renormalized Fermi velocity) plotted as a function of momentum along the Fermi surface for $\ell_0=10$.
}
\label{figRENORMVF}
\end{figure*}

In the Fermi liquid region
with $\lor \gg \ellThree$,
the spectral function exhibits a well-defined quasiparticle peak
and it has the Lorentzian form,
\begin{align}
\mc A \left(
\omega,\vec k
\right) 
&= \FR{  
\mfZ_k
}{\pi}
\FR{  \gamma({\vec k})}{
( \omega- \omega_{\vec k} )^2 
+ \gamma({\vec k})^2}.
\label{eq:defnSpecFcn}
\end{align}
Here, 
$\omega_{\vec k} =\mfVFk e_N[\vec k; {\bf v}_k]$ is the quasiparticle dispersion;
${\bf v}_k =  v_k(\ellTwoLoopX)$ is the renormalized nesting angle
and
$\mfVFk  =  \mc E_1( \ellTwoLoopX)^{-1}$ is
the Fermi velocity.
\begin{align}
\mfZ_k  &=   \mc E_1( \ellTwoLoopX
,0)^{-1},\\
\label{mcT}
\gamma(\vec{k}) &=
\omega_{\vec{k}} \alpha_3(\ellTwoLoopX)
e^{-
\ell^{\mathrm{Re}}_{\omega_{\vec{k}}}
+\ellThree} \mc{E}_0(\ell^{\mathrm{Re}}_{\omega_{\vec{k}}},\ellThree)^2
\end{align}
correspond to the quasiparticle weight (which is identical to the renormalized Fermi velocity to the leading order in $v$)
and the quasiparticle decay rate, respectively.
The fully renormalized Fermi surface is given by
$k_y = - v_{k_x}(\infty) k_x$ near hot spot $1$.
The renormalized nesting angle
and the deformed shape of the Fermi surface
for a UV theory 
whose bare Fermi surface is straight 
with a constant nesting angle 
is shown in  \fig{figRENORMFERMISURFACE}.
The quasiparticle weight
and the renormalized Fermi velocity
are plotted as functions of momentum along the Fermi surface in  \fig{figRENORMVF}\footnote{
It is noted that a reduction of the quasiparticle weight and the Fermi velocity near the hot spots has been already observed in earlier numerical simulations\cite{GERLACH}.
However, one may have to use larger system sizes and small bare nesting angles to probe the low-energy dynamics dictated by the $z\approx 1$ scaling\cite{2022arXiv220414241L}. 
}.
Far away from the hot spots ($k > k_c$),
$\mfZ_k = 1$
and
$v_k(\infty) =v_0$
because electrons in the cold region 
are not renormalized.
Electrons in $0 < k < k_c$ 
are subject to 
renormalization caused by spin fluctuations above crossover energy scales below which they
decouple from spin fluctuations.
Since electrons closer to the hot spots remain coupled with spin fluctuations over longer length scales,
the quasiparticle weight, the Fermi velocity  and the nesting angle all decreases as momentum approaches the hot spots.

\section{Conclusion}
\label{sec:conclusion}

In this paper, a field-theoretic functional renormalization group formalism is developed for  full low-energy effective field theories 
 of non-Fermi liquids that include all gapless modes  around the Fermi surface.
The formalism, which is beyond the traditional patch description,
captures the universal low-energy physics of the entire Fermi surface in the minimal way through renormalizable field theories.
Due to the Fermi momentum that does not generally decouple from the low-energy physics,
the usual notions of 
renormalizable field theories and 
scale invariance 
need to be generalized.
We use this functional renormalization group formalism to 
understand the non-Fermi liquid state 
realized at the 2+1 dimensional antiferromagnetic quantum critical point
and the pathway from the non-Fermi liquid to superconductivity.

The low-energy effective field theory of the antiferromagnetic quantum critical metal 
is characterized by couplings 
that are functions of momentum along the Fermi surface.
The full functional renormalization group flow,  
which is controlled in the limit that the bare nesting angle is small for any $N_f$ and $N_c$, 
allows us to identify the non-Fermi liquid fixed point in the space of coupling functions and extract universal low-energy physics controlled by the fixed point.
At low energies,
the renormalized coupling functions acquire universal profiles as functions of momentum along the Fermi surface.
Those coupling functions, in turn, 
fix physical observables that depend on momentum along the Fermi surface such as 
the shape of the renormalized Fermi surface,
the Fermi velocity,
the quasiparticle weight,
the single-electron decay rate
and
the pairing channel 
for the superconducting instability.
When the bare four-fermion coupling is weak, the superconducting instability is controlled by the universal attractive interaction generated by the gapless spin fluctuations.
In the limit that the bare nesting angle is small, there is a large window of energy controlled by the non-Fermi liquid fixed point above the superconducting transition temperature. 
Below the superconducting transition temperature,
which is exponentially suppressed in the limit that the nesting angle  is small,
the non-Fermi liquid state becomes unstable against the spin-singlet d-wave pairing instability.
The pairing wavefunction includes significant weight of electrons away from the hot spot region defined at the scale of the superconducting transition temperature.
Therefore, the hot spot theory can not capture the superconducting instability that involves large-angle scatterings.
We close with some discussions.

\begin{itemize}

\item
Beyond the small nesting angle limit:
For the antiferromagnetic quantum critical metal, we use the nesting angle as a small parameter to access the low-energy physics of the theory in a controlled way.
We can not exclude the possibility of new strongly interacting fixed points in the region where the nesting angle is not small.
However, a recent quantum Monte Carlo study suggests that
the one parameter family of the quasi-fixed points that is connected to the true fixed point at zero nesting angle may still govern the low-energy physics of systems with general nesting angle\cite{2022arXiv220414241L}.
It will be of great interest to understand the fate of the theories with general nesting angles in more detail.

\item Hot Fermi surfaces :
In this work, the functional renormalization group formalism has been applied to the theory 
in which only the hot spots on the Fermi surface remain strongly coupled in the low-energy limit.
In this theory, a superconducting instability is unavoidable at low energies as the majority of the Fermi surface is left with  largely coherent electrons at low energies, which become unstable in the presence of the attractive interaction that has been generated at high energies.
In hot Fermi surfaces, on the other hand,
the critical properties vary more smoothly along the Fermi surface
and electrons can stay incoherent across the  Fermi surface.
It will be of great interest to examine a possibility in which the pair breaking effect due to incoherence prevents  superconductivity even at zero temperature
based on the functional renormalization group formalism which can be applied to theories for hot Fermi surfaces.

\item
Resilience of low-energy effective field theories :
Theories with continuously many gapless modes are prone to `UV/IR mixing' 
as the scale associated with the manifold of the gapless modes in the momentum space 
is not generally decoupled from the low-energy physics\cite{IPSITA,PhysRevB.104.235116,PhysRevLett.128.106402}.
This does not imply a breakdown of low-energy effective theory.
However, the content and the scope of low-energy effective theories need to be adjusted in order for the theories to be predictable.
First, low-energy effective field theories  in general have to include data associated with large momentum scatterings 
if such scatterings occur 
within the manifold of low-energy modes.
All coupling functions
that give rise to infrared singularities
should be parts of low-energy theories 
even if they are formally `irrelevant' 
by power-counting 
and include `microscopic' information.
The fact that one has to include such data means that the low-energy physics retains more information than conventional theories with finitely many gapless modes.
In this context, what UV/IR mixing really means is 
an {\it enrichment of IR physics}.
Second, some observables should be dropped 
from the scope of the low-energy effective theory 
if predicting them requires including physics of high-energy modes.
For example, 
for the antiferromagnetic quantum critical metal,
the 1PI fermion four-point functions 
can not be predicted without including high-energy physics 
as is discussed in Sec \ref{sec:fixedpoint}.
Instead, only the net two-body interaction can be predicted within the low-energy effective field theory.
However, the set of low-energy observables that can be predicted within a low-energy effective field theory in general depend on  the theory.

\end{itemize}



%

\newpage
\appendix
%
%
%


\section{Quantum corrections for the two and three-point functions}
\label{app:QC}

In this appendix, 
we compute the quantum corrections
for the fermion two-point function
and the Yukawa vertex.
In this computation, 
we assume that all coupling functions satisfy the adiabaticity condition 
in \eq{eq:adiabatic_condition}.
In this case, the singular parts of the quantum corrections for the two and three-point functions can be computed by replacing the coupling functions with the values at which the integrand is peaked.
Later we will verify that the coupling functions that arise from the beta functionals obtained under the assumption of adiabaticity at a UV energy scale satisfy the adiabaticity at all energy scales.

\subsection{Fermion self-energy}
\label{Sec:OneLoopFSE}

%
%
%

\subsubsection{One-loop}
\label{eq:QC_FSE_1L}

The real and imaginary part of the one-loop self-energy 
in \eq{eq:1LoopFSE_full} reads
\begin{align}
\hspace{-0.5cm}\mathrm{Im}\left[\Sigma^{\mathrm{1L}}_{N}({\bf k})\right] &= \frac{2g^{(N)}_k g^{(\bar N)}_k (N^2_c-1)}{N_cN_f} \int\dd \bm{q}\left[\frac{(k_0+q_0)}{(k_0+q_0)^2+(V_{F,k}^{(\bar N)})^2e_{\bar{N}}[\vec{k}+\vec{q};v_k^{(\bar N)}]^2}\frac{1}{|q_0|+|cq_x|+|cq_y|}\right],\label{eq:ImPart}\\
\hspace{-0.5cm}\mathrm{Re}\left[\Sigma^{\mathrm{1L}}_{N}({\bf k})\right] &=-\frac{2 g^{(N)}_k g^{(\bar N)}_k (N^2_c-1)}{N_cN_f} \int\dd \bm{q}\left[\frac{V_{F,k}^{(\bar N)} e_{\bar{N}}[\vec{k}+\vec{q};v_k^{(\bar N)} ]}{(k_0+q_0)^2+(V_{F,k}^{(\bar N)})^2 e_{\bar{N}}[\vec{k}+\vec{q};v_k^{(\bar N)}]^2}\frac{1}{|q_0|+|cq_x|+|cq_y|}\right].
\label{eq:RePart}
\end{align}
Without loss of generality,
we can consider the quantum correction for $N=1$. 
The self-energy at other hot spots can be obtained from this through a $C_4$-transformation.



\vspace{0.5cm}
\centerline{ $\mathrm{Im}\left[\Sigma^{\mathrm{1L}}_{N}(\textbf{k})\right]$ }
\vspace{0.5cm}

We first compute the imaginary part,
\begin{align}\label{eq:ImDelta_Ashu}
\hspace{-1.7cm}\mathrm{Im}\left[\Sigma^{\mathrm{1L}}_{1}({\bf k})\right] 
&= 
\frac{2g_k^2(N^2_c-1)}{N_cN_f} 
\int\dd {\bf q}\left[\frac{(k_0+q_0)}{(k_0+q_0)^2+V_{F,k}^2(v_kq_x-q_y+e_4[\vec k,v_k])^2}\frac{1}{|q_0|+|cq_x|+|cq_y|}\right],
\end{align}
where \eq{eq:vVgN_vVg} is used 
to express coupling functions 
in terms of the generic coupling functions.
We shift the internal momentum as $q_y \to q_y+v_kq_x+
e_4[\vec k,v_k]$ so that the internal fermion has zero energy at $q_y=0$. 
The integration over $q_y$ is convergent even if we drop the $cq_y$ term in the boson propagator in the small $v$ limit. 
We further simply the expression by replacing $|cq_x| + |cv_k q_x + c e_4[\vec k,v_k]|$ 
with $c|q_x| + |c e_4[\vec k,v_k]|$ in the small $v$ limit, 
\begin{align}\label{eq:ImNew1}
\hspace{-1.7cm}\mathrm{Im}\left[\Sigma^{\mathrm{1L}}_{1}({\bf k})\right] 
&= 
\frac{2g_k^2(N^2_c-1)}{N_cN_f} 
\int\dd{\bf q}\left[\frac{(k_0+q_0)}{(k_0+q_0)^2+V_{F,k}^2q_y^2}\frac{1}{|q_0|+|cq_x|+
| \bar{\Delta}(\vec{k};v_k) |
}\right],
\end{align}
where 
$\bar{\Delta}(\vec{k};v_k)
= c e_4[\vec k, v_k]$.
The integration over $q_y$ leads to 
\begin{align}\label{eq:ImNew2}
\hspace{-1.7cm}\mathrm{Im}\left[\Sigma^{\mathrm{1L}}_{1}({\bf k})\right] 
&= 
\frac{g_k^2(N^2_c-1)}{V_{F,k}N_cN_f}
\int_{\mathbb{R}}\FR{\dd q_0}{2\pi}
\int_{-\Lambda/c}^{\Lambda/c}\FR{\dd q_x}{2\pi}
\left[\frac{\sgn(k_0+q_0)}{|q_0|+|cq_x|+
| \bar{\Delta}(\vec{k};v_k) |
}\right].
\end{align}
Here $\Lambda$ is a UV energy cutoff,
which is translated to the momentum cutoff for the boson, $\Lambda/c$.
The subsequent integrations over $q_0$ and $q_x$  yield
\begin{align}\label{eq:ImNew4}
\hspace{-1.7cm}\mathrm{Im}\left[\Sigma^{\mathrm{1L}}_{1}({\bf k})\right] 
&= 
\frac{g_k^2(N^2_c-1)}{V_{F,k}N_cN_f}
\FR{1}{\pi^2 c}
k_0
\log\left(\FR{\Lambda}{
| \bar{\Delta}(\vec{k};v_k) |
+|k_0|}\right)
+ reg.,
\end{align}
where $reg.$ represents terms that are regular in the large $\Lambda$ limit.
To remove the singular part of 
$\left.
\frac{ \partial \mathrm{Im}\left[\Sigma^{\mathrm{1L}}_{1}({\bf k})\right] }{\partial k_0} \right|_{{\bf k}=(\mu,k_x, -v_k k_x)}$
in the small $\mu$ limit,
the counter-term is chosen to be
\begin{align}
A^{(1);\mathrm{1L}}(k)
        =
-\frac{N_c^2-1}{\pi^2 N_cN_f} \FR{g_k^2}{cV_{F,k}}
\log\left( \FR{\Lambda}{\mu+2v_kc|k|} \right),
\label{A1Lashu}
\end{align}
where we use
$ \bar{\Delta}(\vec{k};v_k) 
= 2 v_k c k$ when the external fermion is on the Fermi surface.

\vspace{0.5cm}
\centerline{ $\mathrm{Re}\left[\Sigma^{\mathrm{1L}}_{N}(\textbf{k})\right]$ }
\vspace{0.5cm}

\newcommand{\mfc}{\mathfrak c_k}

We rewrite 
the real part of the self-energy in Eq.~(\ref{eq:RePart}) 
by shifting $q_y$ to $q_y + e_4[\vec{k},v_k] + v_k q_x$,
\begin{align}
\begin{split}
\mathrm{Re}\left[\Sigma^{\mathrm{1L}}_{1}({\bf k})\right] &= 
\frac{2g_k^2(N^2_c-1)}{N_cN_f}
\int\dd {\bf q}
\left[\frac{ V_{F,k}q_y}{(k_0+q_0)^2+V_{F,k}^2q_y^2}\right]
\left[\frac{1}{|q_0|+|cq_x|+|cq_y+\bar{\Delta}(\vec{k};v_k) + c v_k q_x | }\right],
\end{split}
\end{align}
where 
$\bar{\Delta}(\vec{k};v_k)
= c e_4[\vec k, v_k]$.
As in the calculation of the imaginary part, we can neglect $cv_k q_x$ in the boson propagator since the leading $q_x$ dependence comes from $|cq_x|$.
However, $cq_y$ can not be ignored as the integration vanishes without it.
Therefore, we consider
\begin{align}
\label{eq:Re2}
\hspace{-1cm}\mathrm{Re}\left[\Sigma^{\mathrm{1L}}_{1}({\bf k})\right] &=
\frac{2g_k^2(N^2_c-1)}{N_cN_f}
\int\dd {\bf q}\left[\frac{V_{F,k}q_y}{(k_0+q_0)^2+V_{F,k}^2q^2_y}\right]\left[\frac{1}{|q_0|+|cq_x|+|cq_y + \bar{\Delta}(\vec{k};v_k)|}\right].
\end{align}
\noindent 
For the imaginary part of the self-energy, 
the full self-energy has been  obtained at general frequency and momentum.
For the real part, we focus on the asymptotic limits: 
($i$) $|k_0|\gg |\bar{\Delta}(\vec{k};v_k)|$ and ($ii$) $|k_0|\ll |\bar{\Delta}(\vec{k};v_k)|$. 
%
%

For $|k_0|\gg |\bar{\Delta}(\vec{k};v_k)|$,  
we rescale the internal momentum as
$q_0\to k_0q_0$ and 
$\vec q\to k_0\vec q/c$ to rewrite Eq. (\ref{eq:Re2}) as
\begin{align}
\hspace{-1.5cm}\mathrm{Re}\left[\Sigma^{\mathrm{1L}}_{1}({\bf k})\right] &=
\frac{2g_k^2(N^2_c-1)k_0}{cV_{F,k}N_cN_f}
\int\dd {\bf q}
\left[
\frac{q_y}{\mfc^2 (1+q_0)^2+q^2_y}\right]
\left[\frac{1}{|q_0|+|q_x|+|q_y+\textswab{h}(\textbf{k};v_k)|}\right],
\end{align}
where $\textswab{h}(\textbf{k};v_k)=\bar{\Delta}(\vec{k};v_k)/k_0$
and
$\mfc \equiv c/V_{F,k}$.
%
One has to be careful in taking the small $c$ 
and small
$\textswab{h}(\textbf{k};v_k)$ 
limits in this expression.
On the one hand, 
the integration over $q_y$
vanishes if
$\textswab{h}(\textbf{k};v_k)=0$. 
On the other hand, setting $\mfc=0$ inside the integrand makes the
integration logarithmically divergent. 
These imply that the quantum correction is proportional  to
$\textswab{h}(\textbf{k},v_k)$ and diverges logarithmically in $\mfc\ll 1$. 
After the integration over $q_y$ is done for general $\mfc$ and $\textswab{h}(\textbf{k};v_k)$, the
leading order contribution in $\textswab{h}(\textbf{k};v_k)$
is given by
\begin{align}
&\mathrm{Re}\left[\Sigma^{\mathrm{1L}}_{1}({\bf k})\right] = 
\frac{2g_k^2(N^2_c-1)\bar{\Delta}(\vec{k};v_k)}{\pi cV_{F,k}N_cN_f}
\int\limits_{\mathbb{R}}\frac{\dd q_0}{(2\pi)}\int\limits_{\mathbb{R}}\frac{\dd q_x}{(2\pi)}
\left(
\frac{\mfc^2(1+q_0)^2+(|q_0|+|q_x|)^2-\pi \mfc|q_0+1|(|q_x|+|q_0|)}{[\mfc^2(1+q_0)^2+(|q_0|+|q_x|)^2]^2}
\right.
\nonumber
\\
&\hspace{7cm}
\left.
+
\frac{1}{2}\frac{\mfc^2(1+q_0)^2-(|q_0|+|q_x|)^2}{[\mfc^2(1+q_0)^2+(|q_0|+|q_x|)^2]^2}\log\left(\frac{(|q_0|+|q_x|)^2}{\mfc^2(1+q_0)^2}\right)
\right).
\label{eqnMFC}
\end{align}
The integration over $q_x$ can be done exactly and is given by
\begin{align}
&\mathrm{Re}\left[\Sigma^{\mathrm{1L}}_{1}({\bf k})\right] = 
\frac{2g_k^2(N^2_c-1)\bar{\Delta}(\vec{k};v_k)}{\pi cV_{F,k}N_cN_f}
\int\limits_{-\Lambda/|k_0|}^{\Lambda/|k_0|}\frac{\dd q_0}{(2\pi)}
\FR{|q_0|\log(\mfc^2(1+q_0)^2/q_0^2)-\pi\mfc|1+q_0|}{2\pi(q_0^2+\mfc^2(1+q_0)^2)},
\label{eqAfterQX}
\end{align}
where the cutoff for the rescaled frequency becomes $\Lambda/|k_0|$.
The frequency integration gives
\begin{align}
&\mathrm{Re}\left[\Sigma^{\mathrm{1L}}_{1}({\bf k})\right] = -
\frac{2g_k^2(N^2_c-1)\bar{\Delta}(\vec{k};v_k)}{\pi^3 cV_{F,k}N_cN_f}
\log\FR{V_{F,k}}{c}\log\FR{\Lambda}{|k_0|}
+ reg.,
\label{eq:RealLim1}
\end{align}
where $reg.$ represents terms that are regular in the large $\Lambda/k_0$ limit.
For $|\bar{\Delta}(\vec{k};v_k)|\gg |k_0|$,
$\bar{\Delta}(\vec{k};v_k)$ cuts off the IR singularity, and the self-energy becomes
\begin{align}\label{eq:RealLim2}
\mathrm{Re}\left[\Sigma^{\mathrm{1L}}_{1}({\bf k})\right] = -
\frac{2g_k^2(N_c^2-1)}{\pi^3cV_{F,k}N_c N_f}
\log\left(\frac{V_{F,k}}{c}\right)\bar{\Delta}(\vec{k};v_k)
\log\left(\frac{\Lambda}{|\bar{\Delta}(\vec{k};v_k)|}\right) + reg.,
\end{align}
where $reg.$ represents terms that are regular in the large $\Lambda/\bar{\Delta}$ limit.
Using Eqs. (\ref{eq:RealLim1}) and (\ref{eq:RealLim2}), 
we can write the real
part of the one-loop fermion self-energy as
\begin{align}
\label{eq:QC3A}
\mathrm{Re}\left[\Sigma^{\mathrm{1L}}_{1}({\bf k})\right] = -
\frac{2g_k^2(N_c^2-1)}{\pi^3V_{F,k}N_cN_f}
\log\left(\frac{V_{F,k}}{c}\right)e_{4}[\vec{k};v_k]
\log\left(\frac{\Lambda}{\mathscr{H}_1(k_0,ce_{4}[\vec{k};v_k])}\right) + reg..
\end{align}
Here, 
$\mathscr{H}_1(x,y)$ is a crossover function 
that satisfies 
$\mathscr{H}_1(x,y)\sim \max(|x|,|y|)$ 
if $|x|\gg |y|$ or $|y|\gg |x|$.
Counter terms that remove IR divergent parts of 
$\left.  \frac{\partial \mathrm{Re}\left[\Sigma^{\mathrm{1L}}_{1}({\bf k}) \right]}{\partial k_x} \right|_{{\bf k}=(\mu, k_x, -v_k k_x)}$
and
$\left.  \frac{\partial \mathrm{Re}\left[\Sigma^{\mathrm{1L}}_{1}({\bf k}) \right]}{\partial k_y} \right|_{{\bf k}=(\mu, k_x, -v_k k_x)}$
in the small $\mu$ limit are given by
\begin{align}
A^{(2);\mathrm{1L}}(k) &=   \frac{2g_k^2(N_c^2-1)}{\pi^3 
V_{F,k}^2 
N_cN_f} \log\left(\frac{
V_{F,k}
}{c}\right)\log\left(\frac{\Lambda}{\mathscr{H}_1(\mu,2 v_k c k)}\right) 
+  A^{(2);\mathrm{1L}}_{reg}(k)
,\label{eq:A21L}\\ 
A^{(3);\mathrm{1L}}(k)  &= - \frac{2g_k^2(N_c^2-1)}{\pi^3 
V_{F,k}^2
N_cN_f} \log\left(\frac{V_{F,k}}{c}\right)\log\left(
\frac{\Lambda}{\mathscr{H}_1(\mu,2 v_k c k)}
\right)
+ A^{(3);\mathrm{1L}}_{reg}(k),
\label{eq:A31L}
\end{align}
where
$A^{(2);\mathrm{1L}}_{reg}(k)$
and
$A^{(3);\mathrm{1L}}_{reg}(k)$
represent terms that are regular
in the large
$\frac{\Lambda}{\mathscr{H}_2(\mu,2 v_k c k)}$
limit.

\subsubsection{Two-loop}
\label{eq:QC_FSE_2L}

The two-loop fermion self-energy reads
\begin{align}\label{eq:2LoopFSE}
\Sigma^{\mathrm{2L}}_{N}({\bf k}) &= 
\frac{4
(g_k^{(N)})^2
(g_k^{(\bar N)})^2
(N_c^2-1)}{N_c^2N_f^2}
\int\dd {\bf q}\int \dd {\bf p} D({\bf p})D({\bf q}) 
G_{\bar{N}}({\bf k}+{\bf p})
G_{N}({\bf k}+{\bf p}+{\bf q})
G_{\bar{N}}({\bf k}+{\bf q}).
\end{align}
With the shifts
$q_y\rightarrow
q_y+v_kq_x+e_{4}[\vec{k};v_k]$ and $p_y\rightarrow
p_y+v_kp_x+e_{4}[\vec{k};v_k]$,
Eq. (\ref{eq:2LoopFSE}) for $N=1$ is written as
\begin{align}
\begin{split}
\Sigma^{\mathrm{2L}}_{1}({\bf k})&= 
\frac{4g_k^4(N_c^2-1)}{N_c^2N_f^2}
\int\dd {\bf q}\int \dd {\bf p}\left\{\frac{1}{|q_0|+c|q_x|+|cq_y+v_kcq_x+\bar{\Delta}(\vec{k};v_k)|}\right.\\
&\times\frac{1}{|p_0|+c|p_x|+|cp_y+v_kcp_x+\bar{\Delta}(\vec{k};v_k)|}\left[\frac{1}{i(k_0+p_0)-V_{F,k}p_y}\frac{1}{i(k_0+q_0)-V_{F,k}q_y}\right.\\
&\times\left.\left.\frac{1}{i(k_0+p_0+q_0)+(V_{F,k}p_y+V_{F,k}q_y+V_{F,k}\gamma(\vec{k};v_k)+2v_kV_{F,k}(p_x+q_x))}\right]\right\},
\end{split}
\end{align}
where
$\bar{\Delta}(\vec{k};v_k) = c 
e_{4}[\vec{k};v_k]$
and 
$\gamma(\vec{k};v_k) = 2
e_{4}[\vec{k};v_k]
+ e_{1}[\vec{k};v_k]$.
In the small $v$ limit,
the $cq_y$ and $cp_y$ terms can be dropped in the boson propagators 
as the integrations of $q_y$ and $p_y$ are convergent without them.
Furthermore, 
we drop $v_kcq_x$ and $v_kcp_x$  as the IR singular term is unaffected by them
 to the leading order in $v$.
With the rescaling of internal momentum as $(p_x,q_x)\rightarrow (p_x/c, q_x/c)$,
the integration over $p_y$ and $q_y$ yields
\begin{align}
\begin{split}\label{eq:PreviousStartingPoint2Loop}
\Sigma^{\mathrm{2L}}_{1}({\bf k})&=- 
\frac{4g_k^4(N_c^2-1)}{c^2V_{F,k}^2N_c^2N_f^2}
\int\limits_{\mathbb{R}}\frac{\dd q_0}{(2\pi)}\int\limits_{\mathbb{R}}\frac{\dd p_0}{(2\pi)}\int\limits_{\mathbb{R}}\frac{\dd q_x}{(2\pi)}\int\limits_{\mathbb{R}}\frac{\dd p_x}{(2\pi)}\left\{\frac{\left[\Theta(p_0+2q_0+2k_0)-\Theta(-k_0-p_0)\right]}{|q_0|+|q_x|+|\bar{\Delta}(\vec{k};v_k)|}\right.\\
&\times \left.\frac{[\Theta(p_0+q_0+k_0)-\Theta(-k_0-q_0)]}{|p_0|+|p_x|+|\bar{\Delta}(\vec{k};v_k)|}\frac{[2\mfw_k(p_x+q_x)+\mfg(\vec{k};v_k)]-i\left[2(p_0+q_0)+3k_0\right]}{(3k_0+2p_0+2q_0)^2+[2\mfw_k(p_x+q_x)+\mfg(\vec{k};v_k)]^2}\right\},
\end{split}
\end{align}
where $w_k=v_k/c$, 
$\mfw_k = V_{F,k}w_k$, 
$\mfg(\vec k,v_k) =
V_{F,k}\gamma(\vec k;v_k)$ 
and
$\Theta(x)$ denotes the Heaviside function.
Since only the real part of the two-loop self-energy is of the same order as the one-loop self-energy\cite{SCHLIEF},
we only compute the real part of the self-energy.

The real part of the two-loop fermion self-energy reads
\begin{align}
\begin{split}\label{eq:Re2Loop}
\mathrm{Re}\left[\Sigma^{\mathrm{2L}}_{1}({\bf k})\right]&=- 
\frac{4g_k^4(N_c^2-1)}{c^2V_{F,k}^2N_c^2N_f^2}
\int\limits_{\mathbb{R}}\frac{\dd q_0}{(2\pi)}\int\limits_{\mathbb{R}}\frac{\dd p_0}{(2\pi)}\int\limits_{\mathbb{R}}\frac{\dd q_x}{(2\pi)}\int\limits_{\mathbb{R}}\frac{\dd p_x}{(2\pi)}\left\{\frac{\left[\Theta(p_0+2q_0+2k_0)-\Theta(-k_0-p_0)\right]}{|q_0|+|q_x|+|\bar{\Delta}(\vec{k};v_k)|}\right.\\
&\times \left.\frac{[\Theta(p_0+q_0+k_0)-\Theta(-k_0-q_0)]}{|p_0|+|p_x|+|\bar{\Delta}(\vec{k};v_k)|}\frac{[2\mfw_k(p_x+q_x)+\mfg(\vec{k};v_k)]}{(3k_0+2p_0+2q_0)^2+[2\mfw_k(p_x+q_x)+\mfg(\vec{k};v_k)]^2}\right\}.
\end{split}
\end{align}
Let us make a change of variables  as 
$a = (p_x+q_x)/2$ and $b = (p_x-q_x)/2$. 
The integration over $b$ gives
\begin{equation}
\begin{aligned}
&\mathrm{Re}\left[\Sigma^{\mathrm{2L}}_{1}({\bf k})\right]
=- 
\frac{4g_k^4(N_c^2-1)}{\pi c^2V_{F,k}^2N_c^2N_f^2}
\int\limits_{\mathbb{R}}\frac{\dd q_0}{(2\pi)}
\int\limits_{\mathbb{R}}\frac{\dd p_0}{(2\pi)}
\int\limits_{\mathbb{R}}\frac{\dd a}{(2\pi)}
\left\{
\frac{
\Theta(k_0;p_0,q_0)
[4\mfw_ka+\mfg(\vec{k};v_k)]}{(3k_0+2p_0+2q_0)^2+[4\mfw_k a+\mfg(\vec{k};v_k)]^2}
\right.
\\
&\times
\left.
\left(
 \FR{\log\PFR{2|a|+|p_0|+|\bar{\Delta}(\vec{k};v_k)|}{|q_0|+|\bar{\Delta}(\vec{k};v_k)|}}{2|a|+|p_0|-|q_0|}
+\FR{\log\PFR{2|a|+|q_0|+|\bar{\Delta}(\vec{k};v_k)|}{|p_0|+|\bar{\Delta}(\vec{k};v_k)|}}{2|a|+|q_0|-|p_0|}
+\FR{\log\PFR{(2|a|+|p_0|+|\bar{\Delta}(\vec{k};v_k)|)(2|a|+|q_0|+|\bar{\Delta}(\vec{k};v_k)|)}{(|p_0|+|\bar{\Delta}(\vec{k};v_k)|)(|q_0|+|\bar{\Delta}(\vec{k};v_k)|)}}{2|a|+|p_0|+|q_0|+2|\bar{\Delta}(\vec{k};v_k)|}
\right)
\right\},
\end{aligned}
\label{eq:Re2LoopAfterBintegration}
\end{equation}
where
$
\Theta(k_0;p_0,q_0)
=
\left[\Theta(p_0+2q_0+2k_0)-\Theta(-k_0-p_0)\right]
\left[\Theta(p_0+q_0+k_0)-\Theta(-k_0-q_0)\right]$.
%
%
Besides $k_0$,
$\mfg(\vec{k};v_k)$  and $\bar{\Delta}(\vec{k};v_k)$
enter as additional energy scales associated with the external momentum $\vec k$.
If the external fermion is close to the Fermi surface,
$| \mfg(\vec{k};v_k)  | \sim 4 V_{F,k} v_k |k_x| \gg | \bar{\Delta}(\vec{k};v_k)  | \sim 2 v_k c |k_x|$.
Therefore, the crossover is determined by the competition between $k_0$
and
$\mfg(\vec{k};v_k)$.
In the 
$| \mfg(\vec{k};v_k)/k_0  | \ll 1$ limit,
the integration over $a$, $q_0$ and $p_0$ gives
\begin{align}
\mathrm{Re}\left[\Sigma^{\mathrm{2L}}_{1}({\bf k})\right]
&= 
-\frac{4g_k^4(N_c^2-1)\mfg(\vec k,v_k)}{\pi c^2V_{F,k}^2N_c^2N_f^2}
\FR{\log^2 \mfw_k}{8\pi^3}
\log\FR{\Lambda}{k_0}.
\label{eq:Re2LoopLargeFrequencyResult}
\end{align}
In the opposite limit with $ | k_0/\mfg(\vec{k},v_k) |  \ll 1$,
the IR divergence is cutoff by $\mfg(\vec{k},v_k)$ instead of $k_0$,
\begin{align}
\mathrm{Re}\left[\Sigma^{\mathrm{2L}}_{1}({\bf k})\right]
&= 
-\frac{4g_k^4(N_c^2-1)\mfg(\vec k,v_k)}{\pi c^2V_{F,k}^2N_c^2N_f^2}
\FR{\log^2 \mfw_k}{8\pi^3}
\log\FR{\Lambda}{\mfg(\vec{k},v_k)}.
\label{eq:Re2LoopLargeGammaResult}
\end{align}
Collecting the results of \eq{eq:Re2LoopLargeFrequencyResult} and \eq{eq:Re2LoopLargeGammaResult},
we conclude that the logarithmically divergent contribution to the real part of the two-loop fermion self-energy
is given by
\begin{align}
\mathrm{Re}\left[\Sigma^{\mathrm{2L}}_{1}({\bf k})\right]
&=- 
\frac{
g_k^4
(N_c^2-1)}{2\pi^4 c^2
V_{F,k}^2
N_c^2N_f^2}
V_{F,k}
\left[
e_{1}[\vec{k};v_k]+2e_{4}[\vec{k};v_k] \right]
\log^2\PFR{V_{F,k} v_k}{c}
\log\left(
\frac{\Lambda}{\mathscr{H}_1(k_0,
V_{F,k}(2e_{4}[\vec k,v_k]+e_1[\vec k, v_k])) }
\right).
\end{align}
The two-loop counterterms are given by
\begin{align}
\label{eqA22L}
A^{(2);2L}(k)
&= 
\frac{3g_k^4(N_c^2-1)}{2\pi^4 c^2V_{F,k}^2
N_c^2N_f^2}
\log^2\PFR{V_{F,k} v_k}{c}
\log\left(\frac{\Lambda}{\mathscr{H}_1(\mu,4V_{F,k}v_kk)}\right)
+
A^{(2);2L}_{reg}(k),
\\
\label{eqA32L}
A^{(3);2L}(k)
&= 
-\frac{g_k^4(N_c^2-1)}{2\pi^4 c^2V_{F,k}^2
N_c^2N_f^2}
\log^2\PFR{V_{F,k} v_k}{c}
\log\left(\frac{\Lambda}{\mathscr{H}_1(\mu,4V_{F,k}v_kk)}\right)
+ A^{(3);2L}_{reg}(k),
\end{align}
where
$A^{(2);2L}_{reg}(k)$
and
$ A^{(3);2L}_{reg}(k)$
represent the terms that are regular
in the large 
$\frac{\Lambda}{\mathscr{H}_2(\mu,4V_{F,k}v_kk)}$
limit.

\subsection{Fermion-boson vertex correction} \label{sec:Yukawa_app_1L}

The one-loop vertex function is given by:
\begin{align}
\varGamma^{(2,1),\mathrm{1L}}_{N}({\bf k}',{\bf k}) = -
\frac{2}{N_cN^\frac{3}{2}_{f}}
g_{k'}^{(N)}
g_{k,k'}^{(\bar{N})} 
g_k^{(N)}
\int \dd {\bf q}
D({\bf q})
G_{\bar{N}}({\bf k'}+{\bf q})
G_{N}({\bf k}+{\bf q}).
\end{align}
Without loss of generality,
we consider the contribution to interaction vertex for the
$N=1$ hot spot,
\begin{align}
\varGamma^{(2,1),\mathrm{1L}}_{1}({\bf k}',{\bf k}) = 
-\frac{2 g_kg_{k'}g_{k',k}}{N_cN^\frac{3}{2}_{f}}
&\int \FR{\dd q_0}{2\pi}\FR{\dd q_x}{2\pi}\FR{\dd q_y}{2\pi}
\frac{1}{|q_0|+|cq_x|+|cq_y|}
\frac{1}{i(k_0'+q_0)+V_{F,k'}(e_4[\vec{k}',v_{k'}]+v_{k'}q_x-q_y)}
\nonumber\\
&\times\frac{1}{i(k_0+q_0)+V_{F,k}(e_1[\vec{k},v_{k}]+v_{k}q_x+q_y)}.
\label{eqA51}
\end{align}
Let us integrate over $q_y$ using a contour integration. 
To do this, we use $|cq_y| = c\sqrt{q_y^2+0^+}$ 
with the branch cuts 
located at 
$|Im ~q_y| > 0^+$. 
Across the branch cut, 
the square root is discontinuous :
$c\sqrt{q_y^2+0^+} = 
c \sgn(\mathrm{Re}(q_y)) q_y
$. 
To ensure a symmetric expression, 
we close the contours in both the upper and lower-half planes and taking the average of these two expressions. 
In each case, we will consider a semicircular contour with a dip along the imaginary axis that avoids the branch cuts. 
The contour integral of Eq. \eqref{eqA51} 
results in
\begin{align}
& \varGamma^{(2,1),\mathrm{1L}}_{1}({\bf k}',{\bf k}) = \nn
    &
    \frac{ig_kg_{k'}g_{k',k}}{N_c N^\frac{3}{2}_{f}c(V_{F,k}+V_{F,k'})}
    \int \FR{\dd q_0}{2\pi} \FR{\dd q_x}{2\pi} 
    \FR{1}{i(\mathscr{M}+q_0)+\mathscr{R}_{k',k}/2+\mathscr{W}q_x}
    \left(
        \FR{\sgn(k_0'+q_0)}{|q_0|+|q_x|+  c|e_4(\vec k', v_{k'})|}+
        \FR{\sgn(k_0+q_0)} {|q_0|+|q_x|+  c|e_1(\vec k, v_k)|}
    \right)
    \nonumber\\
    &
    -
    \frac{2g_kg_{k'}g_{k',k}c}{N_cN^\frac{3}{2}_{f}}
    \int \FR{\dd q_0}{2\pi} \FR{\dd q_x}{2\pi} 
    \int_{-\infty}^\infty 
    \FR{\dd x}{2\pi}
    \FR{|x|}{(c^2x^2+(|q_0|+|cq_x|)^2}
    \frac{1}{i(k_0'+q_0)+V_{F,k'}(e_4(k',v_{k'})+v_{k'}q_x-ix)}
    \nonumber\\
    &\hspace{6cm}
    \times
    \frac{1}{i(k_0+q_0)+V_{F,k}(e_1(k,v_{k})+v_{k}q_x+ix)},
    \label{eq:A30}
\end{align}
where 
$\mathscr{M} = (V_{F,k'}^{-1} k_0' + V_{F,k}^{-1} k_0)/(V_{F,k'}^{-1} + V_{F,k}^{-1})$, 
$\mathscr{R}_{k',k} = 2(e_4(\vec k',v_{k'})+e_1(\vec k,v_k))/(V_{F,k'}^{-1} + V_{F,k}^{-1})$
and
$\mathscr{W}= (v_k+v_{k'})/(cV_{F,k'}^{-1} + cV_{F,k}^{-1})$.
The first term is the contribution from the residues of the poles of the fermion propagators. 
The second term comes from the branch cut. 
In the small $v$ limit,
the first contribution dominates.
The remaining integrand over $q_0$ and $q_x$ 
leads to
\bqa
    \varGamma^{(2,1),\mathrm{1L}}_{1}({\bf k}',{\bf k})
  \bigg|_{
 \scriptsize
   \begin{array}{l}
  {\bf k}' =(2 \mu, k'_{x},-v_{k'} k'_{x}) \\
  {\bf k}=( \mu, k_{x},v_{k} k_{x})  
 \end{array}
 } 
    &= &
    \frac{g_kg_{k'}g_{k',k}}{\pi^2c(V_{F,k}+V_{F,k'})N_c N^\frac{3}{2}_{f}}
    \log\PFR{c(V_{F,k}^{-1}+V_{F,k'}^{-1})}{v_k+v_{k'}} \times \nn
  && \hspace{-2cm}
    \left(
        \log\PFR{\Lambda}{\mathscr{H}_2\left[\mu,
\frac{4v_{k'}k' + 4v_kk}{V_{F,k'}^{-1}+V_{F,k}^{-1}} 
        , 2 c v_k  k \right]}+
        \log\PFR{\Lambda}{\mathscr{H}_2\left[ \mu,
\frac{4v_{k'}k' + 4v_kk}{V_{F,k'}^{-1}+V_{F,k}^{-1}} 
        , 2 c v_{k'}  k'  \right]}
    \right)
\eqa
to the leading order in $c$.
$\mathscr{H}_2(x,y,z)$ is a crossover function 
that satisfies
$\mathscr{H}_2(x,y,z)\sim \max(|x|,|y|,|z|)$ 
if  $|x|\gg |y|,|z|$, 
$|y|\gg |x|,|z|$ or 
$|z|\gg |x|,|y|$\footnote{
For the future convenience, 
$\mathscr{R}_{k',k}$ 
is chosen as the crossover scale 
although $\mathscr{R}_{k',k}/2$ is 
what appears in \eq{eq:A30}.
This is a freedom associated with the choice of the crossover function 
that only affects the finite part 
of the counter term.
}.
Because $c \ll 1$,
$\left| \frac{4v_{k'}k' + 4v_kk}{V_{F,k'}^{-1}+V_{F,k}^{-1}} \right|
\gg  2 c v_k  |k|, 2 c v_{k'}  |k'|$
for most $k$ and $k'$.
$\left| \frac{4v_{k'}k' + 4v_kk}{V_{F,k'}^{-1}+V_{F,k}^{-1}} \right|$
becomes smaller than 
$2 c v_k  |k|$ or $2 c v_{k'}  |k'|$
only in a small wedge near 
$v_{k'}k' + v_kk =0$.
Since  $|v_{k'}k'| \approx |v_kk|$ within the wedge, 
we can combine the two crossover functions into one as
\bqa
        \log\PFR{\Lambda}{\mathscr{H}_2 \left[ \mu,
\frac{4v_{k'}k' + 4v_kk}{V_{F,k'}^{-1}+V_{F,k}^{-1}} 
        , 2 c v_k  k \right]}+
        \log\PFR{\Lambda}{\mathscr{H}_2 \left[ \mu,
\frac{4v_{k'}k' + 4v_kk}{V_{F,k'}^{-1}+V_{F,k}^{-1}} 
        , 2 c v_{k'}  k'  \right] }
= 
2  
\log\PFR{\Lambda}{\mathscr{H}_3\left[\mu,
\frac{4v_{k'}k' + 4v_kk}{V_{F,k'}^{-1}+V_{F,k}^{-1}},
         2 c v_k  k,
         2 c v_{k'}  k' 
\right]}, \nn
\eqa
where $\mathscr{H}_3(w,x,y,z)$ is a crossover function 
that satisfies
$\mathscr{H}_3(w,x,y,z)\sim \max(|w|,|x|,|y|,|z|)$.
Therefore, we write the counter term as
\begin{align}\label{eqA41L}
A^{(4)}(k',k)
    &=-\frac{2g_kg_{k'}}{\pi^2c(V_{F,k}+V_{F,k'})N_c N_{f}}
    \log\PFR{c(V_{F,k}^{-1}+V_{F,k'}^{-1})}{v_k+v_{k'}}
\log\PFR{\Lambda}{
\mathscr{H}_3\left[\mu,
\frac{4v_{k'}k' + 4v_kk}{V_{F,k'}^{-1}+V_{F,k}^{-1}},
         2 c v_k  k,
         2 c v_{k'}  k' 
\right]}.
\end{align}

\section{Quantum corrections 
for the four-point function}
\label{app:QC2}

In this appendix, we compute the quantum corrections for the four-fermion vertex. 
We denote the fermionic four-point vertex function evaluated at external momenta on the Fermi surface as
\begin{equation}
 \mathbf{\Gamma}^{\spmqty{N_1 & N_2 \\ N_4 & N_3};\spmqty{\sigma_1 & \sigma_2 \\ \sigma_4 & \sigma_3}}_{\spmqty{ {k}_1 & {k}_2 \\ {k}_4 & {k}_3} }
\equiv
\left. \mathbf{\Gamma}^{\spmqty{N_1 & N_2 \\ N_4 & N_3};\spmqty{\sigma_1 & \sigma_2 \\ \sigma_4 & \sigma_3}}\spmqty{ {\bf k}_1 & {\bf k}_2 \\ {\bf k}_4 & {\bf k}_3}   
\right|_{
{\bf k}_i={\bf k}_i^*
}.
\label{eq:B1}
\end{equation}
Here the external frequencies are chosen 
as in  \eq{eq:4fmomenta}.
$k_i$ labels the component of $\vec k_i$ 
that is parallel to the Fermi surface
near hot spot $N_i$
in the small $v$ limit.
The other component of the spatial momentum is chosen 
so that external electrons are on the Fermi surface.

\subsection{Generation of the primary couplings from spin fluctuations}

We first consider the quantum corrections
through which the primary four-fermion coupling  are generated.

\subsubsection{Group 1}
\label{app:QC_lambda_0}

In group 1,
the diagram in Fig. 
\ref{fig:s1} 
exhibits an IR singularity 
only when 
all external fermions are at the hot spots.
In our minimal subtraction scheme,
we don't need to add a counter term for it.
Therefore, we focus on the diagram in Fig. 
\ref{fig:s2}.
Its contribution to the quantum effective action is given by
Eq. (\ref{eq:Gamma_lambda0_PH}),
\begin{equation}
\begin{aligned}
{\bf \Gamma}^{\spmqty{1&1\\1&1};\spmqty{\sigma_1&\sigma_2\\\sigma_4&\sigma_3}}_{(0)PH}\spmqty{{\bf k}+{\bf l}&{\bf p}-{\bf l}\\{\bf k}&{\bf p}}
 = & -\frac{\mathsf{T}^{\beta \sigma_2}_{\sigma_4 \alpha}   \mathsf{T}^{\sigma_1 \alpha}_{\beta \sigma_3}}{2N_f^2}
\int \dd{\bf q}\frac{g^{(1)}_{k+l,k+q}g^{(4)}_{k+q,k}g^{(1)}_{p-l,p-l+q}g^{(4)}_{p-l+q,p}}{\left(|q_0| + c|q_x|+c|q_y|\right)\left(|q_0-l_0| + c|q_x-l_x|+c|q_y-l_y|\right)}\times
\\ & \frac{1}{i(k_0+q_0)+V^{(4)}_{F,k+q}e_{4}[\vec{k}+\vec{q};v^{(4)}_{k+q}]}~\frac{1}{i(p_0-l_0+q_0)+V^{(4)}_{F,p-l+q}e_{4}[\vec{p}-\vec{l}+\vec{q};v^{(4)}_{p-l+q}]}~.
\label{eq:A121}
\end{aligned}
\end{equation}
Since it is possible to put all external fermions on the Fermi surface for 
$\vec k+\vec l = \vec p$,
we focus on the quantum correction
in the vicinity of the plane with  
$\vec k+\vec l = \vec p$.
Near the plane,
we can 
replace $k+l$ with $p$
inside coupling functions.
Shifting $q_y\to q_y + v_{k+q}q_x + \Delta_{1,\lbrace 1,1\rbrace}/c$, we obtain
\begin{equation}
\begin{aligned}
{\bf \Gamma}^{\spmqty{1&1\\1&1};\spmqty{\sigma_1&\sigma_2\\\sigma_4&\sigma_3}}_{(0)PH}\spmqty{{\bf k}+{\bf l}&{\bf p}-{\bf l}\\{\bf k}&{\bf p}}
 = & -\frac{\mathsf{T}^{\beta \sigma_2}_{\sigma_4 \alpha}   \mathsf{T}^{\sigma_1 \alpha}_{\beta \sigma_3}}{2N_f^2}
\int \frac{\dd{\bf q}}{V_{F, k+q}^2}\frac{g^2_{k+l,k+q}g^2_{k+q,k}}{\left(|q_0| + c|q_x|+c|q_y+v_{k+q}q_x+\Delta_{1,\lbrace 1,1\rbrace}/c|\right)}\times
\\ & 
\frac{1}{\left(|q_0-l_0| + c|q_x-l_x|+c|q_y-l_y+v_{k+q}q_x+\Delta_{1,\lbrace 1,1\rbrace}/c|\right)}\times
\\ & \frac{1}{i(k_0+q_0)/V_{F, k+q}-q_y + \Delta_{2,\lbrace 1,1\rbrace}}~\frac{1}{i(p_0-l_0+q_0)/V_{F, k+q}-q_y - \Delta_{2,\lbrace 1,1\rbrace}}~,
\label{eq:11_shift}
\end{aligned}
\end{equation}
where 
$\Delta_{1,\lbrace 1,1\rbrace}\equiv \Delta_{1,\lbrace 1,1\rbrace}(q;k,p-l;v) = c(e_{4}[\vec{k};v^{(4)}_{k+q}]+e_{4}[\vec{p}-\vec{l};v^{(4)}_{p-l+q}])/2$ and 
$\Delta_{2,\lbrace 1,1\rbrace} \equiv \Delta_{2,\lbrace 1,1\rbrace}(q;k,p-l;v) = (e_{4}[\vec{k};v^{(4)}_{k+q}]-e_{4}[\vec{p}-\vec{l};v^{(4)}_{p-l+q}])/2.$
We can drop $cq_y$ from the two boson propagators in the small $v$ limit because the integration is convergent without it. 
Using the RG condition for the frequencies $(k_0 = \mu, ~p_0 = \mu, ~l_0 = 2\mu)$ and doing the $q_y$ integration, we obtain
\begin{equation}
\begin{aligned}
 {\bf \Gamma}^{\spmqty{1&1\\1&1};\spmqty{\sigma_1&\sigma_2\\\sigma_4&\sigma_3}}_{ (0)PH; \spmqty{k+l & p-l\\ k &  p}}
 = & -\frac{\mathsf{T}^{\beta \sigma_2}_{\sigma_4 \alpha}   \mathsf{T}^{\sigma_1 \alpha}_{\beta \sigma_3}}{2N_f^2}
\int \frac{\dd{q_0}\dd{q_x}}{8\pi^2V_{F, k+q}}\frac{g^2_{k+l,k+q}g^2_{k+q,k}}{\left(|q_0| + c|q_x|+c|v_{k+q}q_x+\Delta_{1,\lbrace 1,1\rbrace}/c|\right)}\times
\\ & 
\frac{1}{\left(|q_0-2\mu| + c|q_x-l_x|+c|-l_y+v_{k+q}q_x+\Delta_{1,\lbrace 1,1\rbrace}/c|\right)}~\frac{i({\rm sgn}(q_0+\mu)-{\rm sgn}(q_0-\mu))}{2V_{k+q}\Delta_{2,\lbrace 1,1\rbrace}+2i\mu}~.
\label{eq:11_qyint}
\end{aligned}
\end{equation}
We now construct a local counter term for this quantum correction.
There are two crucial conditions that counter terms must satisfy :
1) counter terms must remove the IR divergence of quantum corrections in the small $\mu$ limit,
and 
2) counter terms should be analytic in external momenta as they  are parts of the local action.
\eq{eq:11_qyint} can not be directly used for the counter term because it is a non-analytic function of external momentum.
We can construct analytic counter terms 
by making the momentum dependence smooth 
around $k=0$ and $p=0$ with energy scale $\mu$.
This smearing can be implemented by replacing 
\bqa
|x|_{\mu} \equiv \sqrt{x^2 + \mu^2}
\label{eq:absxmu}
\eqa 
for momentum $x$.
\footnote{
Such smearing naturally would arise 
in the exact boson propagator evaluated 
at a finite frequency $\mu$.
}.
The modification only introduces a finite correction.
The counter term can be further simplified
by replacing $|q_0|$ and $|q_0-2\mu|$ with $\mu$ 
inside the boson propagators in \eq{eq:11_qyint}.
The latter procedure only affects the finite part of the quantum correction
because $q_0$-integration is bounded by $-\mu$ and $\mu$.
The resulting counter term at energy scale $\mu$ is written as
\begin{equation}
\begin{aligned}
&\tilde {\bf \Gamma}^{\spmqty{1&1\\1&1};\spmqty{\sigma_1&\sigma_2\\\sigma_4&\sigma_3}}_{CT; (0)PH; \spmqty{k+l & p-l\\ k &  p}}
 = \frac{\mathsf{T}^{\beta \sigma_2}_{\sigma_4 \alpha}   \mathsf{T}^{\sigma_1 \alpha}_{\beta \sigma_3}}{2N_f^2}
\int \frac{\dd{q_0}\dd{q_x}}{8\pi^2V_{F, k+q}}\frac{g^2_{k+l,k+q}g^2_{k+q,k}}{\left(\mu + c|q_x|_{\mu}+c|v_{k+q}q_x+\Delta_{1,\lbrace 1,1\rbrace}/c|_{\mu}\right)}\times
\\ & 
\frac{1}{\left(\mu + c|q_x-l_x|_{\mu}+c|-l_y+v_{k+q}q_x+\Delta_{1,\lbrace 1,1\rbrace}/c|_{\mu}\right)}~\frac{i({\rm sgn}(q_0+\mu)-{\rm sgn}(q_0-\mu))}{2V_{k+q}\Delta_{2,\lbrace 1,1\rbrace}+2i\mu}~.
\label{eq:11_counter}
\end{aligned}
\end{equation}
%
As expected, the counter term removes the IR singularity of the quantum correction in the plane ${\vec p } = \vec{k}+\vec{l}$\footnote{
The sum of the quantum correction and the counter term in the region away from the hot-spot 
$\left( |k|, |p| \gg \frac{\mu}{vc} \right)$
is given by
\begin{align}
{\bf \Gamma}^{\spmqty{1&1\\1&1};\spmqty{\sigma_1&\sigma_2\\\sigma_4&\sigma_3}}_{(0)PH; \spmqty{p & k\\ k &  p}}
+ \tilde {\bf \Gamma}^{\spmqty{1&1\\1&1};\spmqty{\sigma_1&\sigma_2\\\sigma_4&\sigma_3}}_{CT; (0)PH; \spmqty{p & k\\ k &  p}} 
\sim
\frac{\mathsf{T}^{\beta \sigma_2}_{\sigma_4 \alpha}   \mathsf{T}^{\sigma_1 \alpha}_{\beta \sigma_3}}{2N_f^2}\frac{1}{8\pi^2}\dfrac{g^2_{p,k}g^2_{k, k}}{c V_{F,k}}\nonumber
\Bigg[
-\frac{2 \mu  \log \left(\frac{(c | k_x-p_x| +c v | k_x| ) (c | k_x-p_x| +c v | p_x| )}{c^2 v^2 | k_x|  | p_x| }\right)}{c^2 (|k_x-p_x| +v | k_x| +v | p_x| )^2}
\\
-\frac{2 \mu  (| k_x| +| p_x| )}{c^2 v | k_x|  | p_x|  (| k_x-p_x| +v | k_x| +v | p_x| )}
\Bigg]
\end{align}
In order to arrive at the above expression, we use momentum independent nesting angle.
}.
%
The $\log\mu$ derivative of the counter term 
that determines the beta functional is given by

\begin{eqnarray}\nonumber
4\mu\frac{\partial}{\partial\log\mu}{\bf \tilde\Gamma}^{\spmqty{1&1\\1&1};\spmqty{\sigma_1&\sigma_2\\\sigma_4&\sigma_3}}_{CT;(0)PH;\spmqty{k+l & p-l\\ k &  p}} = -\frac{\mathsf{T}^{\beta \sigma_2}_{\sigma_4\alpha}   \mathsf{T}^{\sigma_1\alpha}_{\beta \sigma_3}}{\pi N_f^2}
\int \frac{\dd q_x}{2\pi\mu V_{F, k+q}}~ \frac{i\mu}{i\mu+V_{F,k+q}\Delta_{2,\lbrace 1,1\rbrace}}
\Bigg[-\frac{V_{F,k+q}\Delta_{2,\lbrace 1,1\rbrace}}{(i\mu+V_{F,k+q}\Delta_{2,\lbrace 1,1\rbrace})}\\\times\nonumber
\frac{g^2_{k+q,k}~\mu}{\left(\mu + c|q_x|_{\mu}+c|v_{k+q}q_x+\Delta_{1,\lbrace 1,1\rbrace}/c|_{\mu}\right)}
\frac{g^2_{k+l,k+q}~\mu}{\left(\mu + c|q_x-l_x|_{\mu}+c|-l_y+v_{k+q}q_x+\Delta_{1,\lbrace 1,1\rbrace}/c|_{\mu}\right)}\\\nonumber
+\frac{g^2_{k+q,k}~\mu}{\left(\mu + c|q_x|_{\mu}+c|v_{k+q}q_x+\Delta_{1,\lbrace 1,1\rbrace}/c|_{\mu}\right)}
\frac{g^2_{k+l,k+q}~\mu^2}{\left(\mu + c|q_x-l_x|_{\mu}+c|-l_y+v_{k+q}q_x+\Delta_{1,\lbrace 1,1\rbrace}/c|_{\mu}\right)^2}\\
+ \frac{g^2_{k+q,k}~\mu^2}{\left(\mu + c|q_x|_{\mu}+c|v_{k+q}q_x+\Delta_{1,\lbrace 1,1\rbrace}/c|_{\mu}\right)^2}
\frac{g^2_{k+l,k+q}~\mu}{\left(\mu + c|q_x-l_x|_{\mu}+c|-l_y+v_{k+q}q_x+\Delta_{1,\lbrace 1,1\rbrace}/c|_{\mu}\right)}\Bigg].
\label{1515_beta_PP_CT2}
\end{eqnarray}
In the plane with 
${\vec p}= {\vec k}+{\vec l}$, 
we have $\Delta_{1,\lbrace 1,1\rbrace} = c e_{4}[\vec{k};v^{(4)}_{k+q}]$ and $\Delta_{2,\lbrace 1,1\rbrace} = 0$.
Denoting 
$q_x$, $p_x$ and $k_x$ as
$q$, $p$ and $k$, respectively,
\eq{1515_beta_PP_CT2} is written as

\begin{equation}
4\mu\frac{\partial}{\partial\log\mu}{\bf \tilde\Gamma}^{\spmqty{1&1\\1&1};\spmqty{\sigma_1&\sigma_2\\\sigma_4&\sigma_3}}_{CT;(0)PH;\spmqty{p & k\\ k &  p}} = -\frac{\mathsf{T}^{\beta \sigma_2}_{\sigma_4 \alpha}   \mathsf{T}^{\sigma_1 \alpha}_{\beta \sigma_3}}{\pi N_f^2}\int d\rho(q)
\left[\mathsf{D}_\mu(q;k)\frac{\mathsf{D}_\mu(p;q)^2}{g_{p,q}^2} + \mathsf{D}_\mu(p;q)\frac{\mathsf{D}_\mu(q;k)^2}{g_{q,k}^2}\right],
\label{eq:1111_beta_PH_CT}
\end{equation}
where $q$ is shifted to $q - k$
and
\begin{equation}
    \mathsf{D}_\mu(p;k) = g_{k,p}^2\frac{\mu}{\mu +c \left(\abs{p-k}_\mu + \abs{v_p p + v_k k}_\mu \right)}, ~~~~~
    d\rho(q) = \frac{\dd q}{2\pi \mu V_{F,q}}
\label{eq:Dmu}
\end{equation}
represent the interaction mediated by gapless spin fluctuations 
and the phase space integration measure, respectively.


\subsubsection{Group 2}

In group 2, the diagram in Fig. 
\ref{fig:s1}
gives rise to 
a singular quantum correction 
to the couplings given by
\eq{eq:SofIS2}
in which the total momentum of the electron pair is zero.
Eq. (\ref{eq:Gamma_lambda0_PP})  for 
the quantum correction reads
\begin{equation}
\begin{aligned}
{\bf \Gamma}^{\spmqty{1&5\\1&5};\spmqty{\sigma_1&\sigma_2\\\sigma_4&\sigma_3}}_{(0)PP}\spmqty{{\bf k}+{\bf l}&{\bf p}-{\bf l}\\{\bf k}&{\bf p}} =& -\frac{\mathsf{T}^{\sigma_1 \sigma_2}_{\alpha\beta}   \mathsf{T}^{\alpha\beta}_{\sigma_4 \sigma_3}}{2N_f^2}\int \dd{\bf q} ~
\frac{1}{i(k_0+q_0)+V^{(4)}_{F,k+q}e_{4}[\vec{k}+\vec{q};v^{(4)}_{k+q}]}~\frac{1}{i(p_0-q_0)+V^{(8)}_{F,p-q}e_{8}[\vec{p}-\vec{q};v^{(8)}_{p-q}]}
\times\\ &
\frac{g^{(4)}_{k+q, k}g^{(8)}_{p-q, p}g^{(1)}_{k+l, k+q}g^{(5)}_{p-l, p-q}}{\left(|q_0| + c|q_x|+c|q_y|\right)\left(|q_0-l_0| + c|q_x-l_x|+c|q_y-l_y|\right)}.
\end{aligned}
\end{equation}
In the vicinity of the plane 
 with ${\vec p}+{\vec k} = {\vec{0}}$, we can replace $p_x$ with $-k_x$ in the coupling functions. 
%
Shifting $q_y\to q_y + v_{k+q}q_x + \Delta_{1,\lbrace 1,5\rbrace}/c$, we obtain
\begin{equation}
\begin{aligned}
{\bf \Gamma}^{\spmqty{1&5\\1&5};\spmqty{\sigma_1&\sigma_2\\\sigma_4&\sigma_3}}_{(0)PP}\spmqty{{\bf k}+{\bf l}&{\bf p}-{\bf l}\\{\bf k}&{\bf p}}
 = & -\frac{   \mathsf{T}^{\sigma_1 \sigma_2}_{\alpha \beta}\mathsf{T}^{\alpha \beta}_{\sigma_4 \sigma_3}}{2N_f^2}
\int \frac{\dd{\bf q}}{V_{F, k+q}^2}\frac{g^2_{k+l,k+q}g^2_{k+q,k}}{\left(|q_0| + c|q_x|+c|q_y+v_{k+q}q_x+\Delta_{1,\lbrace 1,5\rbrace}/c|\right)}\times
\\ & 
\frac{1}{\left(|q_0-l_0| + c|q_x-l_x|+c|q_y-l_y+v_{k+q}q_x+\Delta_{1,\lbrace 1,5\rbrace}/c|\right)}\times
\\ & \frac{1}{i(k_0+q_0)/V_{F, k+q}-q_y + \Delta_{2,\lbrace 1,5\rbrace}}~\frac{1}{i(p_0-q_0)/V_{F, k+q}-q_y - \Delta_{2,\lbrace 1,5\rbrace}}~,
\label{eq:15_shift}
\end{aligned}
\end{equation}
where 
$\Delta_{1,\lbrace 1,5\rbrace} \equiv \Delta_{1,\lbrace 1,5\rbrace}(q;k,p;v) = c(e_{4}[\vec{k};v^{(4)}_{k+q}]+e_{8}[\vec{p};v^{(8)}_{p-q}])/2$ and 
$\Delta_{2,\lbrace 1,5\rbrace} \equiv \Delta_{2,\lbrace 1,5\rbrace}(q;k,p;v) = (e_{4}[\vec{k};v^{(4)}_{k+q}]-e_{8}[\vec{p};v^{(8)}_{p-q}])/2.$
To the leading order in $v$,
one can perform the $q_y$ integration 
with dropping $c q_y$ in the boson propagator
to obtain
\begin{equation}
\begin{aligned}
& {\bf \Gamma}^{\spmqty{1&5\\1&5};\spmqty{\sigma_1&\sigma_2\\\sigma_4&\sigma_3}}_{(0)PP;\spmqty{k+l & p-l\\ k &  p}}
 = -\frac{\mathsf{T}^{\sigma_1 \sigma_2}_{\alpha\beta}   \mathsf{T}^{\alpha\beta}_{\sigma_4 \sigma_3}}{2N_f^2}
\int \frac{\dd q_0 \dd q_x}{8\pi^2V_{F,k+q}} ~
 \frac{{\rm sgn}(q_0+\mu) + {\rm sgn}(q_0-\mu)}{2(q_0-iV_{F, k+q}\Delta_{2,\lbrace1,5\rbrace})}
\times
\\ & \frac{g^2_{k+q, k}g^2_{k+l, k+q}}{\left(|q_0| + c|q_x|+c|
v_{k+q}
q_x+\Delta_{1,\lbrace1,5\rbrace}/c|\right)\left(|q_0-2\mu| + c|q_x-l_x|+c|-l_y+
v_{k+q}
q_x+\Delta_{1,\lbrace1,5\rbrace}/c|\right)}.
\end{aligned}
\label{1515_q0qx_PP}
\end{equation}
Since the support of $q_0$-integration is 
$(-\infty, -\mu)\bigcup (\mu, \infty)$, 
the IR divergent part of the quantum correction
is not affected by dropping $2\mu$ from 
$\abs{q_0 - 2\mu}$ in the second boson propagator. 
We further smear the non-analyticity in external momenta 
to write the counter term as
\begin{equation}
\begin{aligned}
& \tilde {\bf \Gamma}^{\spmqty{1&5\\1&5};\spmqty{\sigma_1&\sigma_2\\\sigma_4&\sigma_3}}_{CT; (0)PP;\spmqty{k+l & p-l\\ k &  p}} 
= \frac{\mathsf{T}^{\sigma_1 \sigma_2}_{\alpha\beta}   \mathsf{T}^{\alpha\beta}_{\sigma_4 \sigma_3}}{2N_f^2}
\int \frac{\dd q_0 \dd q_x}{8\pi^2V_{F,k+q}} ~
 \frac{{\rm sgn}(q_0+\mu) + {\rm sgn}(q_0-\mu)}{2(q_0-iV_{F, k+q}\Delta_{2,\lbrace1,5\rbrace})}
\times
\\ & \frac{g^2_{k+q, k}g^2_{k+l, k+q}}{\left(|q_0| + c|q_x|_{\mu}+c|v_{k+q}q_x+\Delta_{1,\lbrace1,5\rbrace}/c|_{\mu}\right)\left(|q_0| + c|q_x-l_x|_{\mu}+c|-l_y+v_{k+q}q_x+\Delta_{1,\lbrace1,5\rbrace}/c|_{\mu}\right)}.
\end{aligned}
\label{1515_q0qx_PPIRdiv}
\end{equation}
The counter term removes the IR singularity of the quantum correction in the $\vec{p}=-\vec{k}$ plane\footnote{
The sum of the quantum correction and the counter term in the region away from the hot-spot 
$\left( |k|, |k+l| \gg \frac{\mu}{vc} \right)$
is given by
\begin{align}
{\bf \Gamma}^{\spmqty{1&5\\1&5};\spmqty{\sigma_1&\sigma_2\\\sigma_4&\sigma_3}}_{(0)PP; \spmqty{k+l & -k-l\\ k &  -k}}
+ \tilde {\bf \Gamma}^{\spmqty{1&5\\1&5};\spmqty{\sigma_1&\sigma_2\\\sigma_4&\sigma_3}}_{CT; (0)PP; \spmqty{k+l & -k-l\\ k &  -k}} \sim 
\frac{\mathsf{T}^{\sigma_1 \sigma_2}_{\alpha\beta}   \mathsf{T}^{\alpha\beta}_{\sigma_4 \sigma_3}}{2N_f^2}\frac{1}{8\pi^2}
\dfrac{g^2_{k,k}g^2_{k+l, k}}{c V_{F, k}}\frac{4 \mu}{c v |k_x+l_x|  (c v | k_x| +c | l_x| )}
\end{align}
where we use momentum independent nesting angle for the estimation.
}.
Integrating $q_0$ and taking the $\log\mu$ derivative of the counter term, we obtain 
\begin{equation}
\begin{aligned}
& 4\mu\frac{\partial}{\partial\log\mu}\tilde {\bf \Gamma}^{\spmqty{1&5\\1&5};\spmqty{\sigma_1&\sigma_2\\\sigma_4&\sigma_3}}_{CT; (0)PP;\spmqty{k+l & p-l\\ k &  p}} 
= -\frac{\mathsf{T}^{\sigma_1 \sigma_2}_{\alpha\beta}   \mathsf{T}^{\alpha\beta}_{\sigma_4 \sigma_3}}{\pi N_f^2}
\int \frac{ \dd q_x}{2\pi\mu V_{F,k+q}} ~
 \frac{\mu^2}{(\mu^2+V^2_{F, k+q}\Delta^2_{2,\lbrace1,5\rbrace})}
\times
\\ & \frac{g^2_{k+q, k}~\mu}{\left(\mu + c|q_x|_{\mu}+c|v_{k+q}q_x+\Delta_{1,\lbrace1,5\rbrace}/c|_{\mu}\right)}~\frac{g^2_{k+l,k+q}~\mu}{\left(\mu + c|q_x-l_x|_{\mu}+c|-l_y+v_{k+q}q_x+\Delta_{1,\lbrace1,5\rbrace}/c|_{\mu}\right)}.
\end{aligned}
\label{1515_qx_PPIRdiv}
\end{equation}
Away from the plane with
${\vec p}+{\vec k}\neq 0$,
$\Delta_{2,\lbrace1,5\rbrace} \neq 0$ and the counter term vanishes in the low-energy limit.
Non-vanishing contribution to the beta functional arises only for
$\Delta_{2,\lbrace1,5\rbrace} \ll \mu$.
Within the space of IR singularity, 
\eq{1515_qx_PPIRdiv} becomes
\begin{equation}
4\mu\frac{\partial}{\partial\log\mu}{\bf \tilde\Gamma}^{\spmqty{1&5\\1&5};\spmqty{\sigma_1&\sigma_2\\\sigma_4&\sigma_3}}_{CT;(0)PP;\spmqty{k+l & -k-l\\ k &  -k}} = -\frac{\mathsf{T}^{\sigma_1 \sigma_2}_{\alpha\beta}   \mathsf{T}^{\alpha\beta}_{\sigma_4 \sigma_3}}{\pi N_f^2}
\int \dd\rho(q) ~\mathsf{D}_\mu(q;k)\mathsf{D}_\mu(k+l;q),
\label{1515_beta_PP_CT}
\end{equation}
where $\mathsf{D}_\mu(p;k)$  and $d\rho(q)$ 
are defined in Eq. \eqref{eq:Dmu}.
%


\subsection{Linear mixing}

Once the primary couplings are generated from the spin fluctuations,
the secondary couplings are further generated through the linear mixing.
When external fermions are on the Fermi surface,
Eqs. \eqref{eq:Gamma_lambda1_PP0} and
\eqref{eq:Gamma_lambda1_PH0}
that describe mixing of the four-fermion couplings
can be written as the sum of contributions 
from different parts of the Fermi surface as
\begin{equation}
\begin{aligned}
{\bf \Gamma}_{(1)PP; \spmqty{ k+l & p-l \\  k &  p}}^{\spmqty{N_1 & N_2 \\ N_4 & N_3};\spmqty{\sigma_1 & \sigma_2 \\ \sigma_4 & \sigma_3}}
 =  
\frac{1}{4 \mu N_f} \int \frac{\dd q}{2\pi} &
\Biggl[
g^{(\bar N_4)}_{k+q,k}
g^{(\bar N_3)}_{p-q,p} 
~\mathcal{K}_{N_4,N_3}^{(PP)}(q;\mu, \mu, {k},{p})
\lambda^{\spmqty{N_1 & N_2 \\ \bar N_4 & \bar N_3};\spmqty{\sigma_1 & \sigma_2 \\ \alpha & \beta}}_{\spmqty{k+l  & p-l \\ k+q & p-q }}
 \mathsf{T}^{\alpha \beta}_{\sigma_4 \sigma_3} 
\\ & +
g^{(N_1)}_{k+l,k+l+q}
g^{(N_2)}_{p-l,p-l-q} ~
\mathcal{K}_{N_1,N_2}^{(PP)}(q;3\mu,-\mu, k+l,p-l)
    \mathsf{T}^{\sigma_1 \sigma_2}_{\alpha \beta} 
 \lambda^{\spmqty{\bar N_1 & \bar N_2 \\  N_4 &  N_3};\spmqty{\alpha & \beta \\ \sigma_4 & \sigma_3}}_{\spmqty{k+l+q  & p-l-q \\ k & p }} 
\Biggr]
\end{aligned}
\label{eq:Gamma_lambda1_PP}
\end{equation}
and 
\begin{equation}
\begin{aligned}
 {\bf \Gamma}_{(1)PH; \spmqty{ k+l &  p- l \\  k &  p}}^{\spmqty{N_1 & N_2 \\ N_4 & N_3};\spmqty{\sigma_1 & \sigma_2 \\ \sigma_4 & \sigma_3}}
 = 
\frac{1}{4 \mu N_f} \int  \frac{\dd q}{2\pi}  &
\Biggl[
g^{(\bar N_4)}_{k+q,k}
g^{(N_1)}_{k+l,k+l+q} 
~\mathcal{K}_{N_1,N_4}^{(PH)}(q;3\mu,\mu,k+l,k)
    \mathsf{T}^{\alpha \sigma_1}_{\sigma_4 \beta} 
	  \lambda^{\spmqty{\bar N_1 & N_2 \\  \bar N_4 &  N_3};\spmqty{\beta & \sigma_2 \\ \alpha & \sigma_3}}_{\spmqty{k+l+q  & p-l \\ k+q & p }} \\
&
+ 
%
g^{(\bar N_3)}_{p+q,p}
g^{(N_2)}_{p-l,p-l+q} 
~\mathcal{K}_{N_2,N_3}^{(PH)}(q; -\mu, \mu,p-l, p )
    \mathsf{T}^{\alpha \sigma_2}_{\sigma_3 \beta} 
	 	 \lambda^{\spmqty{N_1 & \bar N_2 \\  N_4 &  \bar N_3};\spmqty{\sigma_1 & \beta \\ \sigma_4 & \alpha}}_{\spmqty{k+l  & p-l+q \\ k & p+q }} \\
&
+ 
g^{(\bar N_4)}_{k+q,k}
g^{(N_2)}_{p-l,p-l+q} 
~\mathcal{K}_{N_2,N_4}^{(PH)}(q; -\mu, \mu,p-l, k )
    \mathsf{T}^{\alpha \sigma_2}_{\sigma_4 \beta} 
 \lambda^{\spmqty{N_1 & \bar N_2 \\  \bar N_4 &   N_3};\spmqty{\sigma_1 & \beta \\ \alpha & \sigma_3}}_{\spmqty{k+l  & p-l+q \\ k+q & p }} \\
&
+ 
g^{(\bar N_3)}_{p+q,p}
g^{(N_1)}_{k+l,k+l+q} 
~\mathcal{K}_{N_1,N_3}^{(PH)}(q; 3\mu, \mu, k+l, p )
    \mathsf{T}^{\alpha \sigma_1}_{\sigma_3 \beta} 
 \lambda^{\spmqty{\bar N_1 &  N_2 \\  N_4 &   \bar N_3};\spmqty{\beta & \sigma_2 \\ \sigma_4 & \alpha}}_{\spmqty{k+l+q  & p-l \\ k & p+q }}
 \Biggr].
\end{aligned}
\label{eq:Gamma_lambda1_PH}
\end{equation}
Here,
\bqa
\mathcal{K}_{N_a,N_b}^{(PP)}(q;k_{a,0},k_{b,0},k_a,k_b) =& \int \frac{\dd{q_\perp}\dd{q_0}}{(2\pi)^2}D({\bf q})G_{\bar{N_a}}({\bf k_a+q})G_{\bar{N_b}}({\bf k_b-q})
\label{eq:k1}
\eqa
is the kernel that determines 
the strength of mixing between 
a particle-particle pair with momenta $(k_a, k_b)$ in hot spots $(N_a,N_b)$
with a particle-particle pair with momenta $(k_a+q,k_b-q)$
in hot spots $(\bar N_a,\bar N_b)$.
$k_a$ and $k_b$ are the external momenta along the Fermi surface.
$k_{a,0}$ and $k_{b,0}$ are the external frequencies. 
$q$ ($q_\perp$) denotes the internal momentum that becomes parallel (perpendicular) to the Fermi surface 
in the small $v$ limit. 
Similarly,
\bqa
\mathcal{K}_{N_a,N_b}^{(PH)}(q;k_{a,0},k_{b,0},k_a,k_b) =& \int \frac{\dd{q_\perp}\dd{q_0}}{(2\pi)^2}D({\bf q})G_{\bar N_a}({\bf k_a+q})G_{\bar{N_b}}({\bf k_b+q})
\label{eq:k2}
\eqa
determines the strength of 
mixing between 
a particle-hole pair with momenta $(k_a, k_b)$ in hot spots $(N_a,N_b)$
with a particle-hole pair with momenta $(k_a+q,k_b+q)$
in hot spots $(\bar N_a,\bar N_b)$.
%

\subsubsection{Group 1}


In group 1, the primary couplings generated from the spin fluctuations takes the form of 
\eq{eq:SofIS1}.
The vertex correction that generates the secondary couplings exhibits IR singularity within the extended space of IR singularity only in the PH channel.
In particular, only the last two terms
in Eq. \eqref{eq:Gamma_lambda1_PH}
exhibit IR singularity for 
$\left\{
\lambda^{\spmqty{1 & 1 \\ 1 & 1}}_{\spmqty{0 & k \\ k & 0}},
\lambda^{\spmqty{1 & 1 \\ 1 & 1}}_{\spmqty{k & 0 \\ 0 & k}}
\right\}$
at general $k$.
This is because the primary coupling has zero total momentum 
only in two of the four PH channels at general $k$.
While we only need 
$\mathcal{K}^{(PH)}_{ 1, 1 }(q;k_{a,0},k_{b,0},k,p)$
at $p=k$ for the last two terms in Eq. \eqref{eq:Gamma_lambda1_PH},
let us compute it for general $k$ and $p$ 
to see how the vertex correction dies out at low energies
when $k$ deviates from $p$.
In this case, $q=q_x$ 
and $q_\perp=q_y$.
The kernel associated with a particle-hole pair at momenta $k$ and $p$ reads
(see Figs. 
\ref{fig:ph1}-\ref{fig:ph4})
\begin{equation}
\begin{aligned}
\mathcal{K}^{(PH)}_{ 1, 1 }(q;3\mu,\mu,k,p) = & \int \frac{\dd{q_0}\dd{q_\perp}}{(2\pi)^2}\frac{1}{\abs{q_0}+c (\abs{q} + \abs{q_\perp})}\frac{1}{i (q_0 + 3\mu) + V^{(4)}_{F,k+q}(v^{(4)}_{k+q}q - q_\perp +e_{4}[\vec{k};v^{(4)}_{k+q}])} \times
\\ & 
\frac{1}{i (q_0 + \mu) + 
V^{(4)}_{F,p+q}(v^{(4)}_{p+q}q - q_\perp + e_{4}[\vec{p};v^{(4)}_{p+q}])}.
\end{aligned}
\end{equation}
From \eq{eq:vVgN_vVg},
we can write
$V^{(4)}_{F,p} = V_{F,p}$
and $v^{(4)}_{p} = v_{p}$. 
Since we are interested in the kernel
near the space of IR singularity 
with $p=k$, 
we set
$V_{F,k+q}=V_{F,p+q}$ and $v_{k+q}=v_{p+q}$
in the integrand. 
We shift the internal momenta as
$q_0 \to q_0 -2\mu$,
$q_\perp \to q_\perp + v_{k+q} q +
 \frac{1}{2}\left(e_4[\vec{k};v^{(4)}_{k+q}] + e_4[\vec{p};v^{(4)}_{p+q}]\right)$
and 
drop $c q_\perp$ from the boson propagator.
The $q_\perp$ integration gives
\begin{equation}
\begin{aligned}
\mathcal{K}^{(PH)}_{ 1, 1 }(q;3\mu,\mu,k,p) = & 
\int \frac{\dd{q_0}}{2\pi}
\frac{1}{4 V_{F,k+q}}
\frac{\text{sgn}(q_0+\mu)-\text{sgn}(q_0-\mu)}{\abs{q_0-2\mu} + c\abs{q} + \abs{c v_{k+q}q + \Delta_{1,\{1,1\}}}}\frac{1}{\mu -
i V_{F,k+q} \Delta_{2,\{1,1\}}},
\label{eq:A100}
\end{aligned}
\end{equation}
where 
\begin{align}
\Delta_{1,\{1,1\}}(q;k,p;v) = \frac{c}{2}\left(e_4[\vec{k};v^{(4)}_{k+q}] + e_4[\vec{p};v^{(4)}_{p+q}]\right), ~~&
\Delta_{2,\{1,1\}}(q;k,p;v) = \frac{1}{2}\left(e_4[\vec{k};v^{(4)}_{k+q}] - e_4[\vec{p};v^{(4)}_{p+q}]\right).
\end{align}
The integration over the frequency results in
\begin{equation}
\mathcal{K}^{(PH)}_{ 1, 1 }(q;3\mu,\mu,k,p) 
= \frac{\log\left(1+\frac{2\mu}{\mu + c\abs{q} + \abs{c v_{k+q}q + \Delta_{1,\{1,1\}}}}\right)}{
4\pi V_{F,k+q}
\left(\mu 
-  i  V_{F,k+q} \Delta_{2,\{1,1\}} \right)}.
\label{eq:knotilde11}
\end{equation}
The subsequent integration over momentum $q$ along the Fermi surface would give rise to a logarithmic IR divergence at $k=p=0$.
A simple local counter term that removes the IR divergent part of the quantum correction can be obtained
by regulating the non-analyticity in the external momenta and replacing $q_0$ with $\mu$ inside the boson propagator in \eq{eq:A100} as
\begin{equation}
\tilde{\mathcal{K}}^{(PH)}_{ 1, 1 }(q;k,p) = 
\frac{1}{2 \pi V_{F,k+q}}
\frac{\mu}{\mu 
-  i  V_{F,k+q} \Delta_{2,\{1,1\}} }
\frac{1}{\mu + c\abs{q}_\mu + 
c \abs{ v_{k+q} q + \frac{1}{c}\Delta_{1,\{1,1\}} }_\mu 
},
\label{eq:ktilde11}
\end{equation}
where $\abs{q}_\mu=\sqrt{q^2+\mu^2}$
smears the non-analyticity of the boson propagator.
One can explicitly check that 
\eq{eq:ktilde11} removes all IR singularity of the quantum correction
\footnote{
To the leading order in the small $v$ limit,
the difference between
\eq{eq:knotilde11} and \eq{eq:ktilde11}
is given by
\begin{equation}
\int\frac{\dd q}{2\pi}\left(\mathcal{K}^{(PH)}_{ 1, 1 }(q;3\mu,\mu,k,p)-\tilde{\mathcal{K}}^{(PH)}_{ 1, 1 }(q;k,p)\right) = \frac{\mu - (3\mu +\abs{\Delta_{1,\{1,1\}}})\text{arctanh}\left(\frac{\mu}{2\mu +\abs{\Delta_{1,\{1,1\}}}}\right)}{2c\pi^2V_{F,k}\left(\mu - i \Delta_{2,\{1,1\}} V_{F,k} \right)},
\label{eq:dvktilde11_pre}
\end{equation}
where the momentum dependence of the coupling functions are ignored,
and $c v_{k} q$ is dropped in the boson propagator. 
}. 
The contribution to the beta functional is given by the derivative 
of Eq. \eqref{eq:ktilde11} 
with respect to $\log \mu$\footnote{In principle, the $\log\mu$ derivative affects the $\abs{q}_\mu$ terms, but these result in sub leading contributions, which we ignore.},
\begin{equation}
\frac{\partial\tilde{\mathcal{K}}^{(PH)}_{ 1, 1 }(q;k,p)}{\partial\log\mu} = -\frac{1}{ 2 \pi  V_{F,k+q}}\frac{\mu\left[\mu^2 + i \Delta_{2,\{1,1\}} V_{F,k+q} \left(c\abs{q}_\mu + c \abs{ v_{k+q} q + \frac{1}{c}\Delta_{1,\{1,1\}} }_\mu \right)\right]}{\left(\mu - i \Delta_{2,\{1,1\}} V_{F,k+q} \right)^2 \left(\mu + c\abs{q}_\mu +  c \abs{ v_{k+q} q + \frac{1}{c}\Delta_{1,\{1,1\}} }_\mu \right)^2}.
\label{eq:dvktilde11}
\end{equation}
\eq{eq:dvktilde11} vanishes
in the small $\mu$ limit
unless 
$q$, $\Delta_{2,\{1,1\}}$ and
$\Delta_{1,\{1,1\}}$ 
all vanish.
This implies that 
(1) the vertex correction 
is non-zero only when the pair of external fermions scattered by the critical boson through quantum fluctuations are at the hot spots
and (2) the non-zero vertex correction 
arises  from the boson 
with small momenta near $q=0$.
We can use 
Eqs.  (\ref{eq:dvktilde11_pre}) and (\ref{eq:dvktilde11})
for the counter terms that cancel singular parts of 
quantum corrections  with different choices of frequencies 
that are order of $\mu$ in \eq{eq:Gamma_lambda1_PH}.



The vertex correction in the particle-hole channel in Fig. 
\ref{fig:4f1}
is given by the last two terms of Eq. \eqref{eq:Gamma_lambda1_PH}.
From Eq. \eqref{eq:ktilde11}, we can directly write the counter term,
\begin{equation}
\begin{aligned}
& {\bf \tilde\Gamma}_{CT;(1)PH;\spmqty{k+l & k\\ k &  k+l}}^{\spmqty{1 & 1 \\ 1 & 1};\spmqty{\sigma_1 & \sigma_2 \\ \sigma_4 & \sigma_3}}
= -\frac{1}{8 \pi N_f} \int d\rho(q)
\left[
\frac{\mathsf{D}_\mu(k;q)}{\mu}
    \mathsf{T}^{\alpha \sigma_2}_{\sigma_4 \beta} 
 \lambda^{\spmqty{1 & 4 \\  4 &   1};\spmqty{\sigma_1 & \beta \\ \alpha & \sigma_3}}_{\spmqty{k+l  & q \\ q & k+l }}
+ \frac{\mathsf{D}_\mu(k+l;q)}{\mu}
  \mathsf{T}^{\sigma_1 \alpha}_{\beta \sigma_3} 
 \lambda^{\spmqty{4 & 1 \\ 1 &   4};\spmqty{\beta & \sigma_2 \\ \sigma_4 & \alpha}}_{\spmqty{q  & k \\ k & q }}
%
%
\right],
\end{aligned}
\end{equation}
where we have shifted $q_x \to q_x - k_x$ and $q_x \to q_x -k_x - l_x$, respectively. 
The contribution to the beta functional is given by the $\log\mu$ derivative of the counter term\footnote{To the leading order, fixing $\lambda_B$ is equivalent to fixing $\lambda/\mu$ under the $\log\mu$ derivative.},
\begin{equation}
\begin{aligned}
4\mu\frac{\partial}{\partial\log\mu}{\bf \tilde\Gamma}_{CT;(1)PH;\spmqty{k+l & k\\ k &  k+l}}^{\spmqty{1 & 1 \\ 1 & 1};\spmqty{\sigma_1 & \sigma_2 \\ \sigma_4 & \sigma_3}}
= &
\frac{1}{2 \pi N_f }
\int d\rho(q) 
\left[
\frac{\mathsf{D}_\mu(k;q)^2}{g_{q,k}^2}
\mathsf{T}^{\alpha \sigma_2}_{\sigma_4 \beta} 
\lambda^{\spmqty{1 & 4 \\  4 &   1};\spmqty{\sigma_1 & \beta \\ \alpha & \sigma_3}}_{\spmqty{k+l  & q \\ q & k+l }}
+ 
\frac{\mathsf{D}_\mu(k+l;q)^2}{g_{k+l,q}^2}
\mathsf{T}^{\sigma_1 \alpha}_{ \beta \sigma_3}
\lambda^{\spmqty{4 & 1 \\ 1 &   4};\spmqty{\beta & \sigma_2 \\ \sigma_4 & \alpha}}_{\spmqty{q  & k \\ k & q }}
\right].
\end{aligned}
\label{eq:gamma1111_CT_PH1}
\end{equation}

\subsubsection{Group 2}

In group 2, the primary couplings generated from the spin fluctuations are given by \eq{eq:SofIS2}.
The vertex correction that is linear in the four-fermion coupling further generates the secondary couplings in the PP channel through Figs. \ref{fig:pp1}-
\ref{fig:pp2}.
The relevant expression for the quantum correction is in Eq. \eqref{eq:Gamma_lambda1_PP}.
The kernel that goes into the quantum correction 
in the PP channel 
is written as
\begin{equation}
\begin{aligned}
\mathcal{K}^{(PP)}_{ 1, 5 }(q;3\mu,-\mu,k,p) = & \int \frac{\dd{q_0}\dd{q_{\perp}}}{(2\pi)^2}\frac{1}{\abs{q_0}+c (\abs{q} + \abs{q_{\perp}})}\frac{1}{i (q_0 + 3\mu) + V^{(4)}_{F,k+q}(v^{(4)}_{k+q}q - q_{\perp}	 +e_{4}[\vec{k};v^{(4)}_{k+q}])}  \times
\\ &\frac{1}{i (- q_0 - \mu) + V^{(8)}_{F,p-q}(v^{(8)}_{p-q}q - q_{\perp} + e_{8}[\vec{p};v^{(8)}_{p-q}])}.
\end{aligned}
\end{equation}
Shifting $q_0$ by $- 2\mu$
and $q_\perp$ by $ v_{k+q}q + \Delta_{1\{ 1, 5 \}}(q;k,p;v)/c$,
and dropping $c q_\perp$ in the boson propagator in the small $c$ limit leads to
\begin{equation}
\begin{aligned}
 \mathcal{K}^{(PP)}_{ 1, 5 }(q;3\mu,-\mu,k,p)
 = & 
\int \frac{\dd{q_0}\dd{q_\perp}}{(2\pi)^2}\frac{1}{\abs{q_0-2\mu} + c\abs{q} +\abs{c v_{k+q}q + \Delta_{1,\{ 1, 5 \} }}}\frac{1}{i (q_0 + \mu) + V_{F,k+q}(- q_\perp + \Delta_{2, \{1, 5\} })} \times
\\ & \frac{1}{i (- q_0 + \mu) + V_{F,-p+q}((v_{-p+q}-v_{k+q})q - q_\perp - \Delta_{2,\{1, 5\}})},
\end{aligned}
\label{eq:k1ppoutintegral}
\end{equation}
where
\bqa
\Delta_{1,\{1,5\}}(q;k,p;v) = \frac{c}{2}\left(e_{4}[\vec{k};v^{(4)}_{k+q}] + e_{8}[\vec{p};v^{(8)}_{p-q}]\right), \quad \Delta_{2,\{1,5\}}(q;k,p;v)=\frac{1}{2}\left(e_4[\vec{k};v^{(4)}_{k+q}] - e_8[\vec{p};v^{(8)}_{p-q}]\right).
\eqa 
Integrating $q_\perp$ gives 
\begin{equation}
\begin{aligned}
& \mathcal{K}^{(PP)}_{ 1, 5 }(q;3\mu,-\mu,k,p) 
 =  \int \frac{\dd{q_0}}{2\pi} \frac{1}{\abs{q_0-2\mu} + c\abs{q} + \abs{c v_{k+q}q+\Delta_{1,\{ 1, 5 \} }}}  \times
\\ & \frac{\frac{1}{2}\left( \text{sgn}(q_0 + \mu) + \text{sgn}(q_0 - \mu) \right)}{(V_{F,k+q}+V_{F,-p+q})q_0-(V_{F,k+q}-V_{F,-p+q})\mu - i V_{F,k+q}V_{F,-p+q}((v_{k+q}-v_{-p+q})q + 2\Delta_{2,\{1, 5 \}})}.
\end{aligned}
\end{equation}
Near the space of IR singularity 
with $k = -  p$,
we can set 
$V_{F,k+q}=V_{F,-p+q}$
and $v_{k+q}=v_{-p+q}$ 
to simplify the above expression.
To construct a simple local and analytic counter term 
that removes the IR divergence,
we can drop $2 \mu$ in the boson propagator and 
regularize the non-analyticity in  the external momenta  as
\begin{equation}
\begin{aligned}
& \tilde{\mathcal{K}}^{(PP)}_{ 1, 5 }(q;k,p)
= \int \frac{\dd{q_0}}{2\pi} \frac{1}{\abs{q_0} + c\abs{q}_\mu + c\abs{ v_{k+q}q+\frac{1}{c}\Delta_{1,\{ 1, 5 \} }}_\mu}\frac{\frac{1}{2}\left( \text{sgn}(q_0 + \mu) + \text{sgn}(q_0 - \mu) \right)}{2 V_{F,k+q} q_0 - 2 i V_{F,k+q}^2\Delta_{2,\{1, 5 \}}}.
\end{aligned}
\label{eq:K1tilde}
\end{equation}
Since $\tilde{\mathcal{K}}^{(pp)}_{ 1, 5 }$ differs from
${\mathcal{K}}^{(pp)}_{ 1, 5 }$
only by non-singular terms\footnote{
To the leading order in $v$,
the difference is given by
\begin{equation}
\begin{aligned}
& (\mathcal{K}^{(PP)}_{ 1, 5 }(q;3\mu,-\mu,k,p)
-  \tilde {\mathcal{K}}^{(PP)}_{ 1, 5 }(q;k,p))
\\ & = 
\int \frac{dq_0}{2\pi}
\frac{\frac{1}{2}(\abs{q_0} - \abs{q_0 - 2\mu})\left( \text{sgn}(q_0 + \mu) + \text{sgn}(q_0 - \mu) \right)}{\left\lbrace\begin{gathered}\left[(V_{F,k+q}+V_{F,-p+q})q_0 - 2 i V_{F,k+q}V_{F,-p+q}\Delta_{2,\{1, 5 \}}\right]
\\ \times (\abs{q_0 - 2\mu} + c\abs{q} + \abs{c v_{k+q}q+\Delta_{1,\{ 1, 5 \} }})(\abs{q_0} + c\abs{q} + \abs{c v_{k+q}q+\Delta_{1,\{ 1, 5 \} }})\end{gathered}\right\rbrace}.
\end{aligned}
\end{equation}
Here $\Delta_2$ only makes the magnitude of the integrand strictly larger, thus we can set $\Delta_2 = 0$. For coupling functions that are weakly momentum dependent,
one can perform the $q$ integration by dropping $c v_{k+q}q$ for $v\ll c\ll 1$
to obtain
\begin{equation}
\int\frac{\dd q}{2\pi} (\mathcal{K}^{(PP)}_{ 1, 5 }(q;3\mu,-\mu,k,p)-\tilde {\mathcal{K}}^{(PP)}_{ 1, 5 }(q;k,p)) =
\left\{ \begin{array}{lcl}
 \frac{1-\log4}{\pi^2 c (V_{F,k} + V_{F,-p})}\frac{\mu}{\abs{\Delta_{1,\{1,5\}}}} + O(\mu^2) &
\mbox{for} & \Delta_1 \neq 0, \Delta_2 = 0 \\
 \frac{\pi^2 - 3 \left(2\log^2(2) + \mathrm{Li}_2(1/4)\right)}{12 c \pi^2 (V_{F,k} + V_{F,-p})} & \mbox{for}&
\Delta_1 = 0, \Delta_2 = 0
\end{array}
\right. .
\label{eq:gammaSlimit3}
\end{equation}
},
we can use \eq{eq:K1tilde} in the counter term. After $q_0$ integration, the kernel becomes
\begin{equation}
\begin{aligned}
& \tilde{\mathcal{K}}^{(PP)}_{1,5}(q;k,p)
 = \frac{\left\lbrace\begin{gathered}
2\Delta_{2;\{1,5\}}V_{F,k+q}
\arctan \left( 
\frac{\Delta_{2;\{1,5\}}V_{F,k+q}}{\mu}
\right) 
\\-\left( c \abs{ v_{k+q}q_x+\Delta_{1,\{1,5\}}/c}_\mu +c\abs{q_x}_\mu \right)\log\left( \frac{\mu^2+ 
\left(\Delta_{2;\{1,5\}} V_{F,k+q}\right)^2}{\left( c \abs{ v_{k+q} q_x+\Delta_{1;\{1,5\}}/c}_\mu +c\abs{q_x}_\mu +\mu\right)^2}\right)
\end{gathered}\right\rbrace}{4\pi V_{F,k+q}
\left[ \left( c\abs{q}_\mu +c\abs{v_{k+q}q + \Delta_{1,\{1,5\} }/c }_\mu \right)^2
+ \left(\Delta_{2;\{1,5\}} V_{F,k+q}\right)^2 \right]
}.
\end{aligned}
\label{eq:ktilde15result}
\end{equation}
The contribution to the beta functional is given by the derivative of the kernel with respect to $\log \mu$,
\begin{equation}
\frac{\partial\tilde{\mathcal{K}}^{(PP)}_{1,5}(q;k,p)}{\partial\log\mu} = -\frac{1}{2 \pi V_{F,k+q}}\frac{\mu^2}{\left( \Delta_{2, \{1, 5 \}} V_{F,k+q} \right)^2 + \mu^2}\frac{1}{\mu + c \abs{q}_\mu + c \abs{ v_{k+q}q +\Delta_{1, \{1, 5 \} }/c}_\mu}.
\label{eq:dvktilde15}
\end{equation}
In the small $\mu$ limit,
\eq{eq:dvktilde15} remains non-zero 
as far as $\Delta_{2, \{1, 5 \}}=0$.
This implies that 
(1) the vertex corrections in the pairing channel 
remains important at low energies
irrespective of the relative momentum of Cooper pairs,
and
(2) the singular vertex correction arises from bosons with all momenta.
In particular, even a high-energy boson creates a singular vertex correction by scatterings Cooper pairs along the Fermi surface with large momentum transfers.




Using Eq. \eqref{eq:ktilde15result},
we can directly write the counter term for the quantum correction in 
Eq. \eqref{eq:Gamma_lambda1_PP}.
From \eqref{eq:dvktilde15},
the contribution to the beta functional is obtained to be
\begin{equation}
\begin{aligned}
4\mu\frac{ \partial}{\partial\log\mu} \mathbf{\tilde{\Gamma}}^{\spmqty{1 & 5 \\ 1 & 5};\spmqty{\sigma_1 & \sigma_2 \\ \sigma_4 & \sigma_3}}_{CT;(1) PP;\spmqty{k+l & -k-l\\ k &  -k}} = &
\frac{1}{2\pi N_f}
\int d\rho(q)\left[
\mathsf{D}_\mu(k+l;q)
\mathsf{T}^{\sigma_1 \sigma_2}_{\alpha \beta}\lambda^{\spmqty{4 & 8 \\ 1 & 5}, \spmqty{\alpha & \beta \\ \sigma_4 & \sigma_3}}_{\spmqty{q & -q \\ k & -k}}
+ \mathsf{D}_\mu(q;k)
\lambda^{\spmqty{1 & 5 \\ 4 & 8}, \spmqty{\sigma_1 & \sigma_2 \\ \alpha & \beta}}_{\spmqty{k+l & -k-l \\ q & -q}}\mathsf{T}^{\alpha  \beta}_{\sigma_4 \sigma_3}\right],
\end{aligned}
\end{equation}
where we have shifted $q_x \to q_x -k_x-l_x$ and $q_x \to q_x -k_x$ in the respective terms.

\subsection{BCS processes}

The vertex corrections quadratic in the four-fermion coupling 
(Eqs. \eqref{eq:Gamma_lambda2_PP0} - \eqref{eq:Gamma_lambda2_PH0}
for Fig. \ref{fig:4f2}) can be written as
\begin{align}
 \label{eq:Gamma_lambda2_PP}
& {\bf \Gamma}_{(2)PP;\spmqty{ k+l & p-l \\ k & p}  }^{\spmqty{N_1 & N_2 \\ N_4 & N_3};\spmqty{\sigma_1 & \sigma_2 \\ \sigma_4 & \sigma_3}}
=  
-\frac{1}{8 \mu^2 }
\int \FR{\dd q}{2\pi}
\mc{Q}^{(PP)}_{M_1M_2}(q;\mu,\mu,k,p)
\lambda^{\spmqty{N_1 & N_2 \\ M_1 & M_2};\spmqty{\sigma_1 & \sigma_2 \\ \beta & \alpha}}_{\spmqty{k+l & p-l \\ k+q & p-q}}
\lambda^{\spmqty{M_1 & M_2 \\ N_4 & N_3};\spmqty{\beta & \alpha \\ \sigma_4 & \sigma_3}}_{\spmqty{k+q & p-q \\ k & p}}, 
\end{align}
and
\begin{equation}
\begin{aligned}
{\bf \Gamma}_{(2)PH;\spmqty{k+l & p-l \\ k & p}}^{\spmqty{N_1 & N_2 \\ N_4 & N_3};\spmqty{\sigma_1 & \sigma_2 \\ \sigma_4 & \sigma_3}}
=&
- \frac{1}{8 \mu^2 }
\int \FR{\dd q}{2\pi}
\Biggl[
\mc{Q}^{(PH)}_{M_1M_2}(q;-2\mu,0,-l,0)
\Biggl(
- N_f 
 \lambda^{\spmqty{N_1 & M_1 \\ N_4 & M_2};\spmqty{\sigma_1 & \alpha \\ \sigma_4 & \beta}}_{\spmqty{k+l & q-l \\ k & q}}
 \lambda^{\spmqty{M_2 & N_2 \\ M_1 & N_3};\spmqty{\beta & \sigma_2 \\ \alpha & \sigma_3}}_{\spmqty{q & p-l \\ q-l & p}}  \\
 & 
 + 
 \lambda^{\spmqty{N_1 & M_1 \\ N_4 & M_2};\spmqty{\sigma_1 & \alpha \\ \sigma_4 & \beta}}_{\spmqty{k+l & q-l \\ k & q}}
 \lambda^{\spmqty{M_2 & N_2 \\ N_3 & M_1};\spmqty{\beta & \sigma_2 \\ \sigma_3 & \alpha}}_{\spmqty{q & p-l \\ p & q-l}} 
  + 
   \lambda^{\spmqty{N_1 &  M_1 \\  M_2 &   N_4};\spmqty{\sigma_1 & \alpha \\ \beta & \sigma_4}}_{\spmqty{k+l  & q-l \\ q & k }}
 \lambda^{\spmqty{M_2 & N_2 \\  M_1 &   N_3};\spmqty{\beta & \sigma_2 \\ \alpha & \sigma_3}}_{\spmqty{q  & p-l \\ q-l & p }}
\Biggr) \\
&
+
\mc{Q}^{(PH)}_{M_1M_2}(q;-2\mu,0,p-l-k,0)
 \lambda^{\spmqty{N_1 & M_1 \\ M_2 & N_3};\spmqty{\sigma_1 & \alpha \\ \beta & \sigma_3} }_{\spmqty{k+l & q+p-l-k \\ q & p}}
 \lambda^{\spmqty{M_2 & N_2 \\ N_4 & M_1};\spmqty{\beta & \sigma_2 \\ \sigma_4 & \alpha}}_{\spmqty{q & p-l \\ k & q+p-l-k}}  
 \Biggr],
\end{aligned}
\label{eq:Gamma_lambda2_PH}
\end{equation}
where 
\begin{subequations}
\begin{align}
\mc{Q}^{(PP)}_{N_1N_2}(q;k_{a,0},k_{b,0},k_a,k_b) &= \int \FR{\dd q_0 \dd q_\perp}{(2\pi)^2}G_{N_1}({\bf k_a+q})G_{N_2}({\bf k_b-q}), 
\label{eq:Qkernel1}
\\
\mc{Q}^{(PH)}_{N_1N_2}(q;k_{a,0},k_{b,0},k_a,k_b) &= \int \FR{\dd q_0 \dd q_\perp}{(2\pi)^2}G_{N_1}({\bf k_a+q})G_{N_2}({\bf k_b+q})
\label{eq:Qkernel}
\end{align}
\end{subequations}
are the kernels that determine the strength of
the operator mixing in which two four-fermion operators
'fuse' into one four-fermion operator
as a function of momentum along the Fermi surface 
and frequencies.
For a generic shape of Fermi surface,
the only kernel that produces an IR singularity in an extended space of external momenta is in the PP channel,
\begin{equation}
\begin{aligned}
\mc{Q}^{(PP)}_{15}(q;\mu,\mu,k,p) 
&=
\int \FR{1}{
\Bigl[ i(\mu+q_0) + V^{(1)}_{F,k+q}e_1[\vec{k}+\vec{q};v^{(1)}_{k+q}] \Bigr]
\Bigl[ i(\mu-q_0) + V^{(5)}_{F,p-q}e_5[\vec{p}-\vec{q};v^{(5)}_{p-q}] \Bigr]
}
\FR{\dd q_\perp}{2\pi}
\FR{\dd q_0}{2\pi}.
\end{aligned}
\end{equation}
This is singular when the center of mass momentum is zero.
For $\vec p = - \vec k$,  
the kernel becomes
\begin{equation}
\begin{aligned}
&\mc{Q}^{(PP)}_{15}(q;\mu,\mu,k,-k) 
=\FR{1}{2\pi V_{F,k+q}} \log\FR{\Lambda}{\mu}.
\end{aligned}
\label{eq:Q15result}
\end{equation}
It is noted that this is IR divergent irrespective of $k$ as far as 
$p=-k$.

\newcommand{\VF}[1]{V_{F,{#1}}}


Using the expressions above,
we write down the quantum corrections that are quadratic in $\lambda$ explicitly.
With the help of Eq. \eqref{eq:Q15result} we obtain the counter term,
\begin{equation}
\begin{aligned}
{\bf \tilde\Gamma}_{CT;(2)PP;\spmqty{k+l & -k-l\\ k &  -k}}^{\spmqty{1 & 5 \\ 1 & 5}; \spmqty{\sigma_1 & \sigma_2 \\ \sigma_4 & \sigma_3}}
=  
&
\frac{1}{8 \mu^2 }
\int \FR{\dd q}{2\pi}
\FR{1}{2\pi \VF{q}}\log\left(\FR{\Lambda}{\mu}\right)
\left[
\lambda^{\spmqty{1 & 5\\ 1 & 5};\spmqty{\sigma_1 & \sigma_2 \\ \beta & \alpha}}_{\spmqty{k+l & -k-l \\ q & -q}}
\lambda^{\spmqty{1 & 5 \\ 1 & 5};\spmqty{\beta & \alpha \\ \sigma_4 & \sigma_3}}_{\spmqty{q & -q \\ k & -k}} 
+
\lambda^{\spmqty{1 & 5\\ 4 & 8};\spmqty{\sigma_1 & \sigma_2 \\ \beta & \alpha}}_{\spmqty{k+l & -k-l \\ q &-q}}
\lambda^{\spmqty{4 & 8 \\ 1 & 5};\spmqty{\beta & \alpha \\ \sigma_4 & \sigma_3}}_{\spmqty{q & -q \\ k & -k}} 
\right],
\end{aligned}
\end{equation}
where the internal momentum is shifted by $-k$.
The contribution to the beta functional becomes
\begin{equation}
\begin{aligned}
4\mu\frac{\partial}{\partial\log\mu} {\bf \tilde{\Gamma}}_{CT;(2)PP;\spmqty{k+l & -k-l\\ k &  -k}}^{\spmqty{1 & 5 \\ 1 & 5}; \spmqty{\sigma_1 & \sigma_2 \\ \sigma_4 & \sigma_3}}
=  
&
-\frac{1}{4 \pi }
\int 
d\rho(q)
\left[
\lambda^{\spmqty{1 & 5\\ 1 & 5};\spmqty{\sigma_1 & \sigma_2 \\ \beta & \alpha}}_{\spmqty{k+l & -k-l \\ q & -q}}
\lambda^{\spmqty{1 & 5 \\ 1 & 5};\spmqty{\beta & \alpha \\ \sigma_4 & \sigma_3}}_{\spmqty{q & -q \\ k & -k}} 
+ \lambda^{\spmqty{1 & 5\\ 4 & 8};\spmqty{\sigma_1 & \sigma_2 \\ \beta & \alpha}}_{\spmqty{k+l & -k-l \\ q &-q}}
\lambda^{\spmqty{4 & 8 \\ 1 & 5};\spmqty{\beta & \alpha \\ \sigma_4 & \sigma_3}}_{\spmqty{q & -q \\ k & -k}}
\right].
\end{aligned}
\end{equation}
\newcommand{\upsA}{\Upsilon^{(a)}(x,x')}
\newcommand{\UpsA}{\Upsilon^{(0)}}
\newcommand{\UpsB}{\Upsilon^{(1)}}
\newcommand{\UpsC}{\Upsilon^{(2)}}

\newcommand{\vxltz}    {v_k(\ellthreezerokkp)}
\newcommand{\vxpltz}   {v_k(\ellthreezerokkp)}
\newcommand{\vxlox}    {v_k(\ellonek)}
\newcommand{\vxploxp}  {v_{k'}(\ellonekp)}
\newcommand{\vzlox}    {v_0(\ellonek)}
\newcommand{\vzploxp}  {v_0(\ellonekp)}
\newcommand{\vxpploxp} {v_0(\ellthreezerokkp)}

\newcommand{\vrm}{\mathsf{v}_0}

\newcommand{\msL}{\mathscr{L}}

\section{RG Flow of
the nesting angle, Fermi velocity and Yukawa coupling functions
}
\label{appendixA_SingleParticle}

\newcommand{\newellthreezerokkp}{\ell^{(3,0)}_{k',k}}
\newcommand{\newellonek}        {\ell^{(2L)}_{k}}
\newcommand{\newellonekp}       {\ell^{(2L)}_{k'}}

\newcommand{\newelltwokf}       {L^{(1L)}(k;\ell)}
\newcommand{\newellonekf}       {L^{(2L)}(k;\ell)}

\newcommand{\newelltwok}        {\ell^{(1L)}_k}
\newcommand{\newelltwokp}       {\ell^{(1L)}_{k'}}
\newcommand{\newelltwokk}       {\ell^{(1L)}_{k,k}}

\newcommand{\newellthreekkp}    {\ell^{(1L)}_{k',k}}
\newcommand{\newellTwoLoopX}{\ell^{(2L)}_k}
\newcommand{\newellOneLoopX}{\ell^{(1L)}_k}


\subsection{Diagonal Coupling functions}
 \label{sec:appendixDiagonalFlow}

We can solve the beta functionals for the diagonal couplings, 
$\{ v_k, V_{F,k}, g_k \equiv g_{k,k} \}$ 
because Eqs. (\ref{betaVgen1})-(\ref{mainTextBetaGgen1})
do not depend on the off-diagonal elements
of $g_{k,k'}$ with $k' \neq k$.
The momentum dependent flow of the diagonal coupling functions is controlled 
by three length scales 
$\ellTwoLoopX$,
$\ellOneLoopX$,
$\ellOneLoopXX$.
defined through 
\eq{SelfConsistentLengthScales}.
The length scales satisfy
$\ellTwoLoopX =  \ellOneLoopXX <  \ellOneLoopX$.
Since these logarithmic length scales 
depend on the scale dependent coupling functions, 
they need to be solved along with the beta functionals.
To be concrete, we consider a UV theory which has momentum dependent coupling functions at scale $\Lambda$ as is shown in \eq{eq:allIncond}.

\subsubsection{Short-distance Regime}

At length scales shorter 
than all crossover scales
($\ell < $
$\ellTwoLoopX$,
$\ellOneLoopX$,
$\ellOneLoopXX$),
the beta functionals become
\begin{align}
\FR{\di v_k(\ell)}{\di\ell}    
&= v_k(\ell) 
\Biggl[ 
        -\FR{4(N_c^2-1)}{\pi^3N_cN_f}\FR{g_k(\ell)^2}{V_{F,k}(\ell)^2}\log\PFR{V_{F,k}(\ell)}{c(\ell)} 
        - \FR{2(N_c^2-1)}{\pi^4N_c^2N_f^2}
        \FR{g_k(\ell)^4}{c(\ell)^2V_{F,k}(\ell)^2}
            \log^2\PFR{V_{F,k}(\ell)v_k(\ell)}{c(\ell)}
\Biggr], 
\label{betaVKhi}\\
\FR{\di V_{F,k}(\ell)}{\di\ell}  &= V_{F,k}(\ell)
\Biggl[ 
        \FR{2(N_c^2-1)}{\pi^3N_cN_f}\FR{g_k(\ell)^2}{V_{F,k}(\ell)^2}\log\PFR{V_{F,k}(\ell)}{c(\ell)}
        -\FR{N_c^2-1}{\pi^2N_cN_f}\FR{g_k(\ell)^2}{c(\ell)V_{F,k}(\ell)}
        -\frac{3}{2} \FR{N_c^2-1}{\pi^2N_cN_f}v_0(\ell)\log\PFR{1}{c(\ell)}
\nn &\hspace{2.0cm}
        +\FR{N_c^2-1}{2\pi N_cN_F}w_0(\ell)
        + \FR{(N_c^2-1)}{2\pi^4N_c^2N_f^2}
        \FR{g_k(\ell)^4}{c(\ell)^2V_{F,k}(\ell)^2}
            \log^2\PFR{V_{F,k}(\ell)v_k(\ell)}{c(\ell)}
\Biggr], 
\label{betaVFKhi}\\
\FR{\di g_k(\ell)}{\di\ell} &= g_k(\ell)
\Biggl[ 
        -\FR{1}{2\pi N_cN_f}w_0(\ell)\log\PFR{1}{w_0(\ell)}
        +\FR{N_c^2-1}{2\pi N_cN_f}w_0(\ell)
        -\FR{N_c^2-1}{\pi^2N_cN_f}v_0(\ell)\log\PFR{1}{v_0(\ell)} 
\nn &\hspace{1.0cm}
        -\FR{(N_c^2-1)g_k(\ell)^2}{\pi^2N_cN_fc(\ell)V_{F,k}(\ell) }
        +\FR{g_k(\ell)^2}{\pi^2N_cN_fV_{F,k}(\ell)c(\ell)}\log\PFR{c(\ell)}{V_{F,k}(\ell)v_k(\ell)}
            %
        \Biggr].
\label{betaGKKhi}
\end{align}
The solution to the beta functional reproduces 
the results of the hot spot theory\cite{SCHLIEF}, 
\begin{align}
        \label{appHiEnSoln1}
        v_k    (\ell) &= \FR{\pi^2 N_c N_f}{2(N_c^2-1)} \FR{1}{(\ell+\ell_0)\log(\ell+\ell_0)},\\
        \label{appHiEnSoln3}V_{F,k}(\ell) &= 1,\\
        \label{appHiEnSoln2}g_{k}  (\ell) &= \sqrt{\FR{\pi^3 N_c N_f}{4(N_c^2-1)} \FR{1}{(\ell+\ell_0)\log(\ell+\ell_0)}}.
\end{align}
The speed of the collective mode is given by \eq{eq:CofV},
\begin{align}
        \label{appCELL}
        c(\ell) = \FR{\pi}{4\sqrt{N_c^2-1}} \FR{1}{\sqrt{\ell+\ell_0}}.
\end{align}
Here $\ell_0$ is the parameter that
sets the nesting angle 
at the UV scale $\ell=0$.
In the limit that the nesting angle is small,
$\ell_0 \gg 1$.

\subsubsection{The crossover scales}

\label{sec:FirstCrossoverScale}

As the length scale increases,
the theory encounters the first crossover 
at $\ellTwoLoopX$.
\eq{SelfConsistentLengthScales} 
that determines the crossover scale reads
\begin{align}
  \ellTwoLoopX = \log\PFR{\Lambda}{4v_0(\ellTwoLoopX) k}.
  \label{app:e24}
\end{align}
For $\ell_0 \gg 1$,
$\log 1/v_0  (\ell) \approx
\log (\ell+\ell_0)$,
and
\eq{app:e24} can be written as
\begin{align}
  \ellTwoLoopX = 
 \log\PFR{\Lambda}{4v_0(0)k}
+ \log\PFR{\ell_0+\ellTwoLoopX}{\ell_0}.
\end{align}
To the leading order in the small $v$ limit, 
its solution is obtained to be

\noindent 
\begin{align}\label{eqA44}
\begin{split}
\ellTwoLoopX&=\log\left(\FR{\Lambda}{4v_0(0)k}\right)+\log \left[ \FR{\ell_0+
 \log\PFR{\Lambda}{4v_0(0)k}
 }{\ell_0} \right].
\end{split}
\end{align}



As $\ell$ increases further,
the theory encounters 
the second crossover length scale at
\begin{align}
\ellOneLoopX &= \log \FR{\Lambda}{2v_k(\ellOneLoopX)c(\ellOneLoopX)k}.
\label{eqAsceqelltwok}
\end{align}
It turns out that $\ellOneLoopX-\ellTwoLoopX$ is not large enough to generate 
any significant  flow of coupling functions 
within this window of length scales. 
%
To see this,
let us first assume that 
the logarithmic change of the coupling functions is negligible between the two length scales, 
\begin{align}
\label{eqAdelta}
|\delta^{(J)}_k| \equiv 
|\log J(\ellOneLoopX) - \log J(\ellTwoLoopX) | \ll 1
\end{align}
for all couplings 
$J = \{ g_k, v_k, V_{F,k} \}$.
The self-consistent equation in \eq{eqAsceqelltwok} can be written as
\begin{align}
\ellOneLoopX -  \ellTwoLoopX
&=
- \log c(\ellTwoLoopX) - \delta^{(v_k)}_k - \delta^{(c)}_k + \log 2
\approx \log\FR{2}{c(\ellTwoLoopX)}
\end{align}
for $c \ll 1$.
%
Since all beta functionals goes to zero in powers of $w$ as
\begin{align}
&\left|\FR{1}{v_k(\ell)}\FR{\di v_k(\ell)}{\di \ell}\right| \lesssim \mc O(\,v\log(1/c)\,), 
&
&\left|\FR{1}{V_{F,k}(\ell)}\FR{\di V_{F,k}(\ell)}{\di \ell}\right| \lesssim  \mc O(\,w\,),
&
&\left|\FR{1}{g_{k,k}(\ell)}\FR{\di g_{k,k'}(\ell)}{\di \ell}\right| \lesssim \mc 
O(\,w\log(1/w)\,),
\label{eqAflowlogV}
\end{align}
the change of the coupling that occurs 
in  $\ellTwoLoopX < \ell <  \ellOneLoopX$ is at most
\begin{align}
\left|\FR{J(\newellOneLoopX) - J(\newellonek)}{J(\newellonek)}\right| 
 \sim O(\,w\log(1/w) \log(1/c)\,)
\ll 1
\end{align}
for all couplings.
This justifies the assumption made in \eq{eqAdelta}.
Therefore, the change of couplings is negligible between 
$\ellTwoLoopX$ and $\ellOneLoopX$,
and we can set
\begin{align}
\label{eqAsolnIntRegime}
J_k(\ell) = J(k;\ellTwoLoopX) \mbox{~~for~~}
\newellonek \le \ell \le \newelltwok.
\end{align}

\subsubsection{Low Energy Regime}

In the long distance limit with 
$\ell > \ellOneLoopX$,
the beta functionals become
\begin{align}
\FR{\di v_k(\ell)}{\di\ell}    
\biggr|_{\ell \ge\ellOneLoopX}
&= 
0
\\
\FR{\di V_{F,k}(\ell)}{\di\ell} 
\biggr|_{\ell\ge\ellOneLoopX}
&= V_{F,k}(\ell)   
\Biggl[ \FR{N_c^2-1}{2\pi N_cN_F}w_0(\ell) \Biggr], 
\\
\FR{\di g_k(\ell)}{\di\ell} 
\biggr|_{\ell\ge\ellOneLoopX}
&= g_k(\ell)
\Biggl[  -\FR{1}{2\pi N_cN_f}w_0(\ell)\log\PFR{1}{w_0(\ell)}  \Biggr]
\label{eqC86}
\end{align}
to the leading order in $v$.
With the quantum corrections turned off,
the flow of $v_k$ freezes out.
$V_{F,k}$ that represents the Fermi velocity measured in the unit of the Fermi velocity at the hot spots
increases with increasing length scale. 
This is because the dynamical critical exponent is chosen to keep $V_{F,0}=1$ at the hot spot. 
Cold electrons which are decoupled from spin fluctuations at low energies appear to be moving increasingly faster 
when the speed is measured with the sluggish clock that is tuned  to keep the speed of hot electrons to be $1$.
Since the deviation of the dynamical critical exponent from $1$ is order of $w$, the flow of $V_{F,k}$ is controlled by $w$.
On the contrary, the Yukawa coupling decays to zero away from the hot spots in the low energy limit.
This is because the vertex correction, which tends to strengthen the coupling through the anti-screening effect, turns off at low energies.
With the anti-screening effect gone,
the large anomalous dimension of the boson,
which is $1+O(w \log1/w)$,
forces the Yukawa coupling to decrease rapidly. 
Since the Yukawa coupling is marginal when the anomalous dimension is $1$, the flow of the Yukawa coupling is proportional to $w \log1/w$.
From $w_0(\ell) = v_0(\ell)/c(\ell)$,
the solutions are readily obtained to be
\begin{align}
\begin{split}\label{appLowEnSolnFinal}
v_k(\ell\ge\newelltwok) &= 
 \FR{\pi^2 N_c N_f}{2(N_c^2-1)}  \FR{1}{(\ellOneLoopX+\ell_0)\log(\ellOneLoopX+\ell_0)}, \\
V_{F,k}(\ell\ge\ellOneLoopX) &= \mc E_1(\ell,\ellOneLoopX), \\
g_k(\ell\ge\ellOneLoopX)&= \sqrt{\FR{\pi}{2}v_0(\ellOneLoopX)} \mc E_0(\ell,\ellOneLoopX), 
\end{split}
\end{align}
where
\begin{align}
\label{E0defn}
\mc E_0(X,Y) &\equiv \exp\left(-\FR{\sqrt{X+\ell_0}-\sqrt{Y+\ell_0}}{\sqrt{N_c^2-1}}\right),\\
\mc E_1(X,Y) &\equiv
\exp\left(\sqrt{N_c^2-1}\left(\mathrm{Ei}(\log\sqrt{X+\ell_0})-\mathrm{Ei}(\log\sqrt{Y+\ell_0})\right)\right).
\label{E1defn}
\end{align}


\noindent 
Because 
$v_k(\ellOneLoopX) \approx v_k(\ellTwoLoopX)$,
$\mathcal{E}_0(\ellOneLoopX,\ellTwoLoopX) \approx 1$,
$\mathcal{E}_1(\ellOneLoopX,\ellTwoLoopX) \approx 1$,
the scale dependent diagonal couplings can be written
in terms of only one crossover as
\begin{align}
\label{appSolnFinalSimplified}
\begin{split}
v_k(\ell) &=
        \begin{cases}
v_0(\ell)        & \ell \le \ellTwoLoopX\\
v_0(\newellonek) & \ell \ge \ellTwoLoopX
        \end{cases}
\\
V_{F,k}(\ell) &=
            \begin{cases}
1                          & \ell \le \ellTwoLoopX \\
\mc E_1(\ell,\ellTwoLoopX)  & \ell \ge \ellTwoLoopX
        \end{cases}
\\
g_k(\ell) &=
         \begin{cases}
\sqrt{\pi v_0(\ell)/2}                                         & \ell \le \ellTwoLoopX\\
\sqrt{\FR{\pi}{2} v_0(\ellTwoLoopX)}\mc E_0(\ell,\ellTwoLoopX)   & \ell \ge \ellTwoLoopX
         \end{cases}.
\end{split}
\end{align}

\subsection{Off-diagonal Yukawa Coupling}

\newcommand{\yukscale}{\ell^{(\mathrm{ver})}_{k',k}}

\noindent The crossover scale for the off-diagonal
Yukawa vertex correction is given by  
\begin{align}
        \elltwokkp &=\min\left( \yukscale, \elltwok, \elltwokp \right),
        \label{e64}
\end{align}
where 
$  \yukscale = L^{(1L)}(k,k';\yukscale)$
is the crossover scale associated with the vertex correction.
Inside \eq{e64},
we can use the expression 
for $\yukscale$ that is valid  for
$\yukscale \le \ell^{(1L)}_k,\ell^{(1L)}_{k'}$\footnote{
If $\yukscale$ is greater than 
 $\ell^{(1L)}_k$ or $\ell^{(1L)}_{k'}$, 
$\yukscale$ drops out from \eq{e64} anyway.}.
Therefore,
we can set $V_{F,k} = V_{F,k'} = 1$ and 
$v_k(\yukscale),
v_{k'}(\yukscale) 
= v_0(\yukscale)$
to estimate $\yukscale$.
In this case, 
$\yukscale$ satisfies
$        \yukscale = \log\PFR{\Lambda}{2v_0(\yukscale)|k+k'|}$.
Since this is of the same form as \eq{app:e24} for $\newellonek$ except that $k$ is replaced with $|k+k'|/2$,
$\yukscale = \ell^{(2L)}_{(k+k')/2}$.
From this, we can write the crossover scale for the off-diagonal Yukawa coupling as
\begin{align}
\label{C187}
\newellthreekkp = \min(\ell^{(2L)}_{(k+k')/2},\ell^{(1L)}_k,\ell^{(1L)}_{k'}).
\end{align}
For $\ell \le \newellthreekkp$,
\eq{mainTextBetaGgen1} takes the same form  as the beta functional for the diagonal Yukawa coupling at high energy in \eq{betaGKKhi}, and 
the solution is given by
\begin{align}
\label{flowGkkpHiEn}
g_{k,k'}(\ell) &= \sqrt{\FR{\pi}{2}v_0(\ell)}.
\end{align}
For $\ell > \newellthreekkp$, 
the off-diagonal Yukawa coupling decreases as
\begin{align}
\FR{\di g_{k',k}(\ell)}{\di\ell} &= g_{k',k}(\ell)
\Biggl[ -\FR{1}{2\pi N_cN_f}w_0(\ell)\log\PFR{1}{w_0(\ell)} \Biggr].
\label{betaGgenLoEn}
\end{align}
All other terms in the beta function are sub-leading.
Integrating this differential equation,
we obtain
\begin{align}
\label{flowGkkpLoEn}
g(k',k;\ell) &= \sqrt{\FR{\pi}{2}v_0(\ell^{(1L)}_{k',k})}\mc E_0(\ell,\ell^{(1L)}_{k',k})
\end{align}
for $\ell > \ell^{(1L)}_{k',k}$.
Combining \cref{flowGkkpHiEn,flowGkkpLoEn} we arrive at
\begin{align}
\label{C207}
g_{k,k'}(\ell) &= 
\begin{cases}
\sqrt{\FR{\pi}{2}v_0(\ell)}                                         & \ell \le \newellthreekkp\\
\sqrt{\FR{\pi}{2}v_0(\newellthreekkp)}\mc E_0(\ell,\newellthreekkp) & \ell \ge \newellthreekkp.
\end{cases}
\end{align}

\section{Electronic Spectral Function}
\label{chap:2PFELE}

In this appendix, we compute the electron spectral function discussed in Sec. \ref{sec:normal}.
\eq{eq:SolRGEq2Point}
for $m=1$ and $n=0$ reads 
\begin{align}
\label{eq:Solution}
\varGamma^{(2,0)}(k_0,\vec k;[\whv,\whg,\whV_{\mathrm{F}}];k_{\mathrm{F}}) =
\exp\left\{\int\limits^{l}_{0}\dd\ell~[
z(\ell) - 2 + 2 \hat \eta^{(\psi)}_{k(\ell)}(\ell) 
]\right\}
\varGamma^{(2,0)}_N
\left( k_0(l),\vec{k}(l);\left[\whv(l), \whg(l), \hat V_F(l)\right];\tilde k_F(l) \right),
\end{align}
where 
$k_0(l)$,
$\vec k(l)$,
$\hat{v}(l)$,
$\hat{g}_{K,K'}(l)$,
$\hat V_{F,K}(l)$, 
$\hat \eta^{(\psi)}_K(l)$
are defined in \eq{eq:SolRGEq2Point} and below. 
In \eq{eq:Solution}, we set $l=\ell_{k_0}$ with 
$|k_0|  \exp\left( \int_0^{\lk} z(\ell) \dd \ell \right) = \Lambda$.
For $\ell_0 \gg 1$, its solution is given by
\begin{align}
\label{d7}
\lk = \log\FR{\Lambda}{|k_0|} - \FR{ \sqrt{N_c^2-1}\sqrt{ \ell_0 + \log\FR{\Lambda}{|k_0|} } }{ \log\sqrt{\ell_0 + \log\FR{\Lambda}{|k_0|}} } + \FR{ \sqrt{N_c^2-1}\sqrt{ \ell_0 } }{ \log\sqrt{\ell_0} }.
\end{align}
At high frequency $\Lambda$, 
we assume that the two-point function takes the form of 
$\varGamma^{(2,0)} \left( 
\Lambda,\vec{k}
;\left[\whv, \whg, \whVF \right]; \whkF \right)
= i\Lambda +\hat V_{F,k} e_{N}[\vec{k}; \hat v_{k}]$.
%
Substituting this into the right hand side  of \eq{eq:Solution}, 
we obtain
\begin{align}
\begin{split}
\varGamma^{(2,0)}(k_0,\vec{k}) &=
\exp\left(\int\limits^{\lk}_{0}\dd\ell~[
z(\ell) -2 +2\hat \eta^{(\psi)}_{k(\ell)}(\ell) 
]\right)
\left\{i\sgn(k_0)\Lambda +\hat V_{F,k(\lk)}(\lk)e_{N}[\vec{k}(\lk);\hat v_{k(\lk)}(\lk)]\right\}.
\end{split}
\end{align}
To the leading order in $v$, 
we obtain
\begin{align}
\label{d212}
\varGamma^{(2,0)}(k_0,k) &= F_z(k_0,k)
\left( ik_0 + F_z(k_0,k)^{-1} e_N[\vec{k},
v_{k}(\lk)] \right),
\end{align}
where 
\begin{align}
        F_z(k_0,k) = 
        \exp\left(
                \int_0^{\lk}
                \dd\ell
\frac{N_c^2-1}{\pi^2 N_cN_f} 
\FR{g_{k}(\ell)^2}{c(\ell)V_{F,k}(\ell)}
\FR{\murg}{\murg+2v_{k}(\ell)c(\ell)|k e^\ell|}
        \right).
        \label{intZ1a}
\end{align}
Here we use
$F_z(k_0,k) = 
\exp\left\{
\int_0^{\lk} \frac{\dd \log Z_1(k;\ell)}{\dd \log \mu} \dd\ell
\right\}$,
$Z_{1}(k) =
1-\frac{N_c^2-1}{\pi^2 N_cN_f} \FR{g_k^2}{cV_{F,k}}
\log\left( \FR{\Lambda}{\mu+2v_kc|k|} \right)$
from \eq{A1Lashu}
and
the identity between
the `hatted' coupling functions
and the `unhatted' coupling functions
in \eq{eq:LambdaAux} :
$\hat v_{k \eell}(\ell) = v_{k}(\ell)$,
$\hat V_{F,k \eell}(\ell) = V_{F,k}(\ell)$,
$\hat g_{k \eell}(\ell) = g_{k}(\ell)$,
$\hat \eta^{(\psi)}_{k \eell}(\ell) =  \eta^{(\psi)}_{k}(\ell)$.
$F_z(k_0,k)$ exhibits a crossover around $\lk \sim \ellOneLoopX$.
At length scale much shorter than $\ellOneLoopX$, 
$F_z$ is given by that of the hot spot.
At length scale much larger than $\ellOneLoopX$, the quantum corrections turn off, 
and the Fermi liquid behaviour is  expected to be restored.
In order to capture the crossover smoothly,
we choose a window of intermediate length scales that contains the crossover scale $\ellOneLoopX$ :
$\ellTwoLoopX < \ell < \ellThree$,
where 
$\ellTwoLoopX = \ellOneLoopX - \log \frac{2}{c(\ellOneLoopX)}$ 
and 
$\ellThree = \ellOneLoopX + \log \frac{2}{c(\ellOneLoopX)}$.
We can compute $F_z(k_0,k)$ 
in the three different ranges of scale as
\begin{align}
    F_z(k_0,k) &=
    \left\{
    \begin{aligned}
    &
      \exp\left(
\frac{N_c^2-1}{\pi^2 N_cN_f} 
                \int_0^{\lk}
                \dd\ell
~
\FR{ g_{k}(\ell)^2 }{c(\ell)}
        \right)
    &&
\lk < \ellTwoLoopX \\
    &
F_z( k_0^{(2L)}, k)
\exp\left(
\frac{N_c^2-1}{\pi^2 N_cN_f} 
\FR{g_{k}(\ellTwoLoopX )^2}{c(\ellTwoLoopX )}
                \int_{\ellTwoLoopX}^{\lk}
                \dd\ell
~ \FR{\murg}{\murg+2v_{k}(\ellTwoLoopX )c(\ellTwoLoopX)|k e^\ell|}
        \right)
         && 
\ellTwoLoopX < \lk < \ellThree \\
    &
F_z( k_0^{(3)}, k)
           \exp\left(
\frac{N_c^2-1}{\pi^2 N_cN_f} 
                \int_{\ellThree}^{\lk}
                \dd\ell
~ 
\FR{  g_{k}(\ell)^2 }{  c(\ellThree)}
\FR{\murg}{2v_{k}(\ellThree)
c(\ellThree)|k e^\ell|}
        \right)
    &&
     \ellThree < \lk \\
    \end{aligned}
    \right.,
    \label{eq:Fzin3}
\end{align}
where $k_0^{(2L)}$ and $k_0^{(3)}$ are the frequencies that satisfy
$\ell_{k_0^{(2L)}} = \ellTwoLoopX$
and
$\ell_{k_0^{(3)}} = \ellThree$, respectively.
In the high-energy region,
$2v_{k}(\ell)c(\ell)|k e^\ell|  \ll \murg$
and  one can approximate 
$\FR{\murg}{\murg+2v_{k}(\ell)c(\ell)|k e^\ell|}$ with $1$.
In the intermediate energy region,
coupling functions change little
and
the scale dependence can be ignored in all coupling functions. 
On the other hand, the crossover function
$\FR{\murg}{\murg+2v_{k}(0)c(0)|k e^\ell|}$
changes by a factor of $c$ 
in the intermediate region.
In the low-energy limit,
one can approximate 
$\FR{\murg}{\murg+2v_{k}(\ell)c(\ell)|k e^\ell|}$ with 
$\FR{\murg}{2v_{k}(\ell)c(\ell)|k e^\ell|}$,
and ignore the scale dependence of 
$v_{k}(\ell)$,
$c(\ell)$ 
which vary much slowly compared to 
$g_k(\ell)$ and $k \eell$.
Here
$V_{F,k}(\ell)$ is set to be $1$ even for 
$\ell > \ell^{(3)}_k$ because the dominant contribution arises around 
$\ell = \ell^{(3)}_k$ in the low-energy region.
An explicit computation of 
    \eq{eq:Fzin3} results in
\begin{align}
    F_z(k_0,k) &=
    \left\{
    \begin{aligned}
    &\mc E_1(\lk,0) 
    && \lk < \ellTwoLoopX\\
    &\mc E_1(\ellTwoLoopX,0) 
    \PFR{1+e^{\ellOneLoopX-\ellTwoLoopX}}{1+e^{\ellOneLoopX-\lk}}^{\alpha_1(\ellTwoLoopX)}
    && 
    \ellTwoLoopX < \lk < \ellThree\\
    &
    \begin{aligned}
    &\mc E_1(\ellTwoLoopX,0) 
    \PFR{1+e^{\ellOneLoopX-\ellTwoLoopX}}
    {1+e^{\ellOneLoopX-\ellThree}}^{\alpha_1(\ellTwoLoopX)}
 \exp\left(
 \alpha_3(\ellTwoLoopX)
    \left[ 
         1 - e^{-\lk+\ellThree}\mc E_0(\lk,\ellThree)^2
    \right]
    \right)
    \end{aligned}
    && 
     \ellThree < \lk \\
    \end{aligned}
    \right.,
\end{align}
where 
$\alpha_1(\ell)$ is defined in 
\eq{eq:alpha01}
and 
$\alpha_3(\ell)  =
\FR{\pi}{4}\FR{1}{(\ell+\ell_0)\log(\ell+\ell_0)}$.

The retarded Green's function can be obtained 
from the thermal Green's function through
the analytic continuation.
For this, we start with the thermal Green's function defined in the $k_0<0$ branch and replace $k_0$ with $i \omega$\footnote{
In our Euclidean sign convention,
the spacetime Fourier transformation is defined as
$\psi( k_0, \vec k)
= \int d \tau d \vec r ~ e^{i(k_0\tau+\vec{k}\cdot\vec{r})}
\psi(\tau,\vec r)$,
and the free thermal Green's function is given by 
$G(k) = \frac{1}{i k_0 + E_{\vec k}}$.
To obtain the retarded Green's function, 
$G_R(k) = \frac{1}{- \omega + E_{\vec k} - i 0^+}$,
we start with the thermal Green's function in $k_0< 0$
and replace $k_0 \rightarrow i \omega - 0^+$.
}.
%
The retarded Green's function is written as
\begin{align}
\label{D38}
G^{\mathrm{R}}_N(\omega,\vec{k})^{-1} 
&= \mc F_z(\omega,k)
 \left( -\omega + \mc F_z(\omega,k)^{-1} e_N[\vec k, v_{k}(\lo)]  \right),
\end{align}
where
\begin{equation}
\begin{aligned}
&   \mathcal{F}_z(\omega,k) = \\
&
    \begin{cases}
    \mc E_1(\lor,0)\left(  1+\FR{i\pi \alpha_1(\lor)}{2}
    \right) 
    &
    \lor < \ellTwoLoopX\\
    \mc E_1(\ellTwoLoopX,0)
    \PFR{  
        1+e^{\ellOneLoopX-\ellTwoLoopX}
    }{   
        \sqrt{ 1 +e^{2\ellOneLoopX-2\lor} }
    }^{\alpha_1(\ellTwoLoopX)}
    \Bigl(
        1 + i \alpha_1(\ellTwoLoopX) \arctan(e^{\ellOneLoopX-\lor})
    \Bigr)
    &
    \ellTwoLoopX < \lor < \ellThree\\
    \begin{aligned}
    &\mc E_1(\ellTwoLoopX,0) 
     \PFR{1+e^{\ellOneLoopX-\ellTwoLoopX}}
    {1+e^{\ellOneLoopX-\ellThree}}^{\alpha_1(\ellTwoLoopX)}
    e^{\alpha_3(\ellTwoLoopX)}
\left[ 
        1 +
        i \alpha_3(\ellTwoLoopX)
        e^{-\lor+\ellThree}
        \mc{E}_0(\lor,\ellThree)^2
    \right]
    \end{aligned}
    &
    \ellThree < \lor \\
    \end{cases}.
\end{aligned}
\end{equation}
Here $\lk$ is analytically continued as
\begin{align}
\label{D36}
\lo &= 
\underbrace{
\left[
 \log\FR{\Lambda}{\omega} - \FR{ \sqrt{N_c^2-1}\sqrt{ \ell_0 + \log\FR{\Lambda}{\omega} } }{ \log\sqrt{\ell_0 + \log\FR{\Lambda}{\omega}} } + \FR{ \sqrt{N_c^2-1}\sqrt{ \ell_0 } }{ \log\sqrt{\ell_0} }
\right]
}_{
\mathrm{Re}(\ell_{\omega})\equiv
\ell_{\omega}^{\mathrm{Re}}
}
+  i \FR{\pi}{2}
\end{align}
to the leading order in the large $\ell_0$ limit
and
$\lor$ represents the real part of 
or $\lo$.
The imaginary part generated from $v(\lk)$ is negligible in the small $v$ limit.

\section{Additional beta functionals in the presence of the particle-hole symmetry}
\label{sec:additional_beta}

%

If the PH symmetry is present,
there exists a perfect nesting for the $2 k_F$ scatterings
both in the PP and PH channels\footnote{
The time-reversal and parity symmetries
guarantee that
$
V^{(N_T)}_{\mathrm{F},-k}e_{N_T}[-\vec{k};v_{-k}^{(N_T)}]
=
V^{(N)}_{\mathrm{F},k}e_{ N}[\vec{k};v_{k}^{(N)}]
$ for $N_T = [N+4]_8$.
This makes it possible to put two electrons with zero center of mass momentum
on the Fermi surface in antipodal patches irrespective of their relative momentum.
In the presence of the PH symmetry, we also have  
$ 
 V^{(N)}_{\mathrm{F},-k}e_{N}[-\vec{k};v^{(N)}_{-k}]
=
-V^{(N)}_{\mathrm{F},k}e_{N}[\vec{k};v^{(N)}_{k}]
$.
This further makes it possible for a pair of electrons 
or an electron-hole pair 
with total momentum $2 \vec k_F$ 
to stay on the Fermi surface irrespective of their relative momentum.
}.
In this case, there are additional channels 
with extended spaces of IR singularity.
In group $1$, one needs to include the interaction 
that describes pairings between electrons 
with total momentum $2 k_F$
as is shown in
\eq{eq:SofIS1_2}.
In group $2$, the $2 k_F$ scatterings of particle-hole pairs
in 
\eq{eq:SofIS2_2}
should be included.
Below, we derive the beta functionals for those additional coupling functions
and discuss they affect 
the flow of the couplings considered in the main text.
In the presence of 
the PH symmetry, 
the coupling functions obey Eq. \eqref{eq:PHcouplings}. Therefore, we can simply write $v^{(N)}_{k} = v_{k}, V^{(N)}_{F,k} = V_{F,k}$ and $g^{(N)}_{k',k}=g_{k',k}$,
where $v_k = v_{-k}$
and $V_{F,k} = V_{F,-k}$. 
This implies that the fermion propagator satisfies
\begin{equation}
    G_N ({\bf q}) = - G_N(-{\bf q}).
    \label{eq:fermiongreenPHSym}
\end{equation}
From this,
we can derive relations between the vertex corrections in the PP and PH channels.
The vertex correction that is independent of the four-fermion coupling in
Eqs. \eqref{eq:Gamma_lambda0_PP} and \eqref{eq:Gamma_lambda0_PH}
is determined by the kernels,
\begin{align}
    \gamma^{(PP)}_{N_1,N_2}(q;k_0,p_0,l_0,k,p,l) = & \int\frac{\dd q_0 \dd q_\perp}{(2\pi)^2} D({\bf q})D({\bf l}- {\bf q})G_{\bar{N}_1}({\bf k}+ {\bf q})G_{\bar{N}_2}({\bf p}-{\bf q}),
    \\ \gamma^{(PH)}_{N_1,N_2}(q;k_0,p_0,l_0,k,p,l) = & \int\frac{\dd q_0 \dd q_\perp}{(2\pi)^2} D({\bf q})D({\bf l}- {\bf q})G_{\bar{N}_1}({\bf k}+ {\bf q})G_{\bar{N}_2}({\bf p}-{\bf l}+{\bf q}).
\end{align}
Due to Eq. \eqref{eq:fermiongreenPHSym}, 
$\gamma^{(PP)}_{N_1,N_2}$ and $\gamma^{(PH)}_{N_1,N_2}$ obey 
\begin{equation}
    \gamma^{(PH)}_{N_1,N_2}(q;k_0,p_0,l_0,k,p,l) = -\gamma^{(PP)}_{N_1,N_2}(q;k_0,-p_0+l_0,l_0,k,-p+l,l).
\end{equation}
Similarly, the kernels that determines the linear mixing in Eqs. \eqref{eq:k1} and \eqref{eq:k2} satisfy
\begin{equation}
\mathcal{K}_{N_a,N_b}^{(PH)}(q;k_{a,0},k_{b,0},k_a,k_b) = -\mathcal{K}_{N_a,N_b}^{(PP)}(q;k_{a,0},-k_{b,0},k_a,-k_b).
\label{eq:nocurvatureproperty}
\end{equation}
Finally, the kernels for the quantum corrections quadratic in the four fermion coupling 
in Eqs. \eqref{eq:Qkernel1} and \eqref{eq:Qkernel} are related to each other through
\begin{equation}
    Q_{M_a,M_b}^{(PH)}(q;k_{a,0},k_{b,0},k_a,k_b) = - Q_{M_a,M_b}^{(PP)}(q;k_{a,0},-k_{b,0},k_a,-k_b).
\end{equation}
From these relations, we can readily compute
the beta functions for the $2k_F$ scatterings using Eq. \eqref{eq:betalambda2_main}.

\subsection{Group 1}

\subsubsection{Beta functional for the $2k_F$ pairing}

Let us first consider $\lambda^{\spmqty{1 & 1 \\ 1 & 1}}_{\spmqty{p & -p \\ k & -k}}$,
where the total momentum of two electrons is $2 k_F$ 
($0$ when measured with respect to the momentum of two electrons located at hot spot $1$)
in the PP channel.
The beta functional for the coupling is
\begin{equation}
\begin{aligned}
& \beta^{(\lambda);\spmqty{1 & 1 \\ 1 & 1};\spmqty{\sigma_1 & \sigma_2 \\ \sigma_4 & \sigma_3}}_{\spmqty{p & -p \\ k & -k}} =
 \left( 1 + 3(z-1) + 2\eta^{(\psi)}_{p} + 2\eta^{(\psi)}_{k} \right)
 \lambda^{\spmqty{1 & 1 \\ 1 & 1};\spmqty{\sigma_1 & \sigma_2 \\ \sigma_4 & \sigma_3}}_{\spmqty{p & -p \\ k & -k}}
\\ & +
\int\dd\rho(q)\left\lbrace 
\frac{\mathsf{D}_{\mu}(q;k)^2}{2 \pi N_f g^2_{k,q}}
\lambda^{\spmqty{1 & 1 \\ 4 & 4}; \spmqty{\sigma_1 & \sigma_2 \\ \alpha & \beta}}_{\spmqty{p & -p \\ q & -q}} 
 \mathsf{T}^{\alpha \beta}_{\sigma_4 \sigma_3} 
 + 
 \frac{\mathsf{D}_{\mu}(p;q)^2}{2 \pi N_f g^2_{p,q}}
  \mathsf{T}^{\sigma_1 \sigma_2}_{\alpha \beta} 
 \lambda^{\spmqty{4 & 4 \\ 1 & 1}; \spmqty{\alpha & \beta \\ \sigma_4 & \sigma_3}}_{\spmqty{q & -q \\ k & -k}}  \right.
\\ & -\left.\frac{ \mathsf{D}_{\mu}(p;q)\mathsf{D}_{\mu}(q;k)}{\pi N_f^2}
 \mathsf{T}^{\sigma_1 \sigma_2}_{\alpha \beta}   \mathsf{T}^{\alpha \beta}_{\sigma_4 \sigma_3} 
\left(\frac{\mathsf{D}_{\mu}(q;k)}{ g^2_{q,k}} + \frac{\mathsf{D}_{\mu}(p;q)}{ g^2_{p,q}}\right)
  \right\rbrace.
 \label{eq:beta1111PP}
\end{aligned}
\end{equation}
Performing the $q$ integration in the adiabatic limit, 
we obtain
\begin{equation}
\begin{aligned}
& \beta^{(\lambda);\spmqty{1 & 1 \\ 1 & 1};\spmqty{\sigma_1 & \sigma_2 \\ \sigma_4 & \sigma_3}}_{\spmqty{p & -p \\ k & -k}} =
 \left( 1 + 3(z-1) + 2\eta^{(\psi)}_{p} + 2\eta^{(\psi)}_{k} \right)
 \lambda^{\spmqty{1 & 1 \\ 1 & 1};\spmqty{\sigma_1 & \sigma_2 \\ \sigma_4 & \sigma_3}}_{\spmqty{p & -p \\ k & -k}}
\\ & + \frac{g^2_{k,k}}{2 \pi^2 c N_f V_{F,k}}\frac{\mu}{\mu + 2vc|k|_\mu}
\lambda^{\spmqty{1 & 1 \\ 4 & 4}; \spmqty{\sigma_1 & \sigma_2 \\ \alpha & \beta}}_{\spmqty{p & -p \\ k & -k}} 
 \mathsf{T}^{\alpha \beta}_{\sigma_4 \sigma_3} 
+ 
\frac{g^2_{p,p}}{2 \pi^2 c N_f V_{F,p}}\frac{\mu}{\mu + 2vc|p|_\mu}
  \mathsf{T}^{\sigma_1 \sigma_2}_{\alpha \beta} 
 \lambda^{\spmqty{4 & 4 \\ 1 & 1}; \spmqty{\alpha & \beta \\ \sigma_4 & \sigma_3}}_{\spmqty{p & -p \\ k & -k}} 
 \\ & -
  \mathsf{T}^{\sigma_1 \sigma_2}_{\alpha \beta}   \mathsf{T}^{\alpha \beta}_{\sigma_4 \sigma_3} 
\frac{ \mathsf{D}_{\mu}(p;k)}{\pi^2 c N_f^2}\left[\frac{\mu g^2_{k,k}}{V_{F,k}(\mu + 2v_kc|k|_\mu)}
+\frac{\mu g^2_{p,p}}{V_{F,p}(\mu + 2v_pc|p|_\mu)}\right].
 \label{eq:beta1111PP1}
\end{aligned}
\end{equation}
Since the coupling
$\lambda^{\spmqty{1 & 1 \\ 1 & 1}}$ mixes with
$\lambda^{\spmqty{4 & 4 \\ 1 & 1}}$,
$\lambda^{\spmqty{1 & 1 \\ 4 & 4}}$,
$\lambda^{\spmqty{4 & 4 \\ 4 & 4}}$,
we need to compute the beta functionals for those couplings as well
to have a closed set of beta functionals.
The beta functionals for the rest of the couplings are obtained to be
\begin{equation}
\begin{aligned}
& \beta^{(\lambda);\spmqty{4 & 4 \\ 1 & 1};\spmqty{\sigma_1 & \sigma_2 \\ \sigma_4 & \sigma_3}}_{\spmqty{p & -p \\ k & -k}} =
 \left( 1 + 3(z-1) + 2\eta^{(\psi)}_{p} + 2\eta^{(\psi)}_{k} \right)\lambda^{\spmqty{4 & 4 \\ 1 & 1};\spmqty{\sigma_1 & \sigma_2 \\ \sigma_4 & \sigma_3}}_{\spmqty{p & -p \\ k & -k}}
\\ & +\frac{g^2_{k,k}}{2\pi^2 c N_f V_{F,k}}\frac{\mu}{\mu + 2v_kc|k|_\mu}
\lambda^{\spmqty{4 & 4 \\ 4 & 4}; \spmqty{\sigma_1 & \sigma_2 \\ \alpha & \beta}}_{\spmqty{p & -p \\ k & -k}} 
 \mathsf{T}^{\alpha \beta}_{\sigma_4 \sigma_3} 
+ \frac{g^2_{p,p}}{2\pi^2 c N_f V_{F,p}}\frac{\mu}{\mu + 2v_pc|p|_\mu}
  \mathsf{T}^{\sigma_1 \sigma_2}_{\alpha \beta} 
\lambda^{\spmqty{1 & 1 \\ 1 & 1}; \spmqty{\alpha & \beta \\ \sigma_4 & \sigma_3}}_{\spmqty{p & -p \\ k & -k}},
 \label{eq:beta1111PP2}
\end{aligned}
\end{equation}
\begin{equation}
\begin{aligned}
& \beta^{(\lambda);\spmqty{1 & 1 \\ 4 & 4};\spmqty{\sigma_1 & \sigma_2 \\ \sigma_4 & \sigma_3}}_{\spmqty{p & -p \\ k & -k}} =
 \left( 1 + 3(z-1) + 2\eta^{(\psi)}_{p} + 2\eta^{(\psi)}_{k} \right)\lambda^{\spmqty{1 & 1 \\ 4 & 4};\spmqty{\sigma_1 & \sigma_2 \\ \sigma_4 & \sigma_3}}_{\spmqty{p & -p \\ k & -k}}
\\ & 
+ \frac{g^2_{k,k}}{2\pi^2 c N_f V_{F,k}}
\frac{\mu}{\mu + 2v_kc|k|_\mu}
 \lambda^{\spmqty{1 & 1 \\ 1 & 1}; \spmqty{\sigma_1 & \sigma_2 \\ \alpha & \beta}}_{\spmqty{p & -p \\ k & -k}}   \mathsf{T}^{\alpha \beta}_{\sigma_4 \sigma_3} 
+ \frac{g^2_{p,p}}{2\pi^2 c N_f V_{F,p}}\frac{\mu}{\mu + 2v_pc|p|_\mu}
  \mathsf{T}^{\sigma_1 \sigma_2}_{\alpha \beta} 
 \lambda^{\spmqty{4 & 4 \\ 4 & 4}; \spmqty{\alpha & \beta \\ \sigma_4 & \sigma_3}}_{\spmqty{p & -p \\ k & -k}},
 \label{eq:beta1111PP3}
\end{aligned}
\end{equation}
\begin{equation}
\begin{aligned}
& \beta^{(\lambda);\spmqty{4 & 4 \\ 4 & 4};\spmqty{\sigma_1 & \sigma_2 \\ \sigma_4 & \sigma_3}}_{\spmqty{p & -p \\ k & -k}} =
 \left( 1 + 3(z-1) + 2\eta^{(\psi)}_{p} + 2\eta^{(\psi)}_{k} \right)\lambda^{\spmqty{4 & 4 \\ 4 & 4};\spmqty{\sigma_1 & \sigma_2 \\ \sigma_4 & \sigma_3}}_{\spmqty{p & -p \\ k & -k}}
\\ & + \frac{g^2_{k,k}}{2 \pi^2 c N_f V_{F,k}}\frac{\mu}{\mu + 2v_kc|k|_\mu}
  \lambda^{\spmqty{4 & 4 \\ 1 & 1}; \spmqty{\sigma_1 & \sigma_2 \\ \alpha & \beta}}_{\spmqty{p & -p \\ k & -k}} 
 \mathsf{T}^{\alpha \beta}_{\sigma_4 \sigma_3} 
+ \frac{g^2_{p,p}}{2\pi^2 c N_f V_{F,p}}\frac{\mu}{\mu + 2v_pc|p|_\mu}
  \mathsf{T}^{\sigma_1 \sigma_2}_{\alpha \beta} 
 \lambda^{\spmqty{1 & 1 \\ 4 & 4}; \spmqty{\alpha & \beta \\ \sigma_4 & \sigma_3}}_{\spmqty{p & -p \\ k & -k}} 
\\ & - 
 \mathsf{T}^{\sigma_1 \sigma_2}_{\alpha \beta} 
 \mathsf{T}^{\alpha \beta}_{\sigma_4 \sigma_3} 
 \frac{\mathsf{D}_{\mu}(p;k)}{\pi^2 c N_f^2}\left[\frac{\mu g^2_{k,k}}{V_{F,k}(\mu + 2v_kc|k|_\mu)}+\frac{\mu g^2_{p,p}}{V_{F,p}(\mu + 2v_pc|p|_\mu)}\right] .
 \label{eq:beta1111PP4}
\end{aligned}
\end{equation}
For $\lambda^{\spmqty{4 & 4 \\ 1 & 1}}$
and
$\lambda^{\spmqty{1 & 1 \\ 4 & 4}}$,
there is no source term 
to the leading order in $v$.

\subsubsection{Solution of the beta functional for the $2k_F$ pairing}

In the space of the rescaled momentum,
the set of four beta functionals 
in Eqs. 
\eqref{eq:beta1111PP1}-
\eqref{eq:beta1111PP4}
can be written as
\begin{align}\nonumber
\left[\frac{\partial}{\partial\ell}+K\frac{\partial}{\partial K} + P\frac{\partial}{\partial P}\right]{\hat\lambda}_{\spmqty{P & -P \\  K &  -K}}^{\spmqty{N_1 & N_2 \\ N_4 & N_3};\spmqty{\sigma_1 & \sigma_2 \\ \sigma_4 & \sigma_3}}
 = -\left(1+\hat\eta_K+\hat\eta_P\right)
{\hat\lambda}_{\spmqty{P & -P \\  K &  -K}}^{\spmqty{N_1 & N_2 \\ N_4 & N_3};\spmqty{\sigma_1 & \sigma_2 \\ \sigma_4 & \sigma_3}}
\\
-\frac{\hat B_K}{2N_f} 
\hat\lambda^{\spmqty{N_1 & N_2 \\ \bar N_4 & \bar N_3};\spmqty{\sigma_1 & \sigma_2 \\ \alpha & \beta}}_{\spmqty{P  & -P \\ K & -K }}
 \mathsf{T}^{\alpha \beta}_{\sigma_4 \sigma_3} 
  - \frac{\hat B_P}{2N_f} 
  \mathsf{T}^{\sigma_1 \sigma_2}_{\alpha \beta} 
 \hat\lambda^{\spmqty{\bar N_1 & \bar N_2 \\ N_4 & N_3};\spmqty{\alpha & \beta \\ \sigma_4 & \sigma_3}}_{\spmqty{P  & -P \\ K & -K }} 
 +\frac{\hat S_{K,P}}{N_f^2}
   \mathsf{T}^{\sigma_1 \sigma_2}_{\alpha \beta} 
 \mathsf{T}^{\alpha \beta}_{\sigma_4 \sigma_3} 
 \delta^{N_1}_{N_4}
 \delta^{N_2}_{N_3},
\end{align}
where
$
\hat {\lambda}^{\spmqty{N_1 & N_2 \\ N_4 & N_3}; \spmqty{\alpha & \beta \\ \gamma & \delta}}_{\spmqty{P & -P \\ K & -K}} 
=
\frac{1}{\sqrt{ V_{F,p} V_{F,k}}}
{\lambda}^{\spmqty{N_1 & N_2 \\ N_4 & N_3}; \spmqty{\alpha & \beta \\ \gamma & \delta}}_{1PH \spmqty{p & -p \\ k & -k}}
$
with $K=k \eell$, $P=p \eell$
for $\spmqty{N_1 & N_2 \\ N_4 & N_3}$ in 
\bqa
H_{1111}^{PP}=\Big\{ 
\spmqty{1 & 1 \\ 1 & 1},
\spmqty{1 & 1 \\ 4 & 4},
\spmqty{4 & 4 \\ 1 & 1},
\spmqty{4 & 4 \\ 4 & 4}
\Bigr\}.
\eqa
$\hat B_{K}$ 
and
$\hat S_{K,P}$ 
are
defined in \eq{eq:VXSKP0} and  \eq{eq:VXSKP1}.
We combine the four coupling functions into a matrix as
\begin{equation}
 \hat {\lambda}^{ \spmqty{\sigma_1 & \sigma_2 \\ \sigma_4 & \sigma_3}}_{1PP\spmqty{P & -P \\  K & -K}} = 
\pmqty{\hat{\lambda}^{\spmqty{1 & 1 \\ 1 & 1};\spmqty{\sigma_1 & \sigma_2 \\ \sigma_4 & \sigma_3}  }_{\spmqty{P & -P \\ K & -K}} & \hat{\lambda}^{\spmqty{1 & 1 \\ 4 & 4} ;\spmqty{\sigma_1 & \sigma_2 \\ \sigma_4 & \sigma_3}}_{\spmqty{P & -P \\ K & -K}} \\ \hat{\lambda}^{\spmqty{4 & 4 \\ 1 & 1};\spmqty{\sigma_1 & \sigma_2 \\ \sigma_4 & \sigma_3}}_{\spmqty{P & -P \\ K & -K}} & \hat{\lambda}^{\spmqty{4 & 4 \\ 4 & 4};\spmqty{\sigma_1 & \sigma_2 \\ \sigma_4 & \sigma_3}}_{\spmqty{P & -P \\ K & -K}}} 
\end{equation}
to write the set of beta functionals in a compact form as
\begin{equation}
\begin{aligned}
& \left[\frac{\partial}{\partial\ell}+K\frac{\partial}{\partial K} + P\frac{\partial}{\partial P}\right]
 \hat {\lambda}^{ \spmqty{\sigma_1 & \sigma_2 \\ \sigma_4 & \sigma_3}}_{1PP\spmqty{P & -P \\  K & -K}} 
 =  -\left( 1+\hat\eta_K+\hat\eta_P\right)
 \hat {\lambda}^{ \spmqty{\sigma_1 & \sigma_2 \\ \sigma_4 & \sigma_3}}_{1PP\spmqty{P & -P \\  K & -K}} 
 \\&
 - 
 \frac{\hat B_{K}}{2N_f} 
 \hat {\lambda}^{ \spmqty{\sigma_1 & \sigma_2 \\ \alpha & \beta}}_{1PP\spmqty{P & -P \\  K & -K}} 
 \mathsf{T}^{\alpha \beta}_{\sigma_4 \sigma_3} 
 \pmqty{0& 1 \\ 1 & 0}
 -
\frac{\hat B_P}{2N_f}
   \mathsf{T}^{\sigma_1 \sigma_2}_{\alpha \beta} 
 \pmqty{0& 1 \\ 1 & 0}
 \hat {\lambda}^{ \spmqty{\alpha & \beta \\ \sigma_4 & \sigma_3}}_{1PP\spmqty{P & -P \\  K & -K}} 
 + \frac{\hat S_{K,P}}{N_f^2}
    \mathsf{T}^{\sigma_1 \sigma_2}_{\alpha \beta} 
 \mathsf{T}^{\alpha \beta}_{\sigma_4 \sigma_3} ~
 \pmqty{1& 0 \\ 0 & 1}.
\end{aligned}
\label{eq:betabigmatrix_10_5PP}
\end{equation}
The matrix coupling function can be decomposed into 
 the spin-symmetric s-wave ($+,s$),
 spin-symmetric d-wave ($+,d$),
 spin-anti-symmetric s-wave ($-,s$)
 and
 spin-anti-symmetric d-wave ($-,d$)
 channels as
 \begin{equation}
 \hat {\lambda}^{ \spmqty{\sigma_1 & \sigma_2 \\ \sigma_4 & \sigma_3}}_{1PP\spmqty{P & -P \\  K & -K}} 
=
\hat{\lambda}^{(+)(s)}_{1PP \spmqty{P & -P \\  K & -K}}\mathsf{S}^{\sigma_1 \sigma_2}_{\sigma_4 \sigma_3} 
\mathscr{P}_s
+ \hat{\lambda}^{(+)(d)}_{1PP\spmqty{P & -P \\  K & -K}}\mathsf{S}^{\sigma_1 \sigma_2}_{\sigma_4 \sigma_3} 
\mathscr{P}_d
+ \hat{\lambda}^{(-)(s)}_{1PP \spmqty{P & -P \\  K & -K}}\mathsf{A}^{\sigma_1 \sigma_2}_{\sigma_4 \sigma_3}
\mathscr{P}_s
+ \hat{\lambda}^{(-)(d)}_{1PP \spmqty{P & -P \\  K & -K}}\mathsf{A}^{\sigma_1 \sigma_2}_{\sigma_4 \sigma_3}
\mathscr{P}_d.
\end{equation}
where $S$ and $A$
defined in \eq{eq:spinSA}
represent the operators that project spin wavefunctions
to the symmetric and anti-symmetric channels, respectively.
$\mathscr{P}_s$ and $\mathscr{P}_d$
defined in \eq{eq:PsPd}
are the operators that project 
hot spot wavefunctions to the s and d wave channels, respectively.
In each channel, the beta functional becomes
\bqa
 \left[
\frac{\partial}{\partial\ell}+K\frac{\partial}{\partial K} + P\frac{\partial}{\partial P}
\right]
\hat\lambda^{(\pm)\spmqty{s \\ d}}_{1PP\spmqty{P & -P \\ K & -K}}
 & = & -\left( 
  1+\hat\eta_K+\hat\eta_P 
  + \frac{1^{\spmqty{s \\ d}} Y_{PP}^{(\pm)}}{ 2 N_f} 
  \left[
  \hat B_{K} + \hat B_{P}
  \right] \right)
  \hat\lambda^{(\pm)\spmqty{s \\ d}}_{1PP\spmqty{P & -P \\ K & -K}}  
   +  Y_{PP}^{(\pm)^2}
\frac{\hat S_{K,P}}{N_f^2}. \nn
\label{eq:beta_s_11all}
\eqa
At the quasi-fixed points,
the coupling function is given by
\begin{align}
 \left.\hat \lambda^{*(\pm)\spmqty{s \\ d}}_{1PP\spmqty{P & -P \\ K & -K}}\right|_{w\ll1} 
= 
\frac{\hat g^2_{P,K}   \left.Y_{PP}^{(\pm)}\right.^2
 }{\pi^2c N_f^2\sqrt{\hat V_{F,K}\hat V_{F,P}}}
 & \left[\frac{\hat g^2_{K,K}}{\hat V_{F,K}}\frac{\Lambda  \log \left(\frac{c |\hat v_K K + \hat v_P P|_\Lambda +c |K-P|_\Lambda +\Lambda }{2\hat v_Kc|K|_\Lambda +\Lambda }\right)}{c (|\hat v_K K + \hat v_P P|_\Lambda +|K-P|_\Lambda -2 \hat v_K| K|_\Lambda )} \right.
\nn
& \left. + \frac{\hat g^2_{P,P}}{\hat V_{F,P}}\frac{\Lambda  \log \left(\frac{c |\hat v_K K + \hat v_P P|_\Lambda +c |K-P|_\Lambda +\Lambda }{2\hat v_Pc|P|_\Lambda +\Lambda }\right)}{c (|\hat v_K K + \hat v_P P|_\Lambda +|K-P|_\Lambda -2 \hat v_P| P|_\Lambda )}\right].
 \label{eq:lambda_fp_1PP_1}
\end{align}

\subsection{Group 2}

\begin{figure}[ht]
\centering
\includegraphics[scale=0.75]{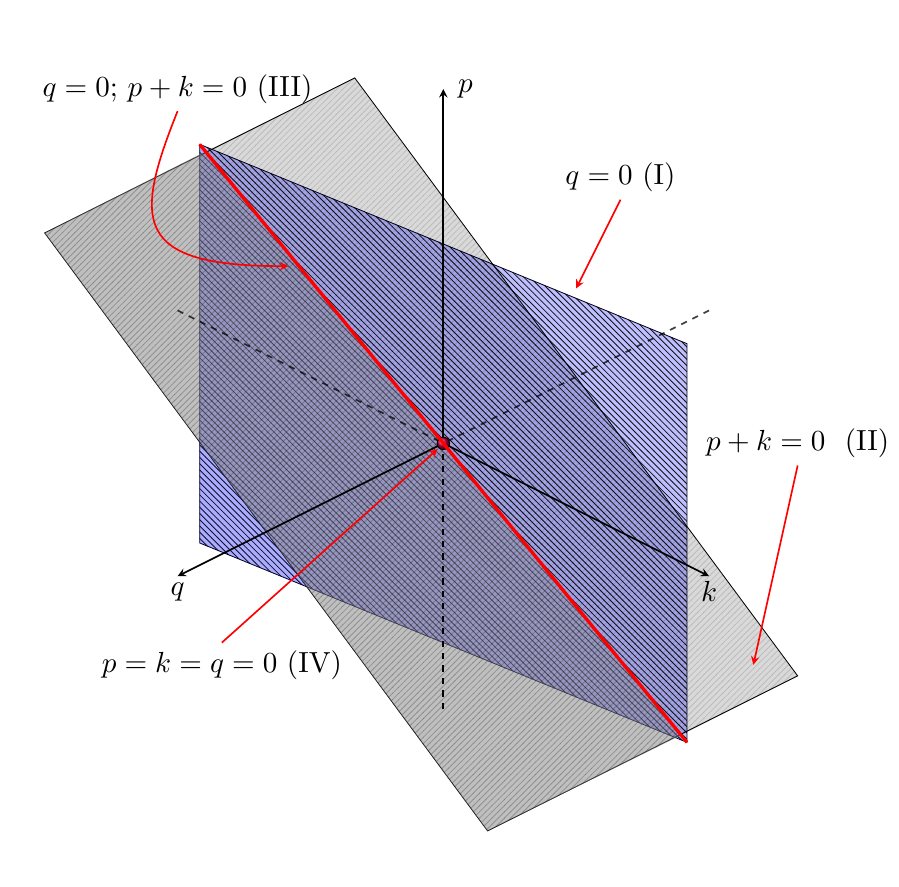}
\caption{
The space of IR singularity for
$\lambda^{\spmqty{1 & 5 \\ 1 & 5}}_{\spmqty{p+q/2 & -p+q/2 \\ k+q/2 & -k+q/2}}$.
Plane I (plane II) represents the set of external momenta at which the vertex correction is singular in the PP (PH) channel.
On line III, where the two planes intersect, the vertex corrections from both channels contribute.
As one deviates from the intersecting line but staying within plane I (II), 
the quantum corrections from plane II (I) dynamically turn off 
 as the deviation of the momentum  from the intersecting line is greater than $\mu/v$.
Since the volume of the intersecting `line' vanishes in the small $\mu$ limit,
one can ignore the contribution of the couplings in the intersecting line to
the flow of the couplings within plane I or II far away from the intersecting line.
}
\label{fig:manifold_singular_1515}
\end{figure}

In group 2, 
the space of IR singularity
consists of two intersecting manifolds
in the presence of the PH symmetry.
The first manifold is the PP-plane in which
the center of mass momentum of two incoming particles is zero.
The second manifold is the PH-plane 
in which a particle-hole pair   with momentum $2k_F$ is formed across anti-podal patches.
(when the momentum is measured with respect to each hot spot the particle-hole pair has zero center of mass momentum).
This space can be divided into three disjoint manifolds.
The first is the PP-plane that excludes
the intersecting line with the PH-plane.
The second is the PH-plane that excludes
the intersecting line with the PP-plane.
The third is the intersecting line. 
Written as a function of three general external momenta as
$\lambda^{\spmqty{1 & 5 \\ 1 & 5}}_{
\spmqty{ p+q/2   &  -p+q/2 \\ 
k+ q/2 & -k+ q/2 }}$,
the space of IR singularity can be divided into four disjoint sub-spaces as
(see \fig{fig:manifold_singular_1515})
\bqa
\begin{array}{lll}
\label{eq:spaceIRsingular2}
 1)~ q = 0, ~p +k \neq  0, ~~&
 2) ~ q \neq 0,~ p +k= 0, ~~& 
 3) ~ q=0, ~p+k=0.
\end{array}
\eqa
The beta functional in the PP-plane is 
computed in Sec. \ref{sec:fullbetalambda_2}.
In this section, we present the beta functionals for the couplings
in the PH-plane and the intersecting line.

\subsubsection{Beta functional for the $2 k_F$ PH interaction}

Within the PH plane, 
the coupling function describes the processes
in which a pair of electron and hole fluctuate 
between different antipodal patches ($1$, $5$ and $4$, $8$ patches)
and different relative momenta. 
The derivation of the beta functional is parallel to the ones for the PP channel.
The beta functional for
$\lambda^{\spmqty{1 & 5 \\ 1 & 5}; \spmqty{\sigma_1 & \sigma_2 \\ \sigma_4 & \sigma_3}}_{\spmqty{p & k \\ k & p}} $
in the PH plane but away from the intersecting line 
is obtained to be 
\begin{equation}
\begin{aligned}
& \beta^{(\lambda);\spmqty{1 & 5 \\ 1 & 5};\spmqty{\sigma_1 & \sigma_2 \\ \sigma_4 & \sigma_3}}_{\spmqty{p & k \\ k & p}} = \left( 1 + 3(z-1) + 2\eta^{(\psi)}_{p} + 2\eta^{(\psi)}_{k} \right)
\lambda^{\spmqty{1 & 5 \\ 1 & 5}; \spmqty{\sigma_1 & \sigma_2 \\ \sigma_4 & \sigma_3}}_{\spmqty{p & k \\ k & p}} \\ &
 + \int \dd\rho(q) \left\lbrace  \frac{\mathsf{D}_{\mu}(q;k)}{2\pi N_f} 
\lambda^{\spmqty{1 & 8 \\ 4 & 5}; \spmqty{\sigma_1 & \alpha \\ \beta & \sigma_3}}_{\spmqty{p & q \\ q & p}}
 \mathsf{T}^{\beta \sigma_2}_{\sigma_4 \alpha} 
+ \frac{\mathsf{D}_{\mu}(p;q) }{2\pi N_f} 
\lambda^{\spmqty{4 & 5 \\ 1 & 8}; \spmqty{\alpha & \sigma_2 \\ \sigma_4 & \beta}}_{\spmqty{q & k \\ k & q}}
 \mathsf{T}^{\sigma_1 \beta}_{\alpha \sigma_3} 
\right.
\\ & -\frac{1}{\pi N_f^2}
 \mathsf{T}^{\beta \sigma_2}_{\sigma_4 \alpha} 
 \mathsf{T}^{\sigma_1 \alpha}_{\beta \sigma_3} 
\mathsf{D}_{\mu}(p;q)
\mathsf{D}_{\mu}(q;k)
\\ & \left. - \frac{1}{4\pi}
\left(\lambda^{\spmqty{1 & 5 \\ 1 & 5}; \spmqty{\sigma_1 & \alpha \\ \beta & \sigma_3}}_{\spmqty{p & q \\ q & p}}\lambda^{\spmqty{1 & 5 \\ 1 & 5}; \spmqty{\beta & \sigma_2 \\ \sigma_4 & \alpha}}_{\spmqty{q & k \\ k & q}} + \lambda^{\spmqty{1 & 8 \\ 4 & 5}; \spmqty{\sigma_1 & \alpha \\ \beta & \sigma_3}}_{\spmqty{p & q \\ q & p}}\lambda^{\spmqty{4 & 5 \\ 1 & 8}; \spmqty{\beta & \sigma_2 \\ \sigma_4 & \alpha}}_{\spmqty{q & k \\ k & q}} \right)\right\rbrace.
\label{eq:beta1515PH1}
\end{aligned}
\end{equation}
Here
$\lambda^{\spmqty{1 & 5 \\ 1 & 5}; \spmqty{\sigma_1 & \sigma_2 \\ \sigma_4 & \sigma_3}}_{\spmqty{p & k \\ k & p}} $
describes the interaction in which
an electron with  momentum $k$ in hot spot $1$
and a hole with  $-k$ in hot spot $5$
are scattered to electron with $p$ in hot spot $1$
and hole with $-p$ in hot spot $5$.
Since the momentum is measured with respect to the hot spots,
the pair of electron and hole in this channel 
actually carry a non-zero momentum, $2 \vec k_F$.
The physical origin of each term in  \eq{eq:beta1515PH1}
can be understood in the same way as in  \eq{eq:beta1515PP1}.
The beta functionals for the other couplings that form a closed set of flow equations
in the PH-plane are given by
\begin{equation}
\begin{aligned}
& \beta^{(\lambda);\spmqty{4 & 5 \\ 1 & 8}; \spmqty{\sigma_1 & \sigma_2 \\ \sigma_4 & \sigma_3}}_{\spmqty{p & k \\ k & p}} = \left( 1 + 3(z-1)+ 2\eta^{(\psi)}_{p} + 2\eta^{(\psi)}_{k} \right)\lambda^{\spmqty{4 & 5 \\ 1 & 8}; \spmqty{\sigma_1 & \sigma_2 \\ \sigma_4 & \sigma_3}}_{\spmqty{p & k \\ k & p}}
\\ & + 
\int \dd\rho(q) 
\left\lbrace
\frac{1}{2 \pi N_f}\mathsf{D}_{\mu}(q;k) 
\lambda^{\spmqty{4 & 8 \\ 4 & 8} ; \spmqty{\sigma_1 & \alpha \\ \beta & \sigma_3}}_{\spmqty{p & q \\ q & p}}
 \mathsf{T}^{\beta \sigma_2}_{\sigma_4 \alpha} 
+ \frac{1}{2 \pi N_f}\mathsf{D}_{\mu}(p;q) 
\lambda^{\spmqty{1 & 5 \\ 1 & 5} ; \spmqty{\alpha & \sigma_2 \\ \sigma_4 & \beta}}_{\spmqty{q & k \\ k & q}}
 \mathsf{T}^{\sigma_1 \beta}_{\alpha \sigma_3} 
 \right.
\\ & \left. - \frac{1}{4\pi } 
\left(\lambda^{\spmqty{4 & 5 \\ 1 & 8}; \spmqty{\sigma_1 & \alpha \\ \beta & \sigma_3}}_{\spmqty{p & q \\ q & p}}\lambda^{\spmqty{1 & 5 \\ 1 & 5}; \spmqty{\beta & \sigma_2 \\ \sigma_4 & \alpha}}_{\spmqty{q & k \\ k & q}} + \lambda^{\spmqty{4 & 8 \\ 4 & 8}; \spmqty{\sigma_1 & \alpha \\ \beta & \sigma_3}}_{\spmqty{p & q \\ q & p}}\lambda^{\spmqty{4 & 5 \\ 1 & 8}; \spmqty{\beta & \sigma_2 \\ \sigma_4 & \alpha}}_{\spmqty{q & k \\ k & q}} \right) \right\rbrace,
\label{eq:beta1515PH2}
\end{aligned}
\end{equation}
\begin{equation}
\begin{aligned}
& \beta^{(\lambda);\spmqty{1 & 8 \\ 4 & 5}; \spmqty{\sigma_1 & \sigma_2 \\ \sigma_4 & \sigma_3}}_{\spmqty{p & k \\ k & p}} = \left( 1 + 3(z-1)+ 2\eta^{(\psi)}_{p} + 2\eta^{(\psi)}_{k} \right)\lambda^{\spmqty{1 & 8 \\ 4 & 5}; \spmqty{\sigma_1 & \sigma_2 \\ \sigma_4 & \sigma_3}}_{\spmqty{p & k \\ k & p}}
\\ & + 
\int \dd\rho(q) 
\left\lbrace
\frac{1}{2 \pi N_f}\mathsf{D}_{\mu}(q;k) 
\lambda^{\spmqty{1 & 5 \\ 1 & 5} ; \spmqty{\sigma_1 & \alpha \\ \beta & \sigma_3}}_{\spmqty{p & q \\ q & p}}
 \mathsf{T}^{\beta \sigma_2}_{\sigma_4 \alpha} 
+ \frac{1}{2 \pi N_f}\mathsf{D}_{\mu}(p;q) 
\lambda^{\spmqty{4 & 8 \\ 4 & 8} ; \spmqty{\alpha & \sigma_2 \\ \sigma_4 & \beta}}_{\spmqty{q & k \\ k & q}}
 \mathsf{T}^{\sigma_1 \beta}_{\alpha \sigma_3} 
 \right.
\\ & \left. - \frac{1}{4\pi } 
\left(\lambda^{\spmqty{1 & 5 \\ 1 & 5}; \spmqty{\sigma_1 & \alpha \\ \beta & \sigma_3}}_{\spmqty{p & q \\ q & p}}\lambda^{\spmqty{1 & 8 \\ 4 & 5}; \spmqty{\beta & \sigma_2 \\ \sigma_4 & \alpha}}_{\spmqty{q & k \\ k & q}} + \lambda^{\spmqty{1 & 8 \\ 4 & 5}; \spmqty{\sigma_1 & \alpha \\ \beta & \sigma_3}}_{\spmqty{p & q \\ q & p}}\lambda^{\spmqty{4 & 8 \\ 4 & 8}; \spmqty{\beta & \sigma_2 \\ \sigma_4 & \alpha}}_{\spmqty{q & k \\ k & q}} \right) \right\rbrace,
\label{eq:beta1515PH3}
\end{aligned}
\end{equation}
\begin{equation}
\begin{aligned}
& \beta^{(\lambda);\spmqty{4 & 8 \\ 4 & 8};\spmqty{\sigma_1 & \sigma_2 \\ \sigma_4 & \sigma_3}}_{\spmqty{p & k \\ k & p}} = \left( 1 + 3(z-1) + 2\eta^{(\psi)}_{p} + 2\eta^{(\psi)}_{k} \right)
\lambda^{\spmqty{4 & 8 \\ 4 & 8}; \spmqty{\sigma_1 & \sigma_2 \\ \sigma_4 & \sigma_3}}_{\spmqty{p & k \\ k & p}} \\ &
 + \int \dd\rho(q) \left\lbrace  \frac{\mathsf{D}_{\mu}(q;k)}{2\pi N_f} 
\lambda^{\spmqty{4 & 5 \\ 1 & 8}; \spmqty{\sigma_1 & \alpha \\ \beta & \sigma_3}}_{\spmqty{p & q \\ q & p}}
 \mathsf{T}^{\beta \sigma_2}_{\sigma_4 \alpha} 
+ \frac{\mathsf{D}_{\mu}(p;q) }{2\pi N_f} 
\lambda^{\spmqty{1 & 8 \\ 4 & 5}; \spmqty{\alpha & \sigma_2 \\ \sigma_4 & \beta}}_{\spmqty{q & k \\ k & q}}
 \mathsf{T}^{\sigma_1 \beta}_{\alpha \sigma_3} 
\right.
\\ & -\frac{1}{\pi N_f^2}
 \mathsf{T}^{\beta \sigma_2}_{\sigma_4 \alpha} 
 \mathsf{T}^{\sigma_1 \alpha}_{\beta \sigma_3} 
\mathsf{D}_{\mu}(p;q)
\mathsf{D}_{\mu}(q;k)
\\ & \left. - \frac{1}{4\pi}
\left(\lambda^{\spmqty{4 & 5 \\ 1 & 8}; \spmqty{\sigma_1 & \alpha \\ \beta & \sigma_3}}_{\spmqty{p & q \\ q & p}}\lambda^{\spmqty{1 & 8 \\ 4 & 5}; \spmqty{\beta & \sigma_2 \\ \sigma_4 & \alpha}}_{\spmqty{q & k \\ k & q}} + \lambda^{\spmqty{4 & 8 \\ 4 & 8}; \spmqty{\sigma_1 & \alpha \\ \beta & \sigma_3}}_{\spmqty{p & q \\ q & p}}\lambda^{\spmqty{4 & 8 \\ 4 & 8}; \spmqty{\beta & \sigma_2 \\ \sigma_4 & \alpha}}_{\spmqty{q & k \\ k & q}} \right)\right\rbrace.
\label{eq:beta1515PH4}
\end{aligned}
\end{equation}

\subsubsection{Beta functional in the intersection between the PP and PH planes}

Within the PP-plane with $k + p \neq 0$,
an operator mixes with other operators only within the plane.
Similarly, an operator at generic momenta within the PH-plane
only mixes with other operators within the PH-plane.
However, an operator at the intersection of the two planes can mix 
with operators in both planes.
Within the one-dimensional manifold in which the PP and PH planes meet,
the coupling function is parameterized by one variable as
$\lambda^{\spmqty{1 & 5 \\ 1 & 5};\spmqty{\sigma_1 & \sigma_2 \\ \sigma_4 & \sigma_3}}_{\spmqty{-k & k \\ k & -k}}$.
While the beta functional takes a more complicated form in the line,
the underlying physics of each term is not different from the ones 
that determine the beta functionals in each of the PP and PH planes.
The beta functional for the couplings 
at a generic momentum point ($k\neq 0$) 
on this line is 
\begin{equation}
\begin{aligned}
& \beta^{(\lambda);\spmqty{N_1 & N_2 \\ N_4 & N_3};\spmqty{\sigma_1 & \sigma_2 \\ \sigma_4 & \sigma_3}}_{\spmqty{-k & k \\ k & -k}} 
= \left( 1 + 3(z-1) + \eta^{(\psi,N_1)}_{-k} + \eta^{(\psi,N_2)}_{k} + \eta^{(\psi,N_3)}_{-k} + \eta^{(\psi,N_4)}_{k} \right)
\lambda^{\spmqty{N_1 & N_2 \\ N_4 & N_3};\spmqty{\sigma_1 & \sigma_2 \\ \sigma_4 & \sigma_3}}_{\spmqty{-k & k \\ k & -k}}
\\ & +\int \dd\rho(q) \left\lbrace 
-\frac{\mathsf{D}_{\mu}(-k;q) }{2\pi N_f}
\left[ 
 \mathsf{T}^{\sigma_1 \sigma_2}_{\alpha \beta} 
\lambda^{\spmqty{\bar N_1 & \bar N_2 \\ N_4 & N_3}; \spmqty{\alpha & \beta \\ \sigma_4 & \sigma_3}}_{\spmqty{q & -q \\ k & -k}} 
 -\lambda^{\spmqty{\bar N_1 & N_2 \\ N_4 & \bar N_3}; \spmqty{\alpha & \sigma_2 \\ \sigma_4 & \beta}}_{\spmqty{q & k \\ k & q}}
  \mathsf{T}^{\sigma_1 \beta}_{\alpha \sigma_3} 
\right]  \right.
\\ & 
\hspace{2.2cm}
\left. -\frac{\mathsf{D}_{\mu}(q;k) }{2\pi N_f}  \left[ 
\lambda^{\spmqty{N_1 & N_2 \\ \bar N_4 & \bar N_3}; \spmqty{\sigma_1 & \sigma_2 \\ \alpha & \beta}}_{\spmqty{-k & k \\ q & -q}}
 \mathsf{T}^{\alpha \beta}_{\sigma_4 \sigma_3} 
 -
\lambda^{\spmqty{N_1 & \bar N_2 \\ \bar N_4 & N_3}; \spmqty{\sigma_1 & \alpha \\ \beta & \sigma_3}}_{\spmqty{-k & q \\ q & -k}}
 \mathsf{T}^{\beta \sigma_2}_{\sigma_4 \alpha} 
\right] \right\rbrace
\\ & +\frac{1}{\pi N_f^2}
\int \dd\rho(q) 
\left[
 \mathsf{T}^{\sigma_1 \sigma_2}_{\alpha \beta} 
 \mathsf{T}^{\alpha \beta}_{\sigma_4 \sigma_3} 
\mathsf{D}_{\mu}(-k;q) 
\mathsf{D}_{\mu}(q;k) 
-
  \mathsf{T}^{\sigma_1 \beta}_{\alpha \sigma_3} 
 \mathsf{T}^{\alpha \sigma_2}_{\sigma_4 \beta} 
\mathsf{D}_{\mu}(-k;q) 
\mathsf{D}_{\mu}(q;k) 
\right]
\delta^{N_1}_{N_4} \delta^{N_2}_{N_3}
\\ & + \frac{1}{4\pi} \int  \dd\rho(q) 
\left[\left(
\lambda^{\spmqty{N_1 & N_2 \\ M & M'}; \spmqty{\sigma_1 & \sigma_2 \\ \beta & \alpha}}_{\spmqty{-k & k \\ q & -q}}\lambda^{\spmqty{M & M' \\ N_4 & N_3}; \spmqty{\beta & \alpha \\ \sigma_4 & \sigma_3}}_{\spmqty{q & -q \\ k & -k}} 
-
\lambda^{\spmqty{N_1 & M \\ M' & N_3}; \spmqty{\sigma_1 & \alpha \\ \beta & \sigma_3}}_{\spmqty{-k & q \\ q & -k}}\lambda^{\spmqty{M' & N_2 \\ N_4 & M}; \spmqty{\beta & \sigma_2 \\ \sigma_4 & \alpha}}_{\spmqty{q & k \\ k & q}} 
\right)\right].
\end{aligned}
\label{eq:PP&PH1}
\end{equation}
Here
$\spmqty{N_1 & N_2 \\ N_4 & N_3}$
represents any of the elements in the set of
\bqa
h^{(1)}_{1515}=\Big\{ 
\spmqty{1 & 5 \\ 1 & 5},
\spmqty{1 & 5 \\ 4 & 8},
\spmqty{4 & 8 \\ 1 & 5},
\spmqty{4 & 8 \\ 4 & 8},
\spmqty{1 & 8 \\ 4 & 5},
\spmqty{4 & 5 \\ 1 & 8} \Bigr\}.
\eqa
$M$ and $M'$ are summed over hot spot indices
for which the four-fermion couplings 
are in $h^{(1)}_{1515}$. 
If $k=0$,
the four-fermion operator mixes with an even larger set of
operators.
However, we don't need to introduce counter terms for
the operators right at the hot spots
because the IR singularity is localized within the measure zero set
in the low-energy limit.
 
How does \eq{eq:PP&PH1} change to
Eqs. (\ref{eq:beta1515PP1})-
(\ref{eq:beta1515PP4})
or 
Eqs. (\ref{eq:beta1515PH1})-
(\ref{eq:beta1515PH4})
as one moves away from the intersecting line staying
either within the PP or PH plane?
To answer this question,
let us examine how the contribution of the PP diagram
to $\lambda^{\spmqty{1 & 5 \\ 1 & 5}; \spmqty{\sigma_1 & \sigma_2 \\ \sigma_4 & \sigma_3}}_{\spmqty{p & k \\ k & p}} $
decays as $p+k$ becomes non-zero away from the intersecting line.
Away from the intersecting line but within the PH plane, 
the total momentum of the electron pair is non-zero,
which makes it impossible to put both internal electrons on the Fermi surface within the loop:
if a pair of electrons with momenta $k$ and $p$ 
on the Fermi surface near hot spots $1$ and $5$
are scattered to hot spots $4$ and $8$,
the minimum energy that the virtual electron pair must carry is order of $v_p p+ v_k k$.
This cuts off the IR divergence in the PP diagram
in the low energy limit.
Therefore, the contribution of the PP diagram to 
$\lambda^{\spmqty{1 & 5 \\ 1 & 5}; \spmqty{\sigma_1 & \sigma_2 \\ \sigma_4 & \sigma_3}}_{\spmqty{p & k \\ k & p}} $
becomes negligible for $\mu \ll |v_p p+ v_k k|$.
This is confirmed through an explicit calculation in 
Appendix \ref{app:QC2}.
Similarly, the contribution of the PH diagram to 
$\lambda^{\spmqty{1 & 5 \\ 1 & 5};\spmqty{\sigma_1 & \sigma_2 \\ \sigma_4 & \sigma_3}}_{\spmqty{p & -p \\ k & -k}}$
becomes negligible for $\mu \ll |v_p p+ v_k k|$.
This implies that  \eq{eq:PP&PH1} crossovers
to Eqs. (\ref{eq:beta1515PP1})-
(\ref{eq:beta1515PP4})
or 
Eqs. (\ref{eq:beta1515PH1})-
(\ref{eq:beta1515PH4})
as the momentum deviates more than $\mu/v$ 
away from the intersecting line in each plane.

\subsubsection{Decoupling between the PP and PH-planes}

 The full beta functionals that describe the coupling functions
defined in this space are given by
Eqs. (\ref{eq:beta1515PP1})-(\ref{eq:beta1515PP4}),
Eqs. (\ref{eq:beta1515PH1})-(\ref{eq:beta1515PH4})
and \eq{eq:PP&PH1}.
The couplings in the PP plane are coupled with
the couplings in the PH plane through the intersection.
However, a simplification arises at low energies.
In the low-energy limit, 
the phase space of the intersection
becomes vanishingly small 
compared to the phase space of the PP and PH planes.
To see this in more detail, 
let us consider the beta functional of 
$\lambda^{\spmqty{1 & 5 \\ 1 & 5};\spmqty{\sigma_1 & \sigma_2 \\ \sigma_4 & \sigma_3}}_{\spmqty{p & -p \\ k & -k}}$
for  $p$ and $k$ far away from the intersecting line, that is $|v_k k+ v_p p| > \mu$.  
The $q$ integration in \eq{eq:beta1515PP1}
can be broken into the contribution
that depends on the couplings in the intersecting line
and the remaining contribution that does not depend on 
the couplings in the intersecting line as
 \begin{equation}
\begin{aligned}
& \beta^{(\lambda);\spmqty{1 & 5 \\ 1 & 5};\spmqty{\sigma_1 & \sigma_2 \\ \sigma_4 & \sigma_3}}_{\spmqty{p & -p \\ k & -k}} =  \left( 1 + 3(z-1) + 2\eta^{(\psi)}_{p} + 2\eta^{(\psi)}_{k} \right)
\lambda^{\spmqty{1 & 5 \\ 1 & 5};\spmqty{\sigma_1 & \sigma_2 \\ \sigma_4 & \sigma_3}}_{\spmqty{p & -p \\ k & -k}}
\\ & +
\int_{C_{k,p}} \dd\rho(q)
~~ V^{(\lambda);\spmqty{1 & 5 \\ 1 & 5};\spmqty{\sigma_1 & \sigma_2 \\ \sigma_4 & \sigma_3}}_{p,k,q}
+ \int_{C'_{k,p}} \dd\rho(q)
~~ V^{(\lambda);\spmqty{1 & 5 \\ 1 & 5};\spmqty{\sigma_1 & \sigma_2 \\ \sigma_4 & \sigma_3}}_{p,k,q},
\label{eq:beta1515PP1_2}
\end{aligned}
\end{equation}
where
\bqa
&&
 V^{(\lambda);\spmqty{1 & 5 \\ 1 & 5};\spmqty{\sigma_1 & \sigma_2 \\ \sigma_4 & \sigma_3}}_{p,k,q}
 =  
 - \frac{ \mathsf{D}_{\mu}(p;q)}{2 \pi N_f}  
 \mathsf{T}^{\sigma_1 \sigma_2}_{\alpha \beta} 
 \lambda^{\spmqty{4 & 8 \\ 1 & 5}; \spmqty{\alpha & \beta \\ \sigma_4 & \sigma_3}}_{\spmqty{q & -q \\ k & -k}}
  -\frac{\mathsf{D}_{\mu}(q;k)}{2\pi N_f} 
 \lambda^{\spmqty{1 & 5 \\ 4 & 8}; \spmqty{\sigma_1 & \sigma_2 \\ \alpha & \beta}}_{\spmqty{p & -p \\ q & -q}}
  \mathsf{T}^{\alpha \beta}_{\sigma_4 \sigma_3}  \nn
&& +\frac{1}{\pi N_f^2}
 \mathsf{T}^{\sigma_1 \sigma_2}_{\alpha \beta} 
 \mathsf{T}^{\alpha \beta}_{\sigma_4 \sigma_3} 
\mathsf{D}_{\mu}(p;q)
\mathsf{D}_{\mu}(q;k)
  + \frac{1}{4\pi}
\left(\lambda^{\spmqty{1 & 5 \\ 1 & 5}; \spmqty{\sigma_1 & \sigma_2 \\ \beta & \alpha}}_{\spmqty{p & -p \\ q & -q}}\lambda^{\spmqty{1 & 5 \\ 1 & 5}; \spmqty{\beta & \alpha \\ \sigma_4 & \sigma_3}}_{\spmqty{q & -q \\ k & -k}} + \lambda^{\spmqty{1 & 5 \\ 4 & 8}; \spmqty{\sigma_1 & \sigma_2 \\ \beta & \alpha}}_{\spmqty{p & -p \\ q & -q}}\lambda^{\spmqty{4 & 8 \\ 1 & 5}; \spmqty{\beta & \alpha \\ \sigma_4 & \sigma_3}}_{\spmqty{q & -q \\ k & -k}} \right),
\eqa
and
\begin{equation}
\begin{aligned}
C_{k,p} =& \left\{ q ~\big|~ |v_q q+ v_k k| > \mu ~\& ~|v_q q + v_p p| > \mu  ~\right\}, \\
C'_{k,p} = & (C_{k,p})^\complement.
\end{aligned}
\end{equation}
$C_{k,p}$ represents the set of $q$ at which
all coupling functions in 
$V^{(\lambda);\spmqty{1 & 5 \\ 1 & 5};\spmqty{\sigma_1 & \sigma_2 \\ \sigma_4 & \sigma_3}}_{p,k,q}$
 are away from the intersection and obey the beta functionals given by
Eqs. (\ref{eq:beta1515PP1})-(\ref{eq:beta1515PP4}).
$C'_{k,p}$, being the complement of $C_{k,p}$, represents the set of $q$
at which at least one coupling function in
$V^{(\lambda);\spmqty{1 & 5 \\ 1 & 5};\spmqty{\sigma_1 & \sigma_2 \\ \sigma_4 & \sigma_3}}_{p,k,q}$
 is in the intersection of the PP and PH planes
 and satisfy \eq{eq:PP&PH1}.
In the small $\mu$ limit, the phase space of 
$C'_{k,p}$ vanishes linearly in $\mu$.
Consequently, 
the contribution of $C'_{k,p}$ 
to the beta functions of  
$\lambda^{\spmqty{1 & 5 \\ 1 & 5};\spmqty{\sigma_1 & \sigma_2 \\ \sigma_4 & \sigma_3}}_{\spmqty{p & -p \\ k & -k}}$
away from the intersection
becomes negligible in the small $\mu$ limit.
Similarly, the contributions of the intersection
to the beta functional of the couplings in the PH plane away from the intersection
are negligible in the low-energy limit.
Therefore, one can ignore the intersection
for the purpose of understanding the RG flow of the coupling functions
in the PP and PH planes 
away from the intersecting line.
%
As a result, the couplings in the PP plane
and the couplings in the PH plane become 
effectively decoupled in the low-energy limit,
and we can study
Eqs. (\ref{eq:beta1515PP1})-(\ref{eq:beta1515PP4})
and
Eqs. (\ref{eq:beta1515PH1})-(\ref{eq:beta1515PH4}),
separately.
The solution of the beta function in the PP-plane is discussed in Sec. \ref{sec:four_fermion_fp2}.
Here, we present the solution of the beta function
for the couplings defined in the PH-plane.

\subsubsection{Solution of the beta functional for the $2k_F$ PH interaction}

The closed set of beta functionals for
$  {\lambda}^{\spmqty{1 & 5 \\ 1 & 5};\spmqty{\sigma_1 & \sigma_2 \\ \sigma_4 & \sigma_3}}_{\spmqty{p & k \\ k & k}} $,
$  {\lambda}^{\spmqty{1 & 8 \\ 4 & 5};\spmqty{\sigma_1 & \sigma_2 \\ \sigma_4 & \sigma_3}}_{\spmqty{p & k \\ k & p}} $,
$  {\lambda}^{\spmqty{4 & 5 \\ 1 & 8};\spmqty{\sigma_1 & \sigma_2 \\ \sigma_4 & \sigma_3}}_{\spmqty{p & k \\ k & p}} $,
$  {\lambda}^{\spmqty{4 & 8 \\ 4 & 8};\spmqty{\sigma_1 & \sigma_2 \\ \sigma_4 & \sigma_3}}_{\spmqty{p & k \\ k & p}} $ 
that describe \ffc in the  PH channel with momentum $2 \vec k_F$ are given by
Eqs.
(\ref{eq:beta1515PH1})-
(\ref{eq:beta1515PH4}).
With
\begin{equation}
{\lambda}^{\spmqty{\sigma_1 & \sigma_2 \\ \sigma_4 & \sigma_3}}_{2PH\spmqty{p & k \\ k & p}} = \pmqty{
{\lambda}^{\spmqty{1 & 5 \\ 1 & 5}; \spmqty{\sigma_1 & \sigma_2 \\ \sigma_4 & \sigma_3}}_{\spmqty{p & k \\ k & p}} & 
{\lambda}^{\spmqty{1 & 8 \\ 4 & 5}; \spmqty{\sigma_1 & \sigma_2 \\ \sigma_4 & \sigma_3}}_{\spmqty{p & k \\ k & p}} \\ 
{\lambda}^{\spmqty{4 & 5 \\ 1 & 8}; \spmqty{\sigma_1 & \sigma_2 \\ \sigma_4 & \sigma_3}}_{\spmqty{p & k \\ k & p}} & 
{\lambda}^{\spmqty{4 & 8 \\ 4 & 8}; \spmqty{\sigma_1 & \sigma_2 \\ \sigma_4 & \sigma_3}}_{\spmqty{p & k \\ k & p}}},
\end{equation}
the set of beta functionals can be combined into
\begin{align}
& \frac{\partial}{\partial\ell}{\lambda}^{\spmqty{\sigma_1 & \sigma_2 \\ \sigma_4 & \sigma_3}}_{2PH \spmqty{p & k \\ k & p}} = 
 -\left( 1 + 3(z-1) + 2\eta^{(\psi)}_{k} + 2\eta^{(\psi)}_{p} \right)
 {\lambda}^{\spmqty{\sigma_1 & \sigma_2 \\ \sigma_4 & \sigma_3}}_{2PH \spmqty{p & k \\ k & p}}
\nn
& + \frac{1}{4\pi}\int \frac{dq}{2 \pi \mu V_{F,q}}  
\left[ 
{\lambda}^{\spmqty{\sigma_1 & \alpha \\ \beta & \sigma_3}}_{2PH \spmqty{p & q \\ q & p}} 
-\frac{ 2 \mathsf{T}^{\sigma_1  \alpha}_{  \beta  \sigma_3 }}{N_f}
\mathsf{D}_{\mu}(q; p)  \pmqty{0 & 1 \\ 1 & 0} \right]
\left[ 
{\lambda}^{\spmqty{\beta & \sigma_2 \\ \sigma_4 & \alpha}}_{2PH \spmqty{q & k \\ k & q}} 
-\frac{2 \mathsf{T}^{ \beta  \sigma_2}_{  \sigma_4  \alpha }}{N_f}\mathsf{D}_{\mu}(q; k)  \pmqty{0 & 1 \\ 1 & 0} \right].
\label{eq:betabigmatrixsimple_main_PH0}
\end{align}
To make the analysis parallel with that of the PH channel,
we define
\begin{align}
\tilde {\lambda}^{ \spmqty{\alpha & \beta \\ \gamma & \delta }}_{2PH \spmqty{P & K \\ K & P}} = & 
\frac{1}{\sqrt{V_{F,p}  V_{F,k}}}
\left[
- {\lambda}^{\spmqty{\alpha & \beta \\ \gamma & \delta}}_{2PH \spmqty{p & k \\ k & p}} 
+ \frac{2  \mathsf{T}^{\alpha  \beta}_{  \gamma  \delta }}{N_f }
\mathsf{D}_{\mu}(p; k)
\pmqty{0 & 1 \\ 1 & 0} 
\right]
\end{align}
with $P=p \eell$, $K=k \eell$
to rewrite the beta functional as
\begin{equation}
\begin{aligned}
\frac{\partial}{\partial\ell}\tilde{\lambda}^{ \spmqty{\sigma_1 & \sigma_2 \\ \sigma_4 & \sigma_3}}_{2PH \spmqty{P & K \\ K & P}} = &
 -\left(1 + K\frac{\partial}{\partial K} + P\frac{\partial}{\partial P} 
 + \hat \eta_K + \hat \eta_P \right)\tilde{\lambda}^{\spmqty{\sigma_1 & \sigma_2 \\ \sigma_4 & \sigma_3}}_{2PH \spmqty{P & K \\ K & P}}
 -  \frac{1}{4 \pi} 
\int \frac{\dd Q}{2\pi\Lambda} 
\tilde{\lambda}^{\spmqty{\sigma_1 & \alpha \\ \beta & \sigma_3}}_{2PH \spmqty{P & Q \\ Q & P}} 
\tilde{\lambda}^{\spmqty{\beta & \sigma_2 \\ \sigma_4 & \alpha}}_{2PH \spmqty{Q & K \\ K & Q}} 
\\&
+\frac{2  \mathsf{T}^{\sigma_1  \sigma_2}_{  \sigma_4  \sigma_3 }}{N_f}\mathsf{R}(K, P) 
(\mathscr{P}_s - \mathscr{P}_d),
\end{aligned}
\label{eq:etabeta_main_PH}
\end{equation}
where
$\mathsf{R}(K, P)$ and
$\hat{\eta}_{K}$,
are defined 
in Eqs.
(\ref{eq:RetaKK}).
%
%
%
The coupling function is decomposed into four different channels as
 \begin{equation}
\tilde{\lambda}^{\spmqty{\sigma_1 & \sigma_2 \\ \sigma_4 & \sigma_3}}_{2PH \{K_i\}} = 
\tilde{\lambda}^{ (t)(s)}_{2PH \{K_i\}}\mathsf{I}^{\sigma_1 \sigma_2}_{\sigma_4 \sigma_3} 
\mathscr{P}_s
+ \tilde{\lambda}^{(t)(d)}_{2PH \{K_i\}}\mathsf{I}^{\sigma_1 \sigma_2}_{\sigma_4 \sigma_3} 
\mathscr{P}_d
+ \tilde{\lambda}^{(a)(s)}_{2PH \{K_i\}}\mathsf{\chi}^{\sigma_1 \sigma_2}_{\sigma_4 \sigma_3}
\mathscr{P}_s
+ \tilde{\lambda}^{(a)(d)}_{2PH \{K_i\}}\mathsf{\chi}^{\sigma_1 \sigma_2}_{\sigma_4 \sigma_3}
\mathscr{P}_d,
\end{equation}
where
$\mathsf{I}^{\sigma_1 \sigma_2}_{\sigma_4 \sigma_3} $
and 
$\mathsf{\chi}^{\sigma_1 \sigma_2}_{\sigma_4 \sigma_3}$
are defined in \eq{eq:IChi}.
Each of the four channels are 
$SU(N_c)$-trivial s-wave,
 $SU(N_c)$-trivial d-wave,
 $SU(N_c)$-adjoint s-wave
and
 $SU(N_c)$-adjoint d-wave channels.
The beta functional in each channel becomes 
\begin{equation}
\begin{aligned} 
 \frac{\dd }{\dd\ell}\tilde{\lambda}^{ \spmqty{t \\ a},\spmqty{s \\ d}}_{2PH \spmqty{P & K \\ K & P}} &= 
 -\left(1 + K\frac{\partial}{\partial K} + P\frac{\partial}{\partial P} + \hat \eta_K + \hat \eta_P \right)
 \tilde{\lambda}^{\spmqty{t \\ a},\spmqty{s \\ d}}_{2PH \spmqty{P & K \\ K & P}}
 -\frac{1}{4\pi} \int \frac{\dd Q}{2\pi\Lambda}
\tilde{\lambda}^{ \spmqty{t \\ a},\spmqty{s \\ d}}_{2PH \spmqty{P & Q \\ Q & P}}
\tilde{\lambda}^{ \spmqty{t \\ a},\spmqty{s \\ d}}_{2PH \spmqty{Q & K \\ K & Q}} 
\\ & + \frac{2}{N_f}
Y_{PH}^{ \spmqty{t \\ a}}
1^{\spmqty{s \\ d}}
 \mathsf{R}(P, K ),
\label{eq:betaallwaves_main0_pH}
\end{aligned}
\end{equation}
where
$1^{\spmqty{s \\ d}}$ is defined in \eq{eq:1sd}.

\eq{eq:betaallwaves_main0_pH}
has the same form as
\eq{eq:betaallwaves_main0}.
The only difference is the change 
in the spin wavefunction and the associated eigenvalue 
determined from the representations
of two fermions in the PP and PH channels.
The spin symmetric and anti-symmetric representations in the PP channel
with eigenvalues $Y_{PP}^{(\pm)}$ in the last term
of \eq{eq:betaallwaves_main0}
are replaced with
the trivial and adjoint representations in the PH channel
with eigenvalues $Y_{PH}^{\spmqty{t \\ a}}$
in Eq. \eqref{eq:Yta} \footnote{
For $N_c=2$, even that difference goes away
because the fundamental and anti-fundamental 
representations are identical for $SU(2)$.}.
All discussions on 
$\tilde{\lambda}_{2PP}$
straightforwardly generalize to 
$\tilde{\lambda}_{2PH}$.
In the PH channel, 
the spin fluctuations gives rise to 
an attractive interaction
in the 
$SU(N_c)$-adjoint s-wave  channel
and the
 $SU(N_c)$-trivial d-wave channel
  with the $SU(N_c)$-trivial d-wave channel 
being the stronger.
The other two channels,
$SU(N_c)$-adjoint d-wave  and  $SU(N_c)$-trivial s-wave,
are repulsive.
Therefore,
$\tilde{\lambda}^{(t)(d)}_{2PH}$
and
$\tilde{\lambda}^{(a)(s)}_{2PH}$
become non-Hermitian (complex)
at quasi-fixed point.



%

\section*{Acknowledgement}

This research was supported by the Natural Sciences 
and Engineering Research Council of
Canada. Research at the Perimeter Institute is supported in part by the
Government of Canada through Industry Canada, and by the Province of
Ontario through the Ministry of Research and Information.

\bibliographystyle{apsrev4-1}

\bibliography{references}

\end{document}